\documentclass[aps,pra,reprint,aps,twocolumn,superscriptaddress]{revtex4}

\usepackage{amsmath,amsthm,amssymb,graphicx,xcolor,float,makecell,multirow}

\usepackage{esint}
\usepackage[unicode=true]{hyperref}
\hypersetup{
     colorlinks=true,       		
     linkcolor=blue,          	
     citecolor=red,            
     urlcolor=magenta,           	
 }

\usepackage{mathrsfs,mathtools,mathdots,dsfont}

\newtheorem{theorem}{Theorem}
\newtheorem*{theorem*}{Theorem}

\newtheorem*{corollary*}{Corollary}

\newtheorem*{lemma*}{Lemma}

\newtheorem*{proposition*}{Example*}

\newtheorem*{conjecture*}{Conjecture}
\theoremstyle{definition}

\newtheorem*{definition*}{Definition}
\theoremstyle{remark}

\newtheorem*{remark*}{Remark}

\newcommand{\ket}[1]{\left|#1\right\rangle}
\newcommand{\bra}[1]{\left\langle#1\right|}

\begin{document}

\title{Steerability of Rank-2 Two-Qubit Entangled States}

\author{Yu-Xuan Zhang}
\affiliation{School of Physics, Nankai University, Tianjin 300071, People's Republic of China}

\author{Jing-Ling Chen}
\email{chenjl@nankai.edu.cn}
\affiliation{Theoretical Physics Division, Chern Institute of Mathematics, Nankai University, Tianjin 300071, People's Republic of
	 	China}

	\date{\today}

	\begin{abstract}
\noindent\textbf{Abstract}\par
\noindent Quantum nonlocality is an essential resource in quantum information and is commonly classified into three distinct forms: quantum entanglement, Einstein-Podolsky-Rosen (EPR) steering, and Bell's nonlocality. Gisin's theorem shows that pure-state entanglement implies Bell nonlocality, and it motivates the question of how entanglement and EPR steering are related beyond pure states. Here we prove that every entangled rank-2 two-qubit state is EPR steerable, with rank-1 states included as a special case. The proof combines a local-unitary parametrization of rank-2 two-qubit states, a necessary and sufficient separability condition, and a state-dependent nonlinear steering inequality. This gives an analytical characterization of EPR steerability for rank-2 two-qubit entangled states. Thus, rank-2 two-qubit entangled states provide certifiable EPR-steering resources in quantum information.
\end{abstract}

\maketitle


\noindent\textbf{INTRODUCTION}

\noindent Quantum nonlocality can be traced back to the famous Einstein-Podolsky-Rosen (EPR) paradox \cite{Einstein1935}. In 1935, EPR published a milestone work by indicating a conflict between quantum mechanics and classical physics, hence initiating the journey of quantum nonlocality. In the same year, as a response to the EPR paradox, Schr{\"o}dinger proposed two fundamental concepts of quantum nonlocality --- \emph{quantum entanglement} and \emph{EPR steering} \cite{Schrodinger35}. In 1964, in a further development of the EPR paradox, Bell proposed the third fundamental concept (i.e., \emph{Bell's nonlocality}) through Bell's theorem for entangled states of two spin-1/2 particles \cite{Bell} \cite{RMP2014}.
In 2007, motivated by quantum information tasks, Wiseman, Jones and Doherty presented a unified framework for these three distinct types of quantum nonlocality \cite{WJD07,WJD07PRA}, and showed that they form a strict hierarchy: quantum entanglement is the most general, Bell's nonlocality is the strongest and most restrictive, and EPR steering lies strictly intermediate between them (i.e., Bell's nonlocality $\subset$ EPR steering $\subset$ quantum entanglement). Nowadays, quantum nonlocalities have become invaluable resources in numerous quantum information protocols \cite{Nielsen2000}, such as quantum teleportation \cite{Bennett1993} and quantum key distribution \cite{Gisin2002}.

According to the hierarchical structure, Bell's nonlocality is the strongest nonlocality, while quantum entanglement is the weakest one. However, before Werner's pioneering work in 1989 \cite{werner89}, the distinction between Bell nonlocality and entanglement was not always explicit. By giving a rigorous mathematical definition of separable states, Werner showed that two-qubit Werner states contain parameter regions that are entangled but do not possess Bell nonlocality. This fact implies that Bell nonlocality and quantum entanglement are two different concepts. Although they are essentially different, they can be equivalent in specific circumstances. In 1991 Gisin presented \emph{Gisin's theorem} on Bell nonlocality \cite{Gisin1991}. The theorem indicates that any two-qubit pure entangled state violates a Bell inequality, namely, all two-qubit pure entangled states have Bell nonlocality. In 2004, the proof of Gisin's theorem was successfully generalized to three qubits \cite{Chen2004}. In 2012, Yu \emph{et al.} finished the proof of Gisin's theorem for all entangled pure states in any quantum system \cite{Yu2012}. It took more than twenty years to complete the proof of Gisin's theorem.

As stated in the review article by the Horodecki family \cite{RMP2009}, Gisin's theorem was one of the central questions at the beginning of the 1990s. The proof of Gisin's theorem for two qubits is quite simple. Given a two-qubit pure state in its Schmidt form, i.e.,
\begin{eqnarray}\label{eq:pure}
\ket{\Psi} &=&\cos \theta \ket{00}+\sin \theta\ket{11},
\end{eqnarray}
with $\theta\in[0,\pi/2]$, by adopting the Clauser-Horne-Shimony-Holt (CHSH) inequality $\mathcal{I}_{\rm CHSH} \leq 2$ \cite{1969CHSH}, one can have the maximal quantum expectation as $\mathcal{I}^{\rm max}_{\rm CHSH} =2 \sqrt{1+\mathcal{C}^2}>2$ for $\mathcal{C}>0$, with $\mathcal{C}=|\sin2\theta|$ being the degree of entanglement (i.e., concurrence) of the pure state $\ket{\Psi}$. Thus all two-qubit pure entangled states violate Bell's inequality. As recalled in his paper \cite{Gisin1991}, Gisin had obtained this simple result for many years and felt that it should be known to many people but (apparently) never published. But surprisingly he found that Bell himself did not know such a simple result during their discussion. This motivated him to publish the work and dedicate it to John Bell. Gisin's theorem therefore provides a benchmark example in which entanglement leads to an experimentally testable form of quantum nonlocality.

Motivated by this relation, it is natural to ask what kind of analytical connection can be established between entanglement and EPR steering when mixed two-qubit states are considered. We address this question for rank-2 two-qubit states and prove that, under local unitary transformations, every entangled state in this class is EPR steerable, with rank-1 states included as a special case. Thus, rank-2 two-qubit entangled states provide certifiable EPR-steering resources in quantum information.

EPR steering is an essential nonlocality resource in quantum information \cite{2017RPPCavalcanti,RMP2020,2013AVN,2014Sun}. It refers to the quantum phenomenon in which one party (Alice) in an entangled pair can prepare the state of the other party (Bob) through local measurements. In practice, steering can be defined via the following two-part information task: Alice prepares a bipartite quantum state and repeatedly sends one subsystem to Bob. Each time, they perform local measurements on their respective parts and exchange classical information. Alice's goal is to convince Bob that the state $\rho$ she prepares is entangled. However, Bob does not trust Alice; Bob worries that she may
send him some unentangled particles and fabricate the results using her knowledge about the local-hidden-state (LHS) of his particles. Bob's task is to prove that no such local hidden states exist. If Bob is convinced, then the state $\rho$ is entangled.

It is known that Bell nonlocality of a quantum state can be detected by violations of Bell inequalities. Similarly, EPR steerability can be certified through violations of steering inequalities. For instance, Saunders \emph{et al.} proposed the $N$-setting linear EPR steering inequality to study the steerability of two qubits \cite{NP2010}. In particular, for $N=3$, we have the 3-setting EPR steering inequality as
\begin{eqnarray}\label{eq:QM-3}
\mathcal{I}_3=\frac{1}{3} \sum_{j=1}^3 A_j (\hat{b}_j\cdot \vec{\sigma}^B) \leq \frac{1}{\sqrt{3}},
\end{eqnarray}
where $A_j=\pm 1$ $(j=1, 2, 3)$ are classical values for Alice, $C_{\rm LHS}=1/\sqrt{3}$ is the classical bound (i.e., the LHS bound), $\vec{\sigma}=(\sigma_x, \sigma_y, \sigma_z)$ is the vector of Pauli matrices, $\vec{\sigma}^B$ is the vector of Pauli matrices for Bob's side, $\hat{b}_1$, $\hat{b}_2$ and $\hat{b}_3$ are 3-dimensional unit vectors for Bob, and they are mutually orthogonal. For example, for the pure state (\ref{eq:pure}), one can have the maximal quantum expectation as $\mathcal{I}_3^{\rm QM}=\frac{2+ \sqrt{1+2\mathcal{C}^2}}{3\sqrt{3}}> \frac{1}{\sqrt{3}}$, thereby revealing the steerability of all pure entangled states.

\vspace{5mm}


\noindent \textbf{RESULTS}

\noindent Our main result gives an analytical characterization of EPR steerability for rank-2 two-qubit entangled states.
\begin{theorem}
Under the local unitary transformations, for any two-qubit quantum state $\rho$ with rank 2 (including rank 1 as a special case), if $\rho$ is entangled, then $\rho$ must be EPR steerable.
\end{theorem}
\noindent Before proving the theorem, we first present the simplified form of the rank-2 quantum state $\rho$ and then state a necessary and sufficient separability condition for $\rho$.

\noindent We consider the parametrization of two-qubit quantum states with rank 2 (including rank 1 as a special case). In general, such states are mixtures of two orthogonal pure states $\ket{\psi_1}$ and $\ket{\psi_2}$. We can write the density matrix as
\begin{eqnarray}\label{eq.1}
\rho=\nu_1  \ket{\psi_1}\bra{\psi_1}+\nu_2 \ket{\psi_2}\bra{\psi_2},
\end{eqnarray}
where $\nu_1$ and $\nu_2$ are two weights; and $\nu_1 \in [0, 1]$, $\nu_2 \in [0, 1]$, $\nu_1+\nu_2=1$. As shown in the Methods section, under the local unitary transformations, the simplified forms of $\ket{\psi_1}$ and $\ket{\psi_2}$ are as follows:
\begin{eqnarray}\label{eq:psi12}
\ket{\psi_1} &=&\cos \theta \ket{00}+\sin \theta\ket{11}, \nonumber\\
\ket{\psi_2}&=&\cos\phi(\cos \alpha \ket{01} +\sin\alpha\ket{10})\nonumber\\
&&+ e^{{\rm i}\beta}\sin\phi(\sin \theta \ket{00}-\cos \theta\ket{11}),
\end{eqnarray}
with $\theta,\phi,\alpha\in [0,\pi/2]$, $\beta \in [0, 2\pi]$.
Besides, under the local unitary transformations, when $\theta=\pi/4$, it is sufficient for us to consider $\ket{\psi_1} =(\ket{00}+\ket{11})/\sqrt{2}$ and $\ket{\psi_2} =(\ket{01}+\ket{10})/\sqrt{2}$, or $(\ket{00}-\ket{11})/\sqrt{2}$.

\noindent The degree of entanglement can be described by  the \emph{concurrence} \cite{Wootters1998}. For the state $\rho$, its concurrence is given by
$ \mathcal{C}=\sqrt{-s_2-2\sqrt{s_1}}$, with
\begin{eqnarray}
 s_2 &=& -\frac{1}{2} \nu_2^2 \sin2 \alpha  \sin 2 \theta \sin ^2(2 \phi ) \cos 2 \beta\nonumber\\
 &&-\nu_2^2 \sin ^2(2 \alpha ) \cos ^4 \phi\nonumber\\
 &&-\sin ^2(2 \theta ) \left[\nu_1^2-\nu_1 \nu_2 \sin ^2\phi +\nu_2^2 \sin ^4\phi \right]\nonumber\\
&&-\frac{1}{2} \nu_1 \nu_2 (\cos4 \theta+3) \sin ^2 \phi, \nonumber\\
 s_1 &=& \frac{1}{2} \nu_1^2 \nu_2^2 [\sin 2 \alpha \sin2 \theta \sin ^2(2 \phi ) \cos 2 \beta \nonumber\\
&&+2 \sin ^2(2 \alpha ) \sin ^2(2 \theta ) \cos ^4\phi +2 \sin ^4\phi ].\nonumber
\end{eqnarray}
The state $\rho$ is separable if and only if $\mathcal{C}=0$.

Let Alice and Bob share the state $\rho$. Alice can perform projective measurement on her qubit along the $\hat{n}$-direction, with $\hat{n}=(\sin\xi \cos\tau, \sin\xi \sin\tau, \cos\xi)$. The density matrix $\rho$ can be recast to the following decomposition form
\begin{eqnarray}\label{eq.3}
 \rho &=& P^{\hat{n}}_{0} \otimes \tilde{\rho}_0^{\hat{n}}+ P^{\hat{n}}_{1} \otimes \tilde{\rho}_1^{\hat{n}}+
|+\hat{n}\rangle\langle-\hat{n}| \otimes \mathcal{M} \nonumber\\
&&+|-\hat{n}\rangle\langle+\hat{n}| \otimes \mathcal{M}^\dagger,
\end{eqnarray}
where $\{\ket{+\hat{n}},\ket{-\hat{n}}\}$ forms Alice's measurement
basis, with
\[
\ket{+\hat{n}}=\cos\frac{\xi}{2}|0\rangle+\sin\frac{\xi}{2}e^{{\rm i}\tau}|1\rangle, \ket{-\hat{n}}=\sin\frac{\xi}{2}|0\rangle-\cos\frac{\xi}{2}e^{{\rm i}\tau}|1\rangle .
\]
The projectors read $P^{\hat{n}}_{0}=|+\hat{n}\rangle\langle+\hat{n}|$ and $P^{\hat{n}}_{1}=|-\hat{n}\rangle\langle-\hat{n}|$. Here $\mathcal{M}$ is a $2\times 2$ matrix, and $\mathcal{M}^\dagger$ is its Hermitian conjugate.
Interestingly, whether the rank-2 state $\rho$ is entangled or not is decided only by the property of $\mathcal{M}$. For this point, we have the following necessary and sufficient (NS) condition of separability: \emph{Under the local unitary transformations, the state $\rho$ is separable if and only if $\mathcal{M}=\mathcal{M}^\dagger$.} The detailed proof is given in the Methods section. Physically, the NS condition means that, if the rank-2 state $\rho$ is separable, then there is always a measurement direction $\hat{n}$ for Alice such that $\mathcal{M}=\mathcal{M}^\dagger$, and vice versa. Explicitly, the measurement direction $\hat{n}$ can be summarized as: when $\sin2\phi \neq 0$, the parameter $\xi$ is arbitrary and $\tau\in\{\beta,-\beta\}$; while $\sin2\phi =0$, the parameter $\xi$ is arbitrary and $\tau=0$, as derived in the Methods section. This property enables us to establish the state-dependent nonlinear steering inequality to detect the steerability of the state $\rho$.

\noindent We now use a steering inequality to detect the steerability of the state $\rho$. The $N$-setting linear steering inequality in Eq.~(\ref{eq:QM-3}) is not sufficient for detecting all entangled regions of $\rho$, so we introduce a state-dependent nonlinear steering inequality. In contrast to the fixed-coefficient linear inequality in Eq.~(\ref{eq:QM-3}), the LHS bound used below depends on the target state and on the measurement parameters entering the construction.
The state-dependent nonlinear steering inequality is given by
\begin{eqnarray}\label{eq.5}
    \langle \mathcal{W} \rangle \leq  C_{\text{LHS}},
\end{eqnarray}
where $C_{\text{LHS}}$ is the classical bound, and $\langle \mathcal{W} \rangle = \mathrm{Tr}(\mathcal{W} \rho)$
is the quantum expectation value of the operator $\mathcal{W}$ acting on $\rho$.

Explicitly, the operator $\mathcal{W}$ is a projector defined by
\begin{eqnarray}\label{eq.6}
 &&\mathcal{W} = |+\rangle\langle+|\otimes |\hat{n}_B \rangle\langle\hat{n}_B|,
\end{eqnarray}
with $|+\rangle=\frac{1}{\sqrt{2}}\left(\ket{+\hat{n}} + e^{{\rm i} \delta} \ket{-\hat{n}}\right)$, $(\delta \in [0, 2\pi])$, whose measurement direction is perpendicular to $\hat{n}$. Similarly, $|\hat{n}_B \rangle\langle\hat{n}_B|$ is Bob's projective measurement, with $  |\hat{n}_B\rangle = \cos \frac{\theta_B}{2} \ket{0} + \sin \frac{\theta_B}{2} e^{{\rm i}\phi_B} \ket{1}$ and $\hat{n}_B=(\sin\theta_B \cos\phi_B, \sin\theta_B \sin\phi_B, \cos\theta_B)$. Here the angles $\theta_B$ and $\phi_B$ are free parameters. The classical bound $C_{\text{LHS}}$ is defined by
\begin{eqnarray}\label{eq.7}
    &&C_{\text{LHS}} =
  \max_{\hat{n}_B} \left\{ {\rm Tr}[  \left( |+\rangle\langle+| \otimes |\hat{n}_B \rangle\langle\hat{n}_B|\right)
  {\rho_{\rm sep}}]\right\},
\end{eqnarray}
where $\rho_{\rm sep}$ represents the rank-2 (including rank-1 as a special case) separable states.

Note that $C_{\text{LHS}}$ depends on the target state $\rho$ and on Alice's measurement parameters, such as $\hat{n}$ and $\delta$. This is the sense in which Eq.~(\ref{eq.5}) is state-dependent and nonlinear. To obtain a tractable bound, we choose measurement directions for which $C_{\text{LHS}}$ can be evaluated explicitly. In this work, we take $\delta=\pm \pi/2$. Then we have $e^{{\rm i} \delta} \mathcal{M} + e^{-{\rm i} \delta} \mathcal{M}^\dagger = \pm{\rm i}( \mathcal{M}- \mathcal{M}^\dagger)=0$ for the separable state ${\rho_{\rm sep}}$. The two choices $\delta=+\pi/2$ and $\delta=-\pi/2$ correspond to the two outcomes of one projective measurement basis on Alice's side. Based on this observation, the classical bound can be simplified to
\begin{eqnarray}\label{eq.8}
    C_{\text{LHS}} &=& \frac{1}{2}\max_{\hat{n}_B} \left[ \mathrm{Tr} \left( |\hat{n}_B \rangle\langle\hat{n}_B| \rho_B\right) \right]= \frac{1}{2}\lambda_{\rm max}(\rho_B),
\end{eqnarray}
where $\rho_B= \tilde{\rho}_0^{\hat{n}} + \tilde{\rho}_1^{\hat{n}}$ is Bob's reduced state, and $\lambda_{\rm max}(\rho_B)$ is the maximal eigenvalue of $\rho_B$. If one expresses $\rho_B$ into its Bloch form, i.e., $\rho_B= (\openone + \vec{\sigma}\cdot \vec{F})/2$ , then the classical bound reads
\begin{eqnarray}\label{eq:lhs-1}
    C_{\text{LHS}} &=& \frac{1}{4}+ \frac{1}{4} \max_{\hat{n}_B}\left(\hat{n}_B \cdot \vec{F}\right)=\frac{1}{4}+ \frac{1}{4} |\vec{F}|.
\end{eqnarray}
Here $\openone$ is the $2\times 2$ identity matrix and $\vec{F}=(F_1,F_2,F_3)$ is the Bloch vector, with
\begin{eqnarray}
F_1 &=&-\nu_2 \cos\beta \sin(\alpha - \theta) \sin2\phi, \nonumber\\
F_2 &=&-\nu_2 \sin\beta \sin(\alpha + \theta) \sin2\phi, \nonumber\\
F_3 &=& -\nu_2 \cos2\alpha \cos^2\phi + \cos2\theta (\nu_1 - \nu_2 \sin^2\phi).\nonumber
\end{eqnarray}

\noindent To prove the theorem, we use the two projectors associated with the phase choices $\delta=+\pi/2$ and $\delta=-\pi/2$. We denote them by $|+\rangle\langle+|$ and $|+'\rangle\langle+'|$, respectively. These two projectors form one measurement basis on Alice's side and lead to the same classical bound $C_{\text{LHS}}$ as shown in Eq. (\ref{eq:lhs-1}). Accordingly, the two corresponding quantum expectations are as follows
\begin{eqnarray}\label{eq:w1-a}
\langle \mathcal{W} \rangle_1^{\rm max}
&=& \max\limits_{\hat{n}_B} \mathrm{Tr} [ (|+\rangle\langle+|\otimes |\hat{n}_B \rangle\langle\hat{n}_B|) \, \rho ],\nonumber\\
\langle \mathcal{W} \rangle_2^{\rm max}
&=& \max\limits_{\hat{n}_B} \mathrm{Tr} [ \left(|+'\rangle\langle+'|\otimes |\hat{n}_B \rangle\langle\hat{n}_B|\right) \, \rho ].
\end{eqnarray}
After direct calculation, we obtain
\begin{eqnarray}\label{eq.10}
\langle \mathcal{W} \rangle_1^{\rm max}
&=&\frac{1}{4}+\frac{1}{4}H_0+\frac{1}{4}|\vec{F}+\vec{H}|,\nonumber\\
\langle \mathcal{W} \rangle_2^{\rm max}
&=& \frac{1}{4}-\frac{1}{4}H_0+\frac{1}{4}|\vec{F}-\vec{H}|,
\end{eqnarray}
with $\vec{H}=(H_1, H_2, H_3)$, and
{\small
$$H_0= \nu_2 \left[ \sin\beta\cos(\alpha - \theta) \cos\tau  - \cos\beta \cos(\alpha + \theta) \sin\tau \right] \sin2\phi,\;\;\;$$
$$H_1 =\sin\tau \left[ \nu_2 \cos^2\phi \sin 2\alpha + \sin2\theta \left( \nu_1 - \nu_2 \sin^2\phi \right) \right],\;\;\;\;\;\;\;\;\;\;\;\;\;\;\;$$
$$H_2 =\cos\tau \left[ -\nu_2 \cos^2\phi \sin2\alpha + \sin2\theta \left( \nu_1 - \nu_2 \sin^2\phi \right) \right],\;\;\;\;\;\;\;\;\;\;\;\;\;\;\;$$
$$H_3 = -\nu_2 \left[ \sin\beta \cos(\alpha + \theta) \cos\tau - \cos\beta \cos(\alpha - \theta) \sin\tau \right] \sin2\phi.$$
}
\noindent Then we can have
\begin{eqnarray}\label{eq.11}
 \langle \mathcal{W} \rangle_1^{\rm max}+ \langle \mathcal{W} \rangle_2^{\rm max}
&=&\frac{1}{2}+\frac{1}{4}|\vec{F}+\vec{H}| +\frac{1}{4}|\vec{F}-\vec{H}| \nonumber\\
&\geq&\frac{1}{2}+{\frac{1}{4}}|\vec{F}+\vec{H}+\vec{F}-\vec{H}|\nonumber\\
&=&\frac{1}{2}+{\frac{1}{2}}|\vec{F}|=2    C_{\text{LHS}},
\end{eqnarray}
which implies that at least one of $\{\langle \mathcal{W} \rangle_1^{\rm max}, \langle \mathcal{W} \rangle_2^{\rm max}\} $ is equal to or larger than the classical bound $C_{\text{LHS}}$. To complete the proof, we need to further show that any entangled state cannot reach the equality. In other words, we only need to prove that reaching the equality must lead to a separable state. To reach the equality, we need
\begin{eqnarray}\label{eq.12}
\vec{H}=k\vec{F}, \;\; \; H_0+k|\vec{F}|=0, \;\;  |H_0|=|\vec{H}|, \;\;\; (|k|\leq 1),
\end{eqnarray}
which leads to the concurrence $\mathcal{C}=0$, thus the corresponding state $\rho$ is separable, as shown in the Methods section. This completes  the proof.

\noindent One may note that, $\langle \mathcal{W} \rangle_1^{\rm max}$ and $\langle \mathcal{W} \rangle_2^{\rm max}$ are the same when $H_0=0$ and $\vec{F} \perp \vec{H}$. In this situation, we have $|\vec{H}|^2=\mathcal{C}^2$. Here we provide some examples.
(i) For $\phi=0$, the concurrence of the state $\rho$ is $\mathcal{C}=\left|\nu_1 \sin2\theta-\nu_2 \sin2\alpha \right|$, the magnitude of Bloch vector is $|\vec{F}|=|\nu_1 \cos2\theta  - \nu_2 \cos2\alpha|$. In this case, the classical bound $C_{\text{LHS}}=(1+|\vec{F}|)/4$, and the quantum expectation $\langle \mathcal{W} \rangle^{\rm max}
 =(1+\sqrt{\vec{F}^2 +\mathcal{C}^2})/4>C_{\text{LHS}}$ for any $\mathcal{C}\neq 0$.
(ii) For $\phi=\pi/2$, one has $\mathcal{C}=|(\nu_1-\nu_2)\sin2\theta|$ and $|\vec{F}|=|(\nu_1 - \nu_2) \cos2\theta|$. In this case, the classical bound $C_{\text{LHS}}=(1+|\vec{F}|)/4$, and the quantum expectation $\langle \mathcal{W} \rangle^{\rm max} =(1+\sqrt{\vec{F}^2 +\mathcal{C}^2})/4>C_{\text{LHS}}$ for any $\mathcal{C}\neq 0$.
(iii) For $0< \phi <\pi/2$, $\beta=\pi/2$ and $\alpha=0$, one has $\mathcal{C}=|\nu_1-\nu_2 \sin^2\phi|\sin 2\theta$ and $|\vec{F}|^2=[\nu_2 \sin (\alpha+\theta) \sin 2 \phi]^2+ \left[-\nu_2 \cos 2 \alpha \cos ^2\phi
+ \cos2\theta (\nu_1- \nu_2  \sin ^2\phi)\right]^2$. In this case, the quantum expectation $\langle \mathcal{W} \rangle^{\rm max} =(1+\sqrt{\vec{F}^2 +\mathcal{C}^2})/4>C_{\text{LHS}}=(1+|\vec{F}|)/4$ for any $\mathcal{C}\neq 0$.

\noindent We now give a concise numerical comparison between the above state-dependent nonlinear criterion and the fixed 3-setting linear steering inequality in Eq.~(\ref{eq:QM-3}). We take the representative family in example (ii) with $\nu_1=0.85$ and $\nu_2=0.15$. Then $\mathcal{C}=0.7|\sin2\theta|$, $C_{\text{LHS}}=(1+0.7|\cos2\theta|)/4$, and $\langle\mathcal{W}\rangle^{\max}=0.425$. Optimizing the fixed 3-setting linear inequality over three orthogonal measurement directions gives $\mathcal{I}_3^{\rm QM}=(1+2\mathcal{C})/3$, with the LHS bound $1/\sqrt{3}$. Table~\ref{tab:numerical-comparison} lists the two violations
\begin{eqnarray}
\Delta_{\rm NL}&=&\langle\mathcal{W}\rangle^{\max}-C_{\text{LHS}},\nonumber\\
\Delta_{\rm lin}&=&\mathcal{I}_3^{\rm QM}-1/\sqrt{3}.
\end{eqnarray}
For all entangled points shown in the table, the state-dependent nonlinear criterion gives a positive violation. By contrast, the fixed 3-setting linear inequality does not show a violation for weakly entangled points close to the separable limits.

\begin{table}[t]
\caption{Numerical comparison for the representative rank-2 two-qubit family in example (ii) with $\nu_1=0.85$ and $\nu_2=0.15$. Here $\Delta_{\rm NL}=\langle\mathcal{W}\rangle^{\max}-C_{\text{LHS}}$ denotes the violation of the state-dependent nonlinear inequality, and $\Delta_{\rm lin}=\mathcal{I}_3^{\rm QM}-1/\sqrt{3}$ denotes the violation of the fixed 3-setting linear inequality in Eq.~(\ref{eq:QM-3}).}
\label{tab:numerical-comparison}
\begin{ruledtabular}
\begin{tabular}{ccccc}
$\theta$ & $\mathcal{C}$ & $\Delta_{\rm NL}$ & $\Delta_{\rm lin}$ & Eq.~(\ref{eq:QM-3}) \\
\hline
0.100 & 0.139 & 0.003 & $-0.151$ & no \\
0.200 & 0.273 & 0.014 & $-0.062$ & no \\
0.300 & 0.395 & 0.031 & 0.019 & yes \\
0.400 & 0.502 & 0.053 & 0.091 & yes \\
$\pi/4$ & 0.700 & 0.175 & 0.223 & yes \\
1.170 & 0.503 & 0.053 & 0.091 & yes \\
1.270 & 0.396 & 0.031 & 0.020 & yes \\
1.370 & 0.274 & 0.014 & $-0.062$ & no \\
1.470 & 0.140 & 0.004 & $-0.151$ & no \\
\end{tabular}
\end{ruledtabular}
\end{table}

\vspace{5mm}


\noindent \textbf{DISCUSSION}

\noindent In this work, we have proved that every entangled rank-2 two-qubit state is EPR steerable, with rank-1 states included as a special case. This gives an analytical characterization of EPR steerability for the complete rank-2 class of mixed two-qubit states. Gisin's theorem serves as a natural motivation for asking how entanglement and EPR steering are related beyond pure states, while the main result is the rank-2 two-qubit theorem proved above.

The proof combines three ingredients. First, we use a local-unitary parametrization of rank-2 two-qubit states. Second, we derive a necessary and sufficient separability condition for this parametrization. Third, we construct a state-dependent nonlinear steering inequality whose LHS bound depends on the target state and on Alice's measurement parameters. The equality analysis shows that the bound can be saturated only by separable states. Hence any entangled rank-2 two-qubit state violates the state-dependent LHS bound and is EPR steerable.

The state-dependent nonlinear form is central to the analytical characterization. Fixed-coefficient linear steering inequalities are useful and experimentally accessible, but they need not detect all weakly entangled mixed states. The numerical comparison in Table~\ref{tab:numerical-comparison} illustrates this point for a representative rank-2 two-qubit family: the state-dependent nonlinear criterion remains positive for weakly entangled examples near the separable limits, while the fixed 3-setting linear inequality in Eq.~(\ref{eq:QM-3}) does not show a violation for those parameters.

Within the hierarchy of quantum nonlocality, this result identifies rank-2 two-qubit states as a natural mixed-state class in which entanglement is sufficient for EPR steerability. Together with known higher-rank examples of entangled states admitting LHS models \cite{WJD07,Nguyen2019}, the result provides a useful analytical benchmark for comparing entanglement, steering criteria, and finite-setting tests in two-qubit systems.

Looking forward, extending this steering theorem to higher-dimensional qudit systems or multipartite configurations presents significant mathematical challenges. For generic bipartite $d \times d$ systems, the increased geometric complexity of the quantum state space makes analytically constructing a tight, state-dependent steering criterion highly intricate. Additionally, multipartite EPR steering involves diverse correlation structures which cannot be straightforwardly accommodated by the current bipartite framework. We leave the rigorous generalization to higher dimensions and multipartite settings as an important direction for future research.

\vspace{5mm}


\noindent \textbf{METHODS}

\noindent \textbf{Parametrization of rank-2 states}

\noindent A general two-qubit state with rank 2, including rank 1 as a special case, can be written as
\begin{eqnarray}
\rho=\nu_1|\psi_1\rangle\langle\psi_1|+\nu_2|\psi_2\rangle\langle\psi_2|,
\end{eqnarray}
where $\nu_1,\nu_2\in[0,1]$, $\nu_1+\nu_2=1$, and $\langle\psi_1|\psi_2\rangle=0$. By the Schmidt decomposition and local unitary transformations, $|\psi_1\rangle$ can be chosen as
\[
|\psi_1\rangle=\cos\theta|00\rangle+\sin\theta|11\rangle .
\]
The orthogonality condition then implies that $|\psi_2\rangle$ lies in the subspace spanned by $|01\rangle$, $|10\rangle$, and $\sin\theta|00\rangle-\cos\theta|11\rangle$. After absorbing an overall phase and removing one relative phase by local phase transformations, one obtains
\begin{eqnarray}
|\psi_2\rangle &=& \cos\phi(\cos\alpha|01\rangle+\sin\alpha|10\rangle)\nonumber\\
&&+e^{{\rm i}\beta}\sin\phi(\sin\theta|00\rangle-\cos\theta|11\rangle),
\end{eqnarray}
with $\theta,\phi,\alpha\in[0,\pi/2]$ and $\beta\in[0,2\pi]$.

\vspace{5mm}


\noindent \textbf{Concurrence of rank-2 states}

\noindent The degree of entanglement of the state $\rho$ is quantified by the concurrence $\mathcal{C}$. Using the spin-flip transformation $\tilde{\rho}=(\sigma_y\otimes\sigma_y)\rho^*(\sigma_y\otimes\sigma_y)$, the concurrence is given by $\mathcal{C}=\max\{0,\lambda_1-\lambda_2-\lambda_3-\lambda_4\}$, where $\lambda_i$ are the square roots of the eigenvalues of $\rho\tilde{\rho}$ in decreasing order. For the rank-2 state considered in this work, the characteristic equation $\det[v\openone-\rho\tilde{\rho}]=0$ reduces to
\begin{eqnarray}
v^2(v^2+s_2v+s_1)=0.
\end{eqnarray}
The concurrence is therefore obtained as
\begin{eqnarray}
\mathcal{C}=\sqrt{-s_2-2\sqrt{s_1}},
\end{eqnarray}
where the coefficients $s_1$ and $s_2$ are functions of the state parameters $(\theta,\phi,\alpha,\beta)$ and the weights $(\nu_1,\nu_2)$, as given in the Results section. Since $\rho$ is separable if and only if its concurrence vanishes, the separability condition is equivalently written as
\begin{eqnarray}
-s_2-2\sqrt{s_1}=0.
\end{eqnarray}

\vspace{5mm}


\noindent \textbf{Condition for separability}

\noindent To analyze steerability, Alice performs a projective measurement along the direction $\hat{n}=(\sin\xi\cos\tau,\sin\xi\sin\tau,\cos\xi)$. In Alice's measurement basis $\{|+\hat{n}\rangle,|-\hat{n}\rangle\}$, the shared state can be decomposed as
\begin{eqnarray}
 \rho &=& P^{\hat{n}}_{0}\otimes\tilde{\rho}_0^{\hat{n}}+P^{\hat{n}}_{1}\otimes\tilde{\rho}_1^{\hat{n}}+|+\hat{n}\rangle\langle-\hat{n}|\otimes\mathcal{M} \nonumber\\
&&+|-\hat{n}\rangle\langle+\hat{n}|\otimes\mathcal{M}^\dagger,
\end{eqnarray}
where $P^{\hat{n}}_{0}=|+\hat{n}\rangle\langle+\hat{n}|$ and $P^{\hat{n}}_{1}=|-\hat{n}\rangle\langle-\hat{n}|$ are Alice's projectors, and $\tilde{\rho}_0^{\hat{n}}$ and $\tilde{\rho}_1^{\hat{n}}$ are Bob's unnormalized conditional states. The off-diagonal block $\mathcal{M}$ depends on Alice's measurement basis and on the parameters of $\rho$.

For the parametrized rank-2 state, direct calculation shows that, under local unitary transformations, $\rho$ is separable if and only if there exists a measurement direction $\hat{n}$ such that $\mathcal{M}=\mathcal{M}^\dagger$. Equivalently, solving the Hermiticity constraints of $\mathcal{M}$ gives the same parameter conditions as $\mathcal{C}=0$. Thus the condition $\mathcal{M}=\mathcal{M}^\dagger$ provides a necessary and sufficient criterion for the separability of the rank-2 state considered here.

\vspace{5mm}


\noindent \textbf{Derivation of the LHS bound}

\noindent The state-dependent nonlinear steering inequality is defined through the operator
\begin{eqnarray}
\mathcal{W}=|+\rangle\langle+|\otimes|\hat{n}_B\rangle\langle\hat{n}_B|,
\end{eqnarray}
where $|+\rangle=(|+\hat{n}\rangle+e^{{\rm i}\delta}|-\hat{n}\rangle)/\sqrt{2}$ and $|\hat{n}_B\rangle$ denotes Bob's projective measurement state. The local-hidden-state bound is
\begin{eqnarray}
C_{\text{LHS}}=\max_{\hat{n}_B}\left\{\mathrm{Tr}\left[\left(|+\rangle\langle+|\otimes|\hat{n}_B\rangle\langle\hat{n}_B|\right)\rho_{\rm sep}\right]\right\},
\end{eqnarray}
where $\rho_{\rm sep}$ denotes the corresponding separable rank-2 state. Substituting the block decomposition of $\rho_{\rm sep}$ into the trace, the off-diagonal contribution is proportional to $e^{{\rm i}\delta}\mathcal{M}+e^{-{\rm i}\delta}\mathcal{M}^\dagger$. For the choices $\delta=\pm\pi/2$, this term becomes $\pm{\rm i}(\mathcal{M}-\mathcal{M}^\dagger)$, which vanishes for separable states in the basis satisfying $\mathcal{M}=\mathcal{M}^\dagger$. The LHS bound is then determined only by Bob's reduced state $\rho_B=\tilde{\rho}_0^{\hat{n}}+\tilde{\rho}_1^{\hat{n}}$:
\begin{eqnarray}
C_{\text{LHS}}=\frac{1}{2}\max_{\hat{n}_B}\left[\mathrm{Tr}\left(|\hat{n}_B\rangle\langle\hat{n}_B|\rho_B\right)\right]=\frac{1}{2}\lambda_{\rm max}(\rho_B).
\end{eqnarray}
Writing Bob's reduced state as $\rho_B=\frac{1}{2}(\openone+\vec{F}\cdot\vec{\sigma})$, we obtain
\begin{eqnarray}
C_{\text{LHS}}=\frac{1}{4}(1+|\vec{F}|).
\end{eqnarray}

\vspace{5mm}


\noindent \textbf{Equality condition}

\noindent We now show that any entangled rank-2 state violates the state-dependent nonlinear steering inequality. For the two choices $\delta=\pi/2$ and $\delta=-\pi/2$, maximizing over Bob's measurement direction gives
\begin{eqnarray}
\langle \mathcal{W} \rangle_1^{\rm max} &=& \frac{1}{4}+\frac{1}{4}H_0+\frac{1}{4}|\vec{F}+\vec{H}|,\nonumber\\
\langle \mathcal{W} \rangle_2^{\rm max} &=& \frac{1}{4}-\frac{1}{4}H_0+\frac{1}{4}|\vec{F}-\vec{H}|,
\end{eqnarray}
where $H_0$ is a scalar and $\vec{H}$ is determined by the off-diagonal block $\mathcal{M}$. Taking the sum and using the triangle inequality, we obtain
\begin{eqnarray}
\langle \mathcal{W} \rangle_1^{\rm max}+\langle \mathcal{W} \rangle_2^{\rm max} &=& \frac{1}{2}+\frac{1}{4}\left(|\vec{F}+\vec{H}|+|\vec{F}-\vec{H}|\right)\nonumber\\
&\geq& \frac{1}{2}+\frac{1}{2}|\vec{F}|=2C_{\text{LHS}}.
\end{eqnarray}
Therefore, at least one of $\langle \mathcal{W} \rangle_1^{\rm max}$ and $\langle \mathcal{W} \rangle_2^{\rm max}$ is not smaller than $C_{\text{LHS}}$.

It remains to identify the equality case. Equality requires $\vec{H}=k\vec{F}$ with $|k|\leq1$, together with $H_0+k|\vec{F}|=0$ and $|H_0|=|\vec{H}|$. Substituting these conditions into the explicit expressions of $H_0$, $\vec{H}$ and $\vec{F}$ gives exactly the zero-concurrence condition
\begin{eqnarray}
-s_2-2\sqrt{s_1}=0.
\end{eqnarray}
Thus equality can occur only for separable states. For any entangled state with $\mathcal{C}>0$, the equality case is excluded, and hence
\begin{eqnarray}
\max\left\{\langle \mathcal{W} \rangle_1^{\rm max},\langle \mathcal{W} \rangle_2^{\rm max}\right\}>C_{\text{LHS}}.
\end{eqnarray}
Consequently, every entangled two-qubit state of rank 2, including rank-1 states as a special case, violates the state-dependent LHS bound and possesses EPR steerability.

\vspace{5mm}


\noindent \textbf{DATA AVAILABILITY}

\noindent The numerical comparison data presented in Table~\ref{tab:numerical-comparison} are included in this article and are obtained by direct evaluation of the analytical expressions given in the Results section. The analytical materials necessary to interpret, replicate and build upon the findings are included in this article. Additional information is available from the corresponding author on reasonable request.

\vspace{5mm}


\noindent \textbf{CODE AVAILABILITY}

\noindent No custom code is required to reproduce the analytical results. The numerical values in Table~\ref{tab:numerical-comparison} are obtained by direct evaluation of the formulas displayed in the Results section. The simple numerical evaluation script is available from the corresponding author on reasonable request.

\vspace{5mm}


\noindent \textbf{ACKNOWLEDGEMENTS}

\noindent This work was supported by the Quantum Science and Technology--National Science and Technology Major Project (Grant No. 2024ZD0301000), and the National Natural Science Foundation of China (Grant No. 12275136). The funders had no role in study design, analysis, manuscript preparation, or the decision to submit the work.

\vspace{5mm}

\noindent \textbf{AUTHOR CONTRIBUTIONS}

\noindent J.-L.C. conceived and supervised the project, constructed the steering inequality, and developed the main theoretical framework. Y.-X.Z. carried out the theoretical calculations, including the parametrization of quantum states, the entanglement criterion, and the derivation of the inequality involving $\vec{F}$, and wrote the first draft of the manuscript. Both authors discussed the results and contributed to the writing and revision of the manuscript.


\vspace{5mm}

\noindent \textbf{COMPETING INTERESTS}

\noindent The authors declare no competing financial or non-financial interests.

\vspace{5mm}

\noindent \textbf{ADDITIONAL INFORMATION}

\noindent \textbf{Supplementary information} The online version contains supplementary material available at https://doi.org/.

\noindent \textbf{Correspondence} and requests for materials should be addressed to J.-L.C.

\end{document}



\title{Supplemental Material for\\``Steerability of Rank-2 Two-Qubit Entangled States''}

\author{Yu-Xuan Zhang}
\affiliation{School of Physics, Nankai University, Tianjin 300071, People's Republic of China}

\author{Jing-Ling Chen}
\email{chenjl@nankai.edu.cn}
\affiliation{Theoretical Physics Division, Chern Institute of Mathematics, Nankai University, Tianjin 300071, People's Republic of
	 	China}

\date{\today}

\begin{abstract}
\end{abstract}


\maketitle\tableofcontents\newpage

\section{Quantum State $\rho$ with Rank 2}

In quantum mechanics, quantum mixed states are represented by density matrices. For a two-qubit state, under the local unitary transformations an arbitrary density matrix with rank 2 (including rank 1 as a special case) can be expressed as
\begin{eqnarray}\label{eq:rank2state}
\rho=\nu_1  \ket{\psi_1}\bra{\psi_1}+\nu_2 \ket{\psi_2}\bra{\psi_2},
\end{eqnarray}
where
\begin{eqnarray}\label{eq:psi1}
\ket{\psi_1} =\cos \theta \ket{00}+\sin \theta\ket{11}, \;\;\;\;\; \theta\in [0, \pi/2],
\end{eqnarray}
\begin{eqnarray}\label{eq:psi2}
\ket{\psi_2} =\cos\phi(\cos \alpha\, {e^{{\rm i}\beta_1}} \ket{01} + {e^{{\rm i}\beta_2}} \sin\alpha\ket{10})
+{e^{{\rm i}\beta_3}}\sin\phi(\sin \theta \ket{00}-\cos \theta\ket{11}),
\end{eqnarray}
are two orthogonal pure states of two qubits, i.e.,
\begin{eqnarray}
\langle \psi_1| \psi_2\rangle =\langle \psi_2| \psi_1\rangle=0.
\end{eqnarray}
The parameters $\nu_i$ ($i=1,2$) are the weights of pure states $|\psi_i\rangle$ ($i=1,2$) in $\rho$ respectively, and their regions are
\begin{eqnarray}
\nu_1 \in [0,1], \;\;\; \nu_2 \in [0,1], \;\;\; \nu_1+\nu_2=1.
\end{eqnarray}
The regions of the angles $\phi$, $\alpha$, $\beta_1$, $\beta_2$ and $\beta_3$ are temporarily $[0, 2\pi]$, and we shall simplify these parameters further.

Generally, the state $\rho$ is a rank-2 density matrix. Specially, when one of the two weights is 1 (i.e., $\nu_1=1$ or $\nu_2=1$), then $\rho$ reduces to a density matrix with rank 1 (i.e., the pure state). In the following, let us explain why the state $\rho$ is written in the form of Eq. (\ref{eq:rank2state}).

\subsection{The Form of $|\psi_1\rangle$} \label{eq:secIA}

\begin{remark} \textcolor{blue}{Why the pure state $|\psi_1\rangle$ is in the form of Eq. (\ref{eq:psi1})?}
Let us consider the most general pure state in the two-qubit system, which is given by
\begin{eqnarray}\label{eq:psi}
\ket{\psi} =a \ket{01} +b\ket{10}+c \ket{00}+d\ket{11},
\end{eqnarray}
where $a, b, c, d$ are complex numbers, and they satisfy the following normalization condition
\begin{eqnarray}
|a|^2+ |b|^2+ |c|^2+|d|^2=1.
\end{eqnarray}
It is easy to known that the pure state $\ket{\psi}$ contains 7 (i.e., $4\times 2-1=7$) free parameters. However, under the local unitary transformation $\mathcal{U}_1 \otimes \mathcal{U}_2$, the initial pure state ($|\psi\rangle_{\rm i}\equiv|\psi\rangle $) can be transformed to the following simple form
\begin{eqnarray}
(\mathcal{U}_1 \otimes \mathcal{U}_2)\, \ket{\psi}_{\rm i} \mapsto \ket{\psi}_{\rm f} =\kappa_1 \ket{00}+\kappa_2\ket{11},
\end{eqnarray}
with $\kappa_1 \geq 0$, $\kappa_2 \geq 0$, and $\kappa_1 +\kappa_2 =1$. Here $\ket{\psi}_{\rm f}$ represents the final state after performing the unitary transformation on the initial state $|\psi\rangle_{\rm i}=|\psi\rangle$. This procedure is well-known as the Schmit decomposition of a two-qubit pure state. After denoting
\begin{eqnarray}
&& \kappa_1=\cos\theta, \;\;\;\;\; \kappa_2=\sin\theta,
\end{eqnarray}
then one naturally explains the reason why the state $|\psi_1\rangle$ is in the form of Eq. (\ref{eq:psi1}). In addition, note that $\kappa_1 \geq 0$ and $\kappa_2 \geq 0$, thus the region of the angle $\theta$ is given by
\begin{eqnarray}
&& \theta\in [0, \pi/2].
\end{eqnarray}
By comparing Eq. (\ref{eq:psi1}) and Eq. (\ref{eq:psi}), the merit of $|\psi_1\rangle$ is obvious, namely, it contains only one free parameter. $\blacksquare$
\end{remark}

\begin{remark} \textcolor{blue}{The Schmit decomposition.}
The Schmit decomposition can be found the textbook of quantum information theory \cite{Nielsen2000}. For the convenience of readers, here we would like to provide a simple derivation. Consider a general two-qubit state in Eq. (\ref{eq:psi}), we want to choose Alice's and Bob's bases to simplify it. Let us consider the following unitary transformation
\begin{eqnarray}
&& \ket{0} \mapsto A_0 \ket{0} +B_0 \ket{1}, \nonumber\\
&&\ket{1}  \mapsto  -B_0^* \ket{0} +A_0^* \ket{1},
\end{eqnarray}
for Alice ( $|A_0|^2+|B_0|^2=1$), and the following unitary transformation
\begin{eqnarray}
&&\ket{0}   \mapsto  A_1 \ket{0} +B_1 \ket{1}, \nonumber\\
&&\ket{1}  \mapsto  -B_1^* \ket{0} +A_1^* \ket{1},
\end{eqnarray}
for Bob ( $|A_1|^2+|B_1|^2=1$ ). Here $A_0^*=(A_0)^*$, similar for $B_0^*$, $A_1^*$ and $B_1^*$. Then we can obtain
\begin{eqnarray}
(\mathcal{U}_1 \otimes \mathcal{U}_2)\, \ket{\psi}_{\rm i} \mapsto \ket{\psi}_{\rm f} = \eta_{00} \ket{00} + \eta_{01} \ket{01} + \eta_{10} \ket{10} + \eta_{11} \ket{11},
\end{eqnarray}
with the coefficients
\begin{eqnarray}
&&\eta_{00} = A_0 A_1 c - A_0 B_1^* a - A_1 B_0^* b + B_0^* B_1^* d,\nonumber\\
&&\eta_{01} = A_0 B_1 c + A_0 A_1^* a - B_0^* B_1 b - B_0^* A_1^* d,\nonumber\\
&&\eta_{10} = B_0 A_1 c - B_0 B_1^* a + A_0^* A_1 b - A_0^* B_1^* d,\nonumber\\
&&\eta_{11} = B_0 B_1 c + B_0 A_1^* a + A_0^* B_1 b + A_0^* A_1^* d.
\end{eqnarray}
We now require the coefficients
\begin{eqnarray}\label{eq:eta-1}
&&\eta_{01} = A_0 B_1 c + A_0 A_1^* a - B_0^* B_1 b - B_0^* A_1^* d=0,\nonumber\\
&&\eta_{10} = B_0 A_1 c - B_0 B_1^* a + A_0^* A_1 b - A_0^* B_1^* d=0,
\end{eqnarray}
such that the resultant state $|\psi\rangle_{\rm f}$ turns to the following Schmit form
\begin{eqnarray}
&& \ket{\psi}_{\rm f} = \eta_{00} \ket{00} + \eta_{11} \ket{11}.
\end{eqnarray}
Eq. (\ref{eq:eta-1}) gives
\begin{eqnarray}
&&\frac{B_0^*}{A_0} = \frac{ B_1 c +  A_1^* a }{ B_1 b + A_1^* d},\nonumber\\
&&\frac{B_0}{A_0^*} =\frac{ -A_1 b + B_1^* d}{A_1 c - B_1^* a },
\end{eqnarray}
i.e.,
\begin{eqnarray}
\frac{ -A_1 b + B_1^* d}{A_1 c - B_1^* a } = \frac{ B_1^* c^* +  A_1 a^* }{ B_1^* b^* + A_1d^*}.
\end{eqnarray}
Assume
\begin{eqnarray}
&&\frac{B_1^*}{A_1}=x_1,
\end{eqnarray}
we can get
\begin{eqnarray}
\frac{ - b + x_1 d}{ c - x_1 a } =\frac{ x_1 c^* +   a^* }{ x_1b^* + d^*},
\end{eqnarray}
i.e.,
\begin{eqnarray}
(ac^*+b^*d)x_1^2+(aa^*-bb^*-cc^*+dd^*)x_1-(a^*c+bd^*)=0,
\end{eqnarray}
i.e.,
\begin{eqnarray}
\mu_1 x_1^2+\mu_2 x_1-\mu_1^*=0,
\end{eqnarray}
where
\begin{eqnarray}
&& \mu_1=ac^*+b^*d, \nonumber\\
&& \mu_2=aa^*-bb^*-cc^*+dd^*=2(aa^*+dd^*)-1=2(|a|^2+|d|^2)-1.
\end{eqnarray}
Here $\mu_2$ is a real number and $\mu_1$ can be a complex number.
Hence
\begin{eqnarray}
x_1 = \frac{ - \mu_2 \pm \sqrt{ \mu_2^2 + 4 |\mu_1|^2 } }{2 \mu_1}.
\end{eqnarray}
Based on which, we can have
\begin{eqnarray}
&& B_1 = \frac{|x_1|}{\sqrt{1 + |x_1|^2}} e^{{\rm i} \phi_1}, \quad A_1 = \frac{1}{\sqrt{1 + |x_1|^2}} e^{{\rm i} \phi_2},\nonumber\\
&& \dfrac{x_1}{|x_1|}= \frac{B_1^*/A_1}{|B_1^*/A_1|}= e^{-{\rm i} (\phi_1+\phi_2)}.
\end{eqnarray}
Similarly, we can get
\begin{eqnarray}\label{eq:B0A0}
&& B_0 = \frac{|x_2|}{\sqrt{1 + |x_2|^2}} e^{{\rm i} \phi_1'}, \quad A_0 = \frac{1}{\sqrt{1 + |x_2|^2}} e^{{\rm i} \phi_2'} ,
\end{eqnarray}
with
\begin{eqnarray}
&& x_2=\dfrac{ B_1 c +  A_1^* a }{ B_1 b + A_1^* d}, \;\;\;\; \dfrac{x_2}{|x_2|} =\frac{B_0^*/A_0}{|B_0^*/A_0|}= e^{-{\rm i} (\phi'_1+\phi'_2)}.
\end{eqnarray}
Then we have the coefficients
\begin{eqnarray}
\eta_{00} &=&  A_0 A_1 c - A_0 B_1^* a - A_1 B_0^* b + B_0^* B_1^* d\equiv \kappa_1 e^{{\rm i}\tau_1},\nonumber\\
\eta_{11} &=& B_0 B_1 c + B_0 A_1^* a + A_0^* B_1 b + A_0^* A_1^* d\equiv \kappa_2 e^{{\rm i}\tau_2},
\end{eqnarray}
where
\begin{eqnarray}
\kappa_1=|\eta_{00}| \geq 0,\quad e^{{\rm i}\tau_1}=\frac{\eta_{00}}{|\eta_{00}|},   \nonumber\\
\kappa_2=|\eta_{11}| \geq 0,\quad e^{{\rm i}\tau_2}=\frac{\eta_{11}}{|\eta_{11}|}.
\end{eqnarray}
So the state becomes
\begin{eqnarray}
\ket{\psi}_{\rm f}=\kappa_1 e^{{\rm i}\tau_1}\ket{00}+\kappa_2 e^{{\rm i}\tau_2}\ket{11}.
\end{eqnarray}

Based on which, we can perform an additional local unitary operation
\begin{eqnarray}
\mathcal{U}'_{1}=\begin{bmatrix}
e^{-{\rm i} \frac{\tau_1}{2}} & 0 \\
0 & {e^{-{\rm i} \frac{\tau_2}{2}}} \\
\end{bmatrix}
\end{eqnarray}
on Alice's qubit, and similarly perform the local unitary operation
\begin{eqnarray}
\mathcal{U}'_{2}=\begin{bmatrix}
e^{-{\rm i} \frac{\tau_1}{2}} & 0 \\
0 & {e^{-{\rm i} \frac{\tau_2}{2}}}  \\
\end{bmatrix}
\end{eqnarray}
on Bob's qubit. Then we can have
\begin{eqnarray}
(\mathcal{U}'_{1} \otimes \mathcal{U}'_{2})\ket{\psi}_{\rm f} \mapsto \ket{\psi}_{\rm f}=\kappa_1 \ket{00}+\kappa_2 \ket{11}.
\end{eqnarray}
Notice that
\begin{eqnarray}
\kappa_1,\, \kappa_2\geq 0, \quad \quad \kappa^2_1+\kappa^2_2=1.
\end{eqnarray}
This ends the demonstration of the Schmit decomposition. $\blacksquare$

%
%
%
\end{remark}

\newpage

\subsection{The Form of $|\psi_2\rangle$}

\begin{remark}\label{remark3}
 \textcolor{blue}{Why the state $|\psi_2\rangle$ is in the form of Eq. (\ref{eq:psi2})?}
To construct a density matrix with rank 2, given the the pure state vector $|\psi_1\rangle=\cos \theta \ket{00}+\sin \theta\ket{11}$, we need to find another state vector $| \psi_2\rangle$, which satisfies the following orthogonal condition
\begin{eqnarray}
\langle \psi_1| \psi_2\rangle =\langle \psi_2| \psi_1\rangle=0.
\end{eqnarray}
Let us assume
\begin{eqnarray}
\ket{\psi_2} =a_1 \ket{01} +b_1\ket{10}+c_1 \ket{00}+d_1\ket{11},
\end{eqnarray}
where $a_1$, $b_1$, $c_1$ and $d_1$ are some superposition coefficients.
From the orthogonal condition, we can get
\begin{eqnarray}
c_1 \cos\theta+d_1\sin\theta=0,
\end{eqnarray}
which means
\begin{eqnarray}
&& c_1 =e^{{\rm i}\beta_3}k\sin\theta, \;\;\;\;\;  d_1=-e^{{\rm i} \beta_3}k\cos\theta,
\end{eqnarray}
where $k$ is a real number (with $k\geq 0$). Without loss of generality, we can assume
\begin{eqnarray}
&& a_1 =e^{{\rm i}\beta_1}l\cos\alpha, \;\;\;\;  b_1=e^{{\rm i} \beta_2} l\sin\alpha,
\end{eqnarray}
where $l$ is a real number (with $l\geq 0$). Notice that we have the following normalized condition
\begin{eqnarray}
\langle \psi_2| \psi_2\rangle =1.
\end{eqnarray}
This yields
\begin{eqnarray}
k^2+l^2=1,
\end{eqnarray}
that is
\begin{eqnarray}
&&k=\sin\phi, \;\;\;\; l=\cos\phi.
\end{eqnarray}
Based on the above results, we can determine the state vector $|\psi_2\rangle$ as
\begin{eqnarray}\label{eq:psi2-a}
\ket{\psi_2} &=&e^{{\rm i}\beta_1}l\cos\alpha \ket{01} +e^{{\rm i}\beta_2} l\sin\alpha\ket{10}+e^{{\rm i} \beta_3}k\sin\theta \ket{00}-e^{{\rm i} \beta_3}k\cos\theta \ket{11}\nonumber\\
&=& \cos\phi(\cos \alpha\, {e^{{\rm i}\beta_1}} \ket{01} + {e^{{\rm i}\beta_2}} \sin\alpha\ket{10})
+ {e^{{\rm i}\beta_3}}\sin\phi(\sin \theta \ket{00}-\cos \theta\ket{11}),
\end{eqnarray}
thus explaining why the state $|\psi_2\rangle$ is in the form of Eq. (\ref{eq:psi2}). $\blacksquare$
\end{remark}

\begin{remark}\textcolor{blue}{The region of the angle $\phi$ can be considered as $\phi \in [0, \pi/2]$ and the region of the angle $\alpha$ can be considered as $\alpha \in [0, \pi/2]$.} In general, the region of $\phi$ is $\phi\in [0, 2\pi]$ and the region of $\alpha$ is also $\alpha\in [0, 2\pi]$. However, the parameters $\phi$ and $\alpha$ can be restricted to the region of $[0, \pi/2]$. The reason is as follows. From the viewpoint of physics, for the state vector $\vec{u}_0\equiv |\psi_1\rangle$, there are three vectors that are perpendicular to it, i.e.,
\begin{eqnarray}
&& \vec{u}_1 \equiv \ket{01}, \;\;\;\;\; \vec{u}_2\equiv \ket{10}, \;\;\;\;\; \vec{u}_3\equiv
\sin \theta \ket{00}-\cos \theta\ket{11}.
\end{eqnarray}
Then the most general vector orthogonal to $|\psi_1\rangle$ is given by the linear superposition of these three vectors, i.e.,
\begin{eqnarray}\label{eq:psi2-b}
&& |\psi_2\rangle= f_1 \vec{u}_1+ f_2 \vec{u}_2 +f_3\vec{u}_3.
\end{eqnarray}
However, the problem considered is in the framework of quantum mechanics, thus $f_j$'s $(j=1, 2, 3)$ can be complex numbers, which can be parameterized as
\begin{eqnarray}
&& f_1 =|f_1|{e^{{\rm i}\beta_1}}, \;\;\;\;\; f_2=|f_2|{e^{{\rm i}\beta_2}}, \;\;\;\; f_3=|f_3|{e^{{\rm i}\beta_3}}.
\end{eqnarray}
What we need is just
\begin{eqnarray}
|f_1|^2+|f_2|^2+|f_3|^2=1,
\end{eqnarray}
and notice that
\begin{eqnarray}
|f_j|\geq0.
\end{eqnarray}
After comparing Eq. (\ref{eq:psi2-a}) and Eq. (\ref{eq:psi2-b}), we can parameterize $|f_j|$'s as
\begin{eqnarray}\label{eq:alpha-1}
|f_1|=\cos\phi\cos\alpha, \quad |f_2|=\cos\phi\sin\alpha, \quad |f_3|=\sin\phi,
\end{eqnarray}
where $0\leq\phi\leq \pi/2$, $0\leq\alpha\leq \pi/2$. Such a parameterization is sufficient because it covers all possible nonnegative triples \((|f_1|, |f_2|, |f_3|)\) satisfying the unit sum of squares. Thus the region of $\phi$ can be simplified as $\phi \in [0, \pi/2]$ and the region of $\alpha$ can be simplified as $\alpha \in [0, \pi/2]$. { By the way, based on $k=\sin\phi\geq 0$ and $l=\cos\phi \geq 0$ in Remark \ref{remark3}, one can also have $\phi \in [0, \pi/2]$. And then due to Eq. (\ref{eq:alpha-1}) we have $\alpha \in [0, \pi/2]$.} $\blacksquare$
\end{remark}

\newpage

\subsection{The Simplification of $|\psi_2\rangle$}

The state $|\psi_2\rangle$ be simplified. In this section, let us study the simplification of the state $|\psi_2\rangle$.

\begin{remark} \textcolor{blue}{The phase factor $\beta_1$ can be set as 0.}
For the state $|\psi_2\rangle$, one can have
\begin{eqnarray}
\ket{\psi_2} &=&\cos\phi(\cos \alpha\, {e^{{\rm i}\beta_1}} \ket{01} + {e^{{\rm i}\beta_2}} \sin\alpha\ket{10})
+{e^{{\rm i}\beta_3}}\sin\phi(\sin \theta \ket{00}-\cos \theta\ket{11})\nonumber\\
&=& {e^{{\rm i}\beta_1}} \, \Big[\cos\phi(\cos \alpha\,  \ket{01} + {e^{{\rm i}(\beta_2-\beta_1)}} \sin\alpha\ket{10})
+{e^{{\rm i}(\beta_3-\beta_1)}}\sin\phi(\sin \theta \ket{00}-\cos \theta\ket{11})\Big]\nonumber\\
&=& {e^{{\rm i}\beta_1}} \, \Big[\cos\phi(\cos \alpha\,  \ket{01} + {e^{{\rm i}\beta'_2}} \sin\alpha\ket{10})
+{e^{{\rm i}\beta'_3}}\sin\phi(\sin \theta \ket{00}-\cos \theta\ket{11})\Big],
\end{eqnarray}
with $\beta'_2=\beta_2-\beta_1$ and $\beta'_3=\beta_3-\beta_1$. As a global phase, the phase ${e^{{\rm i}\beta_1}}$ is trivial and can be ignored, for it vanishes when one considers the density-matrix form of the state $|\psi_2\rangle$ (i.e., $|\psi_2\rangle\langle\psi_2|$). Equivalently, the state $|\psi_2\rangle$ can be simplified by setting $\beta_1=0$. Accordingly, the state $|\psi_2\rangle$ becomes
\begin{eqnarray}
\ket{\psi_2} &=& \cos\phi(\cos \alpha\,  \ket{01} + {e^{{\rm i}\beta'_2}} \sin\alpha\ket{10})
+{e^{{\rm i}\beta'_3}}\sin\phi(\sin \theta \ket{00}-\cos \theta\ket{11}).
\end{eqnarray}
$\blacksquare$
\end{remark}

\begin{remark} \textcolor{blue}{The phase factor $\beta'_2$ can be set as 0.}
From above we have known
\begin{eqnarray}
&&\ket{\psi_1} =\cos \theta \ket{00}+\sin \theta\ket{11}, \nonumber\\
&& \ket{\psi_2} =\cos\phi(\cos \alpha \ket{01} + {e^{{\rm i}\beta'_2}}\sin\alpha\ket{10})+ {e^{{\rm i}\beta'_3}}\sin\phi(\sin \theta \ket{00}-\cos \theta\ket{11}).
\end{eqnarray}
We want to use the local unitary transformations to simplify it further.
We can perform the local unitary operation
\begin{eqnarray}
\mathcal{U}_a=\begin{bmatrix}
e^{{\rm i}\frac{\beta'_2}{2}} & 0 \\
0 & 1 \\
\end{bmatrix}
\end{eqnarray}
on Alice's qubit, and similarly perform the local unitary transformation
\begin{eqnarray}
\mathcal{U}_b=\begin{bmatrix}
1 & 0 \\
0 & e^{{\rm i}\frac{\beta'_2}{2}}  \\
\end{bmatrix}
\end{eqnarray}
on Bob's qubit. Then we can find that
\begin{eqnarray}
(\mathcal{U}_a \otimes \mathcal{U}_b) \ket{\psi_1} \mapsto \ket{\psi'_1} &=& \cos \theta \left(e^{{\rm i}\frac{\beta'_2}{2}}\, \ket{00}\right)+\sin \theta \left(e^{{\rm i}\frac{\beta'_2}{2}}\,\ket{11}\right) = e^{{\rm i} \frac{\beta'_2}{2}} \ket{\psi_1}, \nonumber\\
(\mathcal{U}_a \otimes \mathcal{U}_b) \ket{\psi_2} \mapsto \ket{\psi_2'} &=& e^{{\rm i}\beta'_2} \cos\phi (\cos \alpha \ket{01} + \sin \alpha \ket{10}) + e^{{\rm i}(\beta'_3 + \frac{\beta'_2}{2})} \sin\phi (\sin \theta \ket{00} - \cos \theta \ket{11})\nonumber\\
&=& e^{{\rm i}\beta'_2} \left[\cos\phi (\cos \alpha \ket{01} + \sin \alpha \ket{10}) + e^{{\rm i} \beta''_3 } \sin\phi (\sin \theta \ket{00} - \cos \theta \ket{11})\right],
\end{eqnarray}
with $\beta''_3=\beta'_3 - \frac{\beta'_2}{2}$. The global phases (i.e., $e^{{\rm i} \frac{\beta'_2}{2}}$ and $e^{{\rm i}\beta'_2}$ ) can be ignored, for they vanish when the density-matrix forms ((i.e., $|\psi'_1\rangle\langle\psi'_1|$ and $|\psi'_2\rangle\langle\psi'_2|$) are considered.
Equivalently, the state $|\psi_2\rangle$ can be simplified by setting $\beta'_2=0$. Accordingly, the state $|\psi_2\rangle$ becomes
\begin{eqnarray}
\ket{\psi_2} &=& \cos\phi(\cos \alpha\,  \ket{01} + \sin\alpha\ket{10})
+{e^{{\rm i}\beta''_3}}\sin\phi(\sin \theta \ket{00}-\cos \theta\ket{11}).
\end{eqnarray}
For simplicity, let us denote the parameter $\beta''_3$ as $\beta$, then finally the state $|\psi_2\rangle$ becomes
\begin{eqnarray}\label{eq:psi2-c}
\ket{\psi_2} &=& \cos\phi(\cos \alpha\,  \ket{01} + \sin\alpha\ket{10})
+{e^{{\rm i}\beta}}\sin\phi(\sin \theta \ket{00}-\cos \theta\ket{11}).
\end{eqnarray}
$\blacksquare$
\end{remark}


%
%
%

\newpage

\subsection{Further Simplification of Quantum State $\rho$}

Up to now, based on above analysis, for a two-qubit density matrix with rank 2 (including rank 1 as a special case), it is sufficient to consider the following form
\begin{eqnarray}
\rho=\nu_1  \ket{\psi_1}\bra{\psi_1}+\nu_2 \ket{\psi_2}\bra{\psi_2},
\end{eqnarray}
where
\begin{eqnarray}\label{eq:psi12-a}
&&\ket{\psi_1} =\cos \theta \ket{00}+\sin \theta\ket{11}, \nonumber\\
&& \ket{\psi_2} =\cos\phi(\cos \alpha \ket{01} +\sin\alpha\ket{10})+ e^{{\rm i}\beta}\sin\phi(\sin \theta \ket{00}-\cos \theta\ket{11}),
\end{eqnarray}
with
\begin{eqnarray}
&& {\theta\in [0, \pi/2]}, \;\;\;\; {\phi \in [0, \pi/2]}, \;\;\;\;  {\alpha \in [0, \pi/2]}, \;\;\;\; \beta \in [0, 2\pi],   \nonumber\\
&& \nu_1 \in [0,1], \;\;\; \nu_2 \in [0,1], \;\;\; \nu_1+\nu_2=1.
\end{eqnarray}
It is interesting to ask a question: Can the state $\rho$ can be simplified further? The answer is positive. We find that when $\theta=\pi/4$ (i.e., $|\psi_1\rangle$ is a maximally entangled state), and in this situation the density matrix $\rho$ can be simplified further.

\begin{remark} \textcolor{blue}{Further simplification of the state $\rho$ for $\theta=\pi/4$.} It is well-known that for an arbitrary two-qubit pure state
\begin{eqnarray}
\ket{\psi} =a \ket{01} +b\ket{10}+c \ket{00}+d\ket{11},
\end{eqnarray}
its degree of entanglement (i.e., \emph{concurrence}) is given by
\begin{eqnarray}
\mathcal{C}_{\ket{\psi}} = 2|ab-cd|.
\end{eqnarray}
For example, for $|\psi_1\rangle$ as shown in Eq. (\ref{eq:psi12-a}), its concurrence is
\begin{eqnarray}
\mathcal{C}_{\ket{\psi_1}} = |\sin2\theta|=\sin2\theta.
\end{eqnarray}
When $\theta=\pi/4$, $|\psi_1\rangle$ becomes the maximally entangled state (spanned in the subspace of $\{\ket{00}, \ket{11}\}$) as follows
\begin{eqnarray}
&&\ket{\psi_1} =\frac{1}{\sqrt{2}}(\ket{00}+\ket{11}),
\end{eqnarray}
and its concurrence is equal to 1. Accordingly, when $\theta=\pi/4$, the state $|\psi_2\rangle$ as shown in Eq. (\ref{eq:psi12-a}) turns to
\begin{eqnarray}
\ket{\psi_2} &=& \cos\phi(\cos \alpha \ket{01} +\sin\alpha\ket{10})+ e^{{\rm i}\beta}\, \frac{\sin\phi}{\sqrt{2}}( \ket{00}-\ket{11}),
\end{eqnarray}
whose concurrence is given by
\begin{eqnarray}
 \mathcal{C}_{\ket{\psi_2}} &=& 2 \left|e^{{\rm i}\beta}\frac{\sin\phi}{\sqrt{2}} \times \left(-e^{{\rm i}\beta}\frac{\sin\phi}{\sqrt{2}}\right) -\cos\phi\cos \alpha\times \cos\phi\sin \alpha\right|\nonumber\\
 &=& \left|e^{{\rm i} 2\beta}\sin^2 \phi + \cos^2\phi\sin 2 \alpha\right|\nonumber\\
&=& \sqrt{( \cos^2\phi\sin 2 \alpha+\sin^2 \phi \cos 2\beta)^2+ \sin^4 \phi  \sin^2 2\beta}.
\end{eqnarray}
Generally, one has $\mathcal{C}_{\ket{\psi_2}} \neq 1$, i.e., $\ket{\psi_2}$ is a non-maximally entangled state. Here, we find that when $\theta=\pi/4$, i.e., $|\psi_1\rangle$ is a maximally entangled state and in general $|\psi_2\rangle$ is a non-maximally entangled state, then the density matrix $\rho$ can be simplified further.

The procedure of the simplification is shown as follows: for the rank-2 state $\rho$ (with $\theta=\pi/4$) as follows
\begin{eqnarray}
\rho=\nu_1  \ket{\psi_1}\bra{\psi_1}+\nu_2 \ket{\psi_2}\bra{\psi_2},
\end{eqnarray}
where
\begin{eqnarray}
&&\ket{\psi_1} = \frac{1}{\sqrt{2}}(\ket{00}+\ket{11}), \nonumber\\
&& \ket{\psi_2} =
\cos\phi(\cos \alpha \ket{01} +\sin\alpha\ket{10})+ e^{{\rm i}\beta}\, \frac{\sin\phi}{\sqrt{2}}( \ket{00}-\ket{11}),
\end{eqnarray}
we can perform the local unitary transformation ${(\mathcal{V}_a \otimes \mathcal{V}_b)}$ on the rank-2 state $\rho$, such that
\begin{eqnarray}\label{eq:proof1}
&&{(\mathcal{V}_a \otimes \mathcal{V}_b)}\, \ket{\psi_2} \mapsto |\psi'_1\rangle =  {e^{{\rm i}\zeta}} \left(\cos\theta' \ket{00}+\sin\theta'\ket{11}\right), \nonumber\\
&&(\mathcal{V}_a \otimes \mathcal{V}_b)\, \ket{\psi_1} \mapsto |\psi'_2\rangle
= { \frac{1}{\sqrt{2}}(\ket{01}+\ket{10})},
\end{eqnarray}
{where $e^{{\rm i}\zeta}$ is a global phase that does not affect the density matrix $|\psi'_1\rangle\langle \psi'_1|$. }
As a result, under the local unitary transformation $\mathcal{V}_a \otimes \mathcal{V}_b$, the state $\rho$ becomes
\begin{eqnarray}
\rho' &=& (\mathcal{V}_a \otimes \mathcal{V}_b) \rho (\mathcal{V}_a \otimes \mathcal{V}_b)^\dagger \nonumber\\
 &=& \nu_1 (\mathcal{V}_a \otimes \mathcal{V}_b)  \ket{\psi_1}\bra{\psi_1} (\mathcal{V}_a \otimes \mathcal{V}_b)^\dagger +\nu_2 (\mathcal{V}_a \otimes \mathcal{V}_b) \ket{\psi_2}\bra{\psi_2} (\mathcal{V}_a \otimes \mathcal{V}_b)^\dagger\nonumber\\
&=& \nu_1  \ket{\psi'_2}\bra{\psi'_2} +\nu_2  \ket{\psi'_1}\bra{\psi'_1},
\end{eqnarray}
i.e.,
\begin{eqnarray}
\rho' &=& \nu'_1  \ket{\psi'_1}\bra{\psi'_1} +\nu'_2  \ket{\psi'_2}\bra{\psi'_2},
\end{eqnarray}
with
\begin{eqnarray}
&&|\psi'_1\rangle =  {e^{{\rm i}\zeta}} \left(\cos\theta' \ket{00}+\sin\theta'\ket{11}\right), \nonumber\\
&&|\psi'_2\rangle= { \frac{1}{\sqrt{2}}(\ket{01}+\ket{10})}.
\end{eqnarray}
Here we have denoted $\nu_1=\nu'_2$ and $\nu_2=\nu'_1$. In the new state $\rho'$, one may observe that the roles of $|\psi_1\rangle$ and $|\psi_2\rangle$ have been swapped, and the global phase factor ${e^{{\rm i}\zeta}}$ has no contribution to $\rho'$. Obviously, the new state $\rho'$ belongs to the original state $\rho$ for the case of $\theta' \neq \pi/4$ ( i.e., $\ket{\psi'_1}$ is not a maximally entangled state). This fact implies that we need not consider the state $\rho$ separately when $\theta=\pi/4$ and $|\psi_2\rangle$ is non-maximal entangled, thus can simplify some calculations. $\blacksquare$
\end{remark}

\begin{remark}\textcolor{blue}{Proof of Eq. (\ref{eq:proof1}).} Here we would like to provide a proof for Eq. (\ref{eq:proof1}).
The task is to show the explicit form of the unitary transformation $(\mathcal{V}_a \otimes \mathcal{V}_b)$, such that
\begin{eqnarray}\label{eq:unitary-1}
(\mathcal{V}_a \otimes \mathcal{V}_b)\, \ket{\psi_1} \mapsto |\psi'_2\rangle
= {\frac{1}{\sqrt{2}}(\ket{01}+\ket{10})},
\end{eqnarray}
\begin{eqnarray}
{(\mathcal{V}_a \otimes \mathcal{V}_b)}\, \ket{\psi_2} \mapsto |\psi'_1\rangle = {e^{{\rm i}\zeta}}\left( \cos\theta' \ket{00}+\sin\theta'\ket{11}\right),
\end{eqnarray}
with
\begin{eqnarray}
&& {\ket{\psi_1} =\frac{1}{\sqrt{2}}(\ket{00}+\ket{11}),}  \nonumber\\
&& \ket{\psi_2}= \cos\phi(\cos \alpha \ket{01} +\sin\alpha\ket{10})+ e^{{\rm i}\beta}\,\frac{\sin\phi}{\sqrt{2}}( \ket{00}-\ket{11}).
\end{eqnarray}

As it is well-known that the most general $2\times 2$ unitary matrix is given by
\begin{eqnarray}
&& \mathcal{V} =
\left(
                 \begin{array}{cc}
                  e^{{\rm i}\omega_1} & 0 \\
                   0 & e^{{\rm i}\omega_2} \\
                 \end{array}
               \right) \left(
                 \begin{array}{cc}
                   \cos\vartheta & -\sin\vartheta e^{-{\rm i}\varphi} \\
                   \sin\vartheta e^{{\rm i}\varphi} &  \cos\vartheta \\
                 \end{array}
               \right),
\end{eqnarray}
where
\begin{eqnarray}
&& \Omega =
\left(
                 \begin{array}{cc}
                  e^{{\rm i}\omega_1} & 0 \\
                   0 & e^{{\rm i}\omega_2} \\
                 \end{array}
               \right)
\end{eqnarray}
is the usual phase gate transformation, and
\begin{eqnarray}
&& \mathcal{U}= \left(
                 \begin{array}{cc}
                   \cos\vartheta & -\sin\vartheta e^{-{\rm i}\varphi} \\
                   \sin\vartheta e^{{\rm i}\varphi} &  \cos\vartheta \\
                 \end{array}
               \right)
\end{eqnarray}
is the usual SU(2) unitary transformation. Now we let Alice's unitary matrix be
\begin{eqnarray}
&& \mathcal{V}_a = \Omega_1 \mathcal{U}_1=
\left(
                 \begin{array}{cc}
                  e^{{\rm i}\omega_0} & 0 \\
                   0 & e^{{\rm i}\omega_1} \\
                 \end{array}
               \right) \left(
                 \begin{array}{cc}
                   \cos\vartheta_0 & -\sin\vartheta_0 e^{-{\rm i}\varphi_0} \\
                   \sin\vartheta_0 e^{{\rm i}\varphi_0} &  \cos\vartheta_0 \\
                 \end{array}
               \right),
\end{eqnarray}
and Bob's unitary matrix be
\begin{eqnarray}
&& \mathcal{V}_b = \Omega_2 \mathcal{U}_2=
\left(
                 \begin{array}{cc}
                  e^{{\rm i}\omega_2} & 0 \\
                   0 & e^{{\rm i}\omega_3} \\
                 \end{array}
               \right) \left(
                 \begin{array}{cc}
                   \cos\vartheta_1 & -\sin\vartheta_1 e^{-{\rm i}\varphi_1} \\
                   \sin\vartheta_1 e^{{\rm i}\varphi_1} &  \cos\vartheta_1 \\
                 \end{array}
               \right).
\end{eqnarray}
For simplicity, we denote
\begin{eqnarray}
&& \mathcal{U}_1 =\left(
                 \begin{array}{cc}
                   A_0 & -B_0^* \\
                   B_0 & A_0^* \\
                 \end{array}
               \right), \;\;\;\;\;\;
               \mathcal{U}_2 =\left(
                 \begin{array}{cc}
                   A_1 & -B_1^* \\
                   B_1 & A_1^* \\
                 \end{array}
               \right),
\end{eqnarray}
with
\begin{eqnarray}
&& A_0=\cos\vartheta_0, \;\;\;\; B_0=\sin\vartheta_0 e^{{\rm i}\varphi_0},  \nonumber\\
&& A_1=\cos\vartheta_1, \;\;\;\; B_1=\sin\vartheta_1 e^{{\rm i}\varphi_1}.
\end{eqnarray}
The unitary matrix $\mathcal{U}_1$ transforms Alice's qubit states as
\begin{eqnarray}
&& \ket{0} \mapsto A_0 \ket{0} +B_0 \ket{1}, \nonumber\\
&&\ket{1} \mapsto -B_0^* \ket{0} +A_0^* \ket{1},
\end{eqnarray}
and the unitary matrix $\mathcal{U}_2$ transforms Bob's qubit states as
\begin{eqnarray}
&&\ket{0} \mapsto A_1 \ket{0} +B_1 \ket{1}, \nonumber\\
&&\ket{1} \mapsto -B_1^* \ket{0} +A_1^* \ket{1}.
\end{eqnarray}
For example,
\begin{eqnarray}
&& \mathcal{U}_1 |0\rangle=\mathcal{U}_1\left(
                                          \begin{array}{c}
                                            1 \\
                                            0 \\
                                          \end{array}
                                        \right)
=\left( \begin{array}{cc}
                   A_0 & -B_0^* \\
                   B_0 & A_0^* \\
                 \end{array}
               \right) \left( \begin{array}{c}
                                            1 \\
                                            0 \\
                                          \end{array}
                                        \right)=A_0 \left(
                                          \begin{array}{c}
                                            1 \\
                                            0 \\
                                          \end{array}
                                        \right)+B_0\left(
                                          \begin{array}{c}
                                            0 \\
                                            1 \\
                                          \end{array}
                                        \right),\nonumber\\
&& {\mathcal{U}_1 |1\rangle=\mathcal{U}_1\left(
                                          \begin{array}{c}
                                            0 \\
                                            1 \\
                                          \end{array}
                                        \right)
=\left( \begin{array}{cc}
                   A_0 & -B_0^* \\
                   B_0 & A_0^* \\
                 \end{array}
               \right) \left( \begin{array}{c}
                                            0 \\
                                            1 \\
                                          \end{array}
                                        \right)=-B_0^* \left(
                                          \begin{array}{c}
                                            1 \\
                                            0 \\
                                          \end{array}
                                        \right)+A_0^*\left(
                                          \begin{array}{c}
                                            0 \\
                                            1 \\
                                          \end{array}
                                        \right),}
\end{eqnarray}
which correspond to $\ket{0} \mapsto A_0 \ket{0} +B_0 \ket{1}$ and $\ket{1} \mapsto -B_0^* \ket{0} +A_0^* \ket{1}$, respectively.

Now, let us come to determine the explicit form of the unitary transformation $(\mathcal{V}_a \otimes \mathcal{V}_b)$. From Eq. (\ref{eq:unitary-1}), one can have
\begin{eqnarray}
{(\Omega_1 \otimes \Omega_2)}\,(\mathcal{U}_1 \otimes \mathcal{U}_2)\, \frac{1}{\sqrt{2}}(\ket{00}+\ket{11}) ={\frac{1}{\sqrt{2}}(\ket{01}+\ket{10})},
\end{eqnarray}
i.e.,
\begin{eqnarray}
{(\Omega_1 \otimes \Omega_2)}(\mathcal{U}_1 \otimes \mathcal{U}_2)\,(\ket{00}+\ket{11}) ={\ket{01}+\ket{10}},
\end{eqnarray}
i.e.,
\begin{eqnarray}
{(\Omega_1 \otimes \Omega_2)} \Big[(A_0 \ket{0} +B_0 \ket{1})(A_1 \ket{0} +B_1 \ket{1})+(-B_0^* \ket{0} +A_0^* \ket{1})(-B_1^* \ket{0} +A_1^* \ket{1})\Big] ={\ket{01}+\ket{10}},
\end{eqnarray}
i.e.,
\begin{eqnarray}
&& {(\Omega_1 \otimes \Omega_2)} \Big[(A_0A_1+B_0^* B_1^* ) \ket{0}\ket{0}+ (B_0 B_1+A_0^*A_1^*) \ket{1}\ket{1} +(A_0B_1-B_0^*A_1^*) \ket{0}\ket{1}+(B_0A_1-A_0^*B_1^*) \ket{1}\ket{0}\Big] \nonumber\\
& =&{\ket{01}+\ket{10}},
\end{eqnarray}
i.e.,
\begin{eqnarray}
&& {(\Omega_1 \otimes \Omega_2)} \Big[(A_0A_1+B_0^* B_1^* ) \ket{0}\ket{0}+ (A_0A_1+B_0^* B_1^*)^* \ket{1}\ket{1} +(A_0B_1-B_0^*A_1^*) \ket{0}\ket{1}-(A_0B_1-B_0^*A_1^*)^* \ket{1}\ket{0} \Big]\nonumber\\
& =&{\ket{01}-\ket{10}},
\end{eqnarray}
i.e.,
\begin{eqnarray}
&& {e^{{\rm i} (\omega_0+\omega_2)}}(A_0A_1+B_0^* B_1^* ) \ket{0}\ket{0}
+ {e^{{\rm i} (\omega_1+\omega_3)}}(A_0A_1+B_0^* B_1^*)^* \ket{1}\ket{1}
 +{e^{{\rm i} (\omega_0+\omega_3)}}(A_0B_1-B_0^*A_1^*) \ket{0}\ket{1} \nonumber\\
&&-{e^{{\rm i} (\omega_1+\omega_2)}}(A_0B_1-B_0^*A_1^*)^* \ket{1}\ket{0} ={\ket{01}+\ket{10}}.
\end{eqnarray}
Then we have
\begin{eqnarray}
&& {e^{{\rm i} (\omega_0+\omega_2)}}(A_0A_1+B_0^* B_1^* ) =0, \nonumber\\
&& {e^{{\rm i} (\omega_1+\omega_3)}}(A_0A_1+B_0^* B_1^*)^*  =0, \nonumber\\
&& {e^{{\rm i} (\omega_0+\omega_3)}}(A_0B_1-B_0^*A_1^*) =1, \nonumber\\
&& -{e^{{\rm i} (\omega_1+\omega_2)}}(A_0B_1-B_0^*A_1^*)^* =1,
\end{eqnarray}
i.e.,
\begin{eqnarray}
&& {e^{{\rm i} (\omega_0+\omega_2)}}(A_0A_1+B_0^* B_1^* ) =0, \nonumber\\
&& {e^{-{\rm i} (\omega_1+\omega_3)}}(A_0A_1+B_0^* B_1^*)  =0, \nonumber\\
&& {e^{{\rm i} (\omega_0+\omega_3)}}(A_0B_1-B_0^*A_1^*) =1, \nonumber\\
&& -{e^{-{\rm i} (\omega_1+\omega_2)}}(A_0B_1-B_0^*A_1^*) =1,
\end{eqnarray}
i.e.,
\begin{eqnarray}
&& A_0A_1+B_0^* B_1^*  =0, \nonumber\\
&& {e^{{\rm i} (\omega_0+\omega_3)}}(A_0B_1-B_0^*A_1^*) =1, \nonumber\\
&& -{e^{-{\rm i} (\omega_1+\omega_2)}}(A_0B_1-B_0^*A_1^*) =1, \
\end{eqnarray}
i.e.,
\begin{eqnarray}
&& \cos\vartheta_0\cos\vartheta_1  +\sin\vartheta_0\sin\vartheta_1 e^{-{\rm i}(\varphi_0+\varphi_1)} =0, \nonumber\\
&& {e^{{\rm i} (\omega_0+\omega_3)}}(\cos\vartheta_0\sin\vartheta_1 e^{{\rm i}\varphi_1} -\sin\vartheta_0\cos\vartheta_1 e^{-{\rm i}\varphi_0}) =1,\nonumber\\
&& -{e^{-{\rm i} (\omega_1+\omega_2)}}(\cos\vartheta_0\sin\vartheta_1 e^{{\rm i}\varphi_1} -\sin\vartheta_0\cos\vartheta_1 e^{-{\rm i}\varphi_0}) =1,
\end{eqnarray}
i.e.,
\begin{eqnarray}
&& \cos\vartheta_0\cos\vartheta_1  +\sin\vartheta_0\sin\vartheta_1 e^{-{\rm i}(\varphi_0+\varphi_1)} =0, \nonumber\\
&& {e^{{\rm i} (\omega_0+\omega_3)}} e^{{\rm i}\varphi_1}(\cos\vartheta_0\sin\vartheta_1  -\sin\vartheta_0\cos\vartheta_1 e^{-{\rm i}(\varphi_0+\varphi_1)}) =1,\nonumber\\
&& -{e^{-{\rm i} (\omega_1+\omega_2)}} e^{{\rm i}\varphi_1}(\cos\vartheta_0\sin\vartheta_1  -\sin\vartheta_0\cos\vartheta_1 e^{-{\rm i}(\varphi_0+\varphi_1)}) =1,
\end{eqnarray}
i.e.,
\begin{eqnarray}
&& \cos\vartheta_0\cos\vartheta_1  +\sin\vartheta_0\sin\vartheta_1 e^{-{\rm i}(\varphi_0+\varphi_1)} =0, \nonumber\\
&& {e^{{\rm i} (\omega_0+\omega_3)}} e^{{\rm i}\varphi_1}(-\cos\vartheta_0\sin\vartheta_1  +\sin\vartheta_0\cos\vartheta_1 e^{-{\rm i}(\varphi_0+\varphi_1)}) =-1,\nonumber\\
&& -{e^{-{\rm i} (\omega_1+\omega_2)}} e^{{\rm i}\varphi_1}(-\cos\vartheta_0\sin\vartheta_1  +\sin\vartheta_0\cos\vartheta_1 e^{-{\rm i}(\varphi_0+\varphi_1)}) =-1,
\end{eqnarray}
Because the right-hand-side of above equations are real numbers, then we must have
\begin{eqnarray}
&& e^{-{\rm i}(\varphi_0+\varphi_1)} =1, \nonumber\\
&& {e^{{\rm i} (\omega_0+\omega_3)}} e^{{\rm i}\varphi_1}=-1,\nonumber\\
&& -{e^{-{\rm i} (\omega_1+\omega_2)}} e^{{\rm i}\varphi_1}=-1,
\end{eqnarray}
i.e.,
\begin{eqnarray}
&& e^{-{\rm i}(\varphi_0+\varphi_1)} =1, \nonumber\\
&& {e^{{\rm i} (\omega_0+\omega_3)}} e^{{\rm i}\varphi_1}=-1,\nonumber\\
&& {e^{-{\rm i} (\omega_1+\omega_2)}} e^{{\rm i}\varphi_1}=1.
\end{eqnarray}
One can have
\begin{eqnarray}
&& \varphi_0+\varphi_1 =0, \nonumber\\
&& (\omega_0+\omega_3)+\varphi_1 ={\pi},\nonumber\\
&& - (\omega_1+\omega_2)+\varphi_1 =0,
\end{eqnarray}
i.e.,
\begin{eqnarray}
&& \varphi_1=-\varphi_0, \nonumber\\
&& \omega_3={\pi}-\omega_0-\varphi_1 ={\pi} -\omega_0+\varphi_0,\nonumber\\
&& \omega_2=-\omega_1+\varphi_1 = -\omega_1-\varphi_0.
\end{eqnarray}
The above result yields
\begin{eqnarray}
&& \cos\vartheta_0 \cos\vartheta_1  + \sin\vartheta_0\sin\vartheta_1  = \cos(\vartheta_0-\vartheta_1)=0, \nonumber\\
&& \sin\vartheta_0\cos\vartheta_1-\cos\vartheta_0\sin\vartheta_1  =\sin(\vartheta_0-\vartheta_1)=1,
\end{eqnarray}
i.e.,
\begin{eqnarray}
&& \vartheta_1=\vartheta_0-\frac{\pi}{2}.
\end{eqnarray}
In this situation, we have
\begin{eqnarray}
&& A_0=\cos\vartheta, \;\;\;\; B_0=\sin\vartheta e^{{\rm i}\varphi_0},  \nonumber\\
&& A_1=\sin\vartheta, \;\;\;\; B_1=-\cos\vartheta e^{-{\rm i}\varphi_0},
\end{eqnarray}
here we have denoted $\vartheta_0=\vartheta$.

From Sec. \ref{eq:secIA}, we have known that under the unitary transformation $(\mathcal{U}_1 \otimes \mathcal{U}_2)$, the initial two-qubit pure state
\begin{eqnarray}\label{eq:psi-a}
\ket{\psi}_{\rm initial} = \bar{a} \ket{01} + \bar{b}\ket{10}+\bar{c} \ket{00}+\bar{d}\ket{11}
\end{eqnarray}
can be transformed to the following state
\begin{eqnarray}
\ket{\psi}_{\rm middle} = \eta_{00} \ket{00} + \eta_{01} \ket{01} + \eta_{10} \ket{10} + \eta_{11} \ket{11},
\end{eqnarray}
with
\begin{eqnarray}
&&\eta_{00} = A_0 A_1 c - A_0 B_1^* a - A_1 B_0^* b + B_0^* B_1^* d,\nonumber\\
&&\eta_{01} = A_0 B_1 c + A_0 A_1^* a - B_0^* B_1 b - B_0^* A_1^* d,\nonumber\\
&&\eta_{10} = B_0 A_1 c - B_0 B_1^* a + A_0^* A_1 b - A_0^* B_1^* d,\nonumber\\
&&\eta_{11} = B_0 B_1 c + B_0 A_1^* a + A_0^* B_1 b + A_0^* A_1^* d.
\end{eqnarray}
After acting the unitary transformation ${(\Omega_1 \otimes \Omega_2)} $ on $\ket{\psi}_{\rm middle}$, we have the final state as
\begin{eqnarray}
\ket{\psi}_{\rm final} = \gamma_{00} \ket{00} + \gamma_{01} \ket{01} + \gamma_{10} \ket{10} + \gamma_{11} \ket{11},
\end{eqnarray}
with
\begin{eqnarray}
&& \gamma_{00} = {e^{{\rm i} (\omega_0+\omega_2)}} \eta_{00} = {e^{{\rm i} (\omega_0+\omega_2)}}[A_0 A_1 c - A_0 B_1^* a - A_1 B_0^* b + B_0^* B_1^* d],\nonumber\\
&&\gamma_{01} = {e^{{\rm i} (\omega_0+\omega_3)}}\eta_{01} = {e^{{\rm i} (\omega_0+\omega_3)}}[A_0 B_1 c + A_0 A_1^* a - B_0^* B_1 b - B_0^* A_1^* d],\nonumber\\
&&\gamma_{10} = {e^{{\rm i} (\omega_1+\omega_2)}}\eta_{10} = {e^{{\rm i} (\omega_1+\omega_2)}}[B_0 A_1 c - B_0 B_1^* a + A_0^* A_1 b - A_0^* B_1^* d],\nonumber\\
&&\gamma_{11} = {e^{{\rm i} (\omega_1+\omega_3)}}\eta_{11} = {e^{{\rm i} (\omega_1+\omega_3)}}[B_0 B_1 c + B_0 A_1^* a + A_0^* B_1 b + A_0^* A_1^* d].
\end{eqnarray}
We take $|\psi_2\rangle$ as the initial state, then the corresponding coefficients are
\begin{eqnarray}
&& \bar{a}=\cos\phi\cos \alpha, \;\;\;\; \bar{b}=\cos\phi\sin\alpha, \nonumber\\
&&  \bar{c}= e^{{\rm i}\beta}\frac{\sqrt{2}}{2}\sin\phi,\;\;\;\; \bar{d}=- e^{{\rm i}\beta}\frac{\sqrt{2}}{2}\sin\phi.
\end{eqnarray}
We now require
\begin{eqnarray}\label{eq;beta-1}
&&\gamma_{01} = {e^{{\rm i} (\omega_0+\omega_3)}}\eta_{01} =
{e^{{\rm i} (\omega_0+\omega_3)}}[A_0 B_1 c + A_0 A_1^* a - B_0^* B_1 b - B_0^* A_1^* d]=0,\nonumber\\
&&\gamma_{10} = {e^{{\rm i} (\omega_1+\omega_2)}}\eta_{10} = {e^{{\rm i} (\omega_1+\omega_2)}}[B_0 A_1 c - B_0 B_1^* a + A_0^* A_1 b - A_0^* B_1^* d]=0,
\end{eqnarray}
and
\begin{eqnarray}\label{eq;beta-2}
&& \gamma_{00} = {e^{{\rm i} (\omega_0+\omega_2)}} \eta_{00} = {e^{{\rm i} (\omega_0+\omega_2)}}[A_0 A_1 c - A_0 B_1^* a - A_1 B_0^* b + B_0^* B_1^* d]={e^{{\rm i}\zeta}} \cos\theta',\nonumber\\
&&\gamma_{11} = {e^{{\rm i} (\omega_1+\omega_3)}}\eta_{11} = {e^{{\rm i} (\omega_1+\omega_3)}}[B_0 B_1 c + B_0 A_1^* a + A_0^* B_1 b + A_0^* A_1^* d]={e^{{\rm i}\zeta}} \sin\theta',
\end{eqnarray}
such that the resultant state is just $|\psi'_1\rangle$, i.e.,
\begin{eqnarray}
(\mathcal{V}_a \otimes \mathcal{V}_b)\, \ket{\psi_2} \mapsto |\psi'_1\rangle = {e^{{\rm i}\zeta}} \left(\cos\theta' \ket{00}+\sin\theta'\ket{11}\right).
\end{eqnarray}


Based on Eq. (\ref{eq;beta-1}), one obtains
\begin{eqnarray}
\gamma_{01} &=&{e^{{\rm i} (\omega_0+\omega_3)}}[A_0 B_1 c + A_0 A_1^* a - B_0^* B_1 b - B_0^* A_1^* d]\nonumber\\
&=& {e^{{\rm i} (\omega_0+\omega_3)}}\left\{(-\cos^2\vartheta+\sin^2\vartheta) e^{-{\rm i}\varphi_0}\, e^{{\rm i} \beta}\frac{\sqrt{2}}{2}\sin\phi  +\sin\vartheta\cos\vartheta \left[\cos\phi\cos \alpha + e^{-{\rm i}2\varphi_0}\cos\phi\sin\alpha\right]\right\}\nonumber\\
&=& {e^{{\rm i} (\omega_0+\omega_3)}} e^{-{\rm i}\varphi_0} \left\{(-\cos^2\vartheta+\sin^2\vartheta) \, e^{{\rm i}\beta}\frac{\sqrt{2}}{2}\sin\phi  +\sin\vartheta\cos\vartheta \left[\cos\phi\cos \alpha e^{{\rm i}\varphi_0} + e^{-{\rm i}\varphi_0}\cos\phi\sin\alpha\right]\right\}\nonumber\\
&=& (-\cos^2\vartheta+\sin^2\vartheta) \, e^{{\rm i}\beta}\frac{\sqrt{2}}{2}\sin\phi  +\sin\vartheta\cos\vartheta \left[\cos\phi\cos \alpha e^{{\rm i}\varphi_0} + e^{-{\rm i}\varphi_0}\cos\phi\sin\alpha\right]\nonumber\\
&=& (\sin^2\vartheta-\cos^2\vartheta) \, e^{{\rm i}\beta}\frac{\sqrt{2}}{2}\sin\phi  +\sin\vartheta\cos\vartheta \left[\cos\phi\cos \alpha e^{{\rm i}\varphi_0} + \cos\phi\sin\alpha e^{-{\rm i}\varphi_0}\right]=0,\nonumber\\
\gamma_{10} &=&  {e^{{\rm i} (\omega_1+\omega_2)}}[B_0 A_1 c - B_0 B_1^* a + A_0^* A_1 b - A_0^* B_1^* d] \nonumber\\
&=& {e^{{\rm i} (\omega_1+\omega_2)}} \left\{(\sin^2\vartheta- \cos^2\vartheta)e^{{\rm i}\varphi_0} \, e^{{\rm i}\beta}\frac{\sqrt{2}}{2}\sin\phi
+\sin\vartheta\cos\vartheta \left[e^{{\rm i}2\varphi_0}\cos\phi\cos \alpha
+\cos\phi\sin\alpha\right]\right\}\nonumber\\
&=& {e^{{\rm i} (\omega_1+\omega_2)}} e^{{\rm i}\varphi_0} \left\{(\sin^2\vartheta- \cos^2\vartheta) \, e^{{\rm i}\beta}\frac{\sqrt{2}}{2}\sin\phi
+\sin\vartheta\cos\vartheta \left[e^{{\rm i}\varphi_0}\cos\phi\cos \alpha
+\cos\phi\sin\alpha e^{-{\rm i}\varphi_0}\right]\right\}\nonumber\\
&=& (\sin^2\vartheta- \cos^2\vartheta) \, e^{{\rm i}\beta}\frac{\sqrt{2}}{2}\sin\phi
+\sin\vartheta\cos\vartheta \left[\cos\phi\cos \alpha e^{{\rm i}\varphi_0}
+\cos\phi\sin\alpha e^{-{\rm i}\varphi_0}\right]=0,
\end{eqnarray}
i.e.,
\begin{eqnarray}\label{eq:beta-3}
&& (\sin^2\vartheta- \cos^2\vartheta) \, e^{{\rm i}\beta}\frac{\sqrt{2}}{2}\sin\phi
+\sin\vartheta\cos\vartheta \left[\cos\phi\cos \alpha e^{{\rm i}\varphi_0}
+\cos\phi\sin\alpha e^{-{\rm i}\varphi_0}\right]=0,
\end{eqnarray}
i.e.,
\begin{eqnarray}
&& (\cos^2\vartheta- \sin^2\vartheta) \, e^{{\rm i}\beta}\frac{\sqrt{2}}{2}\sin\phi
=\sin\vartheta\cos\vartheta \left[\cos\phi\cos \alpha e^{{\rm i}\varphi_0}
+\cos\phi\sin\alpha e^{-{\rm i}\varphi_0}\right],
\end{eqnarray}
i.e.,
\begin{eqnarray}
&& (\cos^2\vartheta-\sin^2\vartheta)\, e^{{\rm i}\beta}\frac{\sqrt{2}}{2}\sin\phi=\sin\vartheta\cos\vartheta \cos\phi
\left[(\cos\varphi_0+{\rm i}\sin\varphi_0)\cos \alpha
+(\cos\varphi_0-{\rm i}\sin\varphi_0)\sin\alpha\right],
\end{eqnarray}
i.e.,
\begin{eqnarray}
&& (\cos^2\vartheta-\sin^2\vartheta)\, e^{{\rm i}\beta}\frac{\sqrt{2}}{2}\sin\phi
=\sin\vartheta\cos\vartheta \cos\phi \left[\cos\varphi_0(\cos \alpha+\sin \alpha)+{\rm i}\sin\varphi_0(\cos \alpha-\sin\alpha)\right],
\end{eqnarray}
i.e.,
\begin{eqnarray}
&& (\cos^2\vartheta-\sin^2\vartheta)\, \frac{\sqrt{2}}{2}\sin\phi \, e^{{\rm i}\beta}\nonumber\\
&& =\sin\vartheta\cos\vartheta \cos\phi \sqrt{\cos^2\varphi_0(\cos \alpha+\sin \alpha)^2+\sin^2\varphi_0(\cos \alpha-\sin\alpha)^2}\,  e^{{\rm i} \arctan\left[\frac{\sin\varphi_0(\cos \alpha-\sin\alpha)}{\cos\varphi_0(\cos \alpha+\sin \alpha)}\right]},
\end{eqnarray}
i.e.,
\begin{eqnarray}
&& (\cos^2\vartheta-\sin^2\vartheta)\, \frac{\sqrt{2}}{2}\sin\phi \, e^{{\rm i}\beta} =\sin\vartheta\cos\vartheta \cos\phi \sqrt{\cos^2\varphi_0(1+\sin2 \alpha)+\sin^2\varphi_0 (1-\sin2\alpha)}\,  e^{{\rm i} \arctan\left[\frac{\sin\varphi_0(\cos \alpha-\sin\alpha)}{\cos\varphi_0(\cos \alpha+\sin \alpha)}\right]},\nonumber\\
\end{eqnarray}
i.e.,
\begin{eqnarray}
&& (\cos^2\vartheta-\sin^2\vartheta)\, \frac{\sqrt{2}}{2}\sin\phi \, e^{{\rm i}\beta} =\sin\vartheta\cos\vartheta \cos\phi \sqrt{1+\cos2\varphi_0 \sin2 \alpha}\,  e^{{\rm i} \arctan\left[\frac{\sin\varphi_0(\cos \alpha-\sin\alpha)}{\cos\varphi_0(\cos \alpha+\sin \alpha)}\right]},
\end{eqnarray}
i.e.,
\begin{eqnarray}
&& \cos2\vartheta\, \sqrt{2}\sin\phi \, e^{{\rm i}\beta} =\sin2\vartheta \cos\phi \sqrt{1+\cos2\varphi_0\sin2 \alpha}\,  e^{{\rm i} \arctan\left[\frac{\sin\varphi_0(\cos \alpha-\sin\alpha)}{\cos\varphi_0(\cos \alpha+\sin \alpha)}\right]}.
\end{eqnarray}
Then one obtains
\begin{eqnarray}\label{eq:beta-1}
&& \beta= \arctan \left[\frac{\sin\varphi_0(\cos \alpha-\sin\alpha)}{\cos\varphi_0(\cos \alpha+\sin \alpha)}\right],
\end{eqnarray}
and
\begin{eqnarray}\label{eq:beta-2}
&& \cos2\vartheta\, \sqrt{2}\sin\phi  = \sin2\vartheta \cos\phi \sqrt{1+\cos2\varphi_0\,\sin2 \alpha}.
\end{eqnarray}
From Eq. (\ref{eq:beta-1}), we have
\begin{eqnarray}
&& \tan\varphi_0= \frac{\sin\beta(\cos \alpha+\sin\alpha)}{\cos\beta(\cos \alpha-\sin \alpha)},
\end{eqnarray}
and
\begin{eqnarray}
\cos2\varphi_0&=&\frac{1-\tan^ 2\varphi_0}{1+\tan^ 2\varphi_0}=\frac{\cos^2\beta(\cos \alpha-\sin \alpha)^2-\sin^2\beta(\cos \alpha+\sin \alpha)^2}
{\cos^2\beta(\cos \alpha-\sin \alpha)^2-\sin^2\beta(\cos \alpha+\sin \alpha)^2}\nonumber\\
&=&\frac{\cos^2\beta(1-\sin2 \alpha)-\sin^2\beta(1+\sin2 \alpha)}
{\cos^2\beta(1-\sin2 \alpha)+\sin^2\beta(1+\sin2 \alpha)}={ \frac{\cos2\beta-\sin2 \alpha}
{1-\cos2\beta\sin2 \alpha},}
\end{eqnarray}
and
\begin{eqnarray}
1+\cos2\varphi_0\sin2 \alpha &= & 1+{ \frac{\cos2\beta-\sin2 \alpha}
{1-\cos2\beta\sin2 \alpha}} \times \sin2 \alpha\nonumber\\
&=&\frac{(1-\cos2\beta\sin2 \alpha)+(\cos2\beta-\sin2 \alpha)\sin2 \alpha}
{1-\cos2\beta\sin2 \alpha}\nonumber\\
&= & \frac{1-\sin^2 2 \alpha}
{1-\cos2\beta\sin2 \alpha}\nonumber\\
&=&\frac{\cos^2 2 \alpha}
{1-\cos2\beta\sin2 \alpha}.
\end{eqnarray}
From Eq. (\ref{eq:beta-2}), one has
\begin{eqnarray}
\tan2\vartheta &=& \frac{\sqrt{2}\sin\phi}{ \cos\phi \sqrt{1+\cos2\varphi_0\,\sin2 \alpha}}= \sqrt{\frac{1-\cos2\beta\sin2 \alpha}{\cos^2 2 \alpha}} \sqrt{2}\tan\phi.
\end{eqnarray}

From Eq. (\ref{eq;beta-2}), we can have
\begin{eqnarray}
\gamma_{00}& =& {e^{{\rm i} (\omega_0+\omega_2)}}[A_0 A_1 c - A_0 B_1^* a - A_1 B_0^* b + B_0^* B_1^* d]\nonumber\\
&=& {e^{{\rm i} (\omega_0+\omega_2)}} \Big[\sin\vartheta\cos\vartheta(1+1)e^{{\rm i}\beta}\frac{\sqrt{2}}{2}\sin\phi +(\cos^2\vartheta e^{{\rm i}\varphi_0} \cos\phi\cos \alpha-\sin^2\vartheta e^{-{\rm i}\varphi_0}\cos\phi\sin\alpha) \Big]\nonumber\\
&=& {e^{{\rm i} (\omega_0+\omega_2)}} \Big[\sin\vartheta\cos\vartheta e^{{\rm i}\beta}\sqrt{2}\sin\phi +(\cos^2\vartheta e^{{\rm i}\varphi_0} \cos\phi\cos \alpha-\sin^2\vartheta e^{-{\rm i}\varphi_0}\cos\phi\sin\alpha) \Big]={e^{{\rm i}\zeta}} \cos\theta',\nonumber\\
\gamma_{11} &=& {e^{{\rm i} (\omega_1+\omega_3)}}[B_0 B_1 c + B_0 A_1^* a + A_0^* B_1 b + A_0^* A_1^* d] \nonumber\\
&=& {e^{{\rm i} (\omega_1+\omega_3)}}
\Big[\sin\vartheta\cos\vartheta (-1-1)e^{{\rm i}\beta}\frac{\sqrt{2}}{2}\sin\phi+
(\sin^2\vartheta  e^{{\rm i}\varphi_0}\cos\phi\cos \alpha-\cos^2 \vartheta e^{-{\rm i}\varphi_0}\cos\phi\sin\alpha  )\Big]\nonumber\\
&=& {e^{{\rm i} (\omega_1+\omega_3)}}
\Big[-\sin\vartheta\cos\vartheta e^{{\rm i}\beta}\sqrt{2}\sin\phi+
(\sin^2\vartheta  e^{{\rm i}\varphi_0}\cos\phi\cos \alpha-\cos^2 \vartheta e^{-{\rm i}\varphi_0}\cos\phi\sin\alpha  )\Big]
={e^{{\rm i}\zeta}} \sin\theta',
\end{eqnarray}
i.e.,
\begin{eqnarray}
&& {e^{{\rm i} (\omega_0+\omega_2)}} \Big[\sin\vartheta\cos\vartheta e^{{\rm i}\beta}\sqrt{2}\sin\phi +(\cos^2\vartheta e^{{\rm i}\varphi_0} \cos\phi\cos \alpha-\sin^2\vartheta e^{-{\rm i}\varphi_0}\cos\phi\sin\alpha) \Big]={e^{{\rm i}\zeta}}\cos\theta', \nonumber\\
&& {e^{{\rm i} (\omega_1+\omega_3)}}
\Big[-\sin\vartheta\cos\vartheta e^{{\rm i}\beta}\sqrt{2}\sin\phi+
(\sin^2\vartheta  e^{{\rm i}\varphi_0}\cos\phi\cos \alpha-\cos^2 \vartheta e^{-{\rm i}\varphi_0}\cos\phi\sin\alpha  )\Big]
={e^{{\rm i}\zeta}}\sin\theta'.
\end{eqnarray}
From Eq. (\ref{eq:beta-3}), we have
\begin{eqnarray}
&& \, e^{{\rm i}\beta}\sqrt{2}\sin\phi=\frac{2\sin\vartheta\cos\vartheta}{\cos^2\vartheta-\sin^2\vartheta} \left[\cos \alpha e^{{\rm i}\varphi_0} +\sin\alpha e^{-{\rm i}\varphi_0}\right] \cos\phi,
\end{eqnarray}
then
\begin{eqnarray}
&& {e^{{\rm i} (\omega_0+\omega_2)}} \Big[\sin\vartheta\cos\vartheta \times \frac{2\sin\vartheta\cos\vartheta\, [\cos \alpha e^{{\rm i}\varphi_0} +\sin\alpha e^{-{\rm i}\varphi_0}]}{\cos^2\vartheta-\sin^2\vartheta}   +(\cos^2\vartheta e^{{\rm i}\varphi_0} \cos \alpha-\sin^2\vartheta e^{-{\rm i}\varphi_0} \sin\alpha) \Big] \cos\phi ={e^{{\rm i}\zeta}}\cos\theta', \nonumber\\
&& {e^{{\rm i} (\omega_1+\omega_3)}}
\Big[-\sin\vartheta\cos\vartheta  \frac{2\sin\vartheta\cos\vartheta\, [\cos \alpha e^{{\rm i}\varphi_0} +\sin\alpha e^{-{\rm i}\varphi_0}]}{\cos^2\vartheta-\sin^2\vartheta} +
(\sin^2\vartheta  e^{{\rm i}\varphi_0}\cos \alpha-\cos^2 \vartheta e^{-{\rm i}\varphi_0}\sin\alpha  )\Big]\cos\phi
={e^{{\rm i}\zeta}}\sin\theta',\nonumber\\
\end{eqnarray}
i.e.,
\begin{eqnarray}
&& {e^{{\rm i} (\omega_0+\omega_2)}}
\Biggr\{\left[\frac{2\sin^2\vartheta\cos^2 \vartheta}{\cos^2\vartheta-\sin^2\vartheta}+\cos^2\vartheta \right]
\cos \alpha e^{{\rm i}\varphi_0} +
 \left[\frac{2\sin^2\vartheta\cos^2 \vartheta}{\cos^2\vartheta-\sin^2\vartheta}-\sin^2\vartheta \right]\sin\alpha e^{-{\rm i}\varphi_0}\Biggr\}  \cos\phi={e^{{\rm i}\zeta}}\cos\theta', \nonumber\\
&& {e^{{\rm i} (\omega_1+\omega_3)}}
\Biggr\{\left[\frac{-2\sin^2\vartheta\cos^2 \vartheta}{\cos^2\vartheta-\sin^2\vartheta}+\sin^2\vartheta \right]
\cos \alpha e^{{\rm i}\varphi_0} +
 \left[\frac{-2\sin^2\vartheta\cos^2 \vartheta}{\cos^2\vartheta-\sin^2\vartheta}-\cos^2\vartheta \right]\sin\alpha e^{-{\rm i}\varphi_0}\Biggr\}  \cos\phi={e^{{\rm i}\zeta}}\sin\theta', \nonumber\\
\end{eqnarray}
i.e.,
\begin{eqnarray}
&& {e^{{\rm i} (\omega_0+\omega_2)}}
\Biggr\{\left[\frac{\sin^2\vartheta\cos^2 \vartheta+\cos^4 \vartheta}{\cos^2\vartheta+\sin^2\vartheta} \right]
\cos \alpha e^{{\rm i}\varphi_0} +
 \left[\frac{\sin^2\vartheta\cos^2 \vartheta+\sin^4 \vartheta}{\cos^2\vartheta-\sin^2\vartheta} \right]\sin\alpha e^{-{\rm i}\varphi_0}\Biggr\}  \cos\phi={e^{{\rm i}\zeta}}\cos\theta', \nonumber\\
&& {e^{{\rm i} (\omega_1+\omega_3)}}
\Biggr\{\left[\frac{-\sin^2\vartheta\cos^2 \vartheta-\sin^4 \vartheta}{\cos^2\vartheta-\sin^2\vartheta}\right]
\cos \alpha e^{{\rm i}\varphi_0} +
 \left[\frac{-\sin^2\vartheta\cos^2 \vartheta-\cos^4 \vartheta}{\cos^2\vartheta-\sin^2\vartheta} \right]\sin\alpha e^{-{\rm i}\varphi_0}\Biggr\}  \cos\phi={e^{{\rm i}\zeta}}\sin\theta',
\end{eqnarray}
i.e.,
\begin{eqnarray}
&& {e^{{\rm i} (\omega_0+\omega_2)}}\Biggr\{\left[\frac{\cos^2 \vartheta}{\cos^2\vartheta-\sin^2\vartheta} \right]
\cos \alpha e^{{\rm i}\varphi_0} +
 \left[\frac{\sin^2\vartheta}{\cos^2\vartheta-\sin^2\vartheta} \right]\sin\alpha e^{-{\rm i}\varphi_0}\Biggr\}  \cos\phi={e^{{\rm i}\zeta}}\cos\theta', \nonumber\\
&& {e^{{\rm i} (\omega_1+\omega_3)}}\Biggr\{\left[\frac{-\sin^2\vartheta}{\cos^2\vartheta-\sin^2\vartheta}\right]
\cos \alpha e^{{\rm i}\varphi_0} +
 \left[\frac{-\cos^2 \vartheta}{\cos^2\vartheta-\sin^2\vartheta} \right]\sin\alpha e^{-{\rm i}\varphi_0}\Biggr\}  \cos\phi={e^{{\rm i}\zeta}}\sin\theta',
\end{eqnarray}
i.e.,
\begin{eqnarray}
&& {e^{{\rm i} (\omega_0+\omega_2)}}\Big[\cos^2 \vartheta \cos \alpha e^{{\rm i}\varphi_0} +
 \sin^2\vartheta\sin\alpha e^{-{\rm i}\varphi_0}\Big] \frac{ \cos\phi}{\cos^2\vartheta-\sin^2\vartheta}={e^{{\rm i}\zeta}}\cos\theta', \nonumber\\
&& -{e^{{\rm i} (\omega_1+\omega_3)}}\Big[\sin^2\vartheta \cos \alpha e^{{\rm i}\varphi_0} +
 \cos^2 \vartheta \sin\alpha e^{-{\rm i}\varphi_0}\Big]  \frac{ \cos\phi}{\cos^2\vartheta-\sin^2\vartheta}={e^{{\rm i}\zeta}}\sin\theta',
\end{eqnarray}
i.e.,
\begin{eqnarray}
&& {e^{{\rm i} (\omega_0+\omega_2)}}\Big[\cos^2 \vartheta \cos \alpha e^{{\rm i}\varphi_0} +
 \sin^2\vartheta\sin\alpha e^{-{\rm i}\varphi_0}\Big] \frac{ \cos\phi}{\cos2\vartheta}={e^{{\rm i}\zeta}}\cos\theta', \nonumber\\
&& -{e^{{\rm i} (\omega_1+\omega_3)}}\Big[\sin^2\vartheta \cos \alpha e^{{\rm i}\varphi_0} +
 \cos^2 \vartheta \sin\alpha e^{-{\rm i}\varphi_0}\Big]  \frac{ \cos\phi}{\cos2\vartheta}={e^{{\rm i}\zeta}}\sin\theta'.
\end{eqnarray}
By considering
\begin{eqnarray}
&& \omega_2= -\omega_1-\varphi_0,\nonumber\\
&& \omega_3 = {\pi} -\omega_0+\varphi_0,
\end{eqnarray}
we have
\begin{eqnarray}
&& {e^{{\rm i} (\omega_0-\omega_1-\varphi_0)}}\Big[\cos^2 \vartheta \cos \alpha e^{{\rm i}\varphi_0} +
 \sin^2\vartheta\sin\alpha e^{-{\rm i}\varphi_0}\Big] \frac{ \cos\phi}{\cos2\vartheta}={e^{{\rm i}\zeta}}\cos\theta', \nonumber\\
&& -{e^{{\rm i} (\omega_1+{\pi}-\omega_0+\varphi_0)}}\Big[\sin^2\vartheta \cos \alpha e^{{\rm i}\varphi_0} +
 \cos^2 \vartheta \sin\alpha e^{-{\rm i}\varphi_0}\Big]  \frac{ \cos\phi}{\cos2\vartheta}={e^{{\rm i}\zeta}}\sin\theta',
\end{eqnarray}
i.e.,
\begin{eqnarray}
&& {e^{{\rm i} (\omega_0-\omega_1)}}\Big[\cos^2 \vartheta \cos \alpha  +
 \sin^2\vartheta\sin\alpha e^{-{\rm i}2\varphi_0}\Big] \frac{ \cos\phi}{\cos2\vartheta}={e^{{\rm i}\zeta}}\cos\theta', \nonumber\\
&& {e^{-{\rm i} (\omega_0-\omega_1)}}\Big[\sin^2\vartheta \cos \alpha e^{{\rm i}2\varphi_0} +
 \cos^2 \vartheta \sin\alpha \Big]  \frac{ \cos\phi}{\cos2\vartheta}={e^{{\rm i}\zeta}}\sin\theta',
\end{eqnarray}
i.e.,
\begin{eqnarray}\label{eq:real-1}
&& {e^{{\rm i} (\omega_0-\omega_1)}}\Big[(\cos^2 \vartheta \cos \alpha  +
 \sin^2\vartheta\sin\alpha \cos2\varphi_0)-{\rm i} \sin^2\vartheta\sin\alpha \sin2\varphi_0\Big] \frac{ \cos\phi}{\cos2\vartheta}={e^{{\rm i}\zeta}}\cos\theta', \nonumber\\
&& {e^{-{\rm i} (\omega_0-\omega_1)}}\Big[(\sin^2\vartheta \cos \alpha \cos2\varphi_0 +
 \cos^2 \vartheta \sin\alpha )+{\rm i}\sin^2\vartheta \cos \alpha \sin2\varphi_0 \Big]  \frac{ \cos\phi}{\cos2\vartheta}={e^{{\rm i}\zeta}}\sin\theta'.
\end{eqnarray}

We now come to determine the phases in the left-hand side of Eq. (\ref{eq:real-1}). We can have
\begin{eqnarray}
&&{e^{{\rm i} (\omega_0-\omega_1)}}\Big[(\cos^2 \vartheta \cos \alpha  +
 \sin^2\vartheta\sin\alpha \cos2\varphi_0)-{\rm i} \sin^2\vartheta\sin\alpha \sin2\varphi_0\Big] \frac{ \cos\phi}{\cos2\vartheta} \nonumber\\
 &=& \frac{ \cos\phi}{\cos2\vartheta} \left|\Big[(\cos^2 \vartheta \cos \alpha  +
 \sin^2\vartheta\sin\alpha \cos2\varphi_0)-{\rm i} \sin^2\vartheta\sin\alpha \sin2\varphi_0\Big]\right| e^{{\rm i}\zeta_1} \nonumber\\
 && {e^{-{\rm i} (\omega_0-\omega_1)}}\Big[(\sin^2\vartheta \cos \alpha \cos2\varphi_0 +
 \cos^2 \vartheta \sin\alpha )+{\rm i}\sin^2\vartheta \cos \alpha \sin2\varphi_0 \Big]  \frac{ \cos\phi}{\cos2\vartheta}\nonumber\\
 &=&  \frac{ \cos\phi}{\cos2\vartheta} \left|\Big[(\sin^2\vartheta \cos \alpha \cos2\varphi_0 +
 \cos^2 \vartheta \sin\alpha )+{\rm i}\sin^2\vartheta \cos \alpha \sin2\varphi_0 \Big]\right|e^{{\rm i}\zeta_2},
\end{eqnarray}
with
\begin{eqnarray}
&& \zeta_1=+(\omega_0-\omega_1)-\arctan \left[\frac{\sin^2\vartheta\sin\alpha \sin2\varphi_0}{\cos^2 \vartheta \cos \alpha  +
 \sin^2\vartheta\sin\alpha \cos2\varphi_0}\right], \nonumber\\
&&\zeta_2=-(\omega_0-\omega_1)+\arctan\left[\frac{\sin^2\vartheta \cos \alpha \sin2\varphi_0 }{\sin^2\vartheta \cos \alpha \cos2\varphi_0 +
 \cos^2 \vartheta \sin\alpha}\right].
\end{eqnarray}
We now require
\begin{eqnarray}
&& \zeta_1=\zeta_2=\zeta,
\end{eqnarray}
then we can have
\begin{eqnarray}
&& +(\omega_0-\omega_1)-\arctan \left[\frac{\sin^2\vartheta\sin\alpha \sin2\varphi_0}{\cos^2 \vartheta \cos \alpha  +
 \sin^2\vartheta\sin\alpha \cos2\varphi_0}\right]\nonumber\\
 &=&-(\omega_0-\omega_1)+\arctan\left[\frac{\sin^2\vartheta \cos \alpha \sin2\varphi_0 }{\sin^2\vartheta \cos \alpha \cos2\varphi_0 +
 \cos^2 \vartheta \sin\alpha}\right],
\end{eqnarray}
i.e.,
\begin{eqnarray}
 2(\omega_0-\omega_1) &=& {\arctan \left[\frac{\sin^2\vartheta\sin\alpha \sin2\varphi_0}{\cos^2 \vartheta \cos \alpha  +
 \sin^2\vartheta\sin\alpha \cos2\varphi_0}\right]+\arctan\left[\frac{\sin^2\vartheta \cos \alpha \sin2\varphi_0 }{\sin^2\vartheta \cos \alpha \cos2\varphi_0 +
 \cos^2 \vartheta \sin\alpha}\right].}
\end{eqnarray}
One may choose $\omega_1=0$, then we have
\begin{eqnarray}
 \omega_0 &=&{\frac{1}{2}\arctan \left[\frac{\sin^2\vartheta\sin\alpha \sin2\varphi_0}{
 \sin^2\vartheta\sin\alpha \cos2\varphi_0+\cos^2 \vartheta \cos \alpha}\right]+\frac{1}{2}\arctan\left[\frac{\sin^2\vartheta \cos \alpha \sin2\varphi_0 }{\sin^2\vartheta \cos \alpha \cos2\varphi_0 +
 \cos^2 \vartheta \sin\alpha}\right].}
\end{eqnarray}

In conclusion, for the rank-2 state $\rho$ as follows
\begin{eqnarray}
\rho=\nu_1  \ket{\psi_1}\bra{\psi_1}+\nu_2 \ket{\psi_2}\bra{\psi_2},
\end{eqnarray}
where
\begin{eqnarray}
&&\ket{\psi_1} =\cos \theta \ket{00}+\sin \theta\ket{11}, \;\;\;\;\; {\theta\in [0, \pi/2],} \nonumber\\
&& \ket{\psi_2} =\cos\phi(\cos \alpha \ket{01} +\sin\alpha\ket{10})+ e^{{\rm i}\beta}\sin\phi(\sin \theta \ket{00}-\cos \theta\ket{11}),
\end{eqnarray}
when $\theta=\pi/4$, we can perform the local unitary transformation $(\mathcal{V}_a \otimes \mathcal{V}_b)$ on the rank-2 state $\rho$, such that
\begin{eqnarray}
&&(\mathcal{V}_a \otimes \mathcal{V}_b)\, \ket{\psi_2} \mapsto |\psi'_1\rangle =  {e^{{\rm i}\zeta}} \left(\cos\theta' \ket{00}+\sin\theta'\ket{11}\right), \nonumber\\
&&(\mathcal{V}_a \otimes \mathcal{V}_b)\, \ket{\psi_1} \mapsto |\psi'_2\rangle
= { \frac{1}{\sqrt{2}}(\ket{01}+\ket{10})}.
\end{eqnarray}
Here the local unitary transformation $(\mathcal{V}_a \otimes \mathcal{V}_b)$ is given by
\begin{eqnarray}
&& \mathcal{V}_a = \Omega_1 \mathcal{U}_1=
\left(
                 \begin{array}{cc}
                  e^{{\rm i}\omega_0} & 0 \\
                   0 & e^{{\rm i}\omega_1} \\
                 \end{array}
               \right) \left(
                 \begin{array}{cc}
                   \cos\vartheta_0 & -\sin\vartheta_0 e^{-{\rm i}\varphi_0} \\
                   \sin\vartheta_0 e^{{\rm i}\varphi_0} &  \cos\vartheta_0 \\
                 \end{array}
               \right),\nonumber\\
&& \mathcal{V}_b = \Omega_2 \mathcal{U}_2=
\left(
                 \begin{array}{cc}
                  e^{{\rm i}\omega_2} & 0 \\
                   0 & e^{{\rm i}\omega_3} \\
                 \end{array}
               \right) \left(
                 \begin{array}{cc}
                   \cos\vartheta_1 & -\sin\vartheta_1 e^{-{\rm i}\varphi_1} \\
                   \sin\vartheta_1 e^{{\rm i}\varphi_1} &  \cos\vartheta_1 \\
                 \end{array}
               \right),
\end{eqnarray}
with
\begin{eqnarray}
&&\vartheta_0=\vartheta,\nonumber\\
&& \vartheta_1=\vartheta_0-\frac{\pi}{2}, \nonumber\\
&& \varphi_1=-\varphi_0, \nonumber\\
 &&\omega_0 ={\frac{1}{2}\arctan \left[\frac{\sin^2\vartheta\sin\alpha \sin2\varphi_0}{\cos^2 \vartheta \cos \alpha  +
 \sin^2\vartheta\sin\alpha \cos2\varphi_0}\right]+\frac{1}{2}\arctan\left[\frac{\sin^2\vartheta \cos \alpha \sin2\varphi_0 }{\sin^2\vartheta \cos \alpha \cos2\varphi_0 +
 \cos^2 \vartheta \sin\alpha}\right],}\nonumber\\
&& \omega_1=0, \nonumber\\
&& \omega_2=-\omega_1+\varphi_1 =-\varphi_0,\nonumber\\
&& \omega_3={\pi}-\omega_0-\varphi_1 = {\pi}-\omega_0+\varphi_0,\nonumber\\
&& \tan\varphi_0= \frac{\sin\beta(\cos \alpha+\sin\alpha)}{\cos\beta(\cos \alpha-\sin \alpha)},\nonumber\\
&& \tan2\vartheta =\sqrt{\frac{1-\cos2\beta\sin2 \alpha}{\cos^2 2 \alpha}} \sqrt{2}\tan\phi, \nonumber\\
&&\zeta =-\frac{1}{2}\arctan \left[\frac{\sin^2\vartheta\sin\alpha \sin2\varphi_0}{ \sin^2\vartheta\sin\alpha \cos2\varphi_0+\cos^2 \vartheta \cos \alpha}\right]+\frac{1}{2}\arctan\left[\frac{\sin^2\vartheta \cos \alpha \sin2\varphi_0 }{\sin^2\vartheta \cos \alpha \cos2\varphi_0 +
 \cos^2 \vartheta \sin\alpha}\right].
\end{eqnarray}
This ends the proof. $\blacksquare$
\end{remark}

\begin{remark}Let us come to check whether Eq. (\ref{eq:real-1}) gives
\begin{eqnarray}
&& \left|{e^{{\rm i} (\omega_0-\omega_1)}}\Big[\cos^2 \vartheta \cos \alpha  +
 \sin^2\vartheta\sin\alpha e^{-{\rm i}2\varphi_0}\Big] \frac{ \cos\phi}{\cos2\vartheta}\right|^2 \nonumber\\
 &&+
\left|{e^{-{\rm i} (\omega_0-\omega_1)}}\Big[\sin^2\vartheta \cos \alpha e^{{\rm i}2\varphi_0} +
 \cos^2 \vartheta \sin\alpha \Big]  \frac{ \cos\phi}{\cos2\vartheta}\right|^2=|{e^{{\rm i}\zeta}}\cos\theta'|^2+|{e^{{\rm i}\zeta}}\sin\theta'|^2=1,
\end{eqnarray}
i.e.,
\begin{eqnarray}
&& \left|\Big[\cos^2 \vartheta \cos \alpha  +
 \sin^2\vartheta\sin\alpha e^{-{\rm i}2\varphi_0}\Big] \frac{ \cos\phi}{\cos2\vartheta}\right|^2 +
\left|\Big[\sin^2\vartheta \cos \alpha e^{{\rm i}2\varphi_0} +
 \cos^2 \vartheta \sin\alpha \Big]  \frac{ \cos\phi}{\cos2\vartheta}\right|^2\nonumber\\
 &=&\cos^2\theta'+\sin^2\theta'=1.
\end{eqnarray}
We can have
\begin{eqnarray}
&& \left|\Big[\cos^2 \vartheta \cos \alpha  +
 \sin^2\vartheta\sin\alpha e^{-{\rm i}2\varphi_0}\Big] \frac{ \cos\phi}{\cos2\vartheta}\right|^2 +
\left|\Big[\sin^2\vartheta \cos \alpha e^{{\rm i}2\varphi_0} +
 \cos^2 \vartheta \sin\alpha \Big]  \frac{ \cos\phi}{\cos2\vartheta}\right|^2\nonumber\\
&=& \frac{ \cos^2\phi}{\cos^2 2\vartheta}\times \Big\{ \left|\cos^2 \vartheta \cos \alpha +
 \sin^2\vartheta\sin\alpha e^{-{\rm i}2\varphi_0} \right|^2 +
\left| \sin^2\vartheta \cos \alpha e^{{\rm i}2\varphi_0} +
 \cos^2 \vartheta \sin\alpha \right|^2\Big\}\nonumber\\
&=& \frac{ \cos^2\phi}{\cos^2 2\vartheta}\times \Big\{ \left[ \cos^4 \vartheta \cos^2 \alpha+\sin^4\vartheta\sin^2\alpha+2\cos^2 \vartheta \cos \alpha \sin^2\vartheta\sin\alpha \cos 2\varphi_0\right] \nonumber\\
&&+\left[ \sin^4\vartheta \cos^2 \alpha +\cos^4 \vartheta \sin^2\alpha+2\sin^2\vartheta \cos \alpha \cos^2 \vartheta \sin\alpha \cos 2\varphi_0 \right] \Big\}\nonumber\\
&=& \frac{ \cos^2\phi}{\cos^2 2\vartheta}\times
 \left[ \cos^4 \vartheta +\sin^4\vartheta+2 \sin^2\vartheta \cos^2 \vartheta \sin2\alpha \cos 2\varphi_0\right] \nonumber\\
&=& \frac{ \cos^2\phi}{\cos^2 2\vartheta}\times
 \left[(\cos^2 \vartheta-\sin^2\vartheta)^2+2 \sin^2\vartheta \cos^2 \vartheta+2 \sin^2\vartheta \cos^2 \vartheta \sin2\alpha \cos 2\varphi_0\right] \nonumber\\
&=& \frac{ \cos^2\phi}{\cos^2 2\vartheta}\times
 \left[\cos^2 2\vartheta+2 \sin^2\vartheta \cos^2 \vartheta \left(\frac{\cos2\vartheta\, \sqrt{2}\sin\phi}{\sin2\vartheta \cos\phi}\right)^2\right] \nonumber\\
&=& \cos^2\phi \times
 \left[1+4 \sin^2\vartheta \cos^2 \vartheta \frac{ \sin^2 \phi}{\sin^2 2\vartheta \cos^2 \phi}\right]\nonumber\\
 &=&\cos^2\phi \times
 \left[1+\frac{ \sin^2 \phi}{\cos^2 \phi}\right]\nonumber\\
 &=& 1,
\end{eqnarray}
where we have used Eq. (\ref{eq:beta-2}). This ends the verification. $\blacksquare$
\end{remark}

\begin{remark}Let us come to check whether
\begin{eqnarray}
&& 2  \left|{e^{{\rm i} (\omega_0-\omega_1)}}\Big[\cos^2 \vartheta \cos \alpha  +
 \sin^2\vartheta\sin\alpha e^{-{\rm i}2\varphi_0}\Big] \frac{ \cos\phi}{\cos2\vartheta}\right|\times
\left|{e^{-{\rm i} (\omega_0-\omega_1)}}\Big[\sin^2\vartheta \cos \alpha e^{{\rm i}2\varphi_0} +
 \cos^2 \vartheta \sin\alpha \Big]  \frac{ \cos\phi}{\cos2\vartheta}\right|\nonumber\\
\end{eqnarray}
gives the following concurrence
\begin{eqnarray}
  \sin 2\theta' &=& \mathcal{C}_{\ket{\psi_2}}= \left|e^{{\rm i} 2\beta}\sin^2 \phi + \cos^2\phi\sin 2 \alpha\right|\nonumber\\
  &=& \sqrt{( \cos^2\phi\sin 2 \alpha+\sin^2 \phi \cos 2\beta)^2+ \sin^4 \phi  \sin^2 2\beta}.
\end{eqnarray}
We can have
\begin{eqnarray}
&& 2  \left|{e^{{\rm i} (\omega_0-\omega_1)}}\Big[\cos^2 \vartheta \cos \alpha  +
 \sin^2\vartheta\sin\alpha e^{-{\rm i}2\varphi_0}\Big] \frac{ \cos\phi}{\cos2\vartheta}\right|\times
\left|{e^{-{\rm i} (\omega_0-\omega_1)}}\Big[\sin^2\vartheta \cos \alpha e^{{\rm i}2\varphi_0} +
 \cos^2 \vartheta \sin\alpha \Big]  \frac{ \cos\phi}{\cos2\vartheta}\right|\nonumber\\
 &=&  2 \frac{ \cos^2\phi}{\cos^2 2\vartheta} \left|\cos^2 \vartheta \cos \alpha  +
 \sin^2\vartheta\sin\alpha e^{-{\rm i}2\varphi_0}\right|\times
\left|\sin^2\vartheta \cos \alpha e^{{\rm i}2\varphi_0} +
 \cos^2 \vartheta \sin\alpha \right|\nonumber\\
&=&   \frac{2 \cos^2\phi}{\cos^2 2\vartheta} \left|(\cos^2 \vartheta \cos \alpha  +
 \sin^2\vartheta\sin\alpha e^{-{\rm i}2\varphi_0})(\sin^2\vartheta \cos \alpha e^{{\rm i}2\varphi_0} +
 \cos^2 \vartheta \sin\alpha) \right|\nonumber\\
&=&   \frac{2 \cos^2\phi}{\cos^2 2\vartheta} \left|(\cos^4 \vartheta +\sin^4 \vartheta )\sin\alpha\cos \alpha+\sin^2\vartheta \cos^2 \vartheta (\sin^2\alpha e^{-{\rm i}2\varphi_0}+\cos ^2\alpha e^{{\rm i}2\varphi_0}) \right|\nonumber\\
&=&   \frac{2 \cos^2\phi}{\cos^2 2\vartheta} \left|(\cos^4 \vartheta +\sin^4 \vartheta )\sin\alpha\cos \alpha+\sin^2\vartheta \cos^2 \vartheta [(\cos^2\alpha+\sin^2\alpha)\cos2\varphi_0+ {\rm i}(\cos ^2\alpha-\sin^2\alpha) \sin 2\varphi_0] \right|\nonumber\\
&=&   \frac{2 \cos^2\phi}{\cos^2 2\vartheta} \left|(\cos^4 \vartheta +\sin^4 \vartheta )\sin\alpha\cos \alpha+\sin^2\vartheta \cos^2 \vartheta (\cos2\varphi_0+ {\rm i}\cos2\alpha \sin 2\varphi_0) \right|\nonumber\\
&=&   \frac{2 \cos^2\phi}{\cos^2 2\vartheta} \left|[(\cos^2 \vartheta -\sin^2 \vartheta )^2+2\sin^2\vartheta \cos^2 \vartheta]\sin\alpha\cos \alpha+\sin^2\vartheta \cos^2 \vartheta (\cos2\varphi_0+ {\rm i} \cos2\alpha \sin 2\varphi_0) \right|\nonumber\\
&=&   \frac{2 \cos^2\phi}{\cos^2 2\vartheta} \left|\cos^2 2 \vartheta \sin\alpha\cos \alpha +2\sin^2\vartheta \cos^2 \vartheta \sin\alpha\cos \alpha+\sin^2\vartheta \cos^2 \vartheta (\cos2\varphi_0+ {\rm i}\cos2\alpha \sin 2\varphi_0) \right|\nonumber\\
&=&   2 \cos^2\phi \left| \sin\alpha\cos \alpha + \frac{\sin^2\vartheta \cos^2 \vartheta}{\cos^2 2\vartheta}[ 2 \sin\alpha\cos \alpha+\cos2\varphi_0+ {\rm i}\cos2\alpha \sin 2\varphi_0] \right|\nonumber\\
&=&    \cos^2\phi \left| 2\sin\alpha\cos \alpha + \frac{2\sin^2\vartheta \cos^2 \vartheta}{\cos^2 2\vartheta}[ 2 \sin\alpha\cos \alpha+\cos2\varphi_0+ {\rm i}\cos2\alpha \sin 2\varphi_0] \right|\nonumber\\
&=&    \cos^2\phi \left| \sin2\alpha + \frac{2\sin^2\vartheta \cos^2 \vartheta}{\cos^2 2\vartheta}[  \sin2\alpha+\cos2\varphi_0+ {\rm i}\cos2\alpha \sin 2\varphi_0] \right|\nonumber\\
&=&    \cos^2\phi \left[ \sin2\alpha + \frac{1}{2}\tan^2\vartheta [  \sin2\alpha+\cos2\varphi_0+ {\rm i}\cos2\alpha \sin 2\varphi_0] \right]\nonumber\\
&=&    \cos^2\phi \left| \sin2\alpha + \frac{1}{2} \left[\sqrt{\frac{1-\cos2\beta\sin2 \alpha}{\cos^2 2 \alpha}} \sqrt{2}\tan\phi\right]^2 [  \sin2\alpha+\cos2\varphi_0+ {\rm i}\cos2\alpha \sin 2\varphi_0] \right|\nonumber\\
&=&    \cos^2\phi \left| \sin2\alpha + \frac{1-\cos2\beta\sin2 \alpha}{\cos^2 2 \alpha} \tan^2\phi [  \sin2\alpha+\cos2\varphi_0+ {\rm i}\cos2\alpha \sin 2\varphi_0] \right|\nonumber\\
&=&    \left| \sin2\alpha \cos^2\phi + \frac{1-\cos2\beta\sin2 \alpha}{\cos^2 2 \alpha} \sin^2\phi [  \sin2\alpha+\cos2\varphi_0+ {\rm i}\cos2\alpha \sin 2\varphi_0] \right|\nonumber\\
&=&    \sqrt{ \left[ \sin2\alpha \cos^2\phi + \frac{1-\cos2\beta\sin2 \alpha}{\cos^2 2 \alpha} \sin^2\phi (  \sin2\alpha+\cos2\varphi_0)\right]^2
+ \left(\frac{1-\cos2\beta\sin2 \alpha}{\cos^2 2 \alpha} \sin^2\phi\right)^2(\cos2\alpha \sin 2\varphi_0)^2 }\nonumber\\
&=&    \sqrt{ \left[ \sin2\alpha \cos^2\phi + \frac{1-\cos2\beta\sin2 \alpha}{\cos^2 2 \alpha} \sin^2\phi (  \sin2\alpha+\cos2\varphi_0)\right]^2
+ \frac{(1-\cos2\beta\sin2 \alpha)^2}{\cos^2 2 \alpha} \sin^4\phi\sin^2 2\varphi_0 }.
\end{eqnarray}
Because
\begin{eqnarray}
\cos2\varphi_0 &=&{ \frac{\cos2\beta-\sin2 \alpha}
{1-\cos2\beta\sin2 \alpha},}\nonumber\\
\sin^2 2\varphi_0 &=&1-\cos^22\varphi_0=1- \left( \frac{\cos2\beta-\sin2 \alpha}
{1-\cos2\beta\sin2 \alpha}\right)^2=\frac{(1-\cos2\beta\sin2 \alpha)^2 -(\cos2\beta-\sin2 \alpha)^2}
{(1-\cos2\beta\sin2 \alpha)^2}\nonumber\\
&=&\frac{1+(\cos2\beta\sin2 \alpha)^2 -(\cos^22\beta+\sin^22 \alpha)}
{(1-\cos2\beta\sin2 \alpha)^2}=\frac{\sin^2 2\beta+\sin^2 2 \alpha(\cos^22\beta-1)}{(1-\cos2\beta\sin2 \alpha)^2}\nonumber\\
&=&\frac{\sin^2 2\beta-\sin^2 2 \alpha\sin^22\beta}{(1-\cos2\beta\sin2 \alpha)^2}=\frac{\sin^2 2\beta\cos^2 2 \alpha}{(1-\cos2\beta\sin2 \alpha)^2},
\end{eqnarray}
then we have
\begin{eqnarray}
&&  \left[ \sin2\alpha \cos^2\phi + \frac{1-\cos2\beta\sin2 \alpha}{\cos^2 2 \alpha} \sin^2\phi (  \sin2\alpha+\cos2\varphi_0)\right]^2
+ \frac{(1-\cos2\beta\sin2 \alpha)^2}{\cos^2 2 \alpha} \sin^4\phi\sin^2 2\varphi_0 \nonumber\\
&=& \left[ \sin2\alpha \cos^2\phi + \frac{1-\cos2\beta\sin2 \alpha}{\cos^2 2 \alpha} \sin^2\phi \sin2\alpha+\frac{1-\cos2\beta\sin2 \alpha}{\cos^2 2 \alpha} \sin^2\phi \times \frac{\cos2\beta-\sin2 \alpha}
{1-\cos2\beta\sin2 \alpha}\right]^2+ \sin^4\phi \sin^2 2\beta\nonumber\\
&=& \left[ \sin2\alpha \cos^2\phi + \frac{1-\cos2\beta\sin2 \alpha}{\cos^2 2 \alpha} \sin^2\phi \sin2\alpha
+\frac{\cos2\beta-\sin2 \alpha}{\cos^2 2 \alpha} \sin^2\phi \right]^2+\sin^4\phi \sin^2 2\beta\nonumber\\
&=& [ \sin2\alpha \cos^2\phi + \cos2\beta \sin^2\phi ]^2+ +\sin^4\phi \sin^2 2\beta,
\end{eqnarray}
thus we finally have
\begin{eqnarray}
&=&    \sqrt{ \left[ \sin2\alpha \cos^2\phi + \frac{1-\cos2\beta\sin2 \alpha}{\cos^2 2 \alpha} \sin^2\phi (  \sin2\alpha+\cos2\varphi_0)\right]^2
+ \frac{(1-\cos2\beta\sin2 \alpha)^2}{\cos^2 2 \alpha} \sin^4\phi\sin^2 2\varphi_0 }\nonumber\\
&=& \sqrt{[ \sin2\alpha \cos^2\phi + \cos2\beta \sin^2\phi ]^2+ +\sin^4\phi \sin^2 2\beta}.
\end{eqnarray}
This ends the verification. $\blacksquare$
\end{remark}

\begin{remark}\textcolor{blue}{The case of both $\ket{\psi_1}$ and $\ket{\psi_2}$ are maximally entangled states.} The readers may be curious about what happen when both $\ket{\psi_1}$ and $\ket{\psi_2}$ are maximally entangled states. As above, the state $\ket{\psi_1} $ is given by
\begin{eqnarray}
&&\ket{\psi_1} =\cos \theta \ket{00}+\sin \theta\ket{11}= \frac{1}{\sqrt{2}}(\ket{00}+\ket{11}),
\end{eqnarray}
where $\theta$ has been taken as $\theta=\pi/4$. Here $\ket{\psi_1} $ is the maximally entangled state (i.e., $ \mathcal{C}_{\ket{\psi_1}}=1$) spanned in the subspace of $\{\ket{00}, \ket{11}\}$. Accordingly, the pure state $\ket{\psi_2}$ is given by
\begin{eqnarray}
\ket{\psi_2}
&=& \cos\phi(\cos \alpha \ket{01} +\sin\alpha\ket{10})+ e^{{\rm i}\beta}\frac{\sin\phi}{\sqrt{2}}( \ket{00}-\ket{11}),
\end{eqnarray}
whose concurrence is given by
\begin{eqnarray}\label{eq:psicon-1}
 \mathcal{C}_{\ket{\psi_2}} &=& \left|e^{{\rm i} 2\beta}\sin^2 \phi + \cos^2\phi\sin 2 \alpha\right|.
\end{eqnarray}

\emph{Analysis 1.---} The case of $\phi=0$. Because $\alpha\in [0, \pi/2]$, from Eq. (\ref{eq:psicon-1}) we have $\alpha=\pi/4$. Then we obtain
\begin{eqnarray}
&&\ket{\psi_2} =\frac{1}{\sqrt{2}}(\ket{01}+\ket{10}),
\end{eqnarray}
which is a maximally entangled state spanned in the subspace of $\{\ket{01}, \ket{10}\}$. This state $\ket{\psi_2}$ is simple. In the behind, when we study the quantum nonlocality of the density matrix $\rho=\nu_1  \ket{\psi_1}\bra{\psi_1}+\nu_2 \ket{\psi_2}\bra{\psi_2}$, we need to consider this state.

\emph{Analysis 2.---} The case of $\phi=\pi/2$. We can obtain
\begin{eqnarray}
&&\ket{\psi_2} =e^{{\rm i}\beta}\frac{1}{\sqrt{2}}( \ket{00}-\ket{11}),
\end{eqnarray}
which is the other maximally entangled state spanned in the space of $\{\ket{00}, \ket{11}\}$. The phase factor $e^{{\rm i}\beta}$ is trivial and can be ignored. This state $\ket{\psi_2}$ is simple. In the behind, when we study the quantum nonlocality of the density matrix $\rho=\nu_1  \ket{\psi_1}\bra{\psi_1}+\nu_2 \ket{\psi_2}\bra{\psi_2}$, we need to consider this state.

\emph{Analysis 3.---} The case of $\phi\in(0, \pi/2)$. Because $\alpha\in [0, \pi/2]$, based on Eq. (\ref{eq:psicon-1}) the concurrence $\mathcal{C}_{\ket{\psi_2}}=1$ occurs only at
\begin{eqnarray}\label{eq:concu-2}
&& e^{{\rm i} 2\beta}=\sin 2 \alpha= 1,
\end{eqnarray}
otherwise one has  $\mathcal{C}_{\ket{\psi_2}}<1$. Explicitly, one has $\beta =0$ (or $\pi$) and $\alpha= \pi/4 $. And the state $\ket{\psi_2}$ reads
(we take $\beta=0$ as an example; the result is similar when $\beta=\pi$)
\begin{eqnarray}
&&\ket{\psi_2} = \frac{\cos\phi}{\sqrt{2}}( \ket{01} +\ket{10})+ \frac{\sin\phi}{\sqrt{2}}( \ket{00}-\ket{11}).
\end{eqnarray}
One may check that the local unitary transformation ${(\mathcal{V}_a \otimes \mathcal{V}_b)}$ can realize the following transformation
\begin{eqnarray}
&&{(\mathcal{V}_a \otimes \mathcal{V}_b)}\, \ket{\psi_2} \mapsto |\psi'_1\rangle =  {e^{{\rm i}\zeta}} \frac{1}{\sqrt{2}}\left( \ket{00}+ \ket{11}\right), \nonumber\\
&&(\mathcal{V}_a \otimes \mathcal{V}_b)\, \ket{\psi_1} \mapsto |\psi'_2\rangle
= { \frac{1}{\sqrt{2}}(\ket{01}+\ket{10})}.
\end{eqnarray}
Here
\begin{eqnarray}
&& \mathcal{V}_a = \Omega_1 \mathcal{U}_1=
\left(
                 \begin{array}{cc}
                  e^{{\rm i}\omega_0} & 0 \\
                   0 & e^{{\rm i}\omega_1} \\
                 \end{array}
               \right) \left(
                 \begin{array}{cc}
                   \cos\vartheta_0 & -\sin\vartheta_0 e^{-{\rm i}\varphi_0} \\
                   \sin\vartheta_0 e^{{\rm i}\varphi_0} &  \cos\vartheta_0 \\
                 \end{array}
               \right),\nonumber\\
&& \mathcal{V}_b = \Omega_2 \mathcal{U}_2=
\left(
                 \begin{array}{cc}
                  e^{{\rm i}\omega_2} & 0 \\
                   0 & e^{{\rm i}\omega_3} \\
                 \end{array}
               \right) \left(
                 \begin{array}{cc}
                   \cos\vartheta_1 & -\sin\vartheta_1 e^{-{\rm i}\varphi_1} \\
                   \sin\vartheta_1 e^{{\rm i}\varphi_1} &  \cos\vartheta_1 \\
                 \end{array}
               \right),
\end{eqnarray}
with
\begin{eqnarray}
&& \varphi_0=0, \;\;\; \vartheta=\frac{\phi}{2}, \;\;\; \vartheta_0=\vartheta=\frac{\phi}{2}, \nonumber\\ &&\vartheta_1=\vartheta_0-\frac{\pi}{2}=\frac{\phi}{2}-\frac{\pi}{2}, \;\;\;  \varphi_1=-\varphi_0=0, \nonumber\\
 &&\omega_0 ={\frac{1}{2}\arctan \left[\frac{\sin^2\vartheta\sin\alpha \sin2\varphi_0}{\cos^2 \vartheta \cos \alpha  +
 \sin^2\vartheta\sin\alpha \cos2\varphi_0}\right]+\frac{1}{2}\arctan\left[\frac{\sin^2\vartheta \cos \alpha \sin2\varphi_0 }{\sin^2\vartheta \cos \alpha \cos2\varphi_0 +
 \cos^2 \vartheta \sin\alpha}\right]=0,}\nonumber\\
&& \omega_1=0, \nonumber\\
&& \omega_2=-\omega_1+\varphi_1 = -\varphi_0={0},\nonumber\\
&& \omega_3={\pi}-\omega_0-\varphi_1 = {\pi}-\omega_0+\varphi_0={\pi},\nonumber\\
&&\zeta =-\frac{1}{2}\arctan \left[\frac{\sin^2\vartheta\sin\alpha \sin2\varphi_0}{ \sin^2\vartheta\sin\alpha \cos2\varphi_0+\cos^2 \vartheta \cos \alpha}\right]+\frac{1}{2}\arctan\left[\frac{\sin^2\vartheta \cos \alpha \sin2\varphi_0 }{\sin^2\vartheta \cos \alpha \cos2\varphi_0 +
 \cos^2 \vartheta \sin\alpha}\right]={0}.
\end{eqnarray}
Explicitly, the unitary matrices are given by
\begin{eqnarray}
&& \mathcal{V}_a =
\left(
                 \begin{array}{cc}
                  1 & 0 \\
                   0 & 1 \\
                 \end{array}
               \right) \left(
                 \begin{array}{cc}
                   \cos\frac{\phi}{2} & -\sin\frac{\phi}{2}  \\
                   \sin\frac{\phi}{2} &  \cos\frac{\phi}{2} \\
                 \end{array}
               \right)=
\left(
                 \begin{array}{cc}
                   \cos\frac{\phi}{2} & -\sin\frac{\phi}{2}  \\
                   \sin\frac{\phi}{2} &  \cos\frac{\phi}{2} \\
                 \end{array}
               \right),\nonumber\\
&& \mathcal{V}_b =
\left(
                 \begin{array}{cc}
                  1 & 0 \\
                   0 & -1 \\
                 \end{array}
               \right) \left(
                 \begin{array}{cc}
                   \sin\frac{\phi}{2} & \cos\frac{\phi}{2}  \\
                   -\cos\frac{\phi}{2}  &  \sin\frac{\phi}{2} \\
                 \end{array}
               \right)=\left(
                 \begin{array}{cc}
                   \sin\frac{\phi}{2} & \cos\frac{\phi}{2}  \\
                   \cos\frac{\phi}{2}  &  -\sin\frac{\phi}{2} \\
                 \end{array}
               \right).
\end{eqnarray}
$\blacksquare$
\end{remark}

\newpage

\subsection{Summary of Quantum State $\rho$}\label{sum}

In summary, under the local unitary transformations, for a two-qubit density matrix with rank 2 (including rank 1 as a special case), it is sufficient to consider the following form
\begin{eqnarray}
\rho=\nu_1  \ket{\psi_1}\bra{\psi_1}+\nu_2 \ket{\psi_2}\bra{\psi_2},
\end{eqnarray}
where
\begin{eqnarray}
&&\ket{\psi_1} =\cos \theta \ket{00}+\sin \theta\ket{11}, \nonumber\\
&& \ket{\psi_2} =\cos\phi(\cos \alpha \ket{01} +\sin\alpha\ket{10})+ e^{{\rm i}\beta}\sin\phi(\sin \theta \ket{00}-\cos \theta\ket{11}),
\end{eqnarray}
with
\begin{eqnarray}
&& {\theta\in [0, \pi/2]}, \;\;\;\; {\phi \in [0, \pi/2]}, \;\;\;\;  {\alpha \in [0, \pi/2]}, \;\;\;\; \beta \in [0, 2\pi],   \nonumber\\
&& \nu_1 \in [0,1], \;\;\; \nu_2 \in [0,1], \;\;\; \nu_1+\nu_2=1.
\end{eqnarray}
Specially, when $\theta=\pi/4$, we only need to consider the following form
\begin{eqnarray}
\rho=\nu_1  \ket{\psi_1}\bra{\psi_1}+\nu_2 \ket{\psi_2}\bra{\psi_2},
\end{eqnarray}
where
\begin{eqnarray}
&&\ket{\psi_1} =\frac{1}{\sqrt{2}}(\ket{00}+\ket{11}), \nonumber\\
&& \ket{\psi_2} =\frac{1}{\sqrt{2}}(\ket{01}+\ket{10}), \;\;\;\;\; {\rm or}\;\;\;\; \frac{1}{\sqrt{2}}(\ket{00}-\ket{11}).
\end{eqnarray}

By the way, in the matrix form, one can have
\begin{eqnarray}
\rho_1 = |\psi_1\rangle\langle \psi_1|=
 \begin{bmatrix}
  \cos ^2\theta  & 0 & 0 & \sin \theta \cos \theta  \\
 0 & 0 & 0 & 0 \\
 0 & 0 & 0 & 0 \\
 \sin \theta \cos \theta & 0 & 0 & \sin ^2\theta \\
\end{bmatrix},
\end{eqnarray}
\begin{eqnarray}
\rho_2 &=& |\psi_2\rangle\langle \psi_2|\nonumber\\
&=&
 \begin{bmatrix}
  \sin ^2\theta  \sin ^2\phi  & e^{{\rm i} \beta } \cos \alpha \sin \theta  \sin \phi  \cos \phi  &  e^{{\rm i} \beta  } \sin \alpha  \sin \theta  \sin \phi \cos \phi  & -\sin \theta  \cos \theta  \sin ^2\phi  \\
 e^{-{\rm i} \beta } \cos \alpha \sin \theta \sin \phi \cos \phi & \cos ^2\alpha \cos ^2\phi  &  \sin \alpha \cos \alpha  \cos ^2\phi & -e^{-{\rm i} \beta } \cos \alpha  \cos \theta  \sin \phi  \cos \phi \\
 e^{-{\rm i} \beta }\sin \alpha  \sin \theta \sin \phi \cos \phi  &  \sin \alpha \cos \alpha  \cos ^2\phi  & \sin ^2\alpha  \cos ^2\phi  & -e^{-{\rm i} \beta }\sin \alpha   \cos \theta \sin \phi  \cos \phi \\
 -\sin \theta \cos \theta \sin ^2\phi  & -e^{{\rm i} \beta } \cos \alpha  \cos \theta  \sin \phi  \cos \phi  & -e^{{\rm i} \beta }\sin \alpha   \cos \theta  \sin \phi  \cos \phi  & \cos ^2\theta  \sin ^2\phi  \\
\end{bmatrix},\nonumber\\
\end{eqnarray}
which yield the matrix form of the state $\rho$ as
\begin{eqnarray}
\rho &=& \nu_1\rho_1+\nu_2\rho_2\nonumber\\
&=& {\small\begin{bmatrix}
  \nu_1 \cos^2 \theta +\nu_2\sin ^2\theta  \sin ^2\phi  & \nu_2e^{{\rm i} \beta } \cos \alpha \sin \theta  \sin \phi  \cos \phi  &  \nu_2e^{{\rm i} \beta } \sin \alpha  \sin \theta  \sin \phi \cos \phi  & \sin \theta \cos \theta(\nu_1 -\nu_2  \sin ^2\phi)  \\
 \nu_2e^{-{\rm i} \beta } \cos \alpha \sin \theta \sin \phi \cos \phi & \nu_2\cos ^2\alpha \cos ^2\phi  & \nu_2  \sin \alpha \cos \alpha  \cos ^2\phi & -\nu_2e^{-{\rm i} \beta } \cos \alpha  \cos \theta  \sin \phi  \cos \phi \\
 \nu_2e^{-{\rm i} \beta }\sin \alpha  \sin \theta \sin \phi \cos \phi  & \nu_2  \sin \alpha \cos \alpha  \cos ^2\phi  & \nu_2\sin ^2\alpha  \cos ^2\phi  & -\nu_2e^{-{\rm i} \beta }\sin \alpha   \cos \theta \sin \phi  \cos \phi \\
 \sin \theta \cos \theta(\nu_1 -\nu_2  \sin ^2\phi)  & -\nu_2e^{{\rm i} \beta } \cos \alpha  \cos \theta  \sin \phi  \cos \phi  & -\nu_2e^{{\rm i} \beta }\sin \alpha   \cos \theta  \sin \phi  \cos \phi  &  \nu_1 \sin^2 \theta+\nu_2\cos ^2\theta  \sin ^2\phi  \\
\end{bmatrix}.}\nonumber\\
\end{eqnarray}

\newpage

\section{Degree of Entanglement for Quantum State $\rho$}

Pauli matrices are given by
\begin{eqnarray}
 \label{S-3}
  && \sigma_x=
  \left(
    \begin{array}{cc}
      0 & 1 \\
      1 & 0 \\
    \end{array}
  \right),\;\;\;\sigma_y=
  \left(
    \begin{array}{cc}
      0 & -i \\
      i & 0 \\
    \end{array}
  \right),\;\;\;  \sigma_z=
  \left(
    \begin{array}{cc}
      1 & 0 \\
      0 & -1 \\
    \end{array}
  \right).
   \end{eqnarray}
Based on the density matrix $\rho$, we can define the conjugate matrix of $\rho$ as
\begin{eqnarray}
\rho^* &=& (\rho)^* \nonumber\\
&=& {\small\begin{bmatrix}
  \nu_1 \cos^2 \theta +\nu_2\sin ^2\theta  \sin ^2\phi  & \nu_2e^{-{\rm i} \beta } \cos \alpha \sin \theta  \sin \phi  \cos \phi  &  \nu_2e^{-{\rm i} \beta } \sin \alpha  \sin \theta  \sin \phi \cos \phi  & \sin \theta \cos \theta(\nu_1 -\nu_2  \sin ^2\phi)  \\
 \nu_2e^{{\rm i} \beta } \cos \alpha \sin \theta \sin \phi \cos \phi & \nu_2\cos ^2\alpha \cos ^2\phi  & \nu_2  \sin \alpha \cos \alpha  \cos ^2\phi & -\nu_2e^{{\rm i} \beta } \cos \alpha  \cos \theta  \sin \phi  \cos \phi \\
 \nu_2e^{{\rm i} \beta }\sin \alpha  \sin \theta \sin \phi \cos \phi  & \nu_2  \sin \alpha \cos \alpha  \cos ^2\phi  & \nu_2\sin ^2\alpha  \cos ^2\phi  & -\nu_2e^{{\rm i} \beta }\sin \alpha   \cos \theta \sin \phi  \cos \phi \\
 \sin \theta \cos \theta(\nu_1 -\nu_2  \sin ^2\phi)  & -\nu_2e^{-{\rm i} \beta } \cos \alpha  \cos \theta  \sin \phi  \cos \phi  & -\nu_2e^{-{\rm i} \beta }\sin \alpha   \cos \theta  \sin \phi  \cos \phi  &  \nu_1 \sin^2 \theta+\nu_2\cos ^2\theta  \sin ^2\phi  \\
\end{bmatrix}.}\nonumber\\
\end{eqnarray}
Furthermore, we can define the matrix
\begin{eqnarray}
\tilde{\rho}= (\sigma_y\otimes \sigma_y) \rho^*  (\sigma_y\otimes \sigma_y),
\end{eqnarray}
and the Hermitian matrix
\begin{eqnarray}\label{eq:R-1}
R = \sqrt{\sqrt{\rho} \tilde{\rho} \sqrt{\rho}}.
\end{eqnarray}
Let one denote the four eigenvalues of $R$ as $\lambda_i$ ($i=1, 2, 3, 4$) in decreasing order (i.e., $\lambda_1 \geq \lambda_2\geq \lambda_3\geq \lambda_4$), then the degree of entanglement of state $\rho$ is given by
\begin{eqnarray}
\mathcal{C} = {\rm max} \{0, \lambda_1-\lambda_2-\lambda_3-\lambda_4\},
\end{eqnarray}
which is the so-called ``\emph{concurrence}''. Based on Eq. (\ref{eq:R-1}), one easily knows that the four eigenvalues of
\begin{eqnarray}
R^2 = \sqrt{\rho} \tilde{\rho} \sqrt{\rho}
\end{eqnarray}
are $\lambda^2_i$, ($i=1, 2, 3, 4$). Since the matrix $AB$ and the matrix $BA$ share the same eigenvalues, thus, in order to calculate the eigenvalues of $R^2 = \sqrt{\rho} (\tilde{\rho} \sqrt{\rho})$, we turn to calculate the eigenvalues to the following new operator
\begin{eqnarray}
\mathbb{R}^2 =  (\tilde{\rho} \sqrt{\rho})\sqrt{\rho}=\tilde{\rho} (\sqrt{\rho}\sqrt{\rho})= \tilde{\rho} \rho.
\end{eqnarray}

We now come to determine the eigenvalues of $\mathbb{R}^2$. For simplicity, let us denote $v_i=\lambda^2_i$, then we have
\begin{eqnarray}
\det[v \mathbb{I}-\mathbb{R}^2 ]=0,
\end{eqnarray}
where $\mathbb{I}$ is the $4\times 4$ identity matrix. One can have
\begin{eqnarray}
v(s_3 v^2 +s_2 v^2+ s_1 v+s_0)=0,
\end{eqnarray}
with the coefficients
\begin{eqnarray}\label{eq:s-1}
 s_3&=&1, \nonumber\\
 s_2 &=& -\frac{1}{2} \nu_2^2 \sin2 \alpha  \sin 2 \theta \sin ^2(2 \phi ) \cos 2 \beta-\nu_2^2 \sin ^2(2 \alpha ) \cos ^4 \phi -\sin ^2(2 \theta ) \left[\nu_1^2-\nu_1 \nu_2 \sin ^2\phi +\nu_2^2 \sin ^4\phi \right]
\nonumber\\
&&-\frac{1}{2} \nu_1 \nu_2 (\cos4 \theta+3) \sin ^2 \phi, \nonumber\\
 s_1 &=& \frac{1}{2} \nu_1^2 \nu_2^2 \left[\sin 2 \alpha \sin2 \theta \sin ^2(2 \phi ) \cos 2 \beta+2 \sin ^2(2 \alpha ) \sin ^2(2 \theta ) \cos ^4\phi +2 \sin ^4\phi \right], \nonumber\\
 s_0 &=& 0.
\end{eqnarray}
Namely, we have
\begin{eqnarray}\label{eq:s-2}
v^2(v^2 +s_2 v+ s_1)=0.
\end{eqnarray}

\begin{remark}Let us consider a special case of $\nu_1=1$ and $\nu_2=0$. In this case, $\rho=\ket{\psi_1}\bra{\psi_1}$ represents a pure state. On one hand, the \emph{concurrence} of the pure state $\ket{\psi_1}$ is
\begin{eqnarray}\label{eq:C-1a}
\mathcal{C}=2|\cos\theta\sin\theta|=|\sin2\theta|=\sin2\theta.
\end{eqnarray}
On the other hand, based on Eq. (\ref{eq:s-1}), one has
\begin{eqnarray}
 s_2 = -\sin ^2 2 \theta , \;\;\;\;\; s_1=0.
\end{eqnarray}
Then from Eq. (\ref{eq:s-2}), one has
\begin{eqnarray}
v^3[v-\sin ^2 2 \theta ]=0,
\end{eqnarray}
thus
\begin{eqnarray}
 v_1=\sin ^2 2 \theta, \;\;\; v_2=v_3=v_4=0.
\end{eqnarray}
This leads to
\begin{eqnarray}
\lambda_1= \sqrt{v_1}=|\sin 2 \theta|=\sin 2 \theta, \;\;\; \lambda_2=\lambda_3=\lambda_4=0,
\end{eqnarray}
thus the concurrence is given by
\begin{eqnarray}
\mathcal{C} = {\rm max} \{0, \lambda_1-\lambda_2-\lambda_3-\lambda_4\}=|\sin 2 \theta|=\sin 2 \theta,
\end{eqnarray}
which recovers Eq. (\ref{eq:C-1a}). $\blacksquare$
\end{remark}

\begin{remark}Let us consider a special case of $\nu_1=0$ and $\nu_2=1$. In this case, $\rho=\ket{\psi_2}\bra{\psi_2}$ represents a pure state. On one hand, the \emph{concurrence} of the pure state $\ket{\psi_2}$ is
\begin{eqnarray}\label{eq:C-1aa}
\mathcal{C}&=&2|cd-ab|=2\left|(-e^{{\rm i}\beta}\sin\phi\sin \theta e^{{\rm i}\beta}\sin\phi \cos \theta)-(\cos\phi\cos \alpha \cos\phi \sin\alpha)\right|\nonumber\\
&=& 2\left|e^{{\rm i}\beta}\sin\phi\sin \theta e^{{\rm i}\beta}\sin\phi \cos \theta+\cos\phi\cos \alpha \cos\phi \sin\alpha\right|\nonumber\\
&=& 2\left|e^{{\rm i} 2\beta}\sin \theta\cos\theta \sin^2\phi +  \sin\alpha \cos\alpha \cos^2\phi\right|\nonumber\\
&=& \left|e^{{\rm i} 2\beta}\sin 2\theta \sin^2\phi +  \sin 2\alpha \cos^2\phi\right|\nonumber\\
&=& \sqrt{ \left[\cos2\beta \sin 2\theta \sin^2\phi +  \sin2\alpha \cos^2\phi\right]^2+\left[\sin2\beta \sin2\theta \sin^2\phi\right]^2}.
\end{eqnarray}
On the other hand, based on Eq. (\ref{eq:s-1}), one has
\begin{eqnarray}
 s_2 &=& -\frac{1}{2}  \sin 2 \alpha  \sin 2 \theta  \sin ^2(2 \phi ) \cos 2 \beta- \sin ^2(2 \alpha ) \cos ^4 \phi -\sin ^2(2 \theta ) \sin ^4\phi,\nonumber\\
 s_1 &=&0.
\end{eqnarray}
Then from Eq. (\ref{eq:s-2}), one has
\begin{eqnarray}
v^3[v+s_2]=0,
\end{eqnarray}
thus
\begin{eqnarray}
  v_1&=&-s_2= \frac{1}{2}  \sin 2 \alpha \sin 2 \theta \sin ^2(2 \phi ) \cos 2 \beta +\sin ^2(2 \alpha ) \cos ^4 \phi +\sin ^2(2 \theta ) \sin ^4\phi ,\nonumber\\
 &=&  \left[\cos 2\beta\sin 2\theta \sin^2\phi +  \sin2\alpha \cos^2\phi\right]^2+\left[\sin2\beta\sin2\theta \sin^2\phi\right]^2,\nonumber\\
  v_2 &=&v_3=v_4=0.
\end{eqnarray}
This leads to
\begin{eqnarray}
&&\lambda_1= \sqrt{v_1}=\sqrt{ \left[\cos2\beta \sin 2\theta \sin^2\phi +  \sin2\alpha \cos^2\phi\right]^2+\left[\sin2\beta \sin2\theta \sin^2\phi\right]^2}, \nonumber\\
&& \lambda_2=\lambda_3=\lambda_4=0,
\end{eqnarray}
thus the concurrence is given by
\begin{eqnarray}
\mathcal{C} &=& {\rm max} \{0, \lambda_1-\lambda_2-\lambda_3-\lambda_4\} \nonumber\\
&=& \sqrt{ \left[\cos2\beta \sin 2\theta \sin^2\phi +  \sin2\alpha \cos^2\phi\right]^2+\left[\sin2\beta \sin2\theta \sin^2\phi\right]^2},
\end{eqnarray}
which recovers Eq. (\ref{eq:C-1aa}). $\blacksquare$
\end{remark}

\begin{remark}Let us consider the general case where $\nu_1$ and $\nu_2$ are not 0. In this case, $\rho=\nu_1  \ket{\psi_1}\bra{\psi_1}+\nu_2 \ket{\psi_2}\bra{\psi_2}$ represents a quantum mixed state with rank 2. Based on Eq. (\ref{eq:s-1}), one has
\begin{eqnarray}
 s_2 &=& -\frac{1}{2} \nu_2^2 \sin 2 \alpha  \sin 2 \theta  \sin ^2(2 \phi ) \cos 2 \beta-\nu_2^2 \sin ^2(2 \alpha ) \cos ^4 \phi
 -\sin ^2(2 \theta ) \left[\nu_1^2-\nu_1 \nu_2 \sin ^2\phi +\nu_2^2 \sin ^4\phi \right]
\nonumber\\
&&-\frac{1}{2} \nu_1 \nu_2 (\cos 4 \theta+3) \sin ^2 \phi, \nonumber\\
 s_1 &=& \frac{1}{2} \nu_1^2 \nu_2^2 \left[\sin 2 \alpha \sin 2 \theta  \sin ^2(2 \phi ) \cos 2 \beta+2 \sin ^2(2 \alpha ) \sin ^2(2 \theta ) \cos ^4\phi +2 \sin ^4\phi \right].
\end{eqnarray}
Then from Eq. (\ref{eq:s-2}), one has
\begin{eqnarray}
v^2[v^2+s_2v+s_1]=0,
\end{eqnarray}
thus
\begin{eqnarray}
 \lambda_1-\lambda_2=\sqrt{v_1}-\sqrt{v_2}=\sqrt{v_1+v_2-2\sqrt{v_1 v_2}}=\sqrt{-s_2-2\sqrt{s_1}}.
\end{eqnarray}
As a result, the concurrence is 0 iff
\begin{eqnarray}
-s_2-2\sqrt{s_1}=0.
\end{eqnarray}
That is
\begin{eqnarray}
s_2^2-4s_1=0,
\end{eqnarray}
i.e.,
\begin{eqnarray}
& &  \frac{1}{4} \Bigg(
    \nu_2^2 \sin2\alpha \sin2\theta \sin^2(2\phi) \cos2\beta
    + 2 \nu_2^2 \sin^2(2\alpha) \cos^4\phi \nonumber \\
& & + 2 \sin^2(2\theta) \Big(
    \nu_1^2 - \nu_1 \nu_2 \sin^2\phi + \nu_2^2 \sin^4\phi
\Big) + \nu_1 \nu_2 \left( \cos4\theta + 3 \right) \sin^2\phi
\Bigg)^2 \nonumber\\
& & -2 \nu_1^2 \nu_2^2 \Bigg(
    \sin2\alpha \sin2\theta \sin^2(2\phi) \cos2\beta
    + 2 \sin^2(2\alpha) \sin^2(2\theta) \cos^4\phi
    + 2 \sin^4\phi
\Bigg)
=0,
\end{eqnarray}
i.e.,
\begin{eqnarray}\label{eq:c=0-1}
& &  \Bigg(
    \nu_2^2 \sin2\alpha \sin2\theta \sin^2(2\phi) \cos2\beta
    + 2 \nu_2^2 \sin^2(2\alpha) \cos^4\phi \nonumber \\
& & + 2 \sin^2(2\theta) \Big(
    \nu_1^2 - \nu_1 \nu_2 \sin^2\phi + \nu_2^2 \sin^4\phi
\Big) + \nu_1 \nu_2 \left( \cos4\theta + 3 \right) \sin^2\phi
\Bigg)^2 \nonumber\\
& & -8 \nu_1^2 \nu_2^2 \Bigg(
    \sin2\alpha \sin2\theta \sin^2(2\phi) \cos2\beta
    + 2 \sin^2(2\alpha) \sin^2(2\theta) \cos^4\phi
    + 2 \sin^4\phi
\Bigg)
=0.
\end{eqnarray}
There are three possible situations: (i) $\phi=0$; (ii) $\phi=\pi/2$; and (iii) $0< \phi < \pi/2$.

\emph{Case (i).---} $\phi=0$. In this case, from Eq. (\ref{eq:c=0-1}), the concurrence is 0 iff
\begin{eqnarray}
 \left( -\nu_2^2 \sin^2 2 \alpha - \nu_1^2 \sin^2 2 \theta \right)^2-4 \nu_1^2 \nu_2^2 \sin^2 2 \alpha \sin^2 2 \theta =0,
\end{eqnarray}
i.e.,
\begin{eqnarray}
 \left( \nu_2^2 \sin^2 2 \alpha + \nu_1^2 \sin^2 2 \theta \right)^2-4 \nu_1^2 \nu_2^2 \sin^2 2 \alpha \sin^2 2 \theta =0,
\end{eqnarray}
i.e.,
\begin{eqnarray}
  \nu_2^2 \sin^2 2 \alpha + \nu_1^2 \sin^2 2 \theta = 2 \nu_1 \nu_2 |\sin2 \alpha \sin2 \theta|,
\end{eqnarray}
i.e.,
\begin{eqnarray}
  \nu_2^2 \sin^2 2 \alpha + \nu_1^2 \sin^2 2 \theta = { 2 \nu_1 \nu_2 \sin2 \alpha \sin2 \theta,}
\end{eqnarray}
i.e.,
\begin{eqnarray}
 \left(\nu_2 \sin2\alpha - \nu_1 \sin2\theta \right)^2=0,
\end{eqnarray}
i.e.,
\begin{eqnarray}
 \nu_2 \sin2\alpha = \nu_1 \sin2\theta.
\end{eqnarray}
In this case, the concurrence is given by
\begin{eqnarray}
\mathcal{C} &=& {\rm max} \{0, \lambda_1-\lambda_2-\lambda_3-\lambda_4\} = \lambda_1-\lambda_2= \sqrt{-s_2-2\sqrt{s_1}} \nonumber\\
&=& \sqrt{ \left[ \nu_2^2 \sin^2 2 \alpha + \nu_1^2 \sin^2 2 \theta \right]-2 \sqrt{ \nu_1^2 \nu_2^2 \sin^2 2 \alpha \sin^2 2 \theta }}\nonumber\\
&=& \sqrt{ \left[ \nu_2^2 \sin^2 2 \alpha + \nu_1^2 \sin^2 2 \theta\right]-2 \nu_1 \nu_2 |\sin 2 \alpha \sin2 \theta| }\nonumber\\
&=& \sqrt{ \left(\nu_2 \sin2\alpha - \nu_1 \sin2\theta\right)^2 } \nonumber\\
&=&  \left|\nu_2 \sin2\alpha - \nu_1 \sin2\theta\right|.
\end{eqnarray}

\emph{Case (ii).---} $\phi=\pi/2$. Based on Eq. (\ref{eq:c=0-1}), the concurrence is 0 iff
\begin{eqnarray}
 \left[  \sin^2 2\theta \left( \nu_1^2 - \nu_1 \nu_2 + \nu_2^2 \right) +\frac{1}{2} \nu_1 \nu_2
 \left( \cos4\theta + 3 \right) \right]^2 = 4\nu_1^2 \nu_2^2,
\end{eqnarray}
i.e.,
 \begin{eqnarray}
 \sin^2 2\theta \left( \nu_1^2 - \nu_1 \nu_2 + \nu_2^2 \right) +\frac{1}{2} \nu_1 \nu_2 \left( \cos4\theta + 3 \right)  = 2\nu_1 \nu_2,
\end{eqnarray}
i.e.,
 \begin{eqnarray}
   \sin^2 2\theta \left( \nu_1^2 - \nu_1 \nu_2 + \nu_2^2 \right) + \frac{1}{2}\nu_1 \nu_2 \left( 1-2\sin^2 2\theta + 3 \right)  = 2\nu_1 \nu_2,
\end{eqnarray}
i.e.,
 \begin{eqnarray}
   \sin^2 2\theta \left( \nu_1^2 - \nu_1 \nu_2 + \nu_2^2 \right) -  \sin^2 2\theta \, \nu_1 \nu_2=0,
\end{eqnarray}
i.e.,
\begin{eqnarray}
\sin ^2 2 \theta\, (\nu_1-\nu_2)^2=0.
\end{eqnarray}
In this case, the concurrence is given by
\begin{eqnarray}
\mathcal{C} &=& {\rm max} \{0, \lambda_1-\lambda_2-\lambda_3-\lambda_4\} = \lambda_1-\lambda_2= \sqrt{-s_2-2\sqrt{s_1}} \nonumber\\
&=& \sqrt{ \left[\sin ^2 2 \theta \left(\nu_1^2-\nu_1 \nu_2 +\nu_2^2  \right)
+\frac{1}{2} \nu_1 \nu_2 [\cos 4 \theta+3] \right]-2 \sqrt{ \nu_1^2 \nu_2^2 }}\nonumber\\
&=&\sqrt{ \left[\sin ^2 2 \theta  \left(\nu_1^2-\nu_1 \nu_2 +\nu_2^2  \right)
+\frac{1}{2} \nu_1 \nu_2 [\cos 4 \theta +3] \right]-2 \nu_1 \nu_2}\nonumber\\
&=& \sqrt{\sin ^2 2 \theta\, (\nu_1-\nu_2)^2} \nonumber\\
&=& \left|(\nu_1-\nu_2)\sin 2 \theta \right|\nonumber\\
&=& |\nu_1-\nu_2|\sin 2 \theta.
\end{eqnarray}

\emph{Case (iii).---} $0< \phi<\pi/2$. In this case, the concurrence is given by
\begin{eqnarray}
\mathcal{C} &=& {\rm max} \{0, \lambda_1-\lambda_2-\lambda_3-\lambda_4\} = \lambda_1-\lambda_2= \sqrt{-s_2-2\sqrt{s_1}},
\end{eqnarray}
where $s_1$ and $s_2$ are given in Eq. (\ref{eq:s-1}). $\blacksquare$
\end{remark}

\newpage

\section{Necessary and Sufficient Condition of Separability for Quantum State $\rho$}

In this section, we present the following result for the rank-2 (including rank-1 as a special case) quantum state $\rho$:

\textcolor{blue}{\textbf{Result.}---} \emph{Under the local unitary transformations, the rank-2 quantum state $\rho$ is separable if and only if $\mathcal{M}=\mathcal{M}^\dagger$.}

\subsection{The Proof of $\mathcal{M}=\mathcal{M}^\dagger$ as the Necessary Condition}

\textcolor{blue}{\emph{Necessary Condition:}} If the matrix $\mathcal{M}=\mathcal{M}^\dagger$, then the rank-2 state $\rho$ is separable. In the following, let us come to prove the necessary condition.

\begin{proof}
As shown in the previous section, the two-qubit density matrix with rank 2 (including rank 1 as a special case) is given by
\begin{eqnarray}
\rho=\nu_1  \ket{\psi_1}\bra{\psi_1}+\nu_2 \ket{\psi_2}\bra{\psi_2},
\end{eqnarray}
where
\begin{eqnarray}
&&\ket{\psi_1} =\cos \theta \ket{00}+\sin \theta\ket{11}, \nonumber\\
&& \ket{\psi_2} =\cos\phi(\cos \alpha \ket{01} +\sin\alpha\ket{10})+ e^{{\rm i} \beta}\sin\phi(\sin \theta \ket{00}-\cos \theta\ket{11}),
\end{eqnarray}
with
\begin{eqnarray}
&& {\theta\in [0, \pi/2]}, \;\;\;\;  {\phi \in [0, \pi/2]}, \;\;\;\; {\alpha \in [0, \pi/2]}, \;\;\;\; \beta \in [0, 2\pi], \;\;\;\; \nonumber\\
&& \nu_1 \in [0,1], \;\;\; \nu_2 \in [0,1], \;\;\; \nu_1+\nu_2=1.
\end{eqnarray}
When $\nu_1\neq 0$ and $\nu_2\neq 0$, the state $\rho$ is rank 2, while when one of $\nu_j$ $(j=1, 2)$ is 0, then the state $\rho$ is rank 1.

{Let Alice and Bob share the quantum state $\rho$. Alice can perform projective measurement on her qubit.} The density matrix $\rho$ can be recast to the following decomposition form
\begin{eqnarray}
 \rho &=& P^{\hat{n}}_{0} \otimes \tilde{\rho}_0^{\hat{n}}+ P^{\hat{n}}_{1} \otimes \tilde{\rho}_1^{\hat{n}}+
|+\hat{n}\rangle\langle-\hat{n}| \otimes \mathcal{M}+|-\hat{n}\rangle\langle+\hat{n}| \otimes \mathcal{M}^\dag,
\end{eqnarray}
where
\begin{eqnarray}
&& \hat{n}=(\sin\xi \cos\tau, \sin\xi \sin\tau, \cos\xi)
\end{eqnarray}
is the measurement direction of Alice.  With the direction $\hat{n}$, the standard measurement basis is given by $\{\ket{+\hat{n}}, \ket{-\hat{n}}\}$, where
\begin{eqnarray}
&& \ket{+\hat{n}}=\begin{pmatrix}
\cos\frac{\xi}{2}\\
\sin\frac{\xi}{2}\, e^{{\rm i}\tau}\\
\end{pmatrix}=\cos\frac{\xi}{2} |0\rangle+\sin\frac{\xi}{2}\, e^{{\rm i}\tau} |1\rangle, \nonumber\\
&&
{ \ket{-\hat{n}}=\begin{pmatrix}
\sin\frac{\xi}{2}\\
-\cos\frac{\xi}{2} \, e^{{\rm i}\tau} \\
\end{pmatrix}= \sin\frac{\xi}{2}|0\rangle -\cos\frac{\xi}{2} \, e^{{\rm i}\tau} |1\rangle}.
\end{eqnarray}
Accordingly, the corresponding projectors of Alice are
\begin{eqnarray}
&& P^{\hat{n}}_{0}=\frac{1}{2}\left(\openone+\vec{\sigma}\cdot \hat{n}\right)=|+\hat{n}\rangle\langle+\hat{n}|
= \frac{1}{2}\begin{pmatrix}
 1+\cos\xi&
 \sin\xi e^{-{\rm i} \tau}\\
 \sin\xi e^{{\rm i} \tau} &
 1-\cos\xi
\end{pmatrix},\nonumber\\
&&P^{\hat{n}}_{1}=\frac{1}{2}\left(\openone-\vec{\sigma}\cdot \hat{n}\right)=|-\hat{n}\rangle\langle-\hat{n}|
= \frac{1}{2}\begin{pmatrix}
 1-\cos\xi&
 -\sin\xi e^{-{\rm i} \tau}\\
 -\sin\xi e^{{\rm i} \tau} &
 1+\cos\xi
\end{pmatrix}.
\end{eqnarray}

\begin{remark} By the way, $\ket{+\hat{n}}$ and $\ket{-\hat{n}}$ are two eigenstates of operator $\vec{\sigma}\cdot \hat{n}$ with eigenvalues $+1$ and $-1$, respectively. Here $\vec{\sigma}=(\sigma_x, \sigma_y, \sigma_z)$ is the vector of Pauli's matrices. The non-standard measurement basis can be $\{\ket{+\hat{n}}, \ket{-\hat{n}} e^{{\rm i}\chi}\}$ by introducing an extra phase $e^{{\rm i}\chi}$. Obviously, the extra phase $e^{{\rm i}\chi}$ does not affect the projectors $\{P^{\hat{n}}_{0}, P^{\hat{n}}_{1}\}$. $\blacksquare$
\end{remark}

The states $\tilde{\rho}_{0}^{\hat{n}}$ and $\tilde{\rho}_{1}^{\hat{n}}$ are called Bob's conditional states. One can work out them as
\begin{eqnarray}
\tilde{\rho}_{0}^{\hat{n}}
=
\begin{pmatrix}
a_0 &
b_0 \\
c_0 &
d_0
\end{pmatrix}, \;\;\;\;\;\;
\tilde{\rho}_{1}^{\hat{n}}
=
\begin{pmatrix}
a_1 & b_1 \\
c_1 & d_1
\end{pmatrix},
\end{eqnarray}
with
\begin{eqnarray}
a_0 &=& \nu_2 \cos^2 \phi \sin^2 \alpha \sin^2 \frac{\xi}{2} + \cos^2 \frac{\xi}{2} \left( \nu_1 \cos^2 \theta + \nu_2 \sin^2 \theta \sin^2 \phi \right) \nonumber \\
&& + \frac{1}{2} \nu_2 \cos(\beta  + \tau) \sin \alpha \sin \theta \sin \xi \sin2\phi, \nonumber\\
b_0 &=& \frac{1}{4} e^{-{\rm i} (\beta + \tau)} \biggr\{ e^{{\rm i} \beta } \nu_2 \cos^2 \phi \sin2\alpha \sin \xi + e^{{\rm i} \tau} \biggr[ e^{{\rm i} (\beta + \tau)} \sin2\theta \sin \xi (\nu_1 - \nu_2 \sin^2 \phi)  \nonumber \\
&&  + 2 \nu_2 \left( e^{2 {\rm i} \beta} \cos \alpha \cos^2 \frac{\xi}{2} \sin \theta -  \cos \theta \sin \alpha \sin^2 \frac{\xi}{2} \right) \sin2\phi \biggr] \biggr\},\nonumber\\
c_0 &=& \frac{1}{4} e^{- {\rm i} (\beta  + \tau)} \biggr[ e^{{\rm i} (\beta + 2\tau)} \nu_2 \cos^2 \phi \sin2\alpha \sin \xi + e^{{\rm i} \beta } \sin2\theta \sin \xi (\nu_1 - \nu_2 \sin^2 \phi)  \nonumber \\
&&  + 2 e^{{\rm i} \tau} \nu_2 \left(  \cos \alpha \cos^2 \frac{\xi}{2} \sin \theta - e^{2 {\rm i} \beta} \cos \theta \sin \alpha \sin^2 \frac{\xi}{2} \right) \sin2\phi \biggr]=b_0^*,\nonumber\\
d_0 &=& \nu_2 \cos^2 \alpha \cos^2 \frac{\xi}{2} \cos^2 \phi + \sin^2 \frac{\xi}{2} \left( \nu_1 \sin^2 \theta + \nu_2 \cos^2 \theta \sin^2 \phi \right) \nonumber \\
&& - \frac{1}{2} \nu_2 \cos \alpha \cos \theta \cos(\beta - \tau) \sin \xi \sin2\phi,
\end{eqnarray}
and
\begin{eqnarray}
a_1 &=& \nu_2 \cos^2 \frac{\xi}{2} \cos^2 \phi \sin^2 \alpha + \sin^2 \frac{\xi}{2} \left( \nu_1 \cos^2 \theta + \nu_2 \sin^2 \theta \sin^2 \phi \right) \nonumber \\
&& - \frac{1}{2} \nu_2 \cos(\beta  + \tau) \sin \alpha \sin \theta \sin \xi \sin2\phi,\nonumber\\
b_1 &=& \frac{1}{4} e^{-{\rm i} (\beta + \tau)} \biggr\{ - e^{{\rm i} \beta } \nu_2 \cos^2 \phi \sin2\alpha \sin \xi - e^{{\rm i} \tau} \biggr[ e^{{\rm i} (\beta + \tau)} \sin2\theta \sin \xi (\nu_1 - \nu_2 \sin^2 \phi) \nonumber \\
&&  + 2 \nu_2 \left(  \cos \theta \cos^2 \frac{\xi}{2} \sin \alpha - e^{2 i \beta} \cos \alpha \sin \theta \sin^2 \frac{\xi}{2} \right) \sin2\phi \biggr] \biggr\},\nonumber\\
c_1 &=& \frac{1}{4} e^{-{\rm i} (\beta + \tau)} \biggr[ - e^{{\rm i} (\beta + 2\tau)} \nu_2 \cos^2 \phi \sin2\alpha \sin \xi - e^{{\rm i} \beta } \sin2\theta \sin \xi (\nu_1 - \nu_2 \sin^2 \phi)  \nonumber \\
&&  - 2 e^{{\rm i} \tau} \nu_2 \left( e^{2 {\rm i} \beta} \cos \theta \cos^2 \frac{\xi}{2} \sin \alpha -  \cos \alpha \sin \theta \sin^2 \frac{\xi}{2} \right) \sin2\phi \biggr]=b_1^*,\nonumber\\
d_1 &=& \cos^2 \frac{\xi}{2} \left( \nu_1 \sin^2 \theta + \nu_2 \cos^2 \theta \sin^2 \phi \right) \nonumber \\
&& + \frac{1}{2} \nu_2 \cos \alpha \left( 2 \cos \alpha \cos^2 \phi \sin^2 \frac{\xi}{2} + \cos \theta \cos(\beta - \tau) \sin \xi \sin2\phi \right).
\end{eqnarray}
At the same time, we can obtain the matrix $\mathcal{M}$ and its conjugate transpose matrix $\mathcal{M}^\dagger=({\mathcal{M}^*})^T$ as
\begin{eqnarray}
\mathcal{M}= \begin{pmatrix}
 m_1 & m_2 \\
 m_3 & m_4  \\
\end{pmatrix}, \;\;\;\;
\mathcal{M}^\dagger= \begin{pmatrix}
 m_1^* & m_3^* \\
 m_2^* & m_4^*  \\
\end{pmatrix},
\end{eqnarray}
where
\begin{eqnarray}
m_1 &=& \frac{1}{4} \Big\{
   2 \sin \xi \left( \nu_1 \cos^2 \theta - \nu_2 \cos^2 \phi \sin^2 \alpha + \nu_2 \sin^2 \theta \sin^2 \phi \right)\nonumber\\
  &&\quad - \nu_2 \sin \alpha \sin \theta \left[ \cos(\beta - \xi + \tau) + \cos(\beta + \xi + \tau) + 2\mathrm{i} \sin(\beta + \tau) \right] \sin2\phi \Big\}\nonumber\\
&=& \frac{1}{4} \Big\{
   2 \sin \xi \left( \nu_1 \cos^2 \theta - \nu_2 \cos^2 \phi \sin^2 \alpha + \nu_2 \sin^2 \theta \sin^2 \phi \right)\nonumber\\
  && - \nu_2 \sin \alpha \sin \theta \left[ 2 \cos\xi \cos(\beta + \tau) + 2\mathrm{i} \sin(\beta + \tau) \right] \sin2\phi \Big\}\nonumber\\
&=& \frac{1}{2} \Big\{
    \sin \xi \left( \nu_1 \cos^2 \theta - \nu_2 \sin^2 \alpha \cos^2 \phi + \nu_2 \sin^2 \theta \sin^2 \phi \right)\nonumber\\
  &&- \nu_2 \sin \alpha \sin \theta \left[  \cos\xi \cos(\beta + \tau) + \mathrm{i} \sin(\beta + \tau) \right] \sin2\phi \Big\},
\end{eqnarray}
\begin{eqnarray}
 m_4&=& \frac{1}{2} \Big\{\nu_2 \cos^2 \alpha \cos^2 \phi \sin \xi + \nu_2 \cos \alpha \cos \theta  \left[ \cos \xi \cos(\beta - \tau) - \mathrm{i} \sin(\beta - \tau) \right] \sin2\phi \nonumber\\
  &&-  \sin \xi \left( \nu_1 \sin^2 \theta + \nu_2 \cos^2 \theta \sin^2 \phi \right)\Big\}\nonumber\\
&=& \frac{1}{2} \Big\{ \sin \xi \left( -\nu_1 \sin^2 \theta + \nu_2 \cos^2 \alpha \cos^2 \phi - \nu_2 \cos^2 \theta \sin^2 \phi \right)  \nonumber\\
  &&+ \nu_2 \cos \alpha \cos \theta  \left[ \cos \xi \cos(\beta - \tau) - \mathrm{i} \sin(\beta - \tau) \right] \sin2\phi,
\end{eqnarray}
\begin{eqnarray}
m_2 &=& \frac{1}{4}
     \mathrm{e}^{-\mathrm{i} \tau} \nu_2 (1 - \cos \xi) \cos^2 \phi \sin2\alpha - \frac{1}{8}\mathrm{e}^{\mathrm{i} \tau} (1 + \cos \xi) (2\nu_1 - \nu_2 + \nu_2 \cos2\phi) \sin2\theta\nonumber\\
  && + \frac{1}{4} \mathrm{e}^{-\mathrm{i} \beta } \nu_2 \left( \cos \theta \sin \alpha + \mathrm{e}^{2\mathrm{i}\beta} \cos \alpha \sin \theta \right) \sin \xi \sin2 \phi\nonumber\\
&=& \frac{1}{4}
     \mathrm{e}^{-\mathrm{i} \tau} (1 - \cos \xi) \nu_2 \cos^2 \phi \sin2\alpha - \frac{1}{8}\mathrm{e}^{\mathrm{i} \tau} (1 + \cos \xi) (2\nu_1 - 2\nu_2 {\sin^2 \phi}) \sin2\theta\nonumber\\
  && + \frac{1}{4}  \nu_2 \left( \mathrm{e}^{-\mathrm{i} \beta } \cos \theta \sin \alpha + \mathrm{e}^{\mathrm{i}\beta} \cos \alpha \sin \theta \right) \sin \xi \sin2 \phi\nonumber\\
&=& \frac{1}{4}
     \mathrm{e}^{-\mathrm{i} \tau} (1 - \cos \xi) \nu_2 \cos^2 \phi \sin2\alpha - \frac{1}{4}\mathrm{e}^{\mathrm{i} \tau} (1 + \cos \xi) (\nu_1 - \nu_2 {\sin^2 \phi}) \sin2\theta\nonumber\\
  && + \frac{1}{4}  \nu_2 \left( \mathrm{e}^{-\mathrm{i} \beta } \cos \theta \sin \alpha + \mathrm{e}^{\mathrm{i}\beta} \cos \alpha \sin \theta \right) \sin \xi \sin2 \phi,
\end{eqnarray}
\begin{eqnarray}
m_3 &=&
   - \frac{1}{4} \mathrm{e}^{\mathrm{i} \tau} \nu_2  (1+\cos\xi) \cos^2 \phi \sin2 \alpha+ \frac{1}{8} \mathrm{e}^{-\mathrm{i} \tau} (1-\cos\xi)(2\nu_1 - \nu_2 + \nu_2 \cos2\phi) \sin2\theta \nonumber\\
  && + \frac{1}{4} \mathrm{e}^{-\mathrm{i} \beta} \nu_2 \left( \mathrm{e}^{2\mathrm{i}\beta} \cos \theta \sin \alpha + \cos \alpha \sin \theta \right) \sin \xi \sin2\phi \nonumber\\
&=&
   - \frac{1}{4} \mathrm{e}^{\mathrm{i} \tau} (1+\cos\xi) \nu_2 \cos^2 \phi \sin2 \alpha+ \frac{1}{8} \mathrm{e}^{-\mathrm{i} \tau} (1-\cos\xi)(2\nu_1 - 2\nu_2 {\sin^2 \phi}) \sin2\theta \nonumber\\
  &&+ \frac{1}{4}  \nu_2 \left( \mathrm{e}^{\mathrm{i}\beta} \cos \theta \sin \alpha + \mathrm{e}^{-\mathrm{i} \beta} \cos \alpha \sin \theta \right) \sin \xi \sin2\phi \nonumber\\
&=&
   - \frac{1}{4} \mathrm{e}^{\mathrm{i} \tau} (1+\cos\xi) \nu_2 \cos^2 \phi \sin2 \alpha+ \frac{1}{4} \mathrm{e}^{-\mathrm{i} \tau} (1-\cos\xi)(\nu_1 - \nu_2 {\sin^2 \phi}) \sin2\theta \nonumber\\
  && + \frac{1}{4}  \nu_2 \left( \mathrm{e}^{\mathrm{i}\beta} \cos \theta \sin \alpha + \mathrm{e}^{-\mathrm{i} \beta} \cos \alpha \sin \theta \right) \sin \xi \sin2\phi.
\end{eqnarray}
Based on which, we obtain
\begin{eqnarray}\label{eq:cond-1a}
m_1-m_1^*&=& -{\rm i} \, \nu_2 \sin\alpha  \sin\theta  \sin(\beta+\tau) \sin2\phi,
\end{eqnarray}
\begin{eqnarray}\label{eq:cond-2a}
m_4-m_4^*&=& -{\rm i} \, \nu_2 \cos\alpha  \cos\theta \sin(\beta-\tau)  \sin2\phi,
\end{eqnarray}
\begin{eqnarray}\label{eq:cond-3a}
m_2 -m_3^{*}&=& \Big[\frac{1}{4}
     \mathrm{e}^{-\mathrm{i} \tau} (1 - \cos \xi) \nu_2 \cos^2 \phi \sin2\alpha - \frac{1}{4}\mathrm{e}^{\mathrm{i} \tau} (1 + \cos \xi) (\nu_1 - \nu_2 {\sin^2 \phi}) \sin2\theta\nonumber\\
  && + \frac{1}{4}  \nu_2 \left( \mathrm{e}^{-\mathrm{i} \beta } \cos \theta \sin \alpha + \mathrm{e}^{\mathrm{i}\beta} \cos \alpha \sin \theta \right) \sin \xi \sin2 \phi\Big]\nonumber\\
&&-\Big[ - \frac{1}{4} \mathrm{e}^{\mathrm{i} \tau} (1+\cos\xi) \nu_2 \cos^2 \phi \sin2 \alpha+ \frac{1}{4} \mathrm{e}^{-\mathrm{i} \tau} (1-\cos\xi)(\nu_1 - \nu_2 {\sin^2 \phi}) \sin2\theta \nonumber\\
  &&+ \frac{1}{4}  \nu_2 \left( \mathrm{e}^{\mathrm{i}\beta} \cos \theta \sin \alpha + \mathrm{e}^{-\mathrm{i} \beta} \cos \alpha \sin \theta \right) \sin \xi \sin2\phi\Big]^*\nonumber\\
&= &\frac{1}{4} e^{-{\rm i}\tau}  2 \, \nu_2 \cos^2\phi \sin2\alpha
-\frac{1}{4} e^{{\rm i}\tau} 2 (\nu_1 - \nu_2 {\sin^2 \phi}) \sin2\theta\nonumber\\
&= &\frac{1}{2} e^{-i\tau}  \nu_2 \cos^2\phi \sin2\alpha
-\frac{1}{2} e^{{\rm i}\tau} (\nu_1 - \nu_2 {\sin^2 \phi}) \sin2\theta\nonumber\\
&= &\frac{1}{2} e^{-{\rm i}\tau} \left[ \nu_2 \cos^2\phi \sin2\alpha
-e^{2 {\rm i} \tau} (\nu_1 - \nu_2 {\sin^2 \phi}) \sin2\theta\right].
\end{eqnarray}
\begin{eqnarray}
m_3-m_2^* &=&  -(m_2 -m_3^{*})^* \nonumber\\
&= & -\frac{1}{2} e^{{\rm i}\tau} \left[ \nu_2 \cos^2\phi \sin2\alpha
-e^{-2 {\rm i} \tau} (\nu_1 - \nu_2 {\sin^2 \phi}) \sin2\theta\right].
\end{eqnarray}

Here, we would like to prove the necessary condition. From the requirement $\mathcal{M}=\mathcal{M}^\dagger$, we can have
\begin{eqnarray}
&& m_1=m_1^*, \;\;\;\; m_4=m_4^*, \;\;\;\; m_2=m_3^*, \;\;\;\; m_3=m_2^*.
\end{eqnarray}
Because $e^{-{\rm i}\tau}\neq 0$, then $\mathcal{M}=\mathcal{M}^\dagger$ is equivalent to the following conditions
\begin{subequations}
\begin{eqnarray}
&&\nu_2 \sin\alpha \sin\theta  \sin(\beta+\tau) \sin2\phi=0, \label{eq:ss-1}\\
&& \nu_2 \cos\alpha \cos\theta \sin(\beta-\tau) \sin2\phi=0,  \label{eq:ss-2}\\
&& \nu_2 \cos^2\phi \sin2\alpha - e^{2{\rm i}\tau} (\nu_1 - \nu_2 \sin^2\phi) \sin2\theta=0.\label{eq:ss-3}
\end{eqnarray}
\end{subequations}
Note that the condition $m_2^*=m_3$ is equivalent to the condition $m_3^*=m_2$.

Let us study the first two conditions (i.e., Eq. (\ref{eq:ss-1}) and Eq. (\ref{eq:ss-2})). Their solutions can classified into 9 possible cases:

\begin{enumerate}
  \item $\nu_2=0$.
  \item $\sin2\phi=0$.
  \item $\sin\alpha=0$ and $\cos\theta=0$.
  \item $\sin\theta=0$ and $\cos\alpha=0$.
  \item $\sin\alpha=0$ and $\sin(\beta-\tau)=0$.
  \item $\cos\alpha=0$ and $\sin(\beta+\tau)=0$.
  \item $\sin\theta=0$ and $\sin(\beta-\tau)=0$.
  \item $\cos\theta=0$ and $\sin(\beta+\tau)=0$.
  \item $\sin(\beta+\tau)=0$ and $\sin(\beta-\tau)=0$.
\end{enumerate}
In the following, let us analysis the condition (\ref{eq:ss-3}) based on these 9 cases.

\textcolor{blue}{\emph{Case 1.---} $\nu_2=0$.} One has $\nu_1=1-\nu_2=1$. In this case $\rho=|\psi_1\rangle \langle\psi_1|$ is a pure state with rank 1. The condition (\ref{eq:ss-3}) becomes
\begin{eqnarray}
&&\nu_2 \cos^2\phi \sin2\alpha - e^{2{\rm i} \tau} (\nu_1 - \nu_2 \sin^2\phi) \sin2\theta=-e^{2{\rm i}  \tau} \sin2\theta=-e^{2{\rm i}  \tau} 2 \sin\theta \cos\theta=0,
\end{eqnarray}
which yields $\sin\theta=0$ or $\cos\theta=0$, then $|\psi_1\rangle=|00\rangle$ or $|11\rangle$. Hence $\rho=|\psi_1\rangle \langle\psi_1|$ is a separable state.

\textcolor{blue}{\emph{Case 2.---} $\sin2\phi=0$.} Because $\phi\in[0, \pi/2]$, then one has $\phi=0$ or $\phi=\pi/2$. If $\phi=0$, then $\cos^2\phi=1$ and $\sin^2\phi=0$. The condition (\ref{eq:ss-3}) becomes
\begin{eqnarray}\label{eq:r-1}
&&\nu_2 \cos^2\phi \sin2\alpha - e^{2{\rm i} \tau} (\nu_1 - \nu_2 \sin^2\phi) \sin2\theta=\nu_2 \sin2\alpha - e^{2{\rm i} \tau} \nu_1  \sin2\theta =0.
\end{eqnarray}

\emph{Case 2-A.---} If $\nu_1=0$, then $\nu_2=1-\nu_1=1$. From Eq. (\ref{eq:r-1}) one has $\sin2\alpha=0$. In this case, $\rho=|\psi_2\rangle\langle\psi_2|$, with  $\ket{\psi_2} =\ket{01}$ or $\ket{10}$. Thus $\rho$ is a separable state.

\emph{Case 2-B.---} If $\nu_1 \neq 0$ and $\sin2\theta=2\sin\theta\cos\theta=0$, from Eq. (\ref{eq:r-1}) one has $\sin2\alpha=2\sin\alpha\cos\alpha=0$. Namely, we have $\sin\theta=0$ or $\cos\theta=0$, and $\sin\alpha=0$ or $\cos\alpha=0$. There are four possible cases: $\{\sin\theta=0, \sin\alpha=0\}$, $\{\sin\theta=0, \cos\alpha=0\}$, $\{\cos\theta=0, \sin\alpha=0\}$, and $\{\cos\theta=0, \cos\alpha=0\}$.

For the case $\{\sin\theta=0, \sin\alpha=0\}$, one has $\cos\theta= 1$ and $\cos\alpha=1$, then
\begin{eqnarray}
\rho&=&\nu_1  \ket{\psi_1}\bra{\psi_1}+\nu_2 \ket{\psi_2}\bra{\psi_2}\nonumber\\
&=& \nu_1 \cos\theta|00\rangle\langle 00|+\nu_2 \cos\alpha |01\rangle\langle 01|\nonumber \\
&=& |0\rangle\langle 0| \otimes  \left(\nu_1 |0\rangle\langle 0|+\nu_2  |1\rangle\langle 1|\right),
\end{eqnarray}
thus $\rho$ is a separable state. For other three cases, i.e., $\{\sin\theta=0, \cos\alpha=0\}$, $\{\cos\theta=0, \sin\alpha=0\}$, $\{\cos\theta=0, \cos\alpha=0\}$, one can similarly prove that $\rho$ is a separable state.

\emph{Case 2-C.---}If $\nu_1 \neq 0$ and $\sin2\theta \neq 0$, from Eq. (\ref{eq:r-1}) one has
\begin{eqnarray}
&& \left(\nu_2 \sin2\alpha - \cos{2\tau} \times \nu_1  \sin2\theta\right) - {\rm i} \left(\sin{2\tau}\times  \nu_1  \sin2\theta\right) =0,
\end{eqnarray}
i.e.,
\begin{eqnarray}
&& \sin{2\tau}=0, \nonumber\\
&& \nu_2 \sin2\alpha - \cos{2\tau} \times \nu_1  \sin2\theta=0.
\end{eqnarray}
One can choose $\tau=0$, thus
\begin{eqnarray}\label{eq:r-2}
&& \nu_2 \sin2\alpha = \cos{2\tau} \times \nu_1  \sin2\theta= \nu_1  \sin2\theta.
\end{eqnarray}
As we have known that the concurrence of the state $\rho$ is given by
\begin{eqnarray}
\mathcal{C}=\left|\nu_2 \sin2\alpha - \nu_1 \sin2\theta \right|,
\end{eqnarray}
which is obviously $0$ due to Eq. (\ref{eq:r-2}). Thus the state $\rho$ is a separable state.

\emph{Case 2-D.---}If $\phi=\pi/2$, then $\cos^2\phi=0$ and $\sin^2\phi=1$. Accordingly, the condition (\ref{eq:ss-3}) becomes
\begin{eqnarray}
&&\nu_2 \cos^2\phi \sin2\alpha - e^{2{\rm i}\tau} (\nu_1 - \nu_2 \sin^2\phi) \sin2\theta= - e^{2 {\rm i}\tau} (\nu_1 - \nu_2) \sin2\theta=0.
\end{eqnarray}
Because $e^{2{\rm i} \tau} \neq 0$, then one obtains
\begin{eqnarray}\label{eq:r-3}
(\nu_1-\nu_2) \sin2\theta=0.
\end{eqnarray}
In this situation, the state $\rho$ is given by
\begin{eqnarray}
\rho=\nu_1  \ket{\psi_1}\bra{\psi_1}+\nu_2 \ket{\psi_2}\bra{\psi_2},
\end{eqnarray}
with
\begin{eqnarray}
&&\ket{\psi_1} =\cos \theta \ket{00}+\sin \theta\ket{11}, \nonumber\\
&& \ket{\psi_2} =\sin \theta \ket{00}-\cos \theta\ket{11}.
\end{eqnarray}
The concurrence of the state $\rho$ is given by
\begin{eqnarray}
\mathcal{C}=|(\nu_1-\nu_2)\sin2\theta|,
\end{eqnarray}
which is obviously $0$ due to Eq. (\ref{eq:r-3}). Thus the state $\rho$ is a separable state.

\textcolor{blue}{\emph{Case 3.---} $\sin\alpha=0$ and $\cos\theta=0$.} The condition (\ref{eq:ss-3}) becomes
\begin{eqnarray}
&&\nu_2 \cos^2\phi \sin2\alpha - e^{2{\rm i} \tau} (\nu_1 - \nu_2 \sin^2\phi) \sin2\theta= 0,
\end{eqnarray}
which means the condition (\ref{eq:ss-3}) is satisfied. In this situation, the state $\rho$ is given by
\begin{eqnarray}
\rho=\nu_1  \ket{\psi_1}\bra{\psi_1}+\nu_2 \ket{\psi_2}\bra{\psi_2},
\end{eqnarray}
where
\begin{eqnarray}
\ket{\psi_1} &=& \cos \theta \ket{00}+\sin \theta\ket{11}=\sin \theta \ket{11}=\sin\theta |1\rangle \otimes |1\rangle, \nonumber\\
 \ket{\psi_2} &=& \cos\phi(\cos \alpha \ket{01} +\sin\alpha\ket{10})+ e^{{\rm i}\beta}\sin\phi(\sin \theta \ket{00}-\cos \theta\ket{11})\nonumber\\
&=&
\cos\phi \cos\alpha\ket{01}+ e^{{\rm i} \beta}\sin\phi\sin \theta\ket{00}\nonumber\\
&=& |0\rangle \otimes (\cos\phi \cos\alpha\ket{1}+ e^{{\rm i} \beta}\sin\phi\sin \theta\ket{0}),
\end{eqnarray}
with $\cos\alpha=1$ and $\sin\theta= 1$. Because both $\ket{\psi_1}\bra{\psi_1}$ and $\ket{\psi_2}\bra{\psi_2}$ are separable states, thus the state $\rho$ is a separable state.


\textcolor{blue}{\emph{Case 4.---} $\sin\theta=0$ and $\cos\alpha=0$.} Accordingly, The condition (\ref{eq:ss-3}) becomes
\begin{eqnarray}
&&\nu_2 \cos^2\phi \sin2\alpha - e^{2{\rm i}\tau} (\nu_1 - \nu_2 \sin^2\phi) \sin2\theta= 0,
\end{eqnarray}
which means the condition (\ref{eq:ss-3}) is satisfied. In this situation, the state $\rho$ is given by
\begin{eqnarray}
\rho=\nu_1  \ket{\psi_1}\bra{\psi_1}+\nu_2 \ket{\psi_2}\bra{\psi_2},
\end{eqnarray}
where
\begin{eqnarray}
\ket{\psi_1} &=& \cos \theta \ket{00}+\sin \theta\ket{11}=\cos \theta \ket{00}=\cos\theta |0\rangle \otimes |0\rangle, \nonumber\\
 \ket{\psi_2} &=& \cos\phi(\cos \alpha \ket{01} +\sin\alpha\ket{10})+ e^{{\rm i}\beta}\sin\phi(\sin \theta \ket{00}-\cos \theta\ket{11})\nonumber\\
&=&
\cos\phi \sin\alpha\ket{10}- e^{{\rm i}\beta}\sin\phi\cos \theta\ket{11}\nonumber\\
&=& |1\rangle \otimes (\cos\phi \sin\alpha\ket{0}- e^{{\rm i}\beta}\sin\phi\cos \theta\ket{1}),
\end{eqnarray}
with $\cos\theta=1$ and $\sin\alpha= 1$. Because both $\ket{\psi_1}\bra{\psi_1}$ and $\ket{\psi_2}\bra{\psi_2}$ are separable states, thus the state $\rho$ is a separable state.

\textcolor{blue}{\emph{Case 5.---} $\sin\alpha=0$ and $\sin(\beta-\tau)=0$.} Accordingly, The condition (\ref{eq:ss-3}) becomes
\begin{eqnarray}
&&\nu_2 \cos^2\phi \sin2\alpha - e^{2{\rm i} \tau} (\nu_1 - \nu_2 \sin^2\phi) \sin2\theta=- e^{2{\rm i} \tau} (\nu_1 - \nu_2 \sin^2\phi) \sin2\theta=0.
\end{eqnarray}
Because $e^{2{\rm i} \tau} \neq 0$, then one obtains
\begin{eqnarray}\label{eq:r-4}
(\nu_1 - \nu_2 \sin^2\phi) \sin2\theta=0.
\end{eqnarray}
In this situation, the state $\rho$ is given by
\begin{eqnarray}
\rho=\nu_1  \ket{\psi_1}\bra{\psi_1}+\nu_2 \ket{\psi_2}\bra{\psi_2},
\end{eqnarray}
where
\begin{eqnarray}
\ket{\psi_1} &=& \cos \theta \ket{00}+\sin \theta\ket{11}, \nonumber\\
 \ket{\psi_2} &=& \cos\phi(\cos \alpha \ket{01} +\sin\alpha\ket{10})+ e^{{\rm i}\beta}\sin\phi(\sin \theta \ket{00}-\cos \theta\ket{11})\nonumber\\
&=& \cos\phi \cos \alpha \ket{01} + e^{{\rm i} \beta}\sin\phi(\sin \theta \ket{00}-\cos \theta\ket{11}),
\end{eqnarray}
with $\cos\alpha= 1$. The concurrence of the state $\rho$ is given by
\begin{eqnarray}
\mathcal{C} &=& \sqrt{-s_2-2\sqrt{s_1}},
\end{eqnarray}
where
\begin{eqnarray}
 s_2 &=& -\frac{1}{2} \nu_2^2 \sin2 \alpha  \sin 2 \theta \sin ^22 \phi \cos 2 \beta-\nu_2^2 \sin ^22 \alpha  \cos ^4 \phi
 -\sin ^2 2 \theta  \left(\nu_1^2-\nu_1 \nu_2 \sin ^2\phi +\nu_2^2 \sin ^4\phi \right)
\nonumber\\
&&-\frac{1}{2} \nu_1 \nu_2 (\cos 4 \theta+3) \sin ^2 \phi, \nonumber\\
&=&-\sin ^2 2 \theta  \left(\nu_1^2-\nu_1 \nu_2 \sin ^2\phi +\nu_2^2 \sin ^4\phi \right)
-\frac{1}{2} \nu_1 \nu_2 (\cos 4 \theta +3) \sin ^2 \phi, \nonumber\\
&=&-\sin ^2 2 \theta  \left(\nu_1^2-\nu_1 \nu_2 \sin ^2\phi +\nu_2^2 \sin ^4\phi \right)
-\frac{1}{2} \nu_1 \nu_2 [1-2\sin^2 2 \theta +3] \sin ^2 \phi, \nonumber\\
&=&-\sin ^2 2 \theta \left(\nu_1^2-\nu_1 \nu_2 \sin ^2\phi +\nu_2^2 \sin ^4\phi \right)
- \nu_1 \nu_2 [-\sin^2 2 \theta +2] \sin ^2 \phi, \nonumber\\
&=&-\sin ^2 2 \theta \left(\nu_1^2-2\nu_1 \nu_2 \sin ^2\phi +\nu_2^2 \sin ^4\phi \right)
- 2 \nu_1 \nu_2 \sin ^2 \phi\nonumber\\
&=&- \left(\nu_1- \nu_2 \sin ^2\phi\right)^2 \sin ^2 2 \theta
- 2 \nu_1 \nu_2 \sin ^2 \phi,
\end{eqnarray}
\begin{eqnarray}
s_1 &=& \frac{1}{2} \nu_1^2 \nu_2^2 \left(\sin 2 \alpha  \sin 2 \theta  \sin ^2 2 \phi  \cos 2 \beta+2 \sin ^2 2 \alpha \sin ^2 2 \theta  \cos ^4\phi +2 \sin ^4\phi \right) = \nu_1^2 \nu_2^2 \sin ^4\phi.
\end{eqnarray}
Then we have
\begin{eqnarray}
\mathcal{C} &=& \sqrt{-s_2-2\sqrt{s_1}} \nonumber\\
&=& \sqrt{ \left(\nu_1- \nu_2 \sin ^2\phi\right)^2 \sin ^2 2 \theta
+ 2 \nu_1 \nu_2 \sin ^2 \phi-2 \nu_1 \nu_2 \sin ^2 \phi} \nonumber\\
&=& |(\nu_1 - \nu_2 \sin^2\phi) \sin2\theta|,
\end{eqnarray}
which is obviously 0 due to Eq. (\ref{eq:r-4}). Thus the state $\rho$ is a separable state.

\textcolor{blue}{\emph{Case 6.---} $\cos\alpha=0$ and $\sin(\beta+\tau)=0$.} Accordingly, the condition (\ref{eq:ss-3}) becomes
\begin{eqnarray}
&&\nu_2 \cos^2\phi \sin2\alpha - e^{2 {\rm i} \tau} (\nu_1 - \nu_2 \sin^2\phi) \sin2\theta=- e^{2 {\rm i} \tau} (\nu_1 - \nu_2 \sin^2\phi) \sin2\theta=0.
\end{eqnarray}
Because $e^{2 {\rm i} \tau} \neq 0$, then one obtains
\begin{eqnarray}\label{eq:r-5}
(\nu_1 - \nu_2 \sin^2\phi) \sin2\theta=0.
\end{eqnarray}
In this situation, the state $\rho$ is given by
\begin{eqnarray}
\rho=\nu_1  \ket{\psi_1}\bra{\psi_1}+\nu_2 \ket{\psi_2}\bra{\psi_2},
\end{eqnarray}
where
\begin{eqnarray}
\ket{\psi_1} &=& \cos \theta \ket{00}+\sin \theta\ket{11}, \nonumber\\
 \ket{\psi_2} &=& \cos\phi(\cos \alpha \ket{01} +\sin\alpha\ket{10})+ e^{{\rm i}\beta}\sin\phi(\sin \theta \ket{00}-\cos \theta\ket{11})\nonumber\\
&=& \cos\phi \sin \alpha \ket{10} + e^{{\rm i} \beta}\sin\phi(\sin \theta \ket{00}-\cos \theta\ket{11}),
\end{eqnarray}
with $\sin\alpha= 1$. The concurrence of the state $\rho$ is given by
\begin{eqnarray}
\mathcal{C} &=& \sqrt{-s_2-2\sqrt{s_1}},
\end{eqnarray}
where
\begin{eqnarray}
 s_2 &=& -\frac{1}{2} \nu_2^2 \sin 2 \alpha \sin 2 \theta  \sin ^2 2 \phi \cos 2 \beta-\nu_2^2 \sin ^2 2 \alpha \cos ^4 \phi
 -\sin ^2 2 \theta \left(\nu_1^2-\nu_1 \nu_2 \sin ^2\phi +\nu_2^2 \sin ^4\phi \right)
\nonumber\\
&&-\frac{1}{2} \nu_1 \nu_2 (\cos 4 \theta +3) \sin ^2 \phi, \nonumber\\
&=&- \left(\nu_1- \nu_2 \sin ^2\phi\right)^2 \sin ^2 2 \theta
- 2 \nu_1 \nu_2 \sin ^2 \phi,
\end{eqnarray}
\begin{eqnarray}
s_1 &=& \frac{1}{2} \nu_1^2 \nu_2^2 \left(\sin2 \alpha \sin 2 \theta \sin ^2 2 \phi \cos 2 \beta +2 \sin ^2 2 \alpha
 \sin ^2 2 \theta \cos ^4\phi +2 \sin ^4\phi \right) = \nu_1^2 \nu_2^2 \sin ^4\phi.
\end{eqnarray}
Then we have
\begin{eqnarray}
\mathcal{C} &=& \sqrt{-s_2-2\sqrt{s_1}} \nonumber\\
&=& \sqrt{ \left(\nu_1- \nu_2 \sin ^2\phi\right)^2 \sin ^2 2 \theta
+ 2 \nu_1 \nu_2 \sin ^2 \phi-2 \nu_1 \nu_2 \sin ^2 \phi} \nonumber\\
&=& |(\nu_1 - \nu_2 \sin^2\phi) \sin2\theta|,
\end{eqnarray}
which is obviously 0 due to Eq. (\ref{eq:r-5}). Thus the state $\rho$ is a separable state.

\textcolor{blue}{\emph{Case 7.---} $\sin\theta=0$ and $\sin(\beta-\tau)=0$.} Accordingly, the condition (\ref{eq:ss-3}) becomes
\begin{eqnarray}
&&\nu_2 \cos^2\phi \sin2\alpha - e^{2{\rm i}\tau} (\nu_1 - \nu_2 \sin^2\phi) \sin2\theta=\nu_2 \cos^2\phi \sin2\alpha=0.
\end{eqnarray}

\emph{Case 7-A.---} $\nu_2=0$. In this case, $\rho=\ket{\psi_1}\bra{\psi_1}$, with $|\psi_1\rangle=|00\rangle$, thus the state $\rho$ is a separable state.

\emph{Case 7-B.---} $\cos\phi=0$. One can have $\sin\phi= 1$. In this situation, the state $\rho$ is given by
\begin{eqnarray}
\rho=\nu_1  \ket{\psi_1}\bra{\psi_1}+\nu_2 \ket{\psi_2}\bra{\psi_2},
\end{eqnarray}
where
\begin{eqnarray}
\ket{\psi_1} &=& \cos \theta \ket{00}+\sin \theta\ket{11} = \cos \theta \ket{00}= \cos \theta |0\rangle \otimes |0\rangle, \nonumber\\
 \ket{\psi_2} &=& \cos\phi(\cos \alpha \ket{01} +\sin\alpha\ket{10})+ e^{{\rm i}\beta}\sin\phi(\sin \theta \ket{00}-\cos \theta\ket{11})\nonumber\\
&=& -e^{{\rm i} \beta}\sin\phi \cos \theta \ket{11}=-e^{ {\rm i} \beta}\sin\phi \cos \theta |1\rangle\otimes |1\rangle,
\end{eqnarray}
with $\sin\phi= 1$ and $\cos\theta= 1$. Because both $|\psi_1\rangle$ and $|\psi_2\rangle$ are separable states, thus the state $\rho$ is a separable state.

\emph{Case 7-C.---} $\sin2\alpha=2 \sin\alpha\cos\alpha=0$. If $\sin\alpha=0$, the state $\rho$ becomes
\begin{eqnarray}
\rho=\nu_1  \ket{\psi_1}\bra{\psi_1}+\nu_2 \ket{\psi_2}\bra{\psi_2},
\end{eqnarray}
where
\begin{eqnarray}
\ket{\psi_1} &=& \cos \theta \ket{00}+\sin \theta\ket{11} = \cos \theta \ket{00}= \cos \theta |0\rangle \otimes |0\rangle, \nonumber\\
 \ket{\psi_2} &=& \cos\phi(\cos \alpha \ket{01} +\sin\alpha\ket{10})+ e^{{\rm i}\beta}\sin\phi(\sin \theta \ket{00}-\cos \theta\ket{11})\nonumber\\
&=& \cos\phi \cos \alpha \ket{01}- e^{{\rm i}\beta}\sin\phi \cos \theta \ket{11} \nonumber\\
&=& (\cos\phi \cos \alpha \ket{1}- e^{{\rm i} \beta}\sin\phi \cos \theta \ket{1}) \otimes |1\rangle,
\end{eqnarray}
with $\cos\alpha= 1$ and $\cos\theta= 1$. Because both $|\psi_1\rangle$ and $|\psi_2\rangle$ are separable states, thus the state $\rho$ is a separable state. Similarly, if $\cos\alpha=0$, one can prove the state $\rho$ is a separable state.

\textcolor{blue}{\emph{Case 8.---} $\cos\theta=0$ and $\sin(\beta+\tau)=0$.} Accordingly, the condition (\ref{eq:ss-3}) becomes
\begin{eqnarray}
&&\nu_2 \cos^2\phi \sin2\alpha - e^{2{\rm i} \tau} (\nu_1 - \nu_2 \sin^2\phi) \sin2\theta=\nu_2 \cos^2\phi \sin2\alpha=0.
\end{eqnarray}

\emph{Case 8-A.---} $\nu_2=0$. In this case, $\rho=\ket{\psi_1}\bra{\psi_1}$, with $|\psi_1\rangle=|00\rangle$, thus the state $\rho$ is a separable state.

\emph{Case 8-B.---} $\cos\phi=0$. One can have $\sin\phi= 1$. In this situation, the state $\rho$ is given by
\begin{eqnarray}
\rho=\nu_1  \ket{\psi_1}\bra{\psi_1}+\nu_2 \ket{\psi_2}\bra{\psi_2},
\end{eqnarray}
where
\begin{eqnarray}
\ket{\psi_1} &=& \cos \theta \ket{00}+\sin \theta\ket{11} = \sin \theta \ket{11}= \sin \theta |1\rangle \otimes |1\rangle, \nonumber\\
 \ket{\psi_2} &=& \cos\phi(\cos \alpha \ket{01} +\sin\alpha\ket{10})+ e^{{\rm i} \beta}\sin\phi(\sin \theta \ket{00}-\cos \theta\ket{11})\nonumber\\
&=& e^{{\rm i} \beta}\sin\phi \sin \theta \ket{00}=e^{{\rm i} \beta}\sin\phi \sin \theta |0\rangle\otimes |0\rangle,
\end{eqnarray}
with $\sin\phi= 1$ and $\sin\theta= 1$. Because both $|\psi_1\rangle$ and $|\psi_2\rangle$ are separable states, thus the state $\rho$ is a separable state.

\emph{Case 8-C.---} $\sin2\alpha=2 \sin\alpha\cos\alpha=0$. If $\sin\alpha=0$, the state $\rho$ becomes
\begin{eqnarray}
\rho=\nu_1  \ket{\psi_1}\bra{\psi_1}+\nu_2 \ket{\psi_2}\bra{\psi_2},
\end{eqnarray}
where
\begin{eqnarray}
\ket{\psi_1} &=& \cos \theta \ket{00}+\sin \theta\ket{11} = \sin \theta \ket{11}= \sin \theta |1\rangle \otimes |1\rangle, \nonumber\\
 \ket{\psi_2} &=& \cos\phi(\cos \alpha \ket{01} +\sin\alpha\ket{10})+ e^{{\rm i}\beta}\sin\phi(\sin \theta \ket{00}-\cos \theta\ket{11})\nonumber\\
&=& \cos\phi \cos \alpha \ket{01}+ e^{{\rm i}\beta}\sin\phi \sin \theta \ket{00} \nonumber\\
&=& |0\rangle \otimes (\cos\phi \cos \alpha \ket{1}+ e^{{\rm i}\beta}\sin\phi \sin \theta \ket{0}),
\end{eqnarray}
with $\cos\alpha= 1$ and $\sin\theta= 1$. Because both $|\psi_1\rangle$ and $|\psi_2\rangle$ are separable states, thus the state $\rho$ is a separable state. Similarly, if $\cos\alpha=0$, one can prove the state $\rho$ is a separable state.

\textcolor{blue}{\emph{Case 9.---} $\sin(\beta+\tau)=0$ and $\sin(\beta-\tau)=0$.} We can have
\begin{eqnarray}
&& \sin(\beta+\tau) = \sin\beta\cos\tau + \cos\beta\sin\tau=0, \nonumber\\
&& \sin(\beta-\tau) = \sin\beta\cos\tau - \cos\beta\sin\tau=0,
\end{eqnarray}
i.e.,
\begin{eqnarray}
&& \sin\beta\cos\tau =0, \nonumber\\
&& \cos\beta\sin\tau=0,
\end{eqnarray}
i.e., we have two possible cases: $\{\sin\beta=0, \sin\tau=0\}$ and $\{\cos\beta=0, \cos\tau=0\}$.

\emph{Case 9-A.---} $\sin\beta=0$ and $\sin\tau=0$. There are four possible cases: $\{\beta=0, \tau=0\}$, $\{\beta=0, \tau=\pi\}$,
$\{\beta=\pi, \tau=0\}$, and $\{\beta=\pi, \tau=\pi\}$.

For the case of $\{\beta=0, \tau=0\}$, the condition (\ref{eq:ss-3}) becomes
\begin{eqnarray}
&&\nu_2 \cos^2\phi \sin2\alpha - e^{2{\rm i}\tau} (\nu_1 - \nu_2 \sin^2\phi) \sin2\theta=\nu_2 \cos^2\phi \sin2\alpha -(\nu_1 - \nu_2 \sin^2\phi) \sin2\theta=0,
\end{eqnarray}
i.e.,
\begin{eqnarray}\label{eq:r-6}
&& (\nu_1 - \nu_2 \sin^2\phi) \sin2\theta=\nu_2 \cos^2\phi \sin2\alpha.
\end{eqnarray}
In this situation, the state $\rho$ is given by
\begin{eqnarray}
\rho=\nu_1  \ket{\psi_1}\bra{\psi_1}+\nu_2 \ket{\psi_2}\bra{\psi_2},
\end{eqnarray}
where
\begin{eqnarray}
\ket{\psi_1} &=& \cos \theta \ket{00}+\sin \theta\ket{11}, \nonumber\\
 \ket{\psi_2} &=& \cos\phi(\cos \alpha \ket{01} +\sin\alpha\ket{10})+ e^{{\rm i}\beta}\sin\phi(\sin \theta \ket{00}-\cos \theta\ket{11})\nonumber\\
&=& \cos\phi(\cos \alpha \ket{01} +\sin\alpha\ket{10})+ \sin\phi(\sin \theta \ket{00}-\cos \theta\ket{11}).
\end{eqnarray}
The concurrence of the state $\rho$ is given by
\begin{eqnarray}
\mathcal{C} &=& \sqrt{-s_2-2\sqrt{s_1}},
\end{eqnarray}
where
\begin{eqnarray}
 s_2 &=& -\frac{1}{2} \nu_2^2 \sin 2 \alpha  \sin 2 \theta  \sin ^2 2 \phi  \cos 2 \beta-\nu_2^2 \sin ^2 2 \alpha \cos ^4 \phi
 -\sin ^2 2 \theta \left(\nu_1^2-\nu_1 \nu_2 \sin ^2\phi +\nu_2^2 \sin ^4\phi \right)
\nonumber\\
&&-\frac{1}{2} \nu_1 \nu_2 (\cos 4 \theta +3) \sin ^2 \phi, \nonumber\\
&=& -\frac{1}{2} \nu_2^2 \sin 2 \alpha \sin 2 \theta \sin ^2 2 \phi-\nu_2^2 \sin ^2 2 \alpha \cos ^4 \phi
 -\sin ^2 2 \theta \left(\nu_1^2-\nu_1 \nu_2 \sin ^2\phi +\nu_2^2 \sin ^4\phi \right)
\nonumber\\
&&-\frac{1}{2} \nu_1 \nu_2 (-2\sin^2 2 \theta+4) \sin ^2 \phi, \nonumber\\
&=& -\frac{1}{2} \nu_2^2 \sin 2 \alpha \sin 2 \theta \sin ^2 2 \phi-\nu_2^2 \sin ^2 2 \alpha \cos ^4 \phi
 -\sin ^2 2 \theta \left(\nu_1^2-\nu_1 \nu_2 \sin ^2\phi +\nu_2^2 \sin ^4\phi \right)
\nonumber\\
&&+ \nu_1 \nu_2 (\sin^2 2 \theta -2) \sin ^2 \phi, \nonumber\\
&=& -\frac{1}{2} \nu_2^2 \sin 2 \alpha  \sin 2 \theta  \sin ^2 2 \phi-\nu_2^2 \sin ^2 2 \alpha \cos ^4 \phi
 -\sin ^2 2 \theta \left(\nu_1^2-2\nu_1 \nu_2 \sin ^2\phi +\nu_2^2 \sin ^4\phi \right)-2 \nu_1 \nu_2 \sin ^2 \phi
\nonumber\\
&=& -\frac{1}{2} \nu_2^2 \sin 2 \alpha \sin 2 \theta \sin ^2 2 \phi-\nu_2^2 \sin ^2 2 \alpha \cos ^4 \phi
 -\sin ^2 2 \theta \left(\nu_1-\nu_2 \sin ^2\phi \right)^2-2 \nu_1 \nu_2 \sin ^2 \phi,
\end{eqnarray}

\begin{eqnarray}
s_1 &=& \frac{1}{2} \nu_1^2 \nu_2^2 \left(\sin 2 \alpha \sin 2 \theta \sin ^2 2 \phi \cos 2 \beta +2 \sin ^2 2 \alpha  \sin ^2 2 \theta \cos ^4\phi +2 \sin ^4\phi \right)\nonumber\\
&=& \frac{1}{2} \nu_1^2 \nu_2^2 \left(\sin 2 \alpha  \sin 2 \theta  \sin ^2 2 \phi+2 \sin ^2 2 \alpha  \sin ^2 2 \theta \cos ^4\phi +2 \sin ^4\phi \right)\nonumber\\
&=& \frac{1}{2} \nu_1^2 \nu_2^2 \left(4 \sin 2 \alpha \sin 2 \theta  \sin ^2\phi \cos^2\phi +2 \sin ^2 2 \alpha \sin ^2 2 \theta \cos ^4\phi +2 \sin ^4\phi \right)\nonumber\\
&=&\nu_1^2 \nu_2^2 \left(\sin ^2 2 \alpha \sin ^2 2 \theta \cos ^4\phi +2 \sin 2 \alpha  \sin 2 \theta  \sin ^2\phi \cos^2\phi +  \sin ^4\phi \right)\nonumber\\
&=& \nu_1^2 \nu_2^2 \left(\sin 2 \alpha  \sin 2 \theta  \cos^2\phi +  \sin ^2\phi \right)^2.
\end{eqnarray}
Due to Eq. (\ref{eq:r-6}), we have
\begin{eqnarray}
 s_2 &=& -\frac{1}{2} \nu_2^2 \sin 2 \alpha  \sin 2 \theta  \sin ^2 2 \phi -\nu_2^2 \sin ^22 \alpha  \cos ^4 \phi
 -\sin ^22 \theta \left(\nu_1-\nu_2 \sin ^2\phi \right)^2-2 \nu_1 \nu_2 \sin ^2 \phi\nonumber\\
&=& -\frac{1}{2} \nu_2^2 \sin 2 \alpha  \sin 2 \theta  \sin ^22 \phi -2\nu_2^2 \sin ^2 2 \alpha  \cos ^4 \phi
 -2 \nu_1 \nu_2 \sin ^2 \phi\nonumber\\
&=& -2 \nu_2^2 \sin 2 \alpha  \sin 2 \theta  \sin ^2 \phi \cos^2 \phi-2\nu_2^2 \sin ^2 2 \alpha \cos ^4 \phi
 -2 \nu_1 \nu_2 \sin ^2 \phi.
\end{eqnarray}

To prove that the state $\rho$ is a separable state, we only need to prove the concurrence $\mathcal{C}=0$, i.e.,
\begin{eqnarray}
-s_2-2\sqrt{s_1}=0,
\end{eqnarray}
i.e.,
\begin{eqnarray}
s_2^2-4s_1=0,
\end{eqnarray}
i.e.,
\begin{eqnarray}
\left( \nu_2^2 \sin 2 \alpha  \sin 2 \theta  \sin ^2 \phi \cos^2 \phi+\nu_2^2 \sin ^2 2 \alpha  \cos ^4 \phi
 + \nu_1 \nu_2 \sin ^2 \phi\right)^2-\nu_1^2 \nu_2^2 \left(\sin 2 \alpha  \sin 2 \theta \cos^2\phi +  \sin ^2\phi \right)^2=0, \nonumber\\
\end{eqnarray}
i.e.,
\begin{eqnarray}
T_1^2-T_2^2=(T_1+T_2)(T_1-T_2)=0,
\end{eqnarray}
with
\begin{eqnarray}
&& T_1= \nu_2^2 \sin 2 \alpha \sin 2 \theta  \sin ^2 \phi \cos^2 \phi+\nu_2^2 \sin ^2 2 \alpha  \cos ^4 \phi
 + \nu_1 \nu_2 \sin ^2 \phi, \nonumber\\
 &&T_2= \nu_1 \nu_2 \left(\sin 2 \alpha  \sin 2 \theta  \cos^2\phi +  \sin ^2\phi \right).
\end{eqnarray}
We only need to prove
\begin{eqnarray}
T_1-T_2=0.
\end{eqnarray}
We can have
\begin{eqnarray}
&&T_1-T_2 \nonumber\\
&=&\left( \nu_2^2 \sin 2 \alpha  \sin 2 \theta  \sin ^2 \phi \cos^2 \phi+\nu_2^2 \sin ^22 \alpha  \cos ^4 \phi
 + \nu_1 \nu_2 \sin ^2 \phi\right)-\nu_1 \nu_2( \sin 2 \alpha  \sin 2 \theta \cos^2\phi +  \sin ^2\phi) \nonumber\\
&=&\left( \nu_2^2 \sin 2 \alpha \sin 2 \theta \sin ^2 \phi \cos^2 \phi+\nu_2^2 \sin ^2 2 \alpha  \cos ^4 \phi
\right)-\nu_1 \nu_2 (\sin 2 \alpha  \sin 2 \theta  \cos^2\phi ) \nonumber\\
&=& \left(\nu_2^2 \sin 2 \alpha  \sin 2 \theta \sin ^2 \phi +\nu_2^2 \sin ^2 2 \alpha \cos ^2 \phi
-\nu_1 \nu_2 \sin 2 \alpha \sin 2 \theta \right) \cos^2\phi  \nonumber\\
&=& \left(\nu_2 \sin 2 \alpha \sin 2 \theta \sin ^2 \phi +\nu_2 \sin ^2 2 \alpha \cos ^2 \phi
-\nu_1 \sin 2 \alpha \sin 2 \theta \right) \nu_2 \cos^2\phi  \nonumber\\
&=& \left(\nu_2 \sin ^22 \alpha  \cos ^2 \phi
-\nu_1 \sin 2 \alpha \sin 2 \theta +\nu_2 \sin 2 \alpha  \sin 2 \theta  \sin ^2 \phi\right) \nu_2 \cos^2\phi  \nonumber\\
&=& \left(\nu_2 \sin ^2 2 \alpha  \cos ^2 \phi
- \sin 2 \alpha  \sin 2 \theta  (\nu_1-\nu_2  \sin ^2 \phi) \right) \nu_2 \cos^2\phi  \nonumber\\
&=& \left(\nu_2 \sin ^2 2 \alpha \cos ^2 \phi
- \sin 2 \alpha  \times (\nu_1-\nu_2  \sin ^2 \phi) \sin 2 \theta \right) \nu_2 \cos^2\phi  \nonumber\\
&=& \left(\nu_2 \sin ^2 2 \alpha \cos ^2 \phi
- \sin 2 \alpha \times \nu_2 \cos^2\phi \sin2\alpha \right) \nu_2 \cos^2\phi  \nonumber\\
&=& 0,
\end{eqnarray}
which means the concurrence $\mathcal{C}=0$. Thus the state $\rho$ is a separable state. Similarly, for other three cases (i.e., $\{\beta=0, \tau=\pi\}$, $\{\beta=\pi, \tau=0\}$, and $\{\beta=\pi, \tau=\pi\}$), one can prove that
the state $\rho$ is a separable state.

\emph{Case 9-B.---} $\cos\beta=0$ and $\cos\tau=0$. There are four possible cases: $\{\beta=\pi/2, \tau=\pi/2\}$, $\{\beta=\pi/2, \tau=3\pi/2\}$,
$\{\beta=3\pi/2, \tau=\pi/2\}$, and $\{\beta=3\pi/2, \tau=3\pi/2\}$.

For the case of $\{\beta=\pi/2, \tau=\pi/2\}$, the condition (\ref{eq:ss-3}) becomes
\begin{eqnarray}
&&\nu_2 \cos^2\phi \sin2\alpha - e^{2{\rm i}\tau} (\nu_1 - \nu_2 \sin^2\phi) \sin2\theta=\nu_2 \cos^2\phi \sin2\alpha +(\nu_1 - \nu_2 \sin^2\phi) \sin2\theta=0,
\end{eqnarray}
i.e.,
\begin{eqnarray}\label{eq:r-7}
&& (\nu_1 - \nu_2 \sin^2\phi) \sin2\theta=-\nu_2 \cos^2\phi \sin2\alpha.
\end{eqnarray}
In this situation, the state $\rho$ is given by
\begin{eqnarray}
\rho=\nu_1  \ket{\psi_1}\bra{\psi_1}+\nu_2 \ket{\psi_2}\bra{\psi_2},
\end{eqnarray}
where
\begin{eqnarray}
\ket{\psi_1} &=& \cos \theta \ket{00}+\sin \theta\ket{11}, \nonumber\\
 \ket{\psi_2} &=& \cos\phi(\cos \alpha \ket{01} +\sin\alpha\ket{10})+ e^{{\rm i}\beta}\sin\phi(\sin \theta \ket{00}-\cos \theta\ket{11})\nonumber\\
&=& \cos\phi(\cos \alpha \ket{01} +\sin\alpha\ket{10})+ {\rm i} \sin\phi(\sin \theta \ket{00}-\cos \theta\ket{11}).
\end{eqnarray}
The concurrence of the state $\rho$ is given by
\begin{eqnarray}
\mathcal{C} &=& \sqrt{-s_2-2\sqrt{s_1}},
\end{eqnarray}
where
\begin{eqnarray}
 s_2 &=& -\frac{1}{2} \nu_2^2 \sin 2 \alpha  \sin 2 \theta  \sin ^2 2 \phi \cos 2 \beta-\nu_2^2 \sin ^2 2 \alpha \cos ^4 \phi
 -\sin ^2 2 \theta \left(\nu_1^2-\nu_1 \nu_2 \sin ^2\phi +\nu_2^2 \sin ^4\phi \right)
\nonumber\\
&&-\frac{1}{2} \nu_1 \nu_2 (\cos 4 \theta +3) \sin ^2 \phi, \nonumber\\
&=& \frac{1}{2} \nu_2^2 \sin 2 \alpha  \sin 2 \theta  \sin ^2 2 \phi -\nu_2^2 \sin ^2 2 \alpha  \cos ^4 \phi
 -\sin ^2 2 \theta \left(\nu_1^2-\nu_1 \nu_2 \sin ^2\phi +\nu_2^2 \sin ^4\phi \right)
\nonumber\\
&&-\frac{1}{2} \nu_1 \nu_2 (-2\sin^2 2 \theta +4) \sin ^2 \phi, \nonumber\\
&=& \frac{1}{2} \nu_2^2 \sin 2 \alpha \sin 2 \theta \sin ^2 2 \phi -\nu_2^2 \sin ^2 2 \alpha \cos ^4 \phi
 -\sin ^2 2 \theta \left(\nu_1^2-\nu_1 \nu_2 \sin ^2\phi +\nu_2^2 \sin ^4\phi \right)
\nonumber\\
&&+ \nu_1 \nu_2 (\sin^2 2 \theta -2) \sin ^2 \phi, \nonumber\\
&=& \frac{1}{2} \nu_2^2 \sin 2 \alpha \sin 2 \theta \sin ^2 2 \phi-\nu_2^2 \sin ^2 2 \alpha \cos ^4 \phi
 -\sin ^2 2 \theta \left(\nu_1^2-2\nu_1 \nu_2 \sin ^2\phi +\nu_2^2 \sin ^4\phi \right)-2 \nu_1 \nu_2 \sin ^2 \phi
\nonumber\\
&=& \frac{1}{2} \nu_2^2 \sin 2 \alpha \sin 2 \theta \sin ^2 2 \phi -\nu_2^2 \sin ^2 2 \alpha \cos ^4 \phi
 -\sin ^2 2 \theta \left(\nu_1-\nu_2 \sin ^2\phi \right)^2-2 \nu_1 \nu_2 \sin ^2 \phi,
\end{eqnarray}
\begin{eqnarray}
s_1 &=& \frac{1}{2} \nu_1^2 \nu_2^2 \left(\sin 2 \alpha \sin 2 \theta \sin ^2 2 \phi \cos 2 \beta+2 \sin ^2 2 \alpha \sin ^2 2 \theta \cos ^4\phi +2 \sin ^4\phi \right)\nonumber\\
&=& \frac{1}{2} \nu_1^2 \nu_2^2 \left(-\sin 2 \alpha \sin 2 \theta \sin ^2 2 \phi+2 \sin ^2 2 \alpha  \sin ^2 2 \theta \cos ^4\phi +2 \sin ^4\phi \right)\nonumber\\
&=& \frac{1}{2} \nu_1^2 \nu_2^2 \left(-4 \sin 2 \alpha \sin 2 \theta \sin ^2\phi \cos^2\phi +2 \sin ^2 2 \alpha \sin ^2 2 \theta \cos ^4\phi +2 \sin ^4\phi \right)\nonumber\\
&=&\nu_1^2 \nu_2^2 \left(\sin ^2 2 \alpha \sin ^2 2 \theta \cos ^4\phi -2 \sin 2 \alpha \sin 2 \theta \sin ^2\phi \cos^2\phi +  \sin ^4\phi \right)\nonumber\\
&=& \nu_1^2 \nu_2^2 \left(\sin 2 \alpha \sin 2 \theta  \cos^2\phi - \sin ^2\phi \right)^2.
\end{eqnarray}
Due to Eq. (\ref{eq:r-7}), we have
\begin{eqnarray}
 s_2 &=& \frac{1}{2} \nu_2^2 \sin 2 \alpha \sin 2 \theta \sin ^2 2 \phi-\nu_2^2 \sin ^2 2 \alpha \cos ^4 \phi
 -\sin ^2 2 \theta \left(\nu_1-\nu_2 \sin ^2\phi \right)^2-2 \nu_1 \nu_2 \sin ^2 \phi\nonumber\\
&=& \frac{1}{2} \nu_2^2 \sin 2 \alpha \sin 2 \theta \sin ^2 2 \phi-2\nu_2^2 \sin ^2 2 \alpha \cos ^4 \phi
 -2 \nu_1 \nu_2 \sin ^2 \phi\nonumber\\
&=& 2 \nu_2^2 \sin 2 \alpha \sin 2 \theta  \sin ^2 \phi \cos^2 \phi-2\nu_2^2 \sin ^2 2 \alpha \cos ^4 \phi
 -2 \nu_1 \nu_2 \sin ^2 \phi.
\end{eqnarray}

To prove that the state $\rho$ is a separable state, we only need to prove the concurrence $\mathcal{C}=0$, i.e.,
\begin{eqnarray}
-s_2-2\sqrt{s_1}=0,
\end{eqnarray}
i.e.,
\begin{eqnarray}
s_2^2-4s_1=0,
\end{eqnarray}
i.e.,
\begin{eqnarray}
\left(- \nu_2^2 \sin 2 \alpha  \sin 2 \theta \sin ^2 \phi \cos^2 \phi+\nu_2^2 \sin ^2 2 \alpha \cos ^4 \phi
 + \nu_1 \nu_2 \sin ^2 \phi\right)^2-\nu_1^2 \nu_2^2 \left(\sin 2 \alpha \sin 2 \theta \cos^2\phi -  \sin ^2\phi \right)^2=0,
\end{eqnarray}
i.e.,
\begin{eqnarray}
{T'_1}^2-{T'_2}^2=({T'_1}+{T'_2})({T'_1}-{T'_2})=0,
\end{eqnarray}
with
\begin{eqnarray}
&& {T'_1}= -\nu_2^2 \sin 2 \alpha \sin 2 \theta \sin ^2 \phi \cos^2 \phi+\nu_2^2 \sin ^2 2 \alpha  \cos ^4 \phi
 + \nu_1 \nu_2 \sin ^2 \phi, \nonumber\\
 && {T'_2}= \nu_1 \nu_2 \left(\sin 2 \alpha \sin 2 \theta \cos^2\phi -  \sin ^2\phi \right).
\end{eqnarray}
We only need to prove
\begin{eqnarray}
{T'_1}+{T'_2}=0.
\end{eqnarray}

We can have
\begin{eqnarray}
&&{T'_1}+{T'_2}\nonumber\\
&=&\left(- \nu_2^2 \sin 2 \alpha  \sin 2 \theta  \sin ^2 \phi \cos^2 \phi+\nu_2^2 \sin ^2 2 \alpha  \cos ^4 \phi
 + \nu_1 \nu_2 \sin ^2 \phi\right)+\nu_1 \nu_2( \sin 2 \alpha \sin 2 \theta \cos^2\phi -  \sin ^2\phi) \nonumber\\
&=&\left(- \nu_2^2 \sin 2 \alpha \sin 2 \theta \sin ^2 \phi \cos^2 \phi+\nu_2^2 \sin ^2 2 \alpha \cos ^4 \phi
\right)+\nu_1 \nu_2 (\sin 2 \alpha  \sin 2 \theta  \cos^2\phi ) \nonumber\\
&=& \left(-\nu_2^2 \sin 2 \alpha  \sin 2 \theta  \sin ^2 \phi +\nu_2^2 \sin ^2 2 \alpha \cos ^2 \phi
+\nu_1 \nu_2 \sin 2 \alpha  \sin 2 \theta \right) \cos^2\phi  \nonumber\\
&=& \left(-\nu_2 \sin 2 \alpha  \sin 2 \theta \sin ^2 \phi +\nu_2 \sin ^2 2 \alpha \cos ^2 \phi
+\nu_1 \sin 2 \alpha  \sin 2 \theta \right) \nu_2 \cos^2\phi  \nonumber\\
&=& \left(\nu_2 \sin ^22 \alpha  \cos ^2 \phi
+\nu_1 \sin 2 \alpha  \sin 2 \theta -\nu_2 \sin 2 \alpha  \sin 2 \theta \sin ^2 \phi\right) \nu_2 \cos^2\phi  \nonumber\\
&=& \left(\nu_2 \sin ^2 2 \alpha \cos ^2 \phi
+ \sin 2 \alpha  \sin 2 \theta (\nu_1-\nu_2  \sin ^2 \phi) \right) \nu_2 \cos^2\phi  \nonumber\\
&=& \left(\nu_2 \sin ^2 2 \alpha \cos ^2 \phi
+ \sin 2 \alpha  \times (\nu_1-\nu_2  \sin ^2 \phi) \sin 2 \theta \right) \nu_2 \cos^2\phi  \nonumber\\
&=& \left(\nu_2 \sin ^2 2 \alpha \cos ^2 \phi
- \sin 2 \alpha  \times \nu_2 \cos^2\phi \sin2\alpha \right) \nu_2 \cos^2\phi  \nonumber\\
&=& 0,
\end{eqnarray}
which means $\mathcal{C}=0$. Thus  the state $\rho$ is a separable state. Similarly, for other three cases (i.e., $\{\beta=\pi/2, \tau=3\pi/2\}$, $\{\beta=3\pi/2, \tau=\pi/2\}$, and $\{\beta=3\pi/2, \tau=3\pi/2\}$), one can prove that the state $\rho$ is a separable state.

In summary, for all 9 possible cases, we have successfully prove if the matrix $\mathcal{M}=\mathcal{M}^\dagger$, then the state $\rho$ is separable. This ends the proof of necessary condition.

\end{proof}

\newpage


\subsection{The Proof of $\mathcal{M}=\mathcal{M}^\dagger$ as the Sufficient Condition} \label{sec:IIIB}

\textcolor{blue}{\emph{Sufficient Condition:}} If the rank-2 state $\rho$ is separable, then the matrix $\mathcal{M}=\mathcal{M}^\dagger$. In the following, let us come to prove the sufficient condition.

\begin{proof}
 The two-qubit quantum state with rank 2 (including rank 1 as a special case) is given by
\begin{eqnarray}\label{eq:rho-1}
\rho=\nu_1  \ket{\psi_1}\bra{\psi_1}+\nu_2 \ket{\psi_2}\bra{\psi_2},
\end{eqnarray}
where
\begin{eqnarray}
&&\ket{\psi_1} =\cos \theta \ket{00}+\sin \theta\ket{11}, \nonumber\\
&& \ket{\psi_2} =\cos\phi(\cos \alpha \ket{01} +\sin\alpha\ket{10})+ e^{{\rm i} \beta}\sin\phi(\sin \theta \ket{00}-\cos \theta\ket{11}),
\end{eqnarray}
with
\begin{eqnarray}
&& {\theta\in [0, \pi/2]}, \;\;\;\;  {\phi \in [0, \pi/2]}, \;\;\;\; {\alpha \in [0, \pi/2]}, \;\;\;\; \beta \in [0, 2\pi], \;\;\;\; \nonumber\\
&& \nu_1 \in [0,1], \;\;\; \nu_2 \in [0,1], \;\;\; \nu_1+\nu_2=1.
\end{eqnarray}
When $\nu_1\neq 0$ and $\nu_2\neq 0$, the state $\rho$ is rank 2, while when one of $\nu_j$ $(j=1, 2)$ is 0, then the state $\rho$ is rank 1.

Generally, the concurrence of the state $\rho$ is given by
\begin{eqnarray}\label{eq:concur}
\mathcal{C} &=& \sqrt{-s_2-2\sqrt{s_1}},
\end{eqnarray}
where
\begin{eqnarray}
 s_2 &=& -\frac{1}{2} \nu_2^2 \sin 2 \alpha \sin 2 \theta  \sin ^2 2 \phi \cos 2 \beta-\nu_2^2 \sin ^2 2 \alpha \cos ^4 \phi
 -\sin ^2 2 \theta \left(\nu_1^2-\nu_1 \nu_2 \sin ^2\phi +\nu_2^2 \sin ^4\phi \right)\nonumber\\
&&-\frac{1}{2} \nu_1 \nu_2 (\cos 4 \theta +3) \sin ^2 \phi \nonumber\\
&=& -\frac{1}{2} \nu_2^2 \sin 2 \alpha \sin 2 \theta  \sin ^2 2 \phi \left( 2\cos^ 2 \beta-1\right)-\nu_2^2 \sin ^2 2 \alpha \cos ^4 \phi
 -\sin ^2 2 \theta \left(\nu_1^2-\nu_1 \nu_2 \sin ^2\phi +\nu_2^2 \sin ^4\phi \right)\nonumber\\
&&- \nu_1 \nu_2 [1+\cos^2 2 \theta] \sin ^2 \phi, \nonumber\\
&=& -\frac{1}{2} \nu_2^2 \sin 2 \alpha \sin 2 \theta  \sin ^2 2 \phi \left( 2\cos^ 2 \beta-1\right)-\nu_2^2 \sin ^2 2 \alpha \cos ^4 \phi
 -\sin ^2 2 \theta \left(\nu_1^2 +\nu_2^2 \sin ^4\phi \right)- 2 \nu_1 \nu_2 \cos^2 2 \theta \sin ^2 \phi, \nonumber\\
s_1 &=& \frac{1}{2} \nu_1^2 \nu_2^2 \left(\sin2 \alpha \sin 2 \theta \sin ^2 2 \phi \cos 2 \beta +2 \sin ^2 2 \alpha
 \sin ^2 2 \theta \cos ^4\phi +2 \sin ^4\phi \right)\nonumber\\
&=& \frac{1}{2} \nu_1^2 \nu_2^2 \left[\sin2 \alpha \sin 2 \theta \sin ^2 2 \phi \left( 2\cos^ 2 \beta-1\right) +2 \sin ^2 2 \alpha
 \sin ^2 2 \theta \cos ^4\phi +2 \sin ^4\phi \right].
\end{eqnarray}
The state $\rho$ is separable implies the concurrence is 0. The condition of concurrence $\mathcal{C}=0$ leads to
\begin{eqnarray}\label{eq:conc-2a}
-s_2-2\sqrt{s_1}=0,
\end{eqnarray}
i.e.,
\begin{eqnarray}
s_2^2-4s_1=0.
\end{eqnarray}
For convenience, let us denote
\begin{eqnarray}
G_1= -s_2, \;\;\; G_2= 2 \sqrt{s_1},
\end{eqnarray}
then the concurrence $\mathcal{C}=0$ implies
\begin{eqnarray} \label{eq:conc-3q}
G_1^2-G_2^2=(G_1+G_2)(G_1-G_2)=0.
\end{eqnarray}
According to the definition of concurrence as shown in Eq. (\ref{eq:concur}), to make the formula to be meaningful, one automatically has $-s_2 \geq 0$ and $s_1\geq 0$, i.e., $G_1\geq 0$ and $G_2\geq 0$. Thus, the condition (\ref{eq:conc-3q}) turns to
\begin{eqnarray} \label{eq:conc-4}
G_1-G_2=0,
\end{eqnarray}
which is just Eq. (\ref{eq:conc-2a}).

The state $\rho$ is shared by Alice and Bob. Alice has a qubit in her hand, and Bob has another qubit in his hand. Now, Alice performs a measurement on her qubit along the measurement direction
\begin{eqnarray}
&& \hat{n}=(\sin\xi \cos\tau, \sin\xi \sin\tau, \cos\xi),
\end{eqnarray}
where $\xi$ and $\tau$ are some measurement angles. The corresponding state vectors read
\begin{eqnarray}
&& \ket{+\hat{n}}=\begin{pmatrix}
\cos\frac{\xi}{2}\\
\sin\frac{\xi}{2}\, e^{{\rm i}\tau}\\
\end{pmatrix}=\cos\frac{\xi}{2} |0\rangle+\sin\frac{\xi}{2}\, e^{{\rm i}\tau} |1\rangle, \nonumber\\
&&
 \ket{-\hat{n}}=\begin{pmatrix}
\sin\frac{\xi}{2}\\
-\cos\frac{\xi}{2} \, e^{{\rm i}\tau} \\
\end{pmatrix}= \sin\frac{\xi}{2}|0\rangle -\cos\frac{\xi}{2} \, e^{{\rm i}\tau} |1\rangle,
\end{eqnarray}
and they form the standard measurement basis $\{\ket{+\hat{n}}, \ket{-\hat{n}}\}$. The corresponding projective measurements projectors of Alice are
\begin{eqnarray}
&& P^{\hat{n}}_{0}=\frac{1}{2}\left(\openone+\vec{\sigma}\cdot \hat{n}\right)=|+\hat{n}\rangle\langle+\hat{n}|
= \frac{1}{2}\begin{pmatrix}
 1+\cos\xi&
 \sin\xi e^{-{\rm i} \tau}\\
 \sin\xi e^{{\rm i} \tau} &
 1-\cos\xi
\end{pmatrix},\nonumber\\
&&P^{\hat{n}}_{1}=\frac{1}{2}\left(\openone-\vec{\sigma}\cdot \hat{n}\right)=|-\hat{n}\rangle\langle-\hat{n}|
= \frac{1}{2}\begin{pmatrix}
 1-\cos\xi&
 -\sin\xi e^{-{\rm i} \tau}\\
 -\sin\xi e^{{\rm i} \tau} &
 1+\cos\xi
\end{pmatrix}.
\end{eqnarray}
Here, $\ket{+\hat{n}}$ and $\ket{-\hat{n}}$ are two eigenstates of the operator $\vec{\sigma}\cdot \hat{n}$ with eigenvalues $+1$ and $-1$, respectively.

Notice that the parameter $\phi\in [0, \pi/2]$, thus we divide the proof into three cases: $\phi=0$, $\phi=\pi/2$, and $0< \phi <\pi/2$.

\subsubsection{$\phi=0$}

In this situation, the state $\rho$ is given by
\begin{eqnarray}
\rho=\nu_1  \ket{\psi_1}\bra{\psi_1}+\nu_2 \ket{\psi_2}\bra{\psi_2},
\end{eqnarray}
where
\begin{eqnarray}
\ket{\psi_1} &=& \cos \theta \ket{00}+\sin \theta\ket{11}, \nonumber\\
 \ket{\psi_2} &=& \cos\phi(\cos \alpha \ket{01} +\sin\alpha\ket{10})+ e^{{\rm i}\beta}\sin\phi(\sin \theta \ket{00}-\cos \theta\ket{11})\nonumber\\
 &=& \cos \alpha \ket{01} +\sin\alpha\ket{10},
\end{eqnarray}
with $\theta\in[0, \pi/2]$, $\alpha\in[0, \pi/2]$. {Note that in this case the state $\rho$ does not contain the parameter $\beta$.} The concurrence of the state $\rho$ is given by
\begin{eqnarray}
\mathcal{C}=\left|\nu_2 \sin2\alpha - \nu_1 \sin2\theta\right|.
\end{eqnarray}
If the state $\rho$ is separable, then it implies that $\mathcal{C}=0$, i.e.,
\begin{eqnarray}\label{eq:althe}
\nu_2 \sin2\alpha = \nu_1 \sin2\theta.
\end{eqnarray}
In this case, in the measurement basis $\{|+\hat{n}\rangle, |-\hat{n}\rangle \}$ one will find that $\mathcal{M}=\mathcal{M}^\dagger$. The proof is shown below.

Let us come to check the three conditions (\ref{eq:ss-1})-(\ref{eq:ss-3}) related to $\mathcal{M}=\mathcal{M}^\dagger$, or these three conditions are
rewritten as
\begin{subequations}
\begin{eqnarray}
m_1-m_1^*&=& -{\rm i} \, \nu_2 \sin\alpha  \sin\theta  \sin(\beta+\tau) \sin2\phi, \label{eq:ss-1u}\\
m_4-m_4^*&=& -{\rm i} \, \nu_2 \cos\alpha  \cos\theta \sin(\beta-\tau)  \sin2\phi, \label{eq:ss-2u}\\
m_2 -m_3^{*}&=& \frac{1}{2} e^{-{\rm i}\tau}  \nu_2 \cos^2\phi \sin2\alpha
-\frac{1}{2} e^{{\rm i}\tau} (\nu_1 - \nu_2 \sin^2 \phi) \sin2\theta. \label{eq:ss-3u}
\end{eqnarray}
\end{subequations}
For $\phi=0$, the first two conditions (i.e., Eq. (\ref{eq:ss-1u}) and Eq. (\ref{eq:ss-2u})) are automatically satisfied. For the third condition (i.e., Eq. (\ref{eq:ss-3u})), we obtain
\begin{eqnarray}
&& m_2 -m_3^{*}= \frac{1}{2} \mathrm{e}^{-\mathrm{i} \tau} \nu_2 \sin2\alpha - \frac{1}{2} \mathrm{e}^{\mathrm{i} \tau} \nu_1 \sin2\theta.
\end{eqnarray}
Based on Eq. (\ref{eq:althe}), one has
\begin{eqnarray}
 m_2 -m_3^{*} &=& \frac{1}{2} \mathrm{e}^{-\mathrm{i} \tau} \nu_2 \sin2\alpha - \frac{1}{2} \mathrm{e}^{\mathrm{i} \tau} \nu_1 \sin2\theta=
 \frac{1}{2} (\mathrm{e}^{-\mathrm{i} \tau}- \mathrm{e}^{-\mathrm{i} \tau}) \nu_1 \sin2\theta =-{\rm i}\sin\tau\, (\nu_1 \sin2\theta).
\end{eqnarray}
{Thus, if Alice chooses her measurement parameter $\tau=0$ (i.e., $\sin\tau=0$, and $\xi$ is arbitrary), then one has $\mathcal{M}=\mathcal{M}^\dagger$.} This ends the proof.

\begin{remark} {The above proof is also valid for the case of $\theta=\pi/4$ and $\alpha=\pi/4$, i.e., both of the states
\begin{eqnarray}
\ket{\psi_1} &=& \frac{1}{\sqrt{2}}(\ket{00}+\ket{11}), \nonumber\\
\ket{\psi_2} &=& \frac{1}{\sqrt{2}}(\ket{01}+\ket{10}),
\end{eqnarray}
are maximally entangled states (i.e., the former is spanned in the subspace of $\{\ket{00}, \ket{11}\}$, while the latter is spanned in the subspace of $\{\ket{01}, \ket{10}\}$). $\blacksquare$}
\end{remark}

\begin{remark} By letting $\nu_2=0$, then one has $\nu_1=1$. The above state $\rho$ reduces a rank-1 quantum state $\rho=|\psi_1\rangle\langle \psi_1|$. Actually, the above proof is also valid for the rank-1 state. For convenience, without loss of generality, hereafter, we only consider the true rank-2 quantum state $\rho$ with $\nu_1\neq 0$ and $\nu_2\neq 0$. $\blacksquare$
\end{remark}

\subsubsection{$\phi=\pi/2$}

In this situation, the state $\rho$ is given by
\begin{eqnarray}
\rho=\nu_1  \ket{\psi_1}\bra{\psi_1}+\nu_2 \ket{\psi_2}\bra{\psi_2},
\end{eqnarray}
where
\begin{eqnarray}
\ket{\psi_1} &=& \cos \theta \ket{00}+\sin \theta\ket{11}, \nonumber\\
\ket{\psi_2} &=& \cos\phi(\cos \alpha \ket{01} +\sin\alpha\ket{10})+ e^{{\rm i}\beta}\sin\phi(\sin \theta \ket{00}-\cos \theta\ket{11})\nonumber\\
&=& e^{{\rm i}\beta}(\sin \theta \ket{00}-\cos \theta\ket{11}),
\end{eqnarray}
where $\theta\in[0, \pi/2]$, $\alpha\in[0, \pi/2]$. {Note that in this case the state $\rho$ does not contain the parameter $\beta$ after the density-matrix form $\ket{\psi_2}\bra{\psi_2}$ is considered.}  The concurrence of the state $\rho$ is given by
\begin{eqnarray}
\mathcal{C}=|(\nu_1-\nu_2)\sin2\theta|.
\end{eqnarray}
If the state $\rho$ is separable, then it implies that $\mathcal{C}=0$, i.e.,
\begin{eqnarray}\label{eq:conc-1}
(\nu_1-\nu_2)\sin2\theta=0.
\end{eqnarray}
In this case, in the measurement basis $\{|+\hat{n}\rangle, |-\hat{n}\rangle \}$ one will find that $\mathcal{M}=\mathcal{M}^\dagger$. The proof is shown below.

Let us come to check the three conditions related to $\mathcal{M}=\mathcal{M}^\dagger$ (i.e., Eq. (\ref{eq:ss-1u})-Eq. (\ref{eq:ss-3u})).
For $\phi=\pi/2$, the first two conditions are automatically satisfied. For the third condition, we obtain
\begin{eqnarray}
&&  {m_2 -m_3^{*}} = -\frac{1}{2} e^{{\rm i}\tau} (\nu_1 - \nu_2) \sin2\theta,
\end{eqnarray}
which is obviously 0 due to Eq. (\ref{eq:conc-1}). {Thus, we can have the matrix $\mathcal{M}=\mathcal{M}^\dagger$ for arbitrary measurement
parameters $\xi$ and $\tau$. For simplicity, we can choose $\tau=0$.} This ends the proof.

\begin{remark} The above proof is also valid for the case of $\theta=\pi/4$, i.e., both of the states
\begin{eqnarray}
\ket{\psi_1} &=& \frac{1}{\sqrt{2}}(\ket{00}+\ket{11}), \nonumber\\
\ket{\psi_2} &=& \frac{1}{\sqrt{2}}(\ket{00}-\ket{11}),
\end{eqnarray}
are maximally entangled states spanned in the subspace of $\{\ket{00}, \ket{11}\}$. $\blacksquare$
\end{remark}

\subsubsection{$0< \phi <\pi/2$}

\vspace{2mm}
\centerline{\emph{3.1. The Case of $\cos\beta=0$}}
\vspace{6mm}

In this subsection, let us consider the simple case of $\cos\beta=0$. In this situation, the state $\rho$ is given by
\begin{eqnarray}
\rho=\nu_1  \ket{\psi_1}\bra{\psi_1}+\nu_2 \ket{\psi_2}\bra{\psi_2},
\end{eqnarray}
where
\begin{eqnarray}
\ket{\psi_1} &=& \cos \theta \ket{00}+\sin \theta\ket{11}, \nonumber\\
 \ket{\psi_2} &=& \cos\phi(\cos \alpha \ket{01} +\sin\alpha\ket{10})+ e^{{\rm i}\beta}\sin\phi(\sin \theta \ket{00}-\cos \theta\ket{11}),
\end{eqnarray}
with $\beta=\pi/2$, $e^{{\rm i}\beta}={\rm i}$, $\theta\in[0, \pi/2]$, and $\alpha\in[0, \pi/2]$. {For simplicity, in the following we have taken $\beta=\pi/2$ as example (for the case of $\beta=3\pi/2$, it will yield the similar result).} The concurrence of the state $\rho$ is given by
\begin{eqnarray}
\mathcal{C} &=& \sqrt{-s_2-2\sqrt{s_1}},
\end{eqnarray}
with
\begin{eqnarray}
 s_2 &=& \frac{1}{2} \nu_2^2 \sin 2 \alpha \sin 2 \theta  \sin ^2 2 \phi -\nu_2^2 \sin ^2 2 \alpha \cos ^4 \phi
 -\sin ^2 2 \theta \left(\nu_1^2 +\nu_2^2 \sin ^4\phi \right)- 2 \nu_1 \nu_2 \cos^2 2 \theta \sin ^2 \phi, \nonumber\\
s_1 &=& \frac{1}{2} \nu_1^2 \nu_2^2 \left(-\sin2 \alpha \sin 2 \theta \sin ^2 2 \phi +2 \sin ^2 2 \alpha
 \sin ^2 2 \theta \cos ^4\phi +2 \sin ^4\phi \right).
\end{eqnarray}
The concurrence $\mathcal{C}=0$ means $G_1^2-G_2^2=0$, i.e.,
\begin{eqnarray}\label{eq:G1}
G_1^2-G_2^2&=& \left\{\nu_2^2 \left( \cos^2\phi \sin2\alpha - \sin2\theta \sin^2\phi \right)^2  + 2 \nu_1 \nu_2 \cos^2 2\theta \sin^2\phi + \nu_1^2 \sin^2 2\theta\right\}^2 \nonumber\\
&&- 4 \nu_1^2 \nu_2^2 \left( \sin^2\phi - \sin2\alpha \sin2\theta \cos^2\phi \right)^2
=0.
\end{eqnarray}
Here, we have used
\begin{eqnarray}
G_1 &=& -s_2= -\frac{1}{2} \nu_2^2 \sin 2 \alpha \sin 2 \theta  \sin ^2 2 \phi +\nu_2^2 \sin ^2 2 \alpha \cos ^4 \phi
 +\sin ^2 2 \theta \left(\nu_1^2 +\nu_2^2 \sin ^4\phi \right)+ 2 \nu_1 \nu_2 \cos^2 2 \theta \sin ^2 \phi\nonumber\\
&=& -2 \nu_2^2 \sin 2 \alpha \sin 2 \theta  \sin^2 \phi \cos^2 \phi +\nu_2^2 \sin ^2 2 \alpha \cos ^4 \phi
 +\sin ^2 2 \theta \left(\nu_1^2 +\nu_2^2 \sin ^4\phi \right)+ 2 \nu_1 \nu_2 \cos^2 2 \theta \sin ^2 \phi\nonumber\\
&=& \nu_2^2  \left( \cos^2\phi \sin2\alpha - \sin2\theta \sin^2\phi \right)^2  + 2 \nu_1 \nu_2 \cos^2 2\theta \sin^2\phi + \nu_1^2 \sin^2 2\theta,
\end{eqnarray}
and
\begin{eqnarray}
G_2^2&=& 4s_1 =4\times \frac{1}{2} \nu_1^2 \nu_2^2 \left(-\sin2 \alpha \sin 2 \theta \sin ^2 2 \phi +2 \sin ^2 2 \alpha
 \sin ^2 2 \theta \cos ^4\phi +2 \sin ^4\phi \right)\nonumber\\
&=& 2 \nu_1^2 \nu_2^2 \left(-4\sin2 \alpha \sin 2 \theta \sin ^2 \phi \cos ^2 \phi+2 \sin ^2 2 \alpha
 \sin ^2 2 \theta \cos ^4\phi +2 \sin ^4\phi \right)\nonumber\\
&=& 4 \nu_1^2 \nu_2^2 \left(-2\sin2 \alpha \sin 2 \theta \sin ^2 \phi \cos ^2 \phi+ \sin ^2 2 \alpha
 \sin ^2 2 \theta \cos ^4\phi + \sin ^4\phi \right)\nonumber\\
 &=&  4 \nu_1^2 \nu_2^2 \left( \sin^2\phi - \sin2\alpha \sin2\theta \cos^2\phi \right)^2.
\end{eqnarray}

Without loss of generality, hereafter, we only consider $\nu_1 \neq 0$ and $\nu_2\neq 0$. Now, let us come to check the three conditions related to $\mathcal{M}=\mathcal{M}^\dagger$ (i.e., Eq. (\ref{eq:ss-1u})-Eq. (\ref{eq:ss-3u})). When $\cos\beta=0$, then these three conditions become
\begin{subequations}
\begin{eqnarray}
m_1-m_1^* &=& -{\rm i} \, \nu_2 \sin\alpha  \sin\theta  \sin\beta \cos\tau \sin2\phi, \label{eq:ss-1v}\\
m_4-m_4^* &=& -{\rm i} \, \nu_2 \cos\alpha  \cos\theta \sin\beta \cos\tau  \sin2\phi,\label{eq:ss-2v}\\
m_2 -m_3^{*} &=& \frac{1}{2} e^{-{\rm i}\tau}  \nu_2 \cos^2\phi \sin2\alpha
-\frac{1}{2} e^{{\rm i} \tau} (\nu_1 - \nu_2 \sin^2 \phi) \sin2\theta. \label{eq:ss-3v}
\end{eqnarray}
\end{subequations}
The region of the angle $\theta$ is $\theta\in [0, \pi/2]$, thus $\sin 2\theta\geq 0$. The region of the angle $\alpha$ is $\alpha\in [0, \pi/2]$, hence $\sin 2 \alpha$ can be equal to or greater than $0$. In the following, we would like to study the problem by two different cases: $\sin 2 \alpha= 0$ and $\sin 2 \alpha> 0$.

\textcolor{blue}{\emph{Case 1.---} $\sin2\alpha=0$.} Based on Eq. (\ref{eq:G1}), one has
\begin{eqnarray}
 &&G_1^2-G_2^2=\left\{\nu_2^2 \left( \sin2\theta \sin^2\phi \right)^2  + 2 \nu_1 \nu_2 \cos^2 2\theta \sin^2\phi + \nu_1^2 \sin^2 2\theta\right\}^2 - 4 \nu_1^2 \nu_2^2 \left( \sin^2\phi \right)^2=0,
\end{eqnarray}
then one directly has $G_1-G_2=0$, i.e.,
\begin{eqnarray}
 &&\nu_2^2 \left( \sin2\theta \sin^2\phi \right)^2  + 2 \nu_1 \nu_2 \cos^2 2\theta \sin^2\phi + \nu_1^2 \sin^2 2\theta - 2 \nu_1 \nu_2 \left( \sin^2\phi \right)=0,
\end{eqnarray}
i.e.,
\begin{eqnarray}
 &&\nu_2^2 \left( \sin2\theta \sin^2\phi \right)^2  - 2 \nu_1 \nu_2 \sin^2 2\theta \sin^2\phi + \nu_1^2 \sin^2 2\theta =0,
\end{eqnarray}
i.e,
\begin{eqnarray}
 && \sin^2 2\theta \times \left[\nu_2^2 \sin^4\phi   - 2 \nu_1 \nu_2 \sin^2\phi + \nu_1^2 \right] =0,
\end{eqnarray}
i.e.,
\begin{eqnarray}\label{eq:conc-5a}
 && \sin^2 2\theta \times \left[\nu_1-\nu_2 \sin^2\phi\right]^2 =0,
\end{eqnarray}
which leads to (i) $\sin2\theta=0$ or (ii) $(\nu_1-\nu_2 \sin^2)\phi=0$. Actually, in this case, the concurrence of the state $\rho$ is equal to
\begin{eqnarray}
\mathcal{C} &=& \sqrt{-s_2-2\sqrt{s_1}}=\sqrt{\sin^2 2\theta \times \left[\nu_1-\nu_2 \sin^2\phi\right]^2}=|\sin 2\theta \left(\nu_1-\nu_2 \sin^2\phi\right)|\nonumber\\
&=& \sin 2\theta |\nu_1-\nu_2 \sin^2\phi|.
\end{eqnarray}
Thus Eq. (\ref{eq:conc-5a}) means $\mathcal{C}^2=0$.

(i) $\sin2\theta=0$. In this situation, by choosing $\tau=\beta=\pi/2$ (i.e., $\cos\tau=0$), then the first two conditions (i.e., Eq. (\ref{eq:ss-1v}) and Eq. (\ref{eq:ss-2v})) are satisfied. Moreover, due to $\sin2\alpha=0$  and $\sin2\theta=0$, then the third condition Eq. (\ref{eq:ss-3v}) is also satisfied. {Thus, one can have the matrix $\mathcal{M}=\mathcal{M}^\dagger$ by adopting the measurement direction $\hat{n}$ with $\tau=\beta=\pi/2$ and $\xi$ is arbitrary,} i.e.,
\begin{eqnarray}
&& \hat{n}=(0, \sin\xi, \cos\xi).
\end{eqnarray}

(ii) $\nu_1-\nu_2 \sin^2\phi=0$. In this situation, by choosing $\tau=\beta=\pi/2$ (i.e., $\cos\tau=0$), then the first two conditions (i.e., Eq. (\ref{eq:ss-1v}) and Eq. (\ref{eq:ss-2v})) are satisfied. Moreover, due to $\sin2\alpha=0$  and $\nu_1-\nu_2 \sin^2\phi=0$, then the third condition Eq. (\ref{eq:ss-3v}) is also satisfied. {Thus, one can have the matrix $\mathcal{M}=\mathcal{M}^\dagger$ by adopting the measurement direction $\hat{n}$ with $\tau=\beta=\pi/2$ and $\xi$ is arbitrary,} i.e.,
\begin{eqnarray}
&& \hat{n}=(0, \sin\xi, \cos\xi).
\end{eqnarray}
This ends the proof for this case.

\textcolor{blue}{\emph{Case 2.---} $\sin2\alpha>0$.} Based on Eq. (\ref{eq:G1}), one has two possible solutions: $G_1=G_2=0$; $G_1\neq 0$, $G_2\neq 0$, but $G_1^2=G_2^2$.

\textcolor{blue}{\emph{Case 2-A.---} $G_1=G_2=0$.} In this case, we have

\begin{eqnarray}
G_1 &=& \nu_2^2 \left[ \left( \cos^2\phi \sin2\alpha - \sin2\theta \sin^2\phi \right)^2 + \cos^2\beta \sin2\alpha \sin2\theta \sin^2 2 \phi
\right] + 2 \nu_1 \nu_2 \cos^2 2\theta \sin^2\phi + \nu_1^2 \sin^2 2\theta\nonumber\\
&=&\left( \cos^2\phi \sin2\alpha - \sin2\theta \sin^2\phi \right)^2 + 2 \nu_1 \nu_2 \cos^2 2\theta \sin^2\phi + \nu_1^2 \sin^2 2\theta=0,\nonumber\\
G_2^2 &=&  4 \nu_1^2 \nu_2^2 \left[\left( \sin^2\phi - \sin2\alpha \sin2\theta \cos^2\phi \right)^2
+ \cos^2\beta \sin2\alpha \sin2\theta \sin^2 2\phi \right]\nonumber\\
&=&4 \nu_1^2 \nu_2^2 \left( \sin^2\phi - \sin2\alpha \sin2\theta \cos^2\phi \right)^2 =0.
\end{eqnarray}
Because $\nu_1 \neq 0$ and $\nu_2 \neq 0$, then we have
\begin{eqnarray}
&&  \cos^2\phi \sin2\alpha - \sin2\theta \sin^2\phi=0, \nonumber\\
&& \cos^2 2\theta \sin^2\phi=0, \nonumber\\
&& \sin^2 2\theta=0,\nonumber\\
&& \sin^2\phi - \sin2\alpha \sin2\theta \cos^2\phi=0,
\end{eqnarray}
i.e.,
\begin{eqnarray}
&& \sin 2\theta=0,\nonumber\\
&&  \cos^2\phi=0, \nonumber\\
&& \sin^2\phi=0.
\end{eqnarray}
However, there is a contradiction between $\cos^2\phi=0$ and $\sin^2\phi=0$, which means the concurrence $\mathcal{C}=0$ does not occur for the case of $G_1=G_2=0$. We can neglect this case.

\textcolor{blue}{\emph{Case 2-B.---} $G_1\neq 0$, $G_2\neq 0$, but $G_1^2=G_2^2$.} In this case, we have
\begin{eqnarray}\label{eq:G1G2-1}
&& \left\{\nu_2^2 \left( \cos^2\phi \sin2\alpha - \sin2\theta \sin^2\phi \right)^2 + 2 \nu_1 \nu_2 \cos^2 2\theta \sin^2\phi + \nu_1^2 \sin^2 2\theta\right\}^2\nonumber\\
&&- 4 \nu_1^2 \nu_2^2 \left( \sin^2\phi - \sin2\alpha \sin2\theta \cos^2\phi \right)^2=0.
\end{eqnarray}

(i) If $\sin^2\phi - \sin2\alpha \sin2\theta \cos^2\phi>0$, we can get
\begin{eqnarray}
&& \left\{\nu_2^2  \left( \cos^2\phi \sin2\alpha - \sin2\theta \sin^2\phi \right)^2 + 2 \nu_1 \nu_2 \cos^2 2\theta \sin^2\phi + \nu_1^2 \sin^2 2\theta\right\}- 2 \nu_1 \nu_2 \left( \sin^2\phi - \sin2\alpha \sin2\theta \cos^2\phi \right)=0,\nonumber\\
\end{eqnarray}
i.e.,
\begin{eqnarray}
&& \nu_2^2  \left( \cos^2\phi \sin2\alpha - \sin2\theta \sin^2\phi \right)^2 +  \nu_1^2 \sin^2 2\theta- 2 \nu_1 \nu_2 \left( \sin^2 2\theta\sin^2\phi - \sin2\alpha \sin2\theta \cos^2\phi \right)=0,
\end{eqnarray}
i.e.,
\begin{eqnarray}
&& (\nu_1 \sin 2\theta)^2 - 2 (\nu_1 \sin 2\theta) \nu_2 \left( \sin 2\theta\sin^2\phi - \sin2\alpha \cos^2\phi \right)+ \nu_2^2  \left( \sin 2\theta\sin^2\phi - \sin2\alpha \cos^2\phi \right)^2 =0,
\end{eqnarray}
i.e.,
\begin{eqnarray}
&& [(\nu_1 \sin 2\theta) - \nu_2 \left( \sin 2\theta\sin^2\phi - \sin2\alpha \cos^2\phi \right)]^2=0,
\end{eqnarray}
i.e.,
\begin{eqnarray}\label{eq:nu1a}
&& \nu_1 \sin 2\theta- \nu_2 \left( \sin 2\theta\sin^2\phi - \sin2\alpha \cos^2\phi \right)=0.
\end{eqnarray}

Let us come to check the three conditions related to $\mathcal{M}=\mathcal{M}^\dagger$. In this situation, by choosing $\tau=\beta=\pi/2$ (i.e., $\cos\tau=0$), then the first two conditions (i.e., Eq. (\ref{eq:ss-1v}) and Eq. (\ref{eq:ss-2v})) are satisfied. Moreover, due to $\tau=\beta=\pi/2$, then the third condition Eq. (\ref{eq:ss-3v}) becomes
\begin{eqnarray}
{m_2 -m_3^{*}}&=& \frac{1}{2} e^{-{\rm i}\tau}  \nu_2 \cos^2\phi \sin2\alpha
-\frac{1}{2} e^{{\rm i}\tau} (\nu_1 - \nu_2 \sin^2 \phi) \sin2\theta\nonumber\\
&=& -\frac{{\rm i}}{2} [ \nu_2 \cos^2\phi \sin2\alpha
+ (\nu_1 - \nu_2 \sin^2 \phi) \sin2\theta]\nonumber\\
&=& -\frac{{\rm i}}{2} [\nu_1 \sin 2\theta - \nu_2 \left( \sin 2\theta\sin^2\phi - \sin2\alpha \cos^2\phi \right)],
\end{eqnarray}
which is obviously zero due to Eq. (\ref{eq:nu1a}). {Thus, one can have the matrix $\mathcal{M}=\mathcal{M}^\dagger$ by adopting the measurement direction $\hat{n}$ with $\tau=\beta=\pi/2$ and $\xi$ is arbitrary,} i.e.,
\begin{eqnarray}
&& \hat{n}=(0, \sin\xi, \cos\xi).
\end{eqnarray}
This ends the proof for this case.

(ii) If $\sin^2\phi - \sin2\alpha \sin2\theta \cos^2\phi<0$, we can get
\begin{eqnarray}
&& \left\{\nu_2^2  \left( \cos^2\phi \sin2\alpha - \sin2\theta \sin^2\phi \right)^2 + 2 \nu_1 \nu_2 \cos^2 2\theta \sin^2\phi + \nu_1^2 \sin^2 2\theta \right\}+ 2 \nu_1 \nu_2 \left( \sin^2\phi - \sin2\alpha \sin2\theta \cos^2\phi \right)=0,\nonumber\\
\end{eqnarray}
i.e.,
\begin{eqnarray}
&& \nu_2^2  \left( \cos^2\phi \sin2\alpha - \sin2\theta \sin^2\phi \right)^2 + 4 \nu_1 \nu_2 \cos^2 2\theta \sin^2\phi + \nu_1^2 \sin^2 2\theta+ 2 \nu_1 \nu_2 \left( \sin^2\phi\sin^2 2\theta - \sin2\alpha \sin2\theta \cos^2\phi \right)=0,\nonumber\\
\end{eqnarray}
i.e.,\begin{eqnarray}
&& [(\nu_1 \sin 2\theta) +\nu_2 \left( \sin 2\theta\sin^2\phi - \sin2\alpha \cos^2\phi \right)]^2+ 4 \nu_1 \nu_2 \cos^2 2\theta \sin^2\phi =0,
\end{eqnarray}
i.e.,
\begin{eqnarray}
&& (\nu_1 \sin 2\theta) +\nu_2 \left( \sin 2\theta\sin^2\phi - \sin2\alpha \cos^2\phi \right) =0,\nonumber\\
&&  4 \nu_1 \nu_2 \cos^2 2\theta \sin^2\phi =0.
\end{eqnarray}
Because $0<\phi<\pi/2$, $\sin^2\phi > 0$, $\nu_1\neq 0$, and $\nu_2\neq 0$, hence we must have
\begin{eqnarray}
&& \cos 2\theta=0,
\end{eqnarray}
which leads to
\begin{eqnarray}
&& \theta=\frac{\pi}{4},
\end{eqnarray}
and $\sin 2 \theta=1$. Then from above equations, we can have
\begin{eqnarray}
&& \nu_1 =-\nu_2 \left( \sin^2\phi - \sin2\alpha \cos^2\phi \right).
\end{eqnarray}
With the help of $\nu_1 +\nu_2 =1$, we have $\nu_i$'s (i=1, 2) as
\begin{eqnarray}\label{eq:nu2aa}
&&\nu_1=\frac{ \sin2\alpha \cos^2\phi-\sin^2\phi }{(1 +\sin2\alpha)\cos^2\phi}, \;\;\;\; \nu_2=\frac{1}{(1 +\sin2\alpha)\cos^2\phi}.
\end{eqnarray}
{However, in Sec. \ref{sum} we have shown that we need not consider separately the case of $\theta=\pi/4$, thus ending the proof for this case.}

\vspace{6mm}
\centerline{\emph{3.2. The Case of $\cos\beta=1$}}
\vspace{6mm}

In this subsection, let us consider the simple case of $\cos\beta=1$. In this situation, the state $\rho$ is given by
\begin{eqnarray}
\rho=\nu_1  \ket{\psi_1}\bra{\psi_1}+\nu_2 \ket{\psi_2}\bra{\psi_2},
\end{eqnarray}
where
\begin{eqnarray}
\ket{\psi_1} &=& \cos \theta \ket{00}+\sin \theta\ket{11}, \nonumber\\
 \ket{\psi_2} &=& \cos\phi(\cos \alpha \ket{01} +\sin\alpha\ket{10})+ e^{{\rm i}\beta}\sin\phi(\sin \theta \ket{00}-\cos \theta\ket{11}),
\end{eqnarray}
with $\beta=0$, $e^{{\rm i}\beta}=1$, $\theta\in[0, \pi/2]$, and $\alpha\in[0, \pi/2]$. For simplicity, here we have taken $\cos\beta=1$ as example (for the case of $\cos\beta=-1$, it will yield the similar result). The concurrence of the state $\rho$ is given by
\begin{eqnarray}
\mathcal{C} &=& \sqrt{-s_2-2\sqrt{s_1}},
\end{eqnarray}
with
\begin{eqnarray}
 s_2 &=& -\frac{1}{2} \nu_2^2 \sin 2 \alpha \sin 2 \theta  \sin ^2 2 \phi \left( 2\cos^ 2 \beta-1\right)-\nu_2^2 \sin ^2 2 \alpha \cos ^4 \phi
 -\sin ^2 2 \theta \left(\nu_1^2 +\nu_2^2 \sin ^4\phi \right)- 2 \nu_1 \nu_2 \cos^2 2 \theta \sin ^2 \phi \nonumber\\
&=& -\frac{1}{2} \nu_2^2 \sin 2 \alpha \sin 2 \theta  \sin ^2 2 \phi -\nu_2^2 \sin ^2 2 \alpha \cos ^4 \phi
 -\sin ^2 2 \theta \left(\nu_1^2 +\nu_2^2 \sin ^4\phi \right)- 2 \nu_1 \nu_2 \cos^2 2 \theta \sin ^2 \phi, \nonumber\\
s_1 &=& \frac{1}{2} \nu_1^2 \nu_2^2 \left[\sin2 \alpha \sin 2 \theta \sin ^2 2 \phi \left( 2\cos^ 2 \beta-1\right) +2 \sin ^2 2 \alpha
 \sin ^2 2 \theta \cos ^4\phi +2 \sin ^4\phi \right]\nonumber\\
&=& \frac{1}{2} \nu_1^2 \nu_2^2 \left(\sin2 \alpha \sin 2 \theta \sin ^2 2 \phi +2 \sin ^2 2 \alpha
 \sin ^2 2 \theta \cos ^4\phi +2 \sin ^4\phi \right).
\end{eqnarray}
The concurrence $\mathcal{C}=0$ means $G_1^2-G_2^2=0$, i.e.,
\begin{eqnarray}\label{eq:G1-c}
G_1^2-G_2^2&=& \left\{\nu_2^2  \left( \cos^2\phi \sin2\alpha + \sin2\theta \sin^2\phi \right)^2  + 2 \nu_1 \nu_2 \cos^2 2\theta \sin^2\phi + \nu_1^2 \sin^2 2\theta\right\}^2 \nonumber\\
&&- 4 \nu_1^2 \nu_2^2 \left( \sin^2\phi + \sin2\alpha \sin2\theta \cos^2\phi \right)^2
=0.
\end{eqnarray}
Here, we have used
\begin{eqnarray}
G_1 &=& -s_2= \frac{1}{2} \nu_2^2 \sin 2 \alpha \sin 2 \theta  \sin ^2 2 \phi +\nu_2^2 \sin ^2 2 \alpha \cos ^4 \phi
 +\sin ^2 2 \theta \left[\nu_1^2 +\nu_2^2 \sin ^4\phi \right]+ 2 \nu_1 \nu_2 \cos^2 2 \theta \sin ^2 \phi\nonumber\\
&=& 2 \nu_2^2 \sin 2 \alpha \sin 2 \theta  \sin^2 \phi \cos^2 \phi +\nu_2^2 \sin ^2 2 \alpha \cos ^4 \phi
 +\sin ^2 2 \theta \left[\nu_1^2 +\nu_2^2 \sin ^4\phi \right]+ 2 \nu_1 \nu_2 \cos^2 2 \theta \sin ^2 \phi\nonumber\\
&=& \nu_2^2  \left( \cos^2\phi \sin2\alpha + \sin2\theta \sin^2\phi \right)^2  + 2 \nu_1 \nu_2 \cos^2 2\theta \sin^2\phi + \nu_1^2 \sin^2 2\theta,
\end{eqnarray}
and
\begin{eqnarray}
G_2^2&=& 4s_1 =4\times \frac{1}{2} \nu_1^2 \nu_2^2 \left[\sin2 \alpha \sin 2 \theta \sin ^2 2 \phi +2 \sin ^2 2 \alpha
 \sin ^2 2 \theta \cos ^4\phi +2 \sin ^4\phi \right]\nonumber\\
&=& 2 \nu_1^2 \nu_2^2 \left[4\sin2 \alpha \sin 2 \theta \sin ^2 \phi \cos ^2 \phi+2 \sin ^2 2 \alpha
 \sin ^2 2 \theta \cos ^4\phi +2 \sin ^4\phi \right]\nonumber\\
&=& 4 \nu_1^2 \nu_2^2 \left[2\sin2 \alpha \sin 2 \theta \sin ^2 \phi \cos ^2 \phi+ \sin ^2 2 \alpha
 \sin ^2 2 \theta \cos ^4\phi + \sin ^4\phi \right]\nonumber\\
 &=&  4 \nu_1^2 \nu_2^2 \left( \sin^2\phi +\sin2\alpha \sin2\theta \cos^2\phi \right)^2.
\end{eqnarray}

Without loss of generality, we only consider $\nu_1 \neq 0$ and $\nu_2\neq 0$. Now, let us come to check the three conditions related to $\mathcal{M}=\mathcal{M}^\dagger$ (i.e., Eq. (\ref{eq:ss-1u}) and Eq. (\ref{eq:ss-3u})). When $\cos\beta=1$ (i.e., $\sin\beta=0$), then the three conditions become
\begin{subequations}
\begin{eqnarray}
m_1-m_1^*&=& -{\rm i} \, \nu_2 \sin\alpha  \sin\theta  \cos\beta \sin\tau \sin2\phi, \label{eq:ss-1w}\\
m_4-m_4^*&=& -{\rm i} \, \nu_2 \cos\alpha  \cos\theta \cos\beta \sin\tau  \sin2\phi,\label{eq:ss-2w}\\
m_2 -m_3^{*} &=& \frac{1}{2} e^{-{\rm i}\tau}  \nu_2 \cos^2\phi \sin2\alpha
-\frac{1}{2} e^{{\rm i}\tau} (\nu_1 - \nu_2 \sin^2 \phi) \sin2\theta. \label{eq:ss-3w}
\end{eqnarray}
\end{subequations}
The region of the angle $\theta$ is $\theta\in [0, \pi/2]$, thus $\sin 2\theta\geq 0$. The region of the angle $\alpha$ is $\alpha\in [0, \pi/2]$, hence $\sin 2 \alpha$ can be equal to or greater than $0$. In the following, we would like to study the problem by two different cases: $\sin 2 \alpha= 0$ and $\sin 2 \alpha> 0$.

\textcolor{blue}{\emph{Case 1.---} $\sin2\alpha=0$.} Based on Eq. (\ref{eq:G1-c}), one has
\begin{eqnarray}
 &&G_1^2-G_2^2=\left\{\nu_2^2 \left( \sin2\theta \sin^2\phi \right)^2  + 2 \nu_1 \nu_2 \cos^2 2\theta \sin^2\phi + \nu_1^2 \sin^2 2\theta\right\}^2 - 4 \nu_1^2 \nu_2^2 \left( \sin^2\phi \right)^2=0,
\end{eqnarray}
then one directly has
\begin{eqnarray}
 &&\nu_2^2 \left( \sin2\theta \sin^2\phi \right)^2  + 2 \nu_1 \nu_2 \cos^2 2\theta \sin^2\phi + \nu_1^2 \sin^2 2\theta - 2 \nu_1 \nu_2 \left( \sin^2\phi \right)=0,
\end{eqnarray}
i.e.,
\begin{eqnarray}
 &&\nu_2^2 \left( \sin2\theta \sin^2\phi \right)^2  - 2 \nu_1 \nu_2 \sin^2 2\theta \sin^2\phi + \nu_1^2 \sin^2 2\theta =0,
\end{eqnarray}
i.e,
\begin{eqnarray}
 && \sin^2 2\theta \times \left(\nu_2^2 \sin^4\phi   - 2 \nu_1 \nu_2 \sin^2\phi + \nu_1^2 \right)=0,
\end{eqnarray}
i.e.,
\begin{eqnarray}\label{eq:conc-5}
 && \sin^2 2\theta \times \left(\nu_1-\nu_2 \sin^2\phi\right)^2 =0,
\end{eqnarray}
which leads to (i) $\sin2\theta=0$ or (ii) $(\nu_1-\nu_2 \sin^2)\phi=0$. Actually, in this case, the concurrence of the state $\rho$ is equal to
\begin{eqnarray}
\mathcal{C} &=& \sqrt{-s_2-2\sqrt{s_1}}=\sqrt{\sin^2 2\theta \times \left[\nu_1-\nu_2 \sin^2\phi\right]^2}=|\sin 2\theta \left(\nu_1-\nu_2 \sin^2\phi\right)|\nonumber\\
&=& \sin 2\theta |\nu_1-\nu_2 \sin^2\phi|.
\end{eqnarray}
Thus Eq. (\ref{eq:conc-5}) means $\mathcal{C}^2=0$.

(i) $\sin2\theta=0$. In this situation, by choosing $\tau=\beta=0$ (i.e., $\sin\tau=0$), then the first two conditions (i.e., Eq. (\ref{eq:ss-1w}) and Eq. (\ref{eq:ss-2w})) are satisfied. Moreover, due to $\sin2\alpha=0$  and $\sin2\theta=0$, then the third condition Eq. (\ref{eq:ss-3w}) is also satisfied. {Thus, one can have the matrix $\mathcal{M}=\mathcal{M}^\dagger$ by adopting the measurement direction $\hat{n}$ with $\tau=\beta=0$ and $\xi$ is arbitrary,} i.e.,
\begin{eqnarray}
&& \hat{n}=(\sin\xi, 0, \cos\xi).
\end{eqnarray}

(ii) $\nu_1-\nu_2 \sin^2\phi=0$. In this situation, by choosing $\tau=\beta=0$ (i.e., $\sin\tau=0$), then the first two conditions (i.e., Eq. (\ref{eq:ss-1w}) and Eq. (\ref{eq:ss-2w})) are satisfied. Moreover, due to $\sin2\alpha=0$  and $\nu_1-\nu_2 \sin^2\phi=0$, then the third condition Eq. (\ref{eq:ss-3w}) is also satisfied. {Thus, one can have the matrix $\mathcal{M}=\mathcal{M}^\dagger$ by adopting the measurement direction $\hat{n}$ with $\tau=\beta=0$ and $\xi$ is arbitrary,} i.e.,
\begin{eqnarray}
&& \hat{n}=\hat{n}=(\sin\xi, 0, \cos\xi).
\end{eqnarray}
This ends the proof for this case.

\textcolor{blue}{\emph{Case 2.---} $\sin2\alpha>0$.} Based on Eq. (\ref{eq:G1-c}), one has two possible solutions: $G_1=G_2=0$; $G_1\neq 0$, $G_2\neq 0$, but $G_1^2=G_2^2$.

\textcolor{blue}{\emph{Case 2-A.---} $G_1=G_2=0$.} Similarly, the concurrence $\mathcal{C}=0$ does not occur for the case of $G_1=G_2=0$. We can neglect this case.

\textcolor{blue}{\emph{Case 2-B.---} $G_1\neq 0$, $G_2\neq 0$, but $G_1^2=G_2^2$.} In this case, we have
\begin{eqnarray}
&&\left\{\nu_2^2  \left( \cos^2\phi \sin2\alpha + \sin2\theta \sin^2\phi \right)^2  + 2 \nu_1 \nu_2 \cos^2 2\theta \sin^2\phi + \nu_1^2 \sin^2 2\theta\right\}^2 \nonumber\\
&&- 4 \nu_1^2 \nu_2^2 \left( \sin^2\phi + \sin2\alpha \sin2\theta \cos^2\phi \right)^2=0
\end{eqnarray}
Because $G_2 \neq 0$, $\sin2\alpha>0$ and $0< \phi <\pi/2$, thus  $\sin^2\phi + \sin2\alpha \sin2\theta \cos^2\phi>0$.

When $\sin^2\phi + \sin2\alpha \sin2\theta \cos^2\phi>0$, we can get
\begin{eqnarray}
&& \left\{\nu_2^2  \left( \cos^2\phi \sin2\alpha + \sin2\theta \sin^2\phi \right)^2 + 2 \nu_1 \nu_2 \cos^2 2\theta \sin^2\phi + \nu_1^2 \sin^2 2\theta\right\}- 2 \nu_1 \nu_2 \left( \sin^2\phi + \sin2\alpha \sin2\theta \cos^2\phi \right)=0,\nonumber\\
\end{eqnarray}
i.e.,
\begin{eqnarray}
&& \nu_2^2  \left( \cos^2\phi \sin2\alpha + \sin2\theta \sin^2\phi \right)^2 +  \nu_1^2 \sin^2 2\theta- 2 \nu_1 \nu_2 \left( \sin^2 2\theta\sin^2\phi +\sin2\alpha \sin2\theta \cos^2\phi \right)=0,
\end{eqnarray}
i.e.,
\begin{eqnarray}
&& (\nu_1 \sin 2\theta)^2 - 2 (\nu_1 \sin 2\theta) \nu_2 \left( \sin 2\theta\sin^2\phi + \sin2\alpha \cos^2\phi \right)+ \nu_2^2  \left( \sin 2\theta\sin^2\phi + \sin2\alpha \cos^2\phi \right)^2 =0,
\end{eqnarray}
i.e.,
\begin{eqnarray}
&& [(\nu_1 \sin 2\theta) - \nu_2 \left( \sin 2\theta\sin^2\phi + \sin2\alpha \cos^2\phi \right)]^2=0,
\end{eqnarray}
i.e.,
\begin{eqnarray}\label{eq:nu1ac}
&& \nu_1 \sin 2\theta- \nu_2 \left( \sin 2\theta\sin^2\phi + \sin2\alpha \cos^2\phi \right)=0.
\end{eqnarray}

Let us come to check the three conditions related to $\mathcal{M}=\mathcal{M}^\dagger$. In this situation, by choosing $\tau=\beta=0$ (i.e., $\sin\tau=0$), then the first two conditions (i.e., Eq. (\ref{eq:ss-1w}) and Eq. (\ref{eq:ss-2w})) are satisfied. Moreover, due to $\tau=\beta=0$, then the third condition Eq. (\ref{eq:ss-3w}) becomes
\begin{eqnarray}
m_2 -m_3^{*}  &=& \frac{1}{2} e^{-{\rm i}\tau}  \nu_2 \cos^2\phi \sin2\alpha
-\frac{1}{2} e^{{\rm i}\tau} (\nu_1 - \nu_2 \sin^2 \phi) \sin2\theta\nonumber\\
&=& \frac{1}{2} [ \nu_2 \cos^2\phi \sin2\alpha
- (\nu_1 - \nu_2 \sin^2 \phi) \sin2\theta]\nonumber\\
&=& \frac{1}{2} [(\nu_1 \sin 2\theta) - \nu_2 \left( \sin 2\theta\sin^2\phi + \sin2\alpha \cos^2\phi \right)],
\end{eqnarray}
which is obviously 0 due to Eq. (\ref{eq:nu1ac}). {Thus, one can have the matrix $\mathcal{M}=\mathcal{M}^\dagger$ by adopting the measurement direction $\hat{n}$ with $\tau=\beta=0$ and $\xi$ is arbitrary,} i.e., $\hat{n}=(\sin\xi, 0, \cos\xi)$. It ends the proof for this case.


\vspace{12mm}
\centerline{\emph{3.3. The Case of General $\cos\beta$}}
\vspace{8mm}

 In this subsection, let us consider the general $\cos\beta$, i.e., it is not equal to special values of 0 and $\pm 1$. In other works, $0<|\cos\beta|<1$ and $0<|\sin\beta|<1$. In this situation, the state $\rho$ is given by
\begin{eqnarray}
\rho=\nu_1  \ket{\psi_1}\bra{\psi_1}+\nu_2 \ket{\psi_2}\bra{\psi_2},
\end{eqnarray}
where
\begin{eqnarray}
\ket{\psi_1} &=& \cos \theta \ket{00}+\sin \theta\ket{11}, \nonumber\\
 \ket{\psi_2} &=& \cos\phi(\cos \alpha \ket{01} +\sin\alpha\ket{10})+ e^{{\rm i}\beta}\sin\phi(\sin \theta \ket{00}-\cos \theta\ket{11}),
\end{eqnarray}
with $0<|\cos\beta|<1$, $\theta\in[0, \pi/2]$, and $\alpha\in[0, \pi/2]$. The concurrence of the state $\rho$ is given by
\begin{eqnarray}
\mathcal{C} &=& \sqrt{-s_2-2\sqrt{s_1}},
\end{eqnarray}
with
\begin{eqnarray}
 s_2 &=& -\frac{1}{2} \nu_2^2 \sin 2 \alpha \sin 2 \theta  \sin ^2 2 \phi \left( 2\cos^ 2 \beta-1\right)-\nu_2^2 \sin ^2 2 \alpha \cos ^4 \phi
 -\sin ^2 2 \theta \left(\nu_1^2 +\nu_2^2 \sin ^4\phi \right)- 2 \nu_1 \nu_2 \cos^2 2 \theta \sin ^2 \phi,\nonumber\\
s_1 &=& \frac{1}{2} \nu_1^2 \nu_2^2 \left[\sin2 \alpha \sin 2 \theta \sin ^2 2 \phi \left( 2\cos^ 2 \beta-1\right) +2 \sin ^2 2 \alpha
 \sin ^2 2 \theta \cos ^4\phi +2 \sin ^4\phi \right].
\end{eqnarray}
The concurrence $\mathcal{C}=0$ means $G_1^2-G_2^2=0$, with
\begin{eqnarray}
G_1 &=& -s_2 \nonumber\\
&=&\frac{1}{2} \nu_2^2 \sin 2 \alpha \sin 2 \theta  \sin ^2 2 \phi \left( 2\cos^ 2 \beta-1\right)+\nu_2^2 \sin ^2 2 \alpha \cos ^4 \phi
 +\sin ^2 2 \theta \left[\nu_1^2 +\nu_2^2 \sin ^4\phi \right]+ 2 \nu_1 \nu_2 \cos^2 2 \theta \sin ^2 \phi,\nonumber\\
&=& \nu_1^2 \sin ^2 2 \theta + \nu_2^2 \left[\frac{1}{2} \sin 2 \alpha \sin 2 \theta  \sin ^2 2 \phi \left( 2\cos^ 2 \beta-1\right)+\sin ^2 2 \alpha \cos ^4 \phi +\sin ^2 2 \theta \sin ^4\phi \right]+ 2 \nu_1 \nu_2 \cos^2 2 \theta \sin ^2 \phi,\nonumber\\
&=& \nu_1^2 \sin ^2 2 \theta + \nu_2^2 \left[2 \sin 2 \alpha \sin 2 \theta  \sin ^2 \phi \cos ^2 \phi \left( 2\cos^ 2 \beta-1\right)+\sin ^2 2 \alpha \cos ^4 \phi +\sin ^2 2 \theta \sin ^4\phi \right]+ 2 \nu_1 \nu_2 \cos^2 2 \theta \sin ^2 \phi,\nonumber\\
&=& \nu_1^2 \sin ^2 2 \theta + \nu_2^2 \left[ (\sin 2 \alpha \cos ^2 \phi -\sin 2 \theta \sin^2\phi)^2+ 4 \sin 2 \alpha \sin 2 \theta  \sin ^2 \phi \cos ^2 \phi \cos^ 2 \beta \right]+ 2 \nu_1 \nu_2 \cos^2 2 \theta \sin ^2 \phi,
\end{eqnarray}
or
\begin{eqnarray}
G_1 &=& \nu_1^2 \sin ^2 2 \theta + \nu_2^2 \left[2 \sin 2 \alpha \sin 2 \theta  \sin ^2 \phi \cos ^2 \phi \left( 2\cos^ 2 \beta-1\right)+\sin ^2 2 \alpha \cos ^4 \phi +\sin ^2 2 \theta \sin ^4\phi \right]+ 2 \nu_1 \nu_2 \cos^2 2 \theta \sin ^2 \phi,\nonumber\\
&=& \nu_1^2 \sin ^2 2 \theta + \nu_2^2 \left[2 \sin 2 \alpha \sin 2 \theta  \sin ^2 \phi \cos ^2 \phi \left(1- 2\sin^ 2 \beta\right)+\sin ^2 2 \alpha \cos ^4 \phi +\sin ^2 2 \theta \sin ^4\phi \right]+ 2 \nu_1 \nu_2 \cos^2 2 \theta \sin ^2 \phi,\nonumber\\
&=& \nu_1^2 \sin ^2 2 \theta + \nu_2^2 \left[ (\sin 2 \alpha \cos ^2 \phi +\sin 2 \theta \sin^2\phi)^2-4 \sin 2 \alpha \sin 2 \theta  \sin ^2 \phi \cos ^2 \phi \sin^ 2 \beta \right]+ 2 \nu_1 \nu_2 \cos^2 2 \theta \sin ^2 \phi,
\end{eqnarray}
and
\begin{eqnarray}
G_2^2&=& 4s_1 =4\times \frac{1}{2} \nu_1^2 \nu_2^2 \left[\sin2 \alpha \sin 2 \theta \sin ^2 2 \phi \left( 2\cos^ 2 \beta-1\right) +2 \sin ^2 2 \alpha
 \sin ^2 2 \theta \cos ^4\phi +2 \sin ^4\phi \right] \nonumber\\
&=& 2 \nu_1^2 \nu_2^2 \left[\sin2 \alpha \sin 2 \theta \sin ^2 2 \phi \left( 2\cos^ 2 \beta-1\right) +2 \sin ^2 2 \alpha
 \sin ^2 2 \theta \cos ^4\phi +2 \sin ^4\phi \right] \nonumber\\
&=& 2 \nu_1^2 \nu_2^2 \left[4 \sin2 \alpha \sin 2 \theta \sin^2 \phi \cos^2 \phi \left( 2\cos^ 2 \beta-1\right) +2 \sin ^2 2 \alpha
 \sin ^2 2 \theta \cos ^4\phi +2 \sin ^4\phi \right] \nonumber\\
&=& 4 \nu_1^2 \nu_2^2 \left[2 \sin2 \alpha \sin 2 \theta \sin^2 \phi \cos^2 \phi \left( 2\cos^ 2 \beta-1\right) + \sin ^2 2 \alpha
 \sin ^2 2 \theta \cos ^4\phi + \sin ^4\phi \right] \nonumber\\
&=& 4 \nu_1^2 \nu_2^2 \left[ \left( \sin^2\phi - \sin2\alpha \sin2\theta \cos^2\phi \right)^2+4 \sin2 \alpha \sin 2 \theta \sin^2 \phi \cos^2 \phi \cos^ 2 \beta \right],
\end{eqnarray}
or
\begin{eqnarray}
G_2^2&=& 4 \nu_1^2 \nu_2^2 \left[2 \sin2 \alpha \sin 2 \theta \sin^2 \phi \cos^2 \phi \left( 2\cos^ 2 \beta-1\right) + \sin ^2 2 \alpha
 \sin ^2 2 \theta \cos ^4\phi + \sin ^4\phi \right] \nonumber\\
&=& 4 \nu_1^2 \nu_2^2 \left[2 \sin2 \alpha \sin 2 \theta \sin^2 \phi \cos^2 \phi \left( 1-2\sin^ 2 \beta\right) + \sin ^2 2 \alpha
 \sin ^2 2 \theta \cos ^4\phi + \sin ^4\phi \right] \nonumber\\
&=& 4 \nu_1^2 \nu_2^2 \left[ \left( \sin^2\phi + \sin2\alpha \sin2\theta \cos^2\phi \right)^2-4 \sin2 \alpha \sin 2 \theta \sin^2 \phi \cos^2 \phi \sin^ 2 \beta \right].
\end{eqnarray}
The above results show that there are two different expressions for $G_1$ and $G^2_2$ in terms of $\cos^2\beta$ and $\sin^2\beta$. From
\begin{eqnarray}
\mathcal{C}^2=G_1^2-G_2^2=0,
\end{eqnarray}
we easily find that $\mathcal{C}^2$ is $\beta$-independent if $\sin2\alpha=0$ or $\sin2\theta=0$. In the following, we would like to discuss the issue by four possible cases: (1) $\sin2\alpha=0$ and $\sin2\theta= 0$; (2) $\sin2\alpha=0$ and $\sin2\theta\neq 0$; (3) $\sin2\alpha\neq 0$ and $\sin2\theta=0$; and (4) $\sin2\alpha\neq 0$ and $\sin2\theta\neq 0$.

\vspace{5mm}
\centerline{ \emph{3.3.A. \;\;\; The Case of $\sin2\alpha=0$ and $\sin2\theta= 0$ }}
\vspace{4mm}

In this case, we have
\begin{eqnarray}
G_1 &=& \nu_1^2 \sin ^2 2 \theta + \nu_2^2 \left[ (\sin 2 \alpha \cos ^2 \phi +\sin 2 \theta \sin^2\phi)^2-4 \sin 2 \alpha \sin 2 \theta  \sin ^2 \phi \cos ^2 \phi \sin^ 2 \beta \right]+ 2 \nu_1 \nu_2 \cos^2 2 \theta \sin ^2 \phi\nonumber\\
&=& \nu_1^2 \sin ^2 2 \theta + \nu_2^2 (\sin 2 \theta \sin^2\phi)^2+ 2 \nu_1 \nu_2 \cos^2 2 \theta \sin ^2 \phi,
\end{eqnarray}
\begin{eqnarray}
G_2^2 &=& 4 \nu_1^2 \nu_2^2 \left[ \left( \sin^2\phi + \sin2\alpha \sin2\theta \cos^2\phi \right)^2-4 \sin2 \alpha \sin 2 \theta \sin^2 \phi \cos^2 \phi \sin^ 2 \beta \right]= 4 \nu_1^2 \nu_2^2 \sin^4\phi.
\end{eqnarray}
Since $\sin2\theta= 0$ and $\cos^2 2\theta= 1$, then we have
\begin{eqnarray}
G_1 &=& 2 \nu_1 \nu_2 \sin ^2 \phi,
\end{eqnarray}
this leads to
\begin{eqnarray}
\mathcal{C}^2 &= & G_1^2-G_2^2=0.
\end{eqnarray}
This means that in this case the state $\rho$ is always separable.

Now, let us come to check the three conditions related to $\mathcal{M}=\mathcal{M}^\dagger$ (i.e., Eq. (\ref{eq:ss-1u})-Eq. (\ref{eq:ss-3u})).
The third condition is automatically satisfied. Because $\sin2\alpha=0$, then it means $\sin\alpha=0$ or $\cos\alpha=0$. {For the former, the first condition is satisfied, and by taking $\tau=\beta$, the second condition is satisfied.  For the latter, the second condition is satisfied, and by taking $\tau=-\beta$, the third condition is satisfied. Thus by choosing appropriate projective measurements (i.e., $\tau\in\{\beta, -\beta\}$) one can prove $\mathcal{M}=\mathcal{M}^\dagger$ for this case.}

\vspace{5mm}
\centerline{ \emph{3.3.B. \;\;\; The Case of $\sin2\alpha=0$ and $\sin2\theta\neq 0$ }}
\vspace{4mm}

In this case, we have
\begin{eqnarray}
G_1 &=& \nu_1^2 \sin ^2 2 \theta + \nu_2^2 \left[ (\sin 2 \alpha \cos ^2 \phi +\sin 2 \theta \sin^2\phi)^2-4 \sin 2 \alpha \sin 2 \theta  \sin ^2 \phi \cos ^2 \phi \sin^ 2 \beta \right]+ 2 \nu_1 \nu_2 \cos^2 2 \theta \sin ^2 \phi\nonumber\\
&=& \nu_1^2 \sin ^2 2 \theta + \nu_2^2 (\sin 2 \theta \sin^2\phi)^2+ 2 \nu_1 \nu_2 \cos^2 2 \theta \sin ^2 \phi,
\end{eqnarray}
\begin{eqnarray}
G_2^2 &=& 4 \nu_1^2 \nu_2^2 \left[ \left( \sin^2\phi + \sin2\alpha \sin2\theta \cos^2\phi \right)^2-4 \sin2 \alpha \sin 2 \theta \sin^2 \phi \cos^2 \phi \sin^ 2 \beta \right]\nonumber\\
&=& 4 \nu_1^2 \nu_2^2 \sin^4\phi.
\end{eqnarray}
Then from
\begin{eqnarray}
\mathcal{C}^2 &= & G_1^2-G_2^2= \left[\nu_1^2 \sin ^2 2 \theta + \nu_2^2 (\sin 2 \theta \sin^2\phi)^2+ 2 \nu_1 \nu_2 \cos^2 2 \theta \sin ^2 \phi\right]^2-4 \nu_1^2 \nu_2^2 \sin^4\phi=0,
\end{eqnarray}
we can have
\begin{eqnarray}
&& \left[\nu_1^2 \sin ^2 2 \theta + \nu_2^2 (\sin 2 \theta \sin^2\phi)^2+ 2 \nu_1 \nu_2 \cos^2 2 \theta \sin ^2 \phi\right]-2 \nu_1 \nu_2 \sin^2\phi=0,
\end{eqnarray}
i.e.,
\begin{eqnarray}
&& \nu_1^2 \sin ^2 2 \theta + \nu_2^2 (\sin 2 \theta \sin^2\phi)^2- 2 \nu_1 \nu_2 \sin^2 2 \theta \sin ^2 \phi=0,
\end{eqnarray}
i.e.,
\begin{eqnarray}
&& \sin^2 2 \theta \times \left(\nu_1 - \nu_2 \sin^2\phi\right)^2=0,
\end{eqnarray}
i.e.,
\begin{eqnarray}
&& \nu_1 - \nu_2 \sin^2\phi=0,
\end{eqnarray}
which is condition for the concurrence is 0.

Now, let us come to check the three conditions related to $\mathcal{M}=\mathcal{M}^\dagger$ (i.e., Eq. (\ref{eq:ss-1u})-Eq. (\ref{eq:ss-3u})).
Because $\sin2\alpha=0$ and $\nu_1 - \nu_2 \sin^2\phi=0$, the third condition is automatically satisfied. Because $\sin2\alpha=0$, then it means $\sin\alpha=0$ or $\cos\alpha=0$. {For the former, the first condition is satisfied, and by taking $\tau=\beta$, the second condition is satisfied.  For the latter, the second condition is satisfied, and by taking $\tau=-\beta$, the third condition is satisfied. Thus by choosing appropriate projective measurements (i.e., $\tau\in\{\beta, -\beta\}$) one can prove $\mathcal{M}=\mathcal{M}^\dagger$ for this case.}

\vspace{5mm}

\centerline{ \emph{3.3.C. \;\;\; The Case of $\sin2\alpha\neq 0$ and $\sin2\theta=0$}}

\vspace{5mm}

In this case, we have
\begin{eqnarray}
G_1 &=& \nu_1^2 \sin ^2 2 \theta + \nu_2^2 \left[ (\sin 2 \alpha \cos ^2 \phi +\sin 2 \theta \sin^2\phi)^2-4 \sin 2 \alpha \sin 2 \theta  \sin ^2 \phi \cos ^2 \phi \sin^ 2 \beta \right]+ 2 \nu_1 \nu_2 \cos^2 2 \theta \sin ^2 \phi\nonumber\\
&=&  \nu_2^2 (\sin 2 \alpha \cos ^2 \phi)^2 + 2 \nu_1 \nu_2 \sin ^2 \phi,
\end{eqnarray}
\begin{eqnarray}
G_2^2 &=& 4 \nu_1^2 \nu_2^2 \left[ \left( \sin^2\phi + \sin2\alpha \sin2\theta \cos^2\phi \right)^2-4 \sin2 \alpha \sin 2 \theta \sin^2 \phi \cos^2 \phi \sin^ 2 \beta \right]= 4 \nu_1^2 \nu_2^2 \sin^4\phi.
\end{eqnarray}
Then from
\begin{eqnarray}
\mathcal{C}^2 &= & G_1^2-G_2^2= \left[\nu_2^2 (\sin 2 \alpha \cos ^2 \phi)^2 + 2 \nu_1 \nu_2 \sin ^2 \phi\right]^2-4 \nu_1^2 \nu_2^2 \sin^4\phi=0,
\end{eqnarray}
we can have
\begin{eqnarray}
G_1-G_2= \left[\nu_2^2 (\sin 2 \alpha \cos ^2 \phi)^2 + 2 \nu_1 \nu_2 \sin ^2 \phi\right]-2 \nu_1 \nu_2 \sin^2\phi=0,
\end{eqnarray}
i.e.,
\begin{eqnarray}
\nu_2^2 (\sin 2 \alpha \cos ^2 \phi)^2=0.
\end{eqnarray}
However, in this subsection, $\nu_2\neq 0$, $\sin 2 \alpha \neq 0$, and $\cos ^2 \phi\neq 0$, then the above equation is not valid. This implies that the concurrence $\mathcal{C}=0$ does not occur in this case. We can neglect it.

\vspace{5mm}
\centerline{ \emph{3.4.D. \;\;\; The Case of $\sin2\alpha\neq 0$ and $\sin2\theta\neq 0$}}
\vspace{5mm}

In this case, we can have
\begin{eqnarray}
G_1
&=& \nu_1^2 \sin ^2 2 \theta + \nu_2^2 \left[ (\sin 2 \alpha \cos ^2 \phi -\sin 2 \theta \sin^2\phi)^2+ 4 \sin 2 \alpha \sin 2 \theta  \sin ^2 \phi \cos ^2 \phi \cos^ 2 \beta \right]+ 2 \nu_1 \nu_2 \cos^2 2 \theta \sin ^2 \phi,\nonumber\\
G_2^2 &=& 4 \nu_1^2 \nu_2^2 \left[ \left( \sin^2\phi - \sin2\alpha \sin2\theta \cos^2\phi \right)^2+4 \sin2 \alpha \sin 2 \theta \sin^2 \phi \cos^2 \phi \cos^ 2 \beta \right].
\end{eqnarray}
The concurrence $\mathcal{C}=0$ is equivalent to
\begin{eqnarray}
G_1-G_2=0.
\end{eqnarray}
Let us study the quantity $G_1-G_2$, which is
\begin{eqnarray}
G_1-G_2 &=& \nu_1^2 \sin ^2 2 \theta + \nu_2^2 \left[ (\sin 2 \alpha \cos ^2 \phi -\sin 2 \theta \sin^2\phi)^2+ 4 \sin 2 \alpha \sin 2 \theta  \sin ^2 \phi \cos ^2 \phi \cos^ 2 \beta \right]+ 2 \nu_1 \nu_2 \cos^2 2 \theta \sin ^2 \phi \nonumber \\
&&- 2 \nu_1 \nu_2 \sqrt{ \left( \sin^2\phi - \sin2\alpha \sin 2\theta \cos^2\phi \right)^2 + \cos^2\beta \sin2\alpha \sin2\theta \sin^2 2\phi}.
\end{eqnarray}
With the help of
\begin{eqnarray}
x^2+y^2\geq 2 xy,
\end{eqnarray}
we have
\begin{eqnarray}
G_1
 &\geq & 2 \nu_1 \nu_2 \sin 2 \theta  \sqrt{ (\sin 2 \alpha \cos ^2 \phi -\sin 2 \theta \sin^2\phi)^2+ 4 \sin 2 \alpha \sin 2 \theta  \sin ^2 \phi \cos ^2 \phi \cos^ 2 \beta }+ 2 \nu_1 \nu_2 \cos^2 2 \theta \sin ^2 \phi\nonumber\\
&= & 2 \nu_1 \nu_2 \left[ \sin 2 \theta  \sqrt{ (\sin 2 \alpha \cos ^2 \phi -\sin 2 \theta \sin^2\phi)^2+ 4 \sin 2 \alpha \sin 2 \theta  \sin ^2 \phi \cos ^2 \phi \cos^ 2 \beta }+ \cos^2 2 \theta \sin ^2 \phi\right],
\end{eqnarray}
here we have used $\sin 2 \theta \geq 0$, because $\theta\in [0, \pi/2]$. For the above equation, the equal sign occurs at $x=y$, i.e.,
\begin{eqnarray}\label{eq:cond-b}
 \nu_1 \sin  2 \theta = \nu_2 \sqrt{ (\sin 2 \alpha \cos ^2 \phi -\sin 2 \theta \sin^2\phi)^2+ 4 \sin 2 \alpha \sin 2 \theta  \sin ^2 \phi \cos ^2 \phi \cos^ 2 \beta}.
\end{eqnarray}
This yields
\begin{eqnarray}\label{eq:G1G2-3}
G_1-G_2 &\geq & 2 \nu_1 \nu_2 \left[ \sin 2 \theta  \sqrt{ (\sin 2 \alpha \cos ^2 \phi -\sin 2 \theta \sin^2\phi)^2+ 4 \sin 2 \alpha \sin 2 \theta  \sin ^2 \phi \cos ^2 \phi \cos^ 2 \beta }+ \cos^2 2 \theta \sin ^2 \phi\right]\nonumber\\
&&-2 \nu_1 \nu_2 \sqrt{\left( \sin^2\phi - \sin2\alpha \sin2\theta \cos^2\phi \right)^2+4 \sin2 \alpha \sin 2 \theta \sin^2 \phi \cos^2 \phi \cos^ 2 \beta}\nonumber\\
&=&2\nu_1\nu_2\sqrt{\sin^2 2\theta\left[ \left( \cos^2\phi \sin2\alpha - \sin2\theta \sin^2\phi \right)^2 + \cos^2\beta \sin2\alpha \sin2\theta \sin^2 2\phi \right] }\nonumber\\
&&- 2 \nu_1 \nu_2 \left( \sqrt{ \left( \sin^2\phi - \sin2\alpha \sin 2\theta \cos^2\phi \right)^2 + \cos^2\beta \sin2\alpha \sin2\theta \sin^2 2\phi }-\cos^2 2\theta \sin^2\phi\right).
\end{eqnarray}
Note that
\begin{eqnarray}
&&\left[ \left( \cos^2\phi \sin2\alpha - \sin2\theta \sin^2\phi \right)^2 + 4 \cos^2\beta \sin2\alpha \sin2\theta \sin^2 \phi
\cos^2 \phi\right]\nonumber\\
&=&\left[ \left( \cos^2\phi \sin2\alpha + \sin2\theta \sin^2\phi \right)^2 + 4 \sin^2\beta \sin2\alpha \sin2\theta \sin^2 \phi
\cos^2 \phi\right]\geq0.
\end{eqnarray}
Let us denote
\begin{eqnarray}
&&Y_1=\sqrt{\sin^2 2\theta\left[ \left( \cos^2\phi \sin2\alpha - \sin2\theta \sin^2\phi \right)^2 + \cos^2\beta \sin2\alpha \sin2\theta \sin^2 2\phi \right] }\nonumber\\
&&Y_2= \left( \sqrt{ \left( \sin^2\phi - \sin2\alpha \sin 2\theta \cos^2\phi \right)^2 + \cos^2\beta \sin2\alpha \sin2\theta \sin^2 2\phi }-\cos^2 2\theta \sin^2\phi\right),
\end{eqnarray}
note that $Y_1\geq 0$, we can have
Let us denote
\begin{eqnarray}
Y_1^2-Y_2^2 &=& \left[\sqrt{\sin^2 2\theta\left[ \left( \cos^2\phi \sin2\alpha - \sin2\theta \sin^2\phi \right)^2 + \cos^2\beta \sin2\alpha \sin2\theta \sin^2 2\phi \right] }\right]^2\nonumber\\
&&- \left( \sqrt{ \left( \sin^2\phi - \sin2\alpha \sin 2\theta \cos^2\phi \right)^2 + \cos^2\beta \sin2\alpha \sin2\theta \sin^2 2\phi }-\cos^2 2\theta \sin^2\phi\right)^2\nonumber\\
&=&\sin^2 2\theta\left[ \left( \cos^2\phi \sin2\alpha - \sin2\theta \sin^2\phi \right)^2 + \cos^2\beta \sin2\alpha \sin2\theta \sin^2 2\phi \right] \nonumber\\
&&- \left( \sqrt{ \left( \sin^2\phi - \sin2\alpha \sin 2\theta \cos^2\phi \right)^2 + \cos^2\beta \sin2\alpha \sin2\theta \sin^2 2\phi }-\cos^2 2\theta \sin^2\phi\right)^2\nonumber\\
&=&\sin^2 2\theta\left[ \left( \cos^2\phi \sin2\alpha - \sin2\theta \sin^2\phi \right)^2 + \cos^2\beta \sin2\alpha \sin2\theta \sin^2 2\phi \right] \nonumber\\
&&- \left[  \left( \sin^2\phi - \sin2\alpha \sin 2\theta \cos^2\phi \right)^2 + \cos^2\beta \sin2\alpha \sin2\theta \sin^2 2\phi +\cos^4 2\theta \sin^4\phi\right]\nonumber\\
&&+ 2 \cos^2 2\theta \sin^2\phi \sqrt{ \left( \sin^2\phi - \sin2\alpha \sin 2\theta \cos^2\phi \right)^2 + \cos^2\beta \sin2\alpha \sin2\theta \sin^2 2\phi }\nonumber\\
&=& 2 \cos^2 2\theta \sin^2\phi \sqrt{ \left( \sin^2\phi - \sin2\alpha \sin 2\theta \cos^2\phi \right)^2 + \cos^2\beta \sin2\alpha \sin2\theta \sin^2 2\phi }\nonumber\\
&&-2 \cos^2 2\theta \sin^2\phi \left[\sin^2\phi + \cos2\beta \sin2\alpha \sin2\theta \cos^2 \phi\right]\nonumber\\
&=& 2 \cos^2 2\theta \sin^2\phi \left[Y_3 - \left(\sin^2\phi + \cos2\beta \sin2\alpha \sin2\theta \cos^2 \phi\right)\right],
\end{eqnarray}
with
\begin{eqnarray}
Y_3 &=& \sqrt{ \left( \sin^2\phi - \sin2\alpha \sin 2\theta \cos^2\phi \right)^2 + \cos^2\beta \sin2\alpha \sin2\theta \sin^2 2\phi }.
\end{eqnarray}
Because
\begin{eqnarray}
&&Y_3^2- \left(\sin^2\phi + \cos2\beta \sin2\alpha \sin2\theta \cos^2 \phi\right)^2\nonumber\\
&=& \left[ \left( \sin^2\phi - \sin2\alpha \sin 2\theta \cos^2\phi \right)^2 + \cos^2\beta \sin2\alpha \sin2\theta \sin^2 2\phi \right]
-\left(\sin^2\phi + \cos2\beta \sin2\alpha \sin2\theta \cos^2 \phi\right)^2\nonumber\\
&=& \cos^4 \phi \sin^2 2\beta \sin^2 2\alpha \sin^2 2\theta,
\end{eqnarray}
thus
\begin{eqnarray}
Y_1^2-Y_2^2 &=& 2 \cos^2 2\theta \sin^2\phi \left[Y_3 - \left(\sin^2\phi + \cos2\beta \sin2\alpha \sin2\theta \cos^2 \phi\right)\right]\nonumber\\
&=& \frac{2 \cos^2 2\theta \sin^2\phi \left[Y_3^2 - \left(\sin^2\phi + \cos2\beta \sin2\alpha \sin2\theta \cos^2 \phi\right)^2\right]}
{\left[Y_3 + \left(\sin^2\phi + \cos2\beta \sin2\alpha \sin2\theta \cos^2 \phi\right)\right]}\nonumber\\
&=& \frac{2 \cos^2 2\theta \sin^2\phi \cos^4 \phi \sin^2 2\beta \sin^2 2\alpha \sin^2 2\theta}
{ \left[Y_3 + \left(\sin^2\phi + \cos2\beta \sin2\alpha \sin2\theta \cos^2 \phi\right)\right]}.
\end{eqnarray}
Then from Eq. (\ref{eq:G1G2-3}), we can have
\begin{eqnarray}
G_1-G_2 &\geq & 2\nu_1\nu_2\sqrt{\sin^2 2\theta\left[ \left( \cos^2\phi \sin2\alpha - \sin2\theta \sin^2\phi \right)^2 + \cos^2\beta \sin2\alpha \sin2\theta \sin^2 2\phi \right] }\nonumber\\
&&- 2 \nu_1 \nu_2 \left( \sqrt{ \left( \sin^2\phi - \sin2\alpha \sin 2\theta \cos^2\phi \right)^2 + \cos^2\beta \sin2\alpha \sin2\theta \sin^2 2\phi }-\cos^2 2\theta \sin^2\phi\right)\nonumber\\
&=& 2\nu_1\nu_2 (Y_1-Y_2)\nonumber\\
&=& \frac{2\nu_1\nu_2 (Y_1^2-Y_2^2)}{Y_1+Y_2}\nonumber\\
&=& \frac{4\nu_1\nu_2  \cos^2 2\theta \sin^2\phi \cos^4 \phi \sin^2 2\beta \sin^2 2\alpha \sin^2 2\theta }{(Y_1+Y_2)\left[Y_3 + \left(\sin^2\phi + \cos2\beta \sin2\alpha \sin2\theta \cos^2 \phi\right)\right]} \geq 0.
\end{eqnarray}
From $G_1-G_2=0$, one must have
\begin{eqnarray}\label{eq:G1G2-4}
\nu_1\nu_2  \cos^2 2\theta \sin^2\phi \cos^4 \phi \sin^2 2\beta \sin^2 2\alpha \sin^2 2\theta =0.
\end{eqnarray}
Because in this subsection $\nu_1\neq 0$, $\nu_2\neq 0$, $\sin^2\beta \neq 0$, $\sin 2\alpha\neq 0$ and $\sin 2\theta\neq 0$, then from Eq. (\ref{eq:G1G2-4}) we have
\begin{eqnarray}\label{eq:G1G2-5}
&&\cos 2\theta=0,
\end{eqnarray}
which leads to
\begin{eqnarray}
&& \theta=\frac{\pi}{4},
\end{eqnarray}
and $\sin 2 \theta=1$. Then from above equations, we can have
\begin{eqnarray}
&& \nu_1 = \nu_2 \sqrt{ (\sin 2 \alpha \cos ^2 \phi -\sin^2\phi)^2+ 4 \sin 2 \alpha \sin ^2 \phi \cos ^2 \phi \cos^ 2 \beta}.
\end{eqnarray}
{However, in Sec. \ref{sum} we have shown that we need not consider separately the case of $\theta=\pi/4$, thus ending the proof for this case.}

This ends the proof of the sufficient condition.
\end{proof}

In summary, we have successfully prove our \textbf{Result:} \emph{Under the local unitary transformations, the rank-2 quantum state $\rho$ is separable if and only if $\mathcal{M}=\mathcal{M}^\dagger$.}

\newpage

\section{The EPR Steering Inequality}

\subsection{The $N$-Setting Linear Steering Inequality}

 It is well-known that Bell's nonlocality of quantum state can be detected by the violations of Bell's inequalities. Similarly, for a quantum state, whose property of EPR steerability can be detected through the violations of EPR steering inequalities. For instance, in \cite{NP2010}, the scientists have presented the $N$-setting linear EPR steering inequality to study the EPR steerability. The $N$-setting linear steering inequality is given by
\begin{eqnarray}\label{eq:QM-N}
\mathcal{I}_N=\frac{1}{N} \left[A_1 (\hat{b}_1\cdot \vec{\sigma}^B)+A_2 (\hat{b}_2\cdot \vec{\sigma}^B)+\cdots+A_N (\hat{b}_N\cdot \vec{\sigma}^B)\right] \leq C_{\rm LHS}.
\end{eqnarray}
In particular, for $N=3$, we have the 3-setting EPR steering inequality as

\begin{eqnarray}\label{eq:QM-3}
\mathcal{I}_3=\frac{1}{3} \left[A_1 (\hat{b}_1\cdot \vec{\sigma}^B)+A_2 (\hat{b}_2\cdot \vec{\sigma}^B)+A_3 (\hat{b}_3\cdot \vec{\sigma}^B)\right] \leq \frac{1}{\sqrt{3}}.
\end{eqnarray}
Here $A_1=\pm 1$, $A_2=\pm1$ and $A_3=\pm1$ are classical values for Alice, $C_{\rm LSH}=1/\sqrt{3}$ is the classical bound (i.e., the local-hidden-state (LHS) bound, or the LHS bound), $\vec{\sigma}=(\sigma_x, \sigma_y, \sigma_z)$ is the vector of Pauli matrices, $\vec{\sigma}^B$ is the vector of Pauli matrices for Bob's side, $\hat{b}_1$ , $\hat{b}_2$ and  $\hat{b}_3$ are unit vectors in the 3-dimensional space for Bob, and they are mutually orthogonal.
 Generally, vectors $\{\hat{b}_1, \hat{b}_2,\hat{b}_3\}$ can be obtained by rotating the simple vectors $\{\hat{e}_x=(1, 0, 0), \hat{e}_y=(0, 1, 0),\hat{e}_z=(0, 0, 1)\}$.

 Let us introduce the $2\times 2$ unitary matrix
\begin{eqnarray}
\mathcal{U}=\left(
              \begin{array}{cc}
                \cos{\eta} & -\sin{\eta}e^{-{\rm i} \tau} \\
                \sin{\eta}e^{-{\rm i} \tau} & \cos{\eta} \\
              \end{array}
            \right), \;\;\;\;\;
            \mathcal{U}^\dagger =\left(
              \begin{array}{cc}
                \cos{\eta} & \sin{\eta}e^{-{\rm i} \tau} \\
                -\sin{\eta}e^{{\rm i} \tau} & \cos{\eta} \\
              \end{array}
            \right),
\end{eqnarray}
then we have
\begin{eqnarray}
\hat{b}_1\cdot \vec{\sigma}&=& \mathcal{U} \,(\hat{e}_x \cdot \vec{\sigma}) \,\mathcal{U}^\dagger= \mathcal{U} \,{\sigma}_x \,\mathcal{U}^\dagger, \nonumber\\
\hat{b}_2\cdot \vec{\sigma}&=& \mathcal{U} \,(\hat{e}_y \cdot \vec{\sigma})\, \mathcal{U}^\dagger= \mathcal{U} \,{\sigma}_y \,\mathcal{U}^\dagger, \nonumber\\
\hat{b}_3\cdot \vec{\sigma}&=& \mathcal{U} \,(\hat{e}_z \cdot \vec{\sigma})\, \mathcal{U}^\dagger= \mathcal{U} \,{\sigma}_z \,\mathcal{U}^\dagger.
\end{eqnarray}
In this situation, one can have the vectors as
\begin{eqnarray}
&& \hat{b}_1=(\cos^2\eta-\cos2\tau\sin^2\eta,  -\sin^2{\eta} \sin2\tau, -\sin2\eta \cos\tau), \nonumber\\
&& \hat{b}_2=(-\sin^2{\eta} \sin2\tau, \cos^2\eta+\cos2\tau\sin^2\eta,   -\sin2\eta \sin\tau)\\
&& \hat{b}_3=(\cos\tau \sin2\eta,\sin\tau \sin2\eta,  \cos^2\eta -\sin^2{\eta}).
\end{eqnarray}
One can verify that
\begin{eqnarray}
&& |\hat{b}_1|=1, \;\;\;  |\hat{b}_2|=1, \;\;\; |\hat{b}_3|=1, \;\;\; \hat{b}_1 \cdot \hat{b}_2=0, \;\;\; \hat{b}_2 \cdot \hat{b}_3=0, \;\;\; \hat{b}_3 \cdot \hat{b}_1=0.
\end{eqnarray}
Quantum-mechanically, for Alice, the measurements are represented by
\begin{eqnarray}
&& \hat{A}_1=\hat{a}_1\cdot \vec{\sigma}, \nonumber\\
&& \hat{A}_2=\hat{a}_2\cdot \vec{\sigma}, \nonumber\\
&& \hat{A}_3=\hat{a}_3\cdot \vec{\sigma},
\end{eqnarray}
where $\hat{a}_1$ ,$\hat{a}_2$, and $\hat{a}_3$ are unit vectors in the 3-dimensional space for Alice, and they can be parameterized as
\begin{eqnarray}
&& \hat{a}_1=(\sin\theta_1 \cos\phi_1, \sin\theta_1 \sin\phi_1, \cos\theta_1), \nonumber\\
&& \hat{a}_2=(\sin\theta_2 \cos\phi_2, \sin\theta_2 \sin\phi_2, \cos\theta_2),\nonumber\\
&& \hat{a}_3=(\sin\theta_3 \cos\phi_3, \sin\theta_3 \sin\phi_3, \cos\theta_3).
\end{eqnarray}
The 3-setting (and also general $N$-setting) linear steering inequality is a state-independent inequality. In quantum mechanics (QM), for a given state $\rho$, its quantum expectation value is calculated by
\begin{eqnarray}
\mathcal{I}_3^{\rm QM}&=&{\rm Tr} \left\{\rho \left[\frac{1}{3} \left[(\hat{a}_1\cdot \vec{\sigma}^A) \otimes(\hat{b}_1\cdot \vec{\sigma}^B)+(\hat{a}_2\cdot \vec{\sigma}^A) \otimes(\hat{b}_2\cdot \vec{\sigma}^B)+(\hat{a}_3\cdot \vec{\sigma}^A) \otimes(\hat{b}_3\cdot \vec{\sigma}^B)\right]\right]\right\}.
\end{eqnarray}

In the following, we would like to provided two examples.

\emph{Example 1.---} \textcolor{blue}{Using the 3-setting steering inequality to prove the pure state $\ket{\psi_1}=\cos \theta \ket{00}+\sin \theta\ket{11}$ (or $\rho=\ket{\psi_1}\bra{\psi_1}$) is steerable in all the entangled region.} In quantum mechanics, the expectation value is calculated by
\begin{eqnarray}\label{eq:QM-1-3setting}
\mathcal{I}_3^{\rm QM}&=&{\rm Tr} \left\{\rho \left[\frac{1}{3} \left[(\hat{a}_1\cdot \vec{\sigma}^A) \otimes(\hat{b}_1\cdot \vec{\sigma}^B)+(\hat{a}_2\cdot \vec{\sigma}^A) \otimes(\hat{b}_2\cdot \vec{\sigma}^B)+(\hat{a}_3\cdot \vec{\sigma}^A) \otimes(\hat{b}_3\cdot \vec{\sigma}^B)\right]\right]\right\}\nonumber\\
&=& \frac{1}{3} \left[\langle\psi_1|(\hat{a}_1\cdot \vec{\sigma}^A) \otimes(\hat{b}_1\cdot \vec{\sigma}^B)|\psi_1\rangle+\langle\psi_1|(\hat{a}_2\cdot \vec{\sigma}^A) \otimes(\hat{b}_2\cdot \vec{\sigma}^B)|\psi_1\rangle+\langle\psi_1|(\hat{a}_3\cdot \vec{\sigma}^A) \otimes(\hat{b}_3\cdot \vec{\sigma}^B)|\psi_1\rangle\right].
\end{eqnarray}
We need to prove that for the entangled region $\theta\in(0, \pi/2)$, one always have the quantum violation
\begin{eqnarray}\label{eq:QM-2-3setting}
\mathcal{I}_3^{\rm QM} > \frac{1}{\sqrt{3}}.
\end{eqnarray}
The following is the proof.
\begin{proof}
After direct calculation, we can have
\begin{eqnarray}
\mathcal{I}_3^{\rm QM}&=&\dfrac{1}{3} \Bigg[
    \cos(\tau + \phi_{3}) \sin2\eta \sin\theta_{3} + \cos^2\eta \Big( \cos\phi_{1} \sin\theta_{1} - \sin\theta_{2} \sin\phi_{2} \Big) \nonumber\\
&&- \sin^2\eta \Big[ \cos(2\tau + \phi_{1}) \sin\theta_{1} + \sin\theta_{2} \sin(2\tau + \phi_{2}) \Big]
\Bigg]\sin2\theta \nonumber\\
&&+ \dfrac{1}{3} \Bigg[
    \cos2\eta \cos\theta_{3}  - \sin2\eta \Big( \cos\tau \cos\theta_{1} + \cos\theta_{2} \sin\tau \Big)
\Bigg].
\end{eqnarray}
Let the angles
\begin{eqnarray}
\tau = \frac{\pi}{4},\;\;\; \theta_{1} = \theta_{2} = \pi,\;\;\;  \phi_{3} = -\frac{\pi}{4},\;\;\; \eta = \frac{1}{2}\arctan\sqrt{2},
\end{eqnarray}
then the quantum expectation becomes
\begin{eqnarray}
\mathcal{I}_3^{\rm QM}=\frac{2 + \cos\theta_{3} + \sqrt{2} \sin 2\theta \sin\theta_{3}}{3\sqrt{3}},
\end{eqnarray}
i.e.,
\begin{eqnarray}
\mathcal{I}_3^{\rm QM}&=& \frac{2 + \sqrt{1+2\sin^2 2\theta}\left( \dfrac{1}{\sqrt{1+2\sin^2 2\theta}}\cos\theta_{3}
+ \dfrac{\sqrt{2} \sin2\theta}{\sqrt{1+2\sin^2 2\theta}} \sin\theta_{3}\right)}{3\sqrt{3}}\nonumber\\
&=& \frac{2 + \sqrt{1+2\sin^2 2\theta}\; \left( \cos\theta_0\cos\theta_{3}
+ \sin\theta_0 \sin\theta_{3}\right)}{3\sqrt{3}}\nonumber\\
&=& \frac{2 + \sqrt{1+2\sin^2 2\theta}\; \cos(\theta_{3}-\theta_0)}{3\sqrt{3}},
\end{eqnarray}
where
\begin{eqnarray}
\cos\theta_0=\dfrac{1}{\sqrt{1+2\sin^2 2\theta}}, \;\;\; \sin\theta_0=\dfrac{\sqrt{2} \sin2\theta}{\sqrt{1+2\sin^2 2\theta}}, \;\;\;\;
\tan\theta_0=\sqrt{2} \sin2\theta.
\end{eqnarray}
By taking
\begin{eqnarray}
\theta_3=\theta_0=\arctan\left[\sqrt{2} \sin2\theta\right],
\end{eqnarray}
then $\cos(\theta_3-\theta_0)=1$, thus we arrive at
\begin{eqnarray}
\mathcal{I}_3^{\rm QM}=\frac{2+ \sqrt{1+2\sin^2 2\theta}}{3\sqrt{3}}=\frac{2+ \sqrt{1+2\mathcal{C}^2}}{3\sqrt{3}}> \frac{1}{\sqrt{3}}.
\end{eqnarray}
{Here $\mathcal{C}=\sin 2\theta$ is the concurrence of the pure state $|\psi_1\rangle$. Obviously, the relation Eq. (\ref{eq:QM-2-3setting}) is valid for all the entangled region of $\theta\in(0, \pi/2)$ (i.e., $\mathcal{C}> 0$). This ends the proof.}
\end{proof}

\emph{Example 2.---} \textcolor{blue}{Using the 3-setting steering inequality to prove a special rank-2 quantum state $\rho_{\rm s}$ (given below) is steerable in all the entangled region.} Here, the special quantum state with rank 2 is given by
\begin{eqnarray}
{\rho_{\rm s}=\nu_1  \ket{\psi_1}\bra{\psi_1}+\nu_2 \ket{\psi_2}\bra{\psi_2},}
\end{eqnarray}
with
\begin{eqnarray}
&& \ket{\psi_1} =\cos \theta \ket{00}+\sin \theta\ket{11},  \nonumber\\
&& \ket{\psi_2} = \sin \theta \ket{00}-\cos \theta\ket{11}, \;\;\;\;\; \theta\in [0, \pi/2].
\end{eqnarray}
In the matrix form, the density matrix reads
\begin{eqnarray}
{\rho_{\rm s}}=
 \begin{bmatrix}
 \nu_1 \cos^2 \theta +\nu_2\sin ^2\theta    & 0  &  0  & (\nu_1-\nu_2) \sin \theta \cos \theta    \\
 0 & 0  & 0  & 0 \\
 0 & 0  & 0  & 0 \\
 (\nu_1-\nu_2) \sin \theta \cos \theta  & 0  &  0  &  \nu_1 \cos^2 \theta +\nu_2\sin ^2\theta  \\
\end{bmatrix},
\end{eqnarray}
whose concurrence is given by
\begin{eqnarray}
\mathcal{C}=|(\nu_1-\nu_2)\sin2\theta|.
\end{eqnarray}
{We need to prove that for the entangled state $\rho_{\rm s}$ (i.e., $\mathcal{C}>0$), one always has the quantum violation as shown in Eq. (\ref{eq:QM-2-3setting}). The following is the proof.}
\begin{proof}
After direct calculation,  the quantum expectation is given by
\begin{eqnarray}
\mathcal{I}_3^{\rm QM} &= &\frac{1}{3} \Bigg[
    \cos(\tau + \phi_{3}) \sin2\eta \sin\theta_{3}  + \cos^2\eta \Big( \cos\phi_{1} \sin\theta_{1} - \sin\theta_{2} \sin\phi_{2} \Big) \nonumber \\
&& - \sin^2\eta \Big[ \cos(2\tau + \phi_{1}) \sin\theta_{1} + \sin\theta_{2} \sin(2\tau + \phi_{2}) \Big]
\Bigg](\nu_1-\nu_2)\sin2\theta \nonumber\\
&&+ \frac{1}{3} \Bigg[
    \cos2\eta \cos\theta_{3}  - \sin2\eta \Big( \cos\tau \cos\theta_{1} + \cos\theta_{2} \sin\tau \Big)
\Bigg].
\end{eqnarray}
By choosing the same measure directions as the pure state case, i.e.,
\begin{eqnarray}
\tau = \frac{\pi}{4},\;\;\; \theta_{1} = \theta_{2} = \pi,\;\;\;  \phi_{3} = -\frac{\pi}{4},\;\;\; \eta = \frac{1}{2}\arctan\sqrt{2},
\end{eqnarray}
then the  quantum expectation becomes
\begin{eqnarray}
\mathcal{I}_3^{\rm QM}=\frac{2 + \cos\theta_{3} + \sqrt{2} (\nu_1 - \nu_2) \sin2\theta \sin\theta_{3}}{3 \sqrt{3}},
\end{eqnarray}
i.e.,
\begin{eqnarray}
\mathcal{I}_3^{\rm QM} &=& \frac{2 + \sqrt{1+2[(\nu_1-\nu_2)\sin2\theta]^2}
 \left(\dfrac{1}{\sqrt{1+2[(\nu_1-\nu_2)\sin2\theta]^2}}\cos\theta_{3} + \dfrac{\sqrt{2} (\nu_1 - \nu_2) \sin2\theta}{\sqrt{1+2[(\nu_1-\nu_2)\sin2\theta]^2}} \sin\theta_{3}\right)}{3 \sqrt{3}}\nonumber\\
&=& \frac{2 + \sqrt{1+2[(\nu_1-\nu_2)\sin2\theta]^2} \;\left(\cos\theta'_0\cos\theta_{3}
+ \sin\theta'_0 \sin\theta_{3}\right)}{3 \sqrt{3}}\nonumber\\
&=& \frac{2 + \sqrt{1+2[(\nu_1-\nu_2)\sin2\theta]^2} \; \cos (\theta_{3}-\theta'_0)}{3 \sqrt{3}},
\end{eqnarray}
where
\begin{eqnarray}
\cos\theta'_0=\dfrac{1}{\sqrt{1+2[(\nu_1-\nu_2)\sin2\theta]^2}}, \;\;\; \sin\theta'_0=\dfrac{\sqrt{2} (\nu_1 - \nu_2) \sin2\theta}{\sqrt{1+2[(\nu_1-\nu_2)\sin2\theta]^2}}, \;\;\;\;
\tan\theta'_0=\sqrt{2} (\nu_1 - \nu_2) \sin2\theta.
\end{eqnarray}
By taking
\begin{eqnarray}
\theta_3=\theta'_0=\arctan\left[\sqrt{2} (\nu_1 - \nu_2) \sin(2\theta)\right],
\end{eqnarray}
then $\cos(\theta_3-\theta'_0)=1$, thus we arrive at
\begin{eqnarray}
\mathcal{I}_3^{\rm QM}=\frac{2+ \sqrt{1+2[(\nu_1-\nu_2)\sin2\theta]^2}}{3\sqrt{3}}=\frac{2+ \sqrt{1+2 \mathcal{C}^2}}{3\sqrt{3}}> \frac{1}{\sqrt{3}}.
\end{eqnarray}
{Obviously, the relation Eq. (\ref{eq:QM-2-3setting}) is valid for the entangled state $\rho'$ with $\mathcal{C}> 0$). This ends the proof.}
\end{proof}

\begin{remark} However, the $N$-setting linear steering inequality is not powerful enough to detect all the rank-2 quantum entangled states. For example, the all-versus-nothing (AVN) state which is given by
\begin{eqnarray}\label{eq:rho-a1}
\rho_{\rm AVN}=\nu_1  \ket{\psi_1}\bra{\psi_1}+\nu_2 \ket{\psi_2}\bra{\psi_2},
\end{eqnarray}
with
\begin{eqnarray}
&& \ket{\psi_1} =\cos \theta \ket{00}+\sin \theta\ket{11},  \nonumber\\
&&\ket{\psi_2} =\cos \theta \ket{10} + \sin\theta\ket{01}, \;\;\;\;\; \theta\in [0, \pi/2],
\end{eqnarray}
and whose concurrence is
\begin{eqnarray}
\mathcal{C}=|\nu_1 - \nu_2| \sin2\theta.
\end{eqnarray}
By using the approach of AVN proof of EPR steering (i.e., EPR steering without inequality), Ref. \cite{2013AVN} has been successfully proved that the state $\rho_{\rm AVN}$ is steerable in all the entangled region (i.e., $\mathcal{C}> 0$). However, the $N$-setting linear steering inequality cannot be violated by the state $\rho_{\rm AVN}$ for all the entangled region even we have increased the measurement setting from $N=3$ to $N=10$.
Due to this reason, it motivates us to find other more efficient steering inequalities to detect the steerability of rank-2 quantum states. In the next subsection, to reach this purpose, we shall present a novel state-dependent steering inequality. $\blacksquare$
\end{remark}

\newpage

\subsection{The State-Dependent Steering Inequality}


The two-qubit density matrix $\rho$ can be written as
\begin{eqnarray}
 \rho &=& P^{\hat{n}}_{0} \otimes \tilde{\rho}_0^{\hat{n}}+ P^{\hat{n}}_{1} \otimes \tilde{\rho}_1^{\hat{n}}+
|+\hat{n}\rangle\langle-\hat{n}| \otimes \mathcal{M}+|-\hat{n}\rangle\langle+\hat{n}| \otimes \mathcal{M}^\dag.
\end{eqnarray}
To detect the steerability of the state $\rho$, we adopt the approach of EPR steering inequality. In this work, we present a state-dependent steering inequality, which is given by
\begin{eqnarray}\label{eq:sdsi-1}
    \langle \mathcal{W} \rangle \leq  C_{\text{LHS}}.
\end{eqnarray}
where $C_{\text{LHS}}$ is the classical bound (i.e., the LHS bound), and
\begin{eqnarray}
 \langle \mathcal{W} \rangle = \mathrm{Tr}(\mathcal{W} \rho)
\end{eqnarray}
is quantum expectation of the operator $\mathcal{W}$ acting on $\rho$.

Explicitly, the operator $\mathcal{W}$ is a projector defined by
\begin{eqnarray}
 &&\mathcal{W} = |+\rangle\langle+|\otimes |\hat{n}_B \rangle\langle\hat{n}_B|,
\end{eqnarray}
with
\begin{eqnarray}\label{eq:proj-1}
|+\rangle=\frac{1}{\sqrt{2}}\left(\ket{+\hat{n}} + e^{{\rm i} \delta} \ket{-\hat{n}}\right), \;\;\; \delta \in [0, 2\pi].
\end{eqnarray}
\begin{eqnarray}
&& \hat{n}=(\sin\xi \cos\tau, \sin\xi \sin\tau, \cos\xi),
\end{eqnarray}
\begin{eqnarray}
&& \ket{+\hat{n}}=\begin{pmatrix}
\cos\frac{\xi}{2}\\
\sin\frac{\xi}{2}\, e^{{\rm i}\tau}\\
\end{pmatrix}=\cos\frac{\xi}{2} |0\rangle+\sin\frac{\xi}{2}\, e^{{\rm i}\tau} |1\rangle
\end{eqnarray}

\begin{eqnarray}
 \ket{-\hat{n}}=\begin{pmatrix}
\sin\frac{\xi}{2}\\
-\cos\frac{\xi}{2} \, e^{{\rm i}\tau} \\
\end{pmatrix}= \sin\frac{\xi}{2}|0\rangle -\cos\frac{\xi}{2} \, e^{{\rm i}\tau} |1\rangle,
\end{eqnarray}
Here $|+\rangle\langle+|$ is the projective measurement of Alice, i.e.,
\begin{eqnarray}
|+\rangle\langle+| =\frac{1}{2}\left(\openone+\vec{\sigma}\cdot \hat{m}\right),
\end{eqnarray}
whose measurement direction (denoted by $\hat{m}$) is perpendicular to the $\hat{n}$ direction (i.e., $\hat{m}\cdot\hat{n}=0$).
For example, let $\delta=0$ and $\hat{n}=\hat{e}_x$, then
\begin{eqnarray}
&& \ket{+\hat{n}}=\frac{1}{\sqrt{2}}(\ket{0}+\ket{1}), \nonumber\\
&&  \ket{-\hat{n}}=\frac{1}{\sqrt{2}}(\ket{0}-\ket{1}),
\end{eqnarray}
\begin{eqnarray}
&& \ket{+}=\frac{1}{\sqrt{2}}\left(\ket{+\hat{n}} + \ket{-\hat{n}}\right)=\ket{0}.
\end{eqnarray}
From
\begin{eqnarray}
|+\rangle\langle+| =\frac{1}{2}\left(\openone+\vec{\sigma}\cdot \hat{z}\right),
\end{eqnarray}
we know that the measurement direction corresponding to the projector $|+\rangle\langle+|$ is just $\hat{m}=\hat{e}_z$, which is indeed perpendicular to $\hat{n}=\hat{e}_x$. For the general $\delta\neq 0$, the measurement direction $\hat{m}$ is a unit vector located in the $yz$-plane.
Similarly, $|\hat{n}_B \rangle\langle\hat{n}_B|$ is the projective measurement of Bob, with
\begin{eqnarray}
    |\hat{n}_B\rangle = \cos \frac{\theta_B}{2} \ket{0} + \sin \frac{\theta_B}{2} e^{{\rm i}\phi_B} \ket{1}.
\end{eqnarray}
and the angles $\theta_B$ and $\phi_B$ are free parameters. Explicitly, $|\hat{n}_B \rangle\langle\hat{n}_B|$ is expressed as
\begin{eqnarray}
 |\hat{n}_B\rangle \langle \hat{n}_B| = \frac{1}{2}\begin{pmatrix}
 1+\cos\theta_B &
 \sin\theta_B e^{-{\rm i} \phi_B}\\
 \sin\theta_B e^{{\rm i} \phi_B} &
 1-\cos\theta_B
\end{pmatrix},
\end{eqnarray}
or
\begin{eqnarray}
 |\hat{n}_B\rangle \langle \hat{n}_B| = \frac{1}{2}\left(\openone+\vec{\sigma}\cdot \hat{n}_B\right),
\end{eqnarray}
with
\begin{eqnarray}
\hat{n}_B=(\sin\theta_B \cos\phi_B, \sin\theta_B \sin\phi_B, \cos\theta_B).
\end{eqnarray}

In this work, the classical bound $C_{\text{LHS}}$ is defined by
\begin{eqnarray}\label{eq:C-1d}
    &&C_{\text{LHS}} =
  \max_{\hat{n}_B} \left\{ {\rm Tr}[  \left( |+\rangle\langle+| \otimes |\hat{n}_B \rangle\langle\hat{n}_B|\right)
  {\rho_{\rm sep}}]\right\},
\end{eqnarray}
where
\begin{eqnarray}
&& {\rho_{\rm sep}}= P^{\hat{n}}_{0} \otimes \tilde{\rho}_0^{\hat{n}}+ P^{\hat{n}}_{1} \otimes \tilde{\rho}_1^{\hat{n}}+
|+\hat{n}\rangle\langle-\hat{n}| \otimes \mathcal{M}+|-\hat{n}\rangle\langle+\hat{n}| \otimes \mathcal{M}^\dag
\end{eqnarray}
 represent the rank-2 (including rank-1 as a special case) separable states. From previous sections, we have known that, under the local unitary transformations, for the separable states ${\rho_{\rm sep}}$ there always exists a measurement direction $\hat{n}=(\sin\xi \cos\tau, \sin\xi \sin\tau, \cos\xi)$ for Alice such that the matrix $\mathcal{M}=\mathcal{M}^\dag$. Due to such a measurement direction $\hat{n}$ and the projector defined by Eq. (\ref{eq:proj-1}), one can have the LHS bound as
\begin{eqnarray}\label{eq:clhs-2}
    C_{\text{LHS}} &=& \max_{\hat{n}_B} \left[ \mathrm{Tr} \left( |\hat{n}_B \rangle\langle\hat{n}_B| \frac{\tilde{\rho}_0^{\hat{n}} + \tilde{\rho}_1^{\hat{n}}+e^{{\rm i} \delta} \mathcal{M} + e^{-{\rm i} \delta} \mathcal{M}^\dagger}{2} \right) \right].
\end{eqnarray}
Note that $C_{\text{LHS}}$ depends on the state $\rho$, thus we call the presented steering inequality as a state-dependent steering inequality. Besides, the expression of $C_{\text{LHS}}$ in Eq. (\ref{eq:clhs-2}) also depends on the measurement parameters of Alice (such as $\hat{n}$ and $\delta$). In practical operation, we can choose appropriate measurement directions such that $C_{\text{LHS}}$ is easy to calculate. For example, we can choose
\begin{eqnarray}
&&\delta=\frac{\pi}{2},
\end{eqnarray}
then we have
\begin{eqnarray}
&& e^{{\rm i} \delta} \mathcal{M} + e^{-{\rm i} \delta} \mathcal{M}^\dagger = {\rm i} \left(\mathcal{M}-\mathcal{M}^\dagger \right).
\end{eqnarray}
Remarkably, as we have proved in the previous sections, under the local unitary transformations, for any rank-2 quantum state $\rho_{\rm sep}$, the necessary and sufficient condition for the separability is just $\mathcal{M}=\mathcal{M}^\dagger$. Then the classical bound can be simplified to
\begin{eqnarray}
    C_{\text{LHS}} &=& \max_{\hat{n}_B} \left[ \mathrm{Tr} \left( |\hat{n}_B \rangle\langle\hat{n}_B| \frac{\tilde{\rho}_0^{\hat{n}} + \tilde{\rho}_1^{\hat{n}}}{2} \right) \right],
\end{eqnarray}
i.e.,
\begin{eqnarray}\label{eq:clhs-3}
    C_{\text{LHS}} &=& \frac{1}{2}\max_{\hat{n}_B} \left[ \mathrm{Tr} \left( |\hat{n}_B \rangle\langle\hat{n}_B| \rho_B\right) \right],
\end{eqnarray}
where
\begin{eqnarray}
&&\rho_B= \tilde{\rho}_0^{\hat{n}} + \tilde{\rho}_1^{\hat{n}}
\end{eqnarray}
is the reduced state of Bob. Because $\rho_B$ is independent of the measurement direction $\hat{n}$, thus one may observe that the expression of $C_{\text{LHS}}$ in Eq. (\ref{eq:clhs-3}) does not depend on the measurement direction $\hat{n}$.

By the way, we can also choose
\begin{eqnarray}
&&\delta=-\frac{\pi}{2},
\end{eqnarray}
then we have
\begin{eqnarray}
&& e^{{\rm i} \delta} \mathcal{M} + e^{-{\rm i} \delta} \mathcal{M}^\dagger = -{\rm i} \left(\mathcal{M}-\mathcal{M}^\dagger \right),
\end{eqnarray}
which yields the same $C_{\text{LHS}}$. The physical meaning of $C_{\text{LHS}}$ is simple, it is half of the maximal eigenvalue of $\rho_B$, i.e.,
\begin{eqnarray}\label{eq:clhs-4}
    C_{\text{LHS}} &=& \frac{1}{2}\max_{\hat{n}_B} \left[ \mathrm{Tr} \left( |\hat{n}_B \rangle\langle\hat{n}_B| \rho_B\right) \right]
    =\frac{1}{2}\lambda_{\rm max}(\rho_B).
\end{eqnarray}

In addition, for the convenience of calculation, we would like to list the matrix forms for the state vector $\ket{+}$ and the projector $|+\rangle\langle +|$, which read
\begin{eqnarray}
\ket{+}&=& \frac{1}{\sqrt{2}}\left(|+\hat{n}\rangle + e^{{\rm i} \delta} |-\hat{n}\rangle\right)\nonumber\\
&=& \frac{1}{\sqrt{2}}\left[\left(\cos\frac{\xi}{2} |0\rangle+\sin\frac{\xi}{2}\, e^{{\rm i}\tau} |1\rangle\right)
+ e^{{\rm i} \delta} \left( \sin\frac{\xi}{2}   |0\rangle {-} \, \, e^{{\rm i}\tau}\cos\frac{\xi}{2} |1\rangle\right)\right]\nonumber\\
&=& \frac{1}{\sqrt{2}}\left[\left(\cos\frac{\xi}{2} {+}\sin\frac{\xi}{2}\, e^{{\rm i}\delta} \right) |0\rangle
+e^{{\rm i}\tau}   \left( \sin\frac{\xi}{2}\, {-} \cos\frac{\xi}{2} e^{{\rm i} \delta} \right)|1\rangle\right]\nonumber\\
&=& \frac{1}{\sqrt{2}} \begin{pmatrix}
\cos\frac{\xi}{2} {+}\sin\frac{\xi}{2}\, e^{{\rm i}\delta}\\
e^{{\rm i}\tau}   \left( \sin\frac{\xi}{2}\, {-} \cos\frac{\xi}{2} e^{{\rm i} \delta} \right)\\
\end{pmatrix},
\end{eqnarray}
\begin{eqnarray}
|+\rangle\langle +|
 &=&  \frac{1}{2}
\left(
  \begin{array}{cc}
   (\cos\frac{\xi}{2} {+}\sin\frac{\xi}{2}\, e^{{\rm i}\delta})
     (\cos\frac{\xi}{2} {+}\sin\frac{\xi}{2}\, e^{{\rm i}\delta})^*  &
   \left(\cos\frac{\xi}{2} {+}\sin\frac{\xi}{2}\, e^{{\rm i}\delta}\right)
   \left[e^{{\rm i}\tau}   \left( \sin\frac{\xi}{2}\, {-} \cos\frac{\xi}{2} e^{{\rm i} \delta} \right)\right]^*\\
    \left(\cos\frac{\xi}{2} {+}\sin\frac{\xi}{2}\, e^{{\rm i}\delta}\right)^*
   \left[e^{{\rm i}\tau}   \left( \sin\frac{\xi}{2}\, {-} \cos\frac{\xi}{2} e^{{\rm i} \delta} \right)\right] &
   \left[e^{{\rm i}\tau}   \left( \sin\frac{\xi}{2}\, {-} \cos\frac{\xi}{2} e^{{\rm i} \delta} \right)\right]\left[e^{{\rm i}\tau}   \left( \sin\frac{\xi}{2}\, {-} \cos\frac{\xi}{2} e^{{\rm i} \delta} \right)\right]^*  \\
  \end{array}
\right)\nonumber\\
 &=&  \frac{1}{2}
\left(
  \begin{array}{cc}
   1 {+}\sin\xi \cos\delta  &
 e^{-{\rm i}\tau}  \left[ {-} \left(\cos\frac{\xi}{2}\right)^2e^{-{\rm i}\delta} {+}\left(\sin\frac{\xi}{2}\right)^2\,e^{{\rm i}\delta}\right]\\
e^{{\rm i}\tau}  \left[ {-} \left(\cos\frac{\xi}{2}\right)^2e^{{\rm i}\delta} {+}\left(\sin\frac{\xi}{2}\right)^2\,e^{-{\rm i}\delta}\right] &
   1 {-}\sin\xi \cos\delta  \\
  \end{array}
\right)\nonumber\\
 &=&  \frac{1}{2}
\left(
  \begin{array}{cc}
   1 {+}\sin\xi \cos\delta  &
e^{-{\rm i}\tau} \left[-\cos\delta\cos\xi+\rm i\sin\delta  \right]\\
 e^{{\rm i}\tau}  \left[-\cos\delta\cos\xi-\rm i\sin\delta  \right] & 1 {-}\sin\xi \cos\delta  \\
  \end{array}
\right).
\end{eqnarray}
Specially, for $\delta=\pi/2$, we can have
\begin{eqnarray}
|+\rangle=\frac{1}{\sqrt{2}}\left(\ket{+\hat{n}} + e^{{\rm i} \pi/2} \ket{-\hat{n}}\right)=\frac{1}{\sqrt{2}}\left(\ket{+\hat{n}} +{\rm i} \ket{-\hat{n}}\right),
\end{eqnarray}
and
\begin{eqnarray}
|+\rangle\langle +|
&=&  \frac{1}{2}
\left(
  \begin{array}{cc}
   1  &
 {\rm i} e^{-{\rm i}\tau} \\
-{\rm i} e^{{\rm i}\tau}   & 1  \\
  \end{array}
\right),
\end{eqnarray}
which is $\xi$-independent, thus simplifying the calculations. For $\delta=-\pi/2$, we can have
\begin{eqnarray}
|+'\rangle=\frac{1}{\sqrt{2}}(\ket{+\hat{n}} + e^{-{\rm i} \pi/2} \ket{-\hat{n}})=\frac{1}{\sqrt{2}}(\ket{+\hat{n}} -{\rm i} \ket{-\hat{n}}),
\end{eqnarray}
and
\begin{eqnarray}
|+'\rangle\langle +'|
&=&  \frac{1}{2}
\left(
  \begin{array}{cc}
   1  &
 -{\rm i} e^{-{\rm i}\tau} \\
{\rm i} e^{{\rm i}\tau}   & 1  \\
  \end{array}
\right),
\end{eqnarray}
which is also $\xi$-independent.

In the next section, we shall adopt the state-dependent steering inequality (\ref{eq:sdsi-1}) to study the steerability of the rank-2 quantum states. As a result, we present a theorem on Einstein-Podolsky-Rosen steering, i.e., all the rank-2 (including rank-1 as a special case) two-qubit entangled states violate the state-dependent steering inequality, thus possessing the EPR steerability.

\newpage


\section{A Theorem on Einstein-Podolsky-Rosen Steering}

\subsection{The Theorem}

Our main result is the following theorem on Einstein-Podolsky-Rosen steering.

\begin{theorem}
For the two-qubit quantum state
\begin{eqnarray}\label{eq:rho-1a}
\rho=\nu_1  \ket{\psi_1}\bra{\psi_1}+\nu_2 \ket{\psi_2}\bra{\psi_2},
\end{eqnarray}
with
\begin{eqnarray}\label{eq:rho-1b}
&&\ket{\psi_1} =\cos \theta \ket{00}+\sin \theta\ket{11}, \nonumber\\
&& \ket{\psi_2} =\cos\phi(\cos \alpha \ket{01} +\sin\alpha\ket{10})+ e^{{\rm i}\beta}\sin\phi(\sin \theta \ket{00}-\cos \theta\ket{11}),
\end{eqnarray}
and
\begin{eqnarray}
&& \theta\in [0, \pi/2], \;\;\;\; \phi \in [0, \pi/2], \;\;\;\;  \alpha \in [0, \pi/2], \;\;\;\; \beta \in [0, 2\pi],   \nonumber\\
&& \nu_1 \in [0,1], \;\;\; \nu_2 \in [0,1], \;\;\; \nu_1+\nu_2=1,
\end{eqnarray}
under the local unitary transformations, if $\rho$ is entangled, then $\rho$ must be steerable. Here $\rho$ is generally a rank-2 mixed state, which reduces to the rank-1 pure state when one of the weights ($\nu_1$ or $\nu_2$) becomes 1.
\end{theorem}

Our approach is the following state-dependent steering inequality
\begin{eqnarray}\label{eq:sdsi-2}
    \langle \mathcal{W} \rangle \leq  C_{\text{LHS}}.
\end{eqnarray}
 To prove \textbf{Theorem 1}, we need to prove any entangled state $\rho$ violates the steering inequality (\ref{eq:sdsi-2}). Let us denote $\langle \mathcal{W} \rangle^{\rm max}$ as the maximal value of quantum expectation. Mathematically, we only need to prove that for any concurrence $\mathcal{C}>0$, we always have the following relation
\begin{eqnarray}\label{eq:sdsi-3}
    \langle \mathcal{W} \rangle^{\rm max} > C_{\text{LHS}}, \;\;\;\; \forall \;\;\mathcal{C}>0.
\end{eqnarray}

\subsection{Proof of the Theorem}

To prove the theorem,  we need to calculate the classical bound $C_{\text{LHS}}$ and the quantum expectation value $\langle \mathcal{W} \rangle$.

\begin{remark}\textcolor{blue}{Calculation of the Classical Bound.}
Based on Eqs. (\ref{eq:rho-1a}) and Eqs. (\ref{eq:rho-1b}), one can have Bob's reduced density matrix as
\begin{eqnarray}
&& \rho_B= \left(
\begin{array}{cc}
 \nu_1 \cos ^2\theta+\nu_2 \sin ^2\alpha \cos ^2\phi+\nu_2 \sin ^2\theta \sin ^2\phi
 & \nu_2 \sin \phi \cos\phi \left(e^{{\rm i} \beta } \cos \alpha \sin \theta-e^{-{\rm i} \beta }  \sin \alpha \cos\theta\right) \\
  \nu_2 \sin \phi \cos\phi \left(e^{-{\rm i} \beta } \cos \alpha \sin \theta-e^{{\rm i} \beta }  \sin \alpha \cos\theta\right) &
  \nu_1 \sin ^2\theta+\nu_2 \cos ^2\alpha \cos ^2\phi+\nu_2 \cos ^2\theta \sin ^2\phi \\
\end{array}
\right).
\end{eqnarray}
There are two methods to calculate the classical bound. The first method is based on Eq. (\ref{eq:clhs-3}), i.e.,
\begin{eqnarray}
    C_{\text{LHS}}
    &=&\frac{1}{2} \max_{\hat{n}_B} \left[ \mathrm{Tr} \left( |\hat{n}_B\rangle \langle \hat{n}_B| \rho_{B} \right) \right]\nonumber\\
    &=& \frac{1}{2} \max_{\hat{n}_B} \biggr\{ \mathrm{Tr} \biggr[ \frac{1}{2}\begin{pmatrix}
 1+\cos\theta_B &
 \sin\theta_B e^{-{\rm i} \phi_B}\\
 \sin\theta_B e^{{\rm i} \phi_B} &
 1-\cos\theta_B
\end{pmatrix} \rho_B \biggr] \biggr\}\nonumber\\
&=& \frac{1}{4}+ \frac{1}{4} \max_{\hat{n}_B}\biggr[-\nu_2 \sin \alpha \cos\theta \sin \theta_B \sin 2 \phi \cos (\beta -\phi_B)
+\nu_2 \cos\alpha \sin\theta \sin \theta_B \sin2 \phi \cos (\beta +\phi_B)\nonumber\\
&&-\nu_2 \cos 2 \alpha \cos \theta_B \cos ^2\phi+\cos ^2\theta \cos \theta_B \left(\nu_1-\nu_2 \sin ^2\phi\right)
-\nu_1 \sin ^2\theta\cos \theta_B+\nu_2 \sin ^2\theta \cos \theta_B \sin ^2\phi\biggr]\nonumber\\
&=& \frac{1}{4}+ \frac{1}{4} \max_{\hat{n}_B}\biggr\{ -\sin \theta_B\left[\cos\phi_B\nu_2 \cos\beta \sin(\alpha - \theta) \sin2\phi + \sin\phi_B \nu_2 \sin\beta \sin(\alpha + \theta) \sin2\phi\right]+\nonumber\\
&&\cos \theta_B\left[-\nu_2 \cos2\alpha \cos^2\phi + \cos2\theta (\nu_1 - \nu_2 \sin^2\phi)\right]\biggr\}.
\end{eqnarray}
Let us define the vector
\begin{eqnarray}
\vec{F}=(F_1,F_2,F_3)
\end{eqnarray}
with
\begin{eqnarray}
F_1 &=&-\nu_2 \cos\beta \sin(\alpha - \theta) \sin2\phi, \nonumber\\
F_2 &=&-\nu_2 \sin\beta \sin(\alpha + \theta) \sin2\phi, \nonumber\\
F_3 &=& -\nu_2 \cos2\alpha \cos^2\phi + \cos2\theta (\nu_1 - \nu_2 \sin^2\phi).
\end{eqnarray}
We then have
\begin{eqnarray}
    C_{\text{LHS}}
&=& \frac{1}{4}+ \frac{1}{4} \max_{\hat{n}_B}\biggr(\hat{n}_B \cdot \vec{F}\biggr),
\end{eqnarray}
with
\begin{eqnarray}
\hat{n}_B=(\sin\theta_B \cos\phi_B, \sin\theta_B \sin\phi_B, \cos\theta_B).
\end{eqnarray}
Thus
\begin{eqnarray}\label{eq:clhs-a}
    C_{\text{LHS}}
&=& \frac{1}{4}+ \frac{1}{4} |\vec{F}|.
\end{eqnarray}
Here the maximal value of $(\hat{n}_B \cdot \vec{F})$ occurs when the vectors $\hat{n}_B$ and $\vec{F}$ are parallel.

The second method is based on Eq. (\ref{eq:clhs-4}). We can rewritten the density matrix $\rho_B$ in its Bloch form, i.e.,
\begin{eqnarray}
&&\rho_B= \frac{1}{2} \left(\openone + \vec{\sigma}\cdot \vec{F}\right),
\end{eqnarray}
where $\vec{F}$ represents the Bloch vector. It is easy to know that two eigenvalues of $\rho_B$ are $(1+|\vec{F}|)/2$ and $(1-|\vec{F}|)/2$, respectively. Thus one immediately has
\begin{eqnarray}
    C_{\text{LHS}}
&=& \frac{1}{2}\lambda_{\rm max}(\rho_B)=\frac{1}{4}+ \frac{1}{4} |\vec{F}|.
\end{eqnarray}
$\blacksquare$
\end{remark}

Now, let us come to prove the theorem.

\begin{proof}
To prove the theorem, we choose two kinds of the projector $|+\rangle\langle+|$. The first one (still denoted by $|+\rangle\langle+|$ for convenience) is obtained from $|+\rangle\langle+|$ by letting $\delta=+\pi/2$, and the second one (denoted by $|+'\rangle\langle+'|$) is obtained from $|+\rangle\langle+|$ by letting $\delta=-\pi/2$. Obviously, these two choices lead to the same classical bound $C_{\text{LHS}}$ as shown in Eq. (\ref{eq:clhs-a}). Accordingly, the two corresponding quantum expectations are as follows
\begin{eqnarray}\label{eq:w1-a}
\langle \mathcal{W} \rangle_1^{\rm max}
&=& \max\limits_{\hat{n}_B} \mathrm{Tr} [ (|+\rangle\langle+|\otimes |\hat{n}_B \rangle\langle\hat{n}_B|) \, \rho ],\nonumber\\
\langle \mathcal{W} \rangle_2^{\rm max}
&=& \max\limits_{\hat{n}_B} \mathrm{Tr} [ \left(|+'\rangle\langle+'|\otimes |\hat{n}_B \rangle\langle\hat{n}_B|\right) \, \rho ].
\end{eqnarray}
We shall prove that at least one of them can exceed the classical bound when the state $\rho$ is entangled, i.e.,
\begin{eqnarray}
\max[ \langle \mathcal{W} \rangle_1^{\rm max}, \langle \mathcal{W} \rangle_2^{\rm max}]\geq C_\text{LHS},
\end{eqnarray}
here the equal sign occurs only when the state $\rho$ is separable. Explicitly, we have
\begin{eqnarray}
 \langle \mathcal{W} \rangle_1 &=&\frac{1}{4}+ \frac{1}{4} \left\{ \nu_2 \left[ \cos(\alpha - \theta) \cos\tau \sin\beta - \cos\beta \cos(\alpha + \theta) \sin\tau \right] \sin2\phi \right\} \nonumber\\
&&+ \frac{1}{4} \cos\theta_B \biggr[ -\nu_2 \cos2\alpha \cos^2\phi + \cos2\theta \left( \nu_1 - \nu_2 \sin^2\phi \right) \nonumber\\
&& + \nu_2 \left[ -\sin\beta \cos(\alpha + \theta) \cos\tau + \cos\beta \cos(\alpha - \theta) \sin\tau \right] \sin2\phi \biggr] \nonumber\\
&&+ \frac{1}{4} \sin\theta_B \sin\phi_B \left\{ \cos\tau \left[ -\nu_2 \cos^2\phi \sin2\alpha + \sin2\theta \left( \nu_1 - \nu_2 \sin^2\phi \right) \right] - \nu_2 \sin\beta \sin(\alpha + \theta) \sin 2\phi \right\} \nonumber\\
&&+ \frac{1}{4} \sin\theta_B \cos\phi_B \left\{ \sin\tau \left[ \nu_2 \cos^2\phi \sin 2\alpha + \sin2\theta \left( \nu_1 - \nu_2 \sin^2\phi \right) \right] - \nu_2 \cos\beta \sin(\alpha - \theta) \sin 2\phi \right\}.
\end{eqnarray}

\begin{eqnarray}
 \langle \mathcal{W} \rangle_2 &=&\frac{1}{4}- \frac{1}{4} \left\{ \nu_2 \left[ \cos(\alpha - \theta) \cos\tau \sin\beta - \cos\beta \cos(\alpha + \theta) \sin\tau \right] \sin2\phi \right\} \nonumber\\
&&+ \frac{1}{4} \cos\theta_B \biggr[ -\nu_2 \cos2\alpha \cos^2\phi + \cos2\theta \left( \nu_1 - \nu_2 \sin^2\phi \right) \nonumber\\
&& - \nu_2 \left[ -\sin\beta \cos(\alpha + \theta) \cos\tau + \cos\beta \cos(\alpha - \theta) \sin\tau \right] \sin2\phi \biggr] \nonumber\\
&&+ \frac{1}{4} \sin\theta_B \sin\phi_B \left\{ -\cos\tau \left[ -\nu_2 \cos^2\phi \sin2\alpha + \sin2\theta \left( \nu_1 - \nu_2 \sin^2\phi \right) \right] - \nu_2 \sin\beta \sin(\alpha + \theta) \sin 2\phi \right\} \nonumber\\
&&+ \frac{1}{4} \sin\theta_B \cos\phi_B \left\{ -\sin\tau \left[ \nu_2 \cos^2\phi \sin 2\alpha + \sin2\theta \left( \nu_1 - \nu_2 \sin^2\phi \right) \right] - \nu_2 \cos\beta \sin(\alpha - \theta) \sin 2\phi \right\}.
\end{eqnarray}
Let
\begin{eqnarray}
H_0= \nu_2 \left[ \cos(\alpha - \theta) \cos\tau \sin\beta - \cos\beta \cos(\alpha + \theta) \sin\tau \right] \sin2\phi,
\end{eqnarray}
and
\begin{eqnarray}
\vec{H}=(H_1, H_2, H_3),
\end{eqnarray}
with
\begin{eqnarray}
H_1 &=&\sin\tau \left[ \nu_2 \cos^2\phi \sin 2\alpha + \sin2\theta \left( \nu_1 - \nu_2 \sin^2\phi \right) \right],\nonumber\\
H_2 &=&\cos\tau \left[ -\nu_2 \cos^2\phi \sin2\alpha + \sin2\theta \left( \nu_1 - \nu_2 \sin^2\phi \right) \right], \nonumber\\
H_3 &=& \nu_2 \left[ -\sin\beta \cos(\alpha + \theta) \cos\tau + \cos\beta \cos(\alpha - \theta) \sin\tau \right] \sin2\phi.
\end{eqnarray}
Thus, the quantum expectations are
\begin{eqnarray}
 \langle \mathcal{W} \rangle_1
 &=& \frac{1}{4}+\frac{1}{4}H_0+
 \frac{1}{4}(\sin\theta_{B} \cos\phi_{B}, \sin\theta_{B} \sin\phi_{B}, \cos\theta_{B})\cdot(\vec{F}+\vec{H})\nonumber\\
&=& \frac{1}{4}+\frac{1}{4}H_0+
 \frac{1}{4}\left[\hat{n}_B\cdot(\vec{F}+\vec{H})\right],
\end{eqnarray}
\begin{eqnarray}
 \langle \mathcal{W} \rangle_2 &=& \frac{1}{4}-\frac{1}{4}H_0+
 \frac{1}{4}(\sin\theta_{B} \cos\phi_{B}, \sin\theta_{B} \sin\phi_{B}, \cos\theta_{B})\cdot(\vec{F}-\vec{H})\nonumber\\
 &=& \frac{1}{4}-\frac{1}{4}H_0+
 \frac{1}{4}\left[\hat{n}_B\cdot(\vec{F}-\vec{H})\right].
\end{eqnarray}
Then we have
\begin{eqnarray}\label{eq:w1-b}
 \langle \mathcal{W} \rangle_1^{\rm max}=\frac{1}{4}+\frac{1}{4}H_0+\frac{1}{4}|\vec{F}+\vec{H}|,
\end{eqnarray}
here the maximum occurs when the vectors $\hat{n}_B$ and $(\vec{F}+\vec{H})$ are parallel. Similarly,
\begin{eqnarray}\label{eq:w2-b}
 \langle \mathcal{W} \rangle_2^{\rm max}=\frac{1}{4}-\frac{1}{4}H_0+\frac{1}{4}|\vec{F}-\vec{H}|,
\end{eqnarray}
the maximum occurs when the vectors $\hat{n}_B$ and $(\vec{F}-\vec{H})$ are parallel.

Based on Eqs. (\ref{eq:w1-b}) and (\ref{eq:w2-b}), we can have
\begin{eqnarray}
 \langle \mathcal{W} \rangle_1^{\rm max}+ \langle \mathcal{W} \rangle_2^{\rm max}
&=&\frac{1}{4}+\frac{1}{4}H_0+\frac{1}{4}|\vec{F}+\vec{H}| + \frac{1}{4}-\frac{1}{4}H_0+\frac{1}{4}|\vec{F}-\vec{H}| \nonumber\\
&=&\frac{1}{4}+\frac{1}{4}|\vec{F}+\vec{H}| + \frac{1}{4}+\frac{1}{4}|\vec{F}-\vec{H}| \nonumber\\
&\geq&\frac{1}{2}+{\frac{1}{4}}|\vec{F}+\vec{H}+\vec{F}-\vec{H}|\nonumber\\
&=&\frac{1}{2}+{\frac{1}{2}}|\vec{F}|=2    C_{\text{LHS}},
\end{eqnarray}
which implies that at least one of $\{\langle \mathcal{W} \rangle_1^{\rm max}, \langle \mathcal{W} \rangle_2^{\rm max}\} $ is equal to or larger than the classical bound $C_{\text{LHS}}$ (for example, $\langle \mathcal{W} \rangle_1^{\rm max} \geq C_{\text{LHS}}$). To complete the proof, we need to further show that any entangle state cannot reach the equality. In other words, we only need to prove that reaching the equality must lead to a separable state.

To reach the equality, i.e.,
\begin{eqnarray}
 \langle \mathcal{W} \rangle_1^{\rm max}+ \langle \mathcal{W} \rangle_2^{\rm max}
&=&2    C_{\text{LHS}},
\end{eqnarray}
one needs
\begin{eqnarray}\label{eq:H-1}
\vec{H}=k\vec{F}, \;\; \quad|k|\leq 1.
\end{eqnarray}
Besides, because
\begin{eqnarray}
 \langle \mathcal{W} \rangle_1^{\rm max}=\frac{1}{4}+\frac{1}{4}H_0+\frac{1+k}{4}|\vec{F}|
\end{eqnarray}
\begin{eqnarray}
 \langle \mathcal{W} \rangle_2^{\rm max}=\frac{1}{4}-\frac{1}{4}H_0+\frac{1-k}{4}|\vec{F}|,
\end{eqnarray}
we still need
\begin{eqnarray}
H_0+k|\vec{F}|=0.
\end{eqnarray}
Due to Eq. (\ref{eq:H-1}), we have
\begin{eqnarray}
|H_0|=|\vec{H}|.
\end{eqnarray}

Let's check all possible cases. From previous sections, we have known that the condition $\mathcal{M}=\mathcal{M}^\dagger$ is equivalent to the following conditions
\begin{subequations}
\begin{eqnarray}
&&\nu_2 \sin\alpha \sin\theta  \sin(\beta+\tau) \sin2\phi=0, \\
&& \nu_2 \cos\alpha \cos\theta \sin(\beta-\tau) \sin2\phi=0,  \\
&& \nu_2 \cos^2\phi \sin2\alpha - e^{2{\rm i}\tau} (\nu_1 - \nu_2 \sin^2\phi) \sin2\theta=0.
\end{eqnarray}
\end{subequations}
{For the rank-2 state $\rho$, when $\sin2\phi \neq 0$, we have known that $\tau\in\{\beta,-\beta\}$; and when $\sin2\phi = 0$, we have known that $\tau=0$. Thus we need to analysis two possible cases: 1. $\tau\in\{\beta,-\beta\}$ (For $\sin2\phi \neq 0$); 2. $\tau=0$ (For $\sin2\phi =0$).}

\textcolor{blue}{\emph{Case 1.---} $\tau\in\{\beta,-\beta\}$ (For $\sin2\phi \neq 0$)}.

Recall that
\begin{eqnarray}
F_1 &=&-\nu_2 \cos\beta \sin(\alpha - \theta) \sin2\phi, \nonumber\\
F_2 &=&-\nu_2 \sin\beta \sin(\alpha + \theta) \sin2\phi, \nonumber\\
F_3 &=& -\nu_2 \cos2\alpha \cos^2\phi + \cos2\theta (\nu_1 - \nu_2 \sin^2\phi),
\end{eqnarray}

\begin{eqnarray}
H_0&=& \nu_2 \left[ \cos(\alpha - \theta) \cos\tau \sin\beta - \cos\beta \cos(\alpha + \theta) \sin\tau \right] \sin2\phi,
\end{eqnarray}

\begin{eqnarray}
H_1 &=&\sin\tau \left[ \nu_2 \cos^2\phi \sin 2\alpha + \sin2\theta \left( \nu_1 - \nu_2 \sin^2\phi \right) \right],\nonumber\\
H_2 &=&\cos\tau \left[ -\nu_2 \cos^2\phi \sin2\alpha + \sin2\theta \left( \nu_1 - \nu_2 \sin^2\phi \right) \right], \nonumber\\
H_3 &=& \nu_2 \left[ -\sin\beta \cos(\alpha + \theta) \cos\tau + \cos\beta \cos(\alpha - \theta) \sin\tau \right] \sin2\phi,
\end{eqnarray}
For $\tau=\beta$ or $\tau=-\beta$, we always have
\begin{eqnarray}
|H_0|=|H_3|.
\end{eqnarray}
Because
\begin{eqnarray}
|H_0|=|\vec{H}|,
\end{eqnarray}
we then have
\begin{eqnarray}
H_1=0, \quad H_2=0,
\end{eqnarray}
i.e.,
\begin{eqnarray}
\vec{H}=(H_1, H_2, H_3)=(0, 0, H_3).
\end{eqnarray}
Because
\begin{eqnarray}
\vec{H}=k\vec{F}, \;\;  \quad|k|\leq 1,
\end{eqnarray}
thus
\begin{eqnarray}
\vec{F}=(F_1, F_2, F_3)=(0, 0, F_3),
\end{eqnarray}
i.e.,
\begin{eqnarray}
F_1=0, \quad F_2=0.
\end{eqnarray}
Based on
\begin{eqnarray}\label{eq:fh-1}
F_1 &=&-\nu_2 \cos\beta \sin(\alpha - \theta) \sin2\phi=0, \nonumber\\
F_2 &=&-\nu_2 \sin\beta \sin(\alpha + \theta) \sin2\phi=0, \nonumber\\
H_1 &=&\sin\tau \left[ \nu_2 \cos^2\phi \sin 2\alpha + \sin2\theta \left( \nu_1 - \nu_2 \sin^2\phi \right) \right]=0,\nonumber\\
H_2 &=&\cos\tau \left[ -\nu_2 \cos^2\phi \sin2\alpha + \sin2\theta \left( \nu_1 - \nu_2 \sin^2\phi \right) \right]=0,
\end{eqnarray}
we can easily determine that the corresponding state is separable.

Let us take $\tau=\beta$ as an example. For the case of $\tau=-\beta$, the result is similar.

\emph{Example 1.---}The case of $\nu_1=1$ and $\nu_2=0$. In this case, the state $\rho=|\psi_1\rangle\langle\psi_1|$ is a rank-1 pure state.
From Eq. (\ref{eq:fh-1}) we have
\begin{eqnarray}
\sin2\theta=0.
\end{eqnarray}
In this situation, $|\psi_1\rangle=|00\rangle$ or $|11\rangle$, thus $\rho$ is obviously separable.

\emph{Example 2.---}The case of $\nu_1\neq 0$ and $\nu_2\neq 0$. From Eq. (\ref{eq:fh-1}) we have
\begin{eqnarray}\label{eq:fh-2}
F_1 &=&-\nu_2 \cos\beta \sin(\alpha - \theta) \sin2\phi=0, \nonumber\\
F_2 &=&-\nu_2 \sin\beta \sin(\alpha + \theta) \sin2\phi=0, \nonumber\\
H_1 &=&\sin\beta \left[ \nu_2 \cos^2\phi \sin 2\alpha + \sin2\theta \left( \nu_1 - \nu_2 \sin^2\phi \right) \right]=0,\nonumber\\
H_2 &=&\cos\beta \left[ -\nu_2 \cos^2\phi \sin2\alpha + \sin2\theta \left( \nu_1 - \nu_2 \sin^2\phi \right) \right]=0,
\end{eqnarray}

(i) If $\sin\beta=0$ or $\cos\beta=0$. Let us take $\sin\beta=0$ as an example. For $\cos\beta=0$, the result is similar.

When $\sin\beta=0$, Eq. (\ref{eq:fh-2}) becomes
\begin{eqnarray}
F_1 &=&-\nu_2 \cos\beta \sin(\alpha - \theta) \sin2\phi=0, \nonumber\\
H_2 &=&\cos\beta \left[ -\nu_2 \cos^2\phi \sin2\alpha + \sin2\theta \left( \nu_1 - \nu_2 \sin^2\phi \right) \right]=0,
\end{eqnarray}
Because $\cos\beta\neq 0$, $\nu_2\neq0$ and $\sin2\phi\neq0$, we then have
\begin{eqnarray}
\alpha=\theta, \;\;\; \quad (\nu_1-\nu_2)\sin2\theta=0.
\end{eqnarray}
In this situation, one can verify that the concurrence of the state $\rho$ is $\mathcal{C}=|(\nu_1-\nu_2)\sin2\theta|$, thus the state is separable.

(ii) If $\sin\beta\neq0$ and $\cos\beta\neq0$. Because $\nu_2\neq0$ and $\sin2\phi\neq0$, from Eq. (\ref{eq:fh-2}) we have
\begin{eqnarray}\label{eq:fh-3}
&&\sin(\alpha - \theta)=0, \nonumber\\
&&\sin(\alpha + \theta)=0, \nonumber\\
&& \nu_2 \cos^2\phi \sin 2\alpha + \sin2\theta \left( \nu_1 - \nu_2 \sin^2\phi \right)=0,\nonumber\\
&&-\nu_2 \cos^2\phi \sin2\alpha + \sin2\theta \left( \nu_1 - \nu_2 \sin^2\phi \right)=0,
\end{eqnarray}
which yields
\begin{eqnarray}
\alpha=\theta, \;\;\; \quad \sin2\theta=0.
\end{eqnarray}
In this situation, the state is separable. Thus only the separable states can reach the equality condition.

\textcolor{blue}{\emph{Case 2.---} $\tau= 0$ (For $\sin2\phi =0$)}.


For $\tau=0$ and $\sin2\phi =0$, we can have
\begin{eqnarray}
F_1 &=&-\nu_2 \cos\beta \sin(\alpha - \theta) \sin2\phi=0, \nonumber\\
F_2 &=&-\nu_2 \sin\beta \sin(\alpha + \theta) \sin2\phi=0, \nonumber\\
F_3 &=& -\nu_2 \cos2\alpha \cos^2\phi + \cos2\theta (\nu_1 - \nu_2 \sin^2\phi),
\end{eqnarray}

\begin{eqnarray}
H_0&=& \nu_2 \left[ \cos(\alpha - \theta) \cos\tau \sin\beta - \cos\beta \cos(\alpha + \theta) \sin\tau \right] \sin2\phi=0,
\end{eqnarray}

\begin{eqnarray}
H_1 &=&\sin\tau \left[ \nu_2 \cos^2\phi \sin 2\alpha + \sin2\theta \left( \nu_1 - \nu_2 \sin^2\phi \right) \right]=0,\nonumber\\
H_2 &=&\cos\tau \left[ -\nu_2 \cos^2\phi \sin2\alpha + \sin2\theta \left( \nu_1 - \nu_2 \sin^2\phi \right) \right], \nonumber\\
H_3 &=& \nu_2 \left[ -\sin\beta \cos(\alpha + \theta) \cos\tau + \cos\beta \cos(\alpha - \theta) \sin\tau \right] \sin2\phi=0,
\end{eqnarray}
Because
\begin{eqnarray}
|H_0|=|\vec{H}|=0,
\end{eqnarray}
then we have $\vec{H}=(0, 0, 0)$, thus
\begin{eqnarray}\label{eq:h-2}
H_2 &=&\cos\tau \left[ -\nu_2 \cos^2\phi \sin2\alpha + \sin2\theta \left( \nu_1 - \nu_2 \sin^2\phi \right) \right]=0.
\end{eqnarray}
Note that when $\vec{H}=(0, 0, 0)$, one automatically has $\vec{H}=k\vec{F}$ with $k=0$.
From Eq. (\ref{eq:h-2}) we have
\begin{eqnarray}
-\nu_2 \sin2\alpha \cos^2\phi  + \sin2\theta ( \nu_1 - \nu_2 \sin^2\phi )=0.
\end{eqnarray}

(i) If $\phi=0$, we have
\begin{eqnarray}
-\nu_2 \sin2\alpha  + \sin2\theta  \nu_1=0.
\end{eqnarray}
In this situation, the concurrence of the state $\rho$ is $\mathcal{C}=|\nu_2 \sin2\alpha  - \sin2\theta  \nu_1|=0$, thus the state $\rho$ is separable.

(ii) If $\phi=\pi/2$, we have
\begin{eqnarray}
(\nu_1 - \nu_2)\sin2\theta=0.
\end{eqnarray}
In this situation, the concurrence of the state $\rho$ is $\mathcal{C}=|(\nu_1 - \nu_2)\sin2\theta|=0$, thus the state $\rho$ is separable.
Thus only the separable states can reach the equality condition.

This ends the proof.
\end{proof}

%
%

\newpage


\section{Some Examples}

In this section, we would like to provide some examples. Notice that the parameter $\phi\in [0, \pi/2]$, thus we can provide examples for these three cases: 1. $\phi=0$; 2. $\phi=\pi/2$; and 3. $0< \phi <\pi/2$.

\subsubsection{$\phi=0$}

\textcolor{blue}{\emph{Example 1.---}} In this situation, the state $\rho$ is given by
\begin{eqnarray}
\rho=\nu_1  \ket{\psi_1}\bra{\psi_1}+\nu_2 \ket{\psi_2}\bra{\psi_2},
\end{eqnarray}
where
\begin{eqnarray}
\ket{\psi_1} &=& \cos \theta \ket{00}+\sin \theta\ket{11}, \nonumber\\
 \ket{\psi_2} &=& \cos \alpha \ket{01} +\sin\alpha\ket{10},
\end{eqnarray}
where $\theta\in[0, \pi/2]$, $\alpha\in[0, \pi/2]$. Note that the state $\rho$ does not contain the parameter $\beta$.

The concurrence of the state $\rho$ is given by
\begin{eqnarray}
\mathcal{C}=\left|\nu_2 \sin2\alpha - \nu_1 \sin2\theta\right|.
\end{eqnarray}
{In Sec. \ref{sec:IIIB} we have known if Alice chooses her measurement parameter $\tau=0$ (i.e., $\sin\tau=0$, and $\xi$ is arbitrary), then one has $\mathcal{M}=\mathcal{M}^\dagger$ for separable states.} In this situation, Alice's measurement direction is given by
\begin{eqnarray}
&& \hat{n}=(\sin\xi \cos\tau, \sin\xi \sin\tau, \cos\xi)=(\sin\xi, 0, \cos\xi).
\end{eqnarray}
Now, we come to prove that the state $\rho$ will violated the state-dependent steering inequality if it is entangled. We can adopt the following the measurement parameters
\begin{eqnarray}\label{eq:tau1-a}
&& \tau=0, \;\;\;\; \xi=\frac{\pi}{2}, \;\;\;\; \delta=\frac{\pi}{2},
\end{eqnarray}
then we have
\begin{eqnarray}
&& \hat{n}=(\sin\xi \cos\tau, \sin\xi \sin\tau, \cos\xi)=(1, 0, 0)=\hat{e}_x,
\end{eqnarray}
\begin{eqnarray}
&& \ket{+\hat{n}}=\begin{pmatrix}
\cos\frac{\xi}{2}\\
\sin\frac{\xi}{2}\, e^{{\rm i}\tau}\\
\end{pmatrix}=\cos\frac{\xi}{2} |0\rangle+\sin\frac{\xi}{2}\, e^{{\rm i}\tau} |1\rangle
=\frac{1}{\sqrt{2}} \left(|0\rangle+ |1\rangle\right),
\end{eqnarray}

\begin{eqnarray}
 \ket{-\hat{n}}=\begin{pmatrix}
\sin\frac{\xi}{2}\\
-\cos\frac{\xi}{2} \, e^{{\rm i}\tau} \\
\end{pmatrix}= \sin\frac{\xi}{2}|0\rangle -\cos\frac{\xi}{2} \, e^{{\rm i}\tau} |1\rangle=\frac{1}{\sqrt{2}} \left(|0\rangle- |1\rangle\right),
\end{eqnarray}

\begin{eqnarray}
P^{\hat{n}}_{0}&=&\frac{1}{2}\left(\openone+\vec{\sigma}\cdot \hat{n}\right)=|+\hat{n}\rangle\langle+\hat{n}|
= \frac{1}{2}\begin{pmatrix}
 1+\cos\xi&
 \sin\xi e^{-{\rm i} \tau}\\
 \sin\xi e^{{\rm i} \tau} &
 1-\cos\xi
\end{pmatrix}= \frac{1}{2}\begin{pmatrix}
 1 &1\\
 1 & 1
\end{pmatrix},
\end{eqnarray}

\begin{eqnarray}
P^{\hat{n}}_{1}=\frac{1}{2}\left(\openone-\vec{\sigma}\cdot \hat{n}\right)=|-\hat{n}\rangle\langle-\hat{n}|
= \frac{1}{2}\begin{pmatrix}
 1-\cos\xi&
 -\sin\xi e^{-{\rm i} \tau}\\
 -\sin\xi e^{{\rm i} \tau} &
 1+\cos\xi
\end{pmatrix}= \frac{1}{2}\begin{pmatrix}
 1 & -1\\
 -1 & 1
\end{pmatrix},
\end{eqnarray}

\begin{eqnarray}
\tilde{\rho}_{0}^{\hat{n}} &=&
\begin{pmatrix}
\nu_1 \cos^2 \theta \cos^2\left(\frac{\xi}{2}\right) + \nu_2 \sin^2 \alpha \sin^2\left(\frac{\xi}{2}\right) &
\dfrac{1}{4} e^{-{\rm i} \tau} \left( \nu_2 \sin 2\alpha + e^{2{\rm i}\tau} \nu_1 \sin 2\theta \right) \sin \xi \\
\dfrac{1}{4} e^{-{\rm i} \tau} \left( e^{2{\rm i}\tau} \nu_2 \sin 2\alpha + \nu_1 \sin 2\theta \right) \sin \xi &
\nu_2 \cos^2 \alpha \cos^2\left(\frac{\xi}{2}\right) + \nu_1 \sin^2 \theta \sin^2\left(\frac{\xi}{2}\right)
\end{pmatrix}\nonumber\\
&=&\frac{1}{2}\begin{pmatrix}
\nu_1 \cos^2 \theta + \nu_2 \sin^2 \alpha  &
\dfrac{1}{2} \left( \nu_2 \sin 2\alpha +  \nu_1 \sin 2\theta \right)  \\
\dfrac{1}{2}  \left( \nu_2 \sin 2\alpha + \nu_1 \sin 2\theta \right)  &
\nu_2 \cos^2 \alpha + \nu_1 \sin^2 \theta
\end{pmatrix},
\end{eqnarray}
\begin{eqnarray}
\tilde{\rho}_{1}^{\hat{n}} &=&
\begin{pmatrix}
\nu_2 \cos^2\left(\frac{\xi}{2}\right) \sin^2 \alpha + \nu_1 \cos^2 \theta \sin^2\left(\frac{\xi}{2}\right) &
-\dfrac{1}{4} e^{-{\rm i} \tau} \left( \nu_2 \sin 2\alpha + e^{2{\rm i}\tau} \nu_1 \sin 2\theta \right) \sin \xi \\
-\dfrac{1}{4} e^{-{\rm i} \tau} \left( e^{2{\rm i}\tau} \nu_2 \sin 2\alpha + \nu_1 \sin 2\theta \right) \sin \xi &
\nu_1 \cos^2\left(\frac{\xi}{2}\right) \sin^2 \theta + \nu_2 \cos^2 \alpha \sin^2\left(\frac{\xi}{2}\right)
\end{pmatrix}\nonumber\\
&=&\frac{1}{2}\begin{pmatrix}
\nu_1 \cos^2 \theta + \nu_2 \sin^2 \alpha  &
-\dfrac{1}{2} \left( \nu_2 \sin 2\alpha +  \nu_1 \sin 2\theta \right)  \\
-\dfrac{1}{2}  \left( \nu_2 \sin 2\alpha + \nu_1 \sin 2\theta \right)  &
\nu_2 \cos^2 \alpha + \nu_1 \sin^2 \theta
\end{pmatrix},
\end{eqnarray}
\begin{eqnarray}
\rho_{B}=\tilde{\rho}_{0}^{\hat{n}}+\tilde{\rho}_{1}^{\hat{n}} =\begin{pmatrix}
\nu_1 \cos^2 \theta + \nu_2 \sin^2 \alpha & 0 \\
0 & \nu_1 \sin^2 \theta + \nu_2 \cos^2 \alpha
\end{pmatrix}.
\end{eqnarray}
Based on the above results, we can have the classical bound as
\begin{eqnarray}\label{eq:class-1}
    C_{\text{LHS}}
        &=& \frac{1}{2} \max_{\hat{n}_B} \left[ \mathrm{Tr}\!\left( |\hat{n}_B\rangle \langle \hat{n}_B|
     \rho_B  \right) \right]\nonumber\\
    &=&\frac{1}{2} \max_{\hat{n}_B} \left[ \mathrm{Tr} \left[ \frac{1}{2}\begin{pmatrix}
 1+\cos\theta_B &
 \sin\theta_B e^{-{\rm i} \phi_B}\\
 \sin\theta_B e^{{\rm i} \phi_B} &
 1-\cos\theta_B
\end{pmatrix} \begin{pmatrix}
\nu_1 \cos^2 \theta  + \nu_2 \sin^2 \alpha  &
0 \\
0 &
\nu_1 \sin^2 \theta + \nu_2 \cos^2 \alpha
\end{pmatrix}\right] \right]\nonumber\\
&=& \frac{1}{4}\max_{\hat{n}_B} \left[(1+\cos\theta_B)(\nu_1 \cos^2 \theta  + \nu_2 \sin^2 \alpha)+
(1-\cos\theta_B)(\nu_1 \sin^2 \theta + \nu_2 \cos^2 \alpha)     \right]\nonumber\\
&=& \frac{1}{4}\max_{\hat{n}_B} \left[1+ (\nu_1 \cos2\theta  - \nu_2 \cos2\alpha)\cos\theta_B     \right]\nonumber\\
&=& \frac{1+|\nu_1 \cos2\theta  - \nu_2 \cos2\alpha|}{4}.
\end{eqnarray}
The quantum expectation is given by (we take $\langle \mathcal{W} \rangle_1$ as $\langle \mathcal{W} \rangle$ and the parameter $\tau=0$)
\begin{eqnarray}
 \langle \mathcal{W} \rangle = \mathrm{Tr}[(|+\rangle\langle+|\otimes |\hat{n}_B \rangle\langle\hat{n}_B|) \rho],
\end{eqnarray}
i.e.,
\begin{eqnarray}
 \langle \mathcal{W} \rangle
  &=& \frac{1}{4}\mathrm{Tr}\biggr\{\biggr[
 { \left(
  \begin{array}{cc}
   1 {+}\sin\xi \cos\delta  &
e^{-{\rm i}\tau} \left[-\cos\delta\cos\xi+\rm i\sin\delta  \right]\\
 e^{{\rm i}\tau}  \left[-\cos\delta\cos\xi-\rm i\sin\delta  \right] & 1 {-}\sin\xi \cos\delta  \\
  \end{array}
\right) }\otimes \begin{pmatrix}
 1+\cos\theta_B &
 \sin\theta_B e^{-{\rm i} \phi_B}\\
 \sin\theta_B e^{{\rm i} \phi_B} &
 1-\cos\theta_B
\end{pmatrix}\biggr] \rho\biggr\},
\end{eqnarray}
i.e.,
\begin{eqnarray}
 \langle \mathcal{W} \rangle
  &=& \frac{1}{4}\mathrm{Tr}\biggr\{\biggr[
 { \frac{1}{2}
\left(
  \begin{array}{cc}
   1  &
 {\rm i} e^{-{\rm i}\tau} \\
-{\rm i} e^{{\rm i}\tau}   & 1  \\
  \end{array}
\right) }\otimes \begin{pmatrix}
 1+\cos\theta_B &
 \sin\theta_B e^{-{\rm i} \phi_B}\\
 \sin\theta_B e^{{\rm i} \phi_B} &
 1-\cos\theta_B
\end{pmatrix}\biggr] \rho\biggr\}\nonumber\\
&=& \frac{1}{4}\mathrm{Tr}\biggr\{\biggr[
 { \frac{1}{2}
\left(
  \begin{array}{cc}
   1  &
 {\rm i}  \\
-{\rm i}    & 1  \\
  \end{array}
\right) }\otimes \begin{pmatrix}
 1+\cos\theta_B &
 \sin\theta_B e^{-{\rm i} \phi_B}\\
 \sin\theta_B e^{{\rm i} \phi_B} &
 1-\cos\theta_B
\end{pmatrix}\biggr] \rho\biggr\}\nonumber\\
\end{eqnarray}
with
\begin{eqnarray}
\rho =
 \begin{bmatrix}
 \nu_1 \cos^2 \theta    & 0  &  0  & \nu_1 \sin \theta \cos \theta    \\
 0 &  \nu_2\cos ^2\alpha  & \nu_2  \sin \alpha \cos \alpha  & 0 \\
 0 & \nu_2 \sin \alpha \cos \alpha  &  \nu_2\sin ^2\alpha   & 0 \\
 \nu_1 \sin \theta \cos \theta  & 0  &  0  &  \nu_1 \sin^2 \theta \\
\end{bmatrix}.
\end{eqnarray}
After direct calculation, we obtain
\begin{eqnarray}
 \langle \mathcal{W} \rangle
&=&  \frac{1}{4}+
 \frac{1}{4} \biggr[\cos\theta_B \left(\nu_1 \cos2\theta -\nu_2 {\cos2\alpha}\right)
 +\sin\theta_B \left(\nu_1 \sin2\theta -\nu_2 {\sin2\alpha}\right) \sin \phi_B \biggr].
\end{eqnarray}
By taking $\sin\phi_B=1$ (i.e., $\phi_B=\pi/2$), we have
\begin{eqnarray}
 \langle \mathcal{W} \rangle &=&  \frac{1}{4}+
 \frac{1}{4} \biggr[\cos\theta_B \left(\nu_1 \cos2\theta -\nu_2 {\cos2\alpha}\right)
 +\sin\theta_B \left(\nu_1 \sin2\theta -\nu_2 {\sin2\alpha}\right) \biggr]\nonumber\\
 &\leq &  \frac{1}{4}+
 \frac{1}{4} \sqrt{\left(\nu_1 \cos2\theta -\nu_2 {\cos2\alpha}\right)^2
 +\left(\nu_1 \sin2\theta -\nu_2 {\sin2\alpha}\right)^2},
\end{eqnarray}
the equal sign occurs at
\begin{eqnarray}
 \tan \theta_B = \frac{\nu_1 \sin2\theta -\nu_2 {\sin2\alpha}}{\nu_1 \cos2\theta -\nu_2 {\cos2\alpha}}.
\end{eqnarray}
Thus
\begin{eqnarray}\label{eq:quatum-1}
 \langle \mathcal{W} \rangle^{\rm max} &=&  \frac{1}{4}+
 \frac{1}{4} \sqrt{\left(\nu_1 \cos2\theta -\nu_2 {\cos2\alpha}\right)^2
 +\left(\nu_1 \sin2\theta -\nu_2 {\sin2\alpha}\right)^2}\nonumber\\
 &=&\frac{1+\sqrt{\left(\nu_1 \cos2\theta -\nu_2 {\cos2\alpha}\right)^2
 +\left(\nu_1 \sin2\theta -\nu_2 {\sin2\alpha}\right)^2}}{4}\nonumber\\
 &=&\frac{1+\sqrt{\left|\nu_1 \cos2\theta -\nu_2 {\cos2\alpha}\right|^2
 +\mathcal{C}^2}}{4}.
\end{eqnarray}
By comparing  the classical bound $C_{\text{LHS}}$ in  Eq. (\ref{eq:class-1}) and the maximal quantum expectation $\langle \mathcal{W} \rangle^{\rm max}$ in Eq. (\ref{eq:quatum-1}), one easily proves the relation (\ref{eq:sdsi-3}). It ends the proof for this case.

\begin{remark} For $\theta=\pi/4$, one has $\cos2\theta=0$, thus the classical bound as
\begin{eqnarray}
    C_{\text{LHS}} &=& \frac{1+\nu_2 \cos2\alpha}{4},
\end{eqnarray}
and the maximal quantum expectation as
\begin{eqnarray}
 \langle \mathcal{W} \rangle^{\rm max}  &=&\frac{1+\sqrt{\left|\nu_2 {\cos2\alpha}\right|^2
 +\mathcal{C}^2}}{4}=\frac{1+\sqrt{\left|\nu_2 \cos2\alpha\right|^2
 +(\nu_1 -\nu_2 \sin2\alpha)^2}}{4}
\end{eqnarray}
Obviously, the relation (\ref{eq:sdsi-3}) is valid. Specially, when $\alpha=\pi/4$, then both $|\psi_1\rangle$ and $|\psi_2\rangle$ are maximally entangled states, and we have
\begin{eqnarray}
 \langle \mathcal{W} \rangle^{\rm max}  &=&\frac{1+\left|\nu_1-\nu_2\right|}{4}>  C_{\text{LHS}} = \frac{1}{4}
 \end{eqnarray}
for any $\nu_1-\nu_2\neq 0$ (i.e., the concurrence $\mathcal{C}\neq 0$). $\blacksquare$
\end{remark}

\begin{remark} By letting $\nu_2=0$, then one has $\nu_1=1$. The above quantum state $\rho$ reduce a rank-1 pure state $\rho=|\psi_1\rangle\langle \psi_1|$. Actually, the above proof is also valid for the rank-1 pure state. $\blacksquare$
\end{remark}

\begin{remark}  One may note that, $\langle \mathcal{W} \rangle_1^{\rm max}$ and $\langle \mathcal{W} \rangle_2^{\rm max}$ are the same when $H_0=0$ and $\vec{F} \perp \vec{H}$. In this situation, we have $|\vec{H}|^2=\mathcal{C}^2$. By the way, for this quantum state $\rho$, the classical bound can be written as
\begin{eqnarray}
    C_{\text{LHS}}
&=& \frac{1+|\nu_1 \cos2\theta  - \nu_2 \cos2\alpha|}{4}=\frac{1+|\vec{F}|}{4}=\frac{1+\sqrt{\vec{F}^2}}{4},
\end{eqnarray}
where $\vec{F}$ is the Bloch vector of the state $\rho_{B}$. Similarly, the quantum expectation can be written as
\begin{eqnarray}
 \langle \mathcal{W} \rangle^{\rm max}
 &=&\frac{1+\sqrt{\vec{F}^2 +\mathcal{C}^2}}{4}.
\end{eqnarray}
$\blacksquare$
\end{remark}

\subsubsection{$\phi=\pi/2$}\label{sec:IIIB2}

\textcolor{blue}{\emph{Example 2.---}} In this situation, the state $\rho$ is given by
\begin{eqnarray}
\rho=\nu_1  \ket{\psi_1}\bra{\psi_1}+\nu_2 \ket{\psi_2}\bra{\psi_2},
\end{eqnarray}
where
\begin{eqnarray}
\ket{\psi_1} &=& \cos \theta \ket{00}+\sin \theta\ket{11}, \nonumber\\
\ket{\psi_2} &=& e^{{\rm i}\beta}(\sin \theta \ket{00}-\cos \theta\ket{11}),
\end{eqnarray}
where $\theta\in[0, \pi/2]$, $\alpha\in[0, \pi/2]$. Note that $\ket{\psi_2}$ contains the phase factor $e^{{\rm i}\beta}$, but $\ket{\psi_2}\bra{\psi_2}$ does not contain the parameter $\beta$, thus the global phase $e^{{\rm i}\beta}$ is trivial and can be ignored.

The concurrence of the state $\rho$ is given by
\begin{eqnarray}
\mathcal{C}=|(\nu_1-\nu_2)\sin2\theta|.
\end{eqnarray}
Now, we come to verify that this specific state $\rho$ will violated the state-dependent steering inequality if it is entangled.
{In Sec. \ref{sec:IIIB} we have known for any measurement parameter $\xi$ and $\tau$ (i.e., $\xi$ and $\tau$ can be arbitrary), one can have $\mathcal{M}=\mathcal{M}^\dagger$ for separable states (for simplicity, one can choose $\tau=0$).} In this situation, Alice's measurement direction is given by
\begin{eqnarray}
&& \hat{n}=(\sin\xi \cos\tau, \sin\xi \sin\tau, \cos\xi).
\end{eqnarray}
For convenience, we can adopt the following the measurement parameters
\begin{eqnarray}
&& \tau=0, \;\;\;\; \xi=\frac{\pi}{2}, \;\;\;\; \delta=\frac{\pi}{2},
\end{eqnarray}
then we obtain
\begin{eqnarray}
&& \hat{n}=(\sin\xi \cos\tau, \sin\xi \sin\tau, \cos\xi)=(1, 0, 0)=\hat{e}_x,
\end{eqnarray}
\begin{eqnarray}
&& \ket{+\hat{n}}=\begin{pmatrix}
\cos\frac{\xi}{2}\\
\sin\frac{\xi}{2}\, e^{{\rm i}\tau}\\
\end{pmatrix}=\cos\frac{\xi}{2} |0\rangle+\sin\frac{\xi}{2}\, e^{{\rm i}\tau} |1\rangle=\frac{1}{\sqrt{2}} \left(|0\rangle+ |1\rangle\right),
\end{eqnarray}
\begin{eqnarray}
 \ket{-\hat{n}}=\begin{pmatrix}
\sin\frac{\xi}{2}\\
-\cos\frac{\xi}{2} \, e^{{\rm i}\tau} \\
\end{pmatrix}= \sin\frac{\xi}{2}|0\rangle -\cos\frac{\xi}{2} \, e^{{\rm i}\tau} |1\rangle=\frac{1}{\sqrt{2}} \left(|0\rangle- |1\rangle\right),
\end{eqnarray}

\begin{eqnarray}
P^{\hat{n}}_{0}&=&\frac{1}{2}\left(\openone+\vec{\sigma}\cdot \hat{n}\right)=|+\hat{n}\rangle\langle+\hat{n}|
= \frac{1}{2}\begin{pmatrix}
 1+\cos\xi&
 \sin\xi e^{-{\rm i} \tau}\\
 \sin\xi e^{{\rm i} \tau} &
 1-\cos\xi
\end{pmatrix}=
{\frac{1}{2}\begin{pmatrix}
 1 &1\\
 1 & 1
\end{pmatrix},}
\end{eqnarray}

\begin{eqnarray}
P^{\hat{n}}_{1}=\frac{1}{2}\left(\openone-\vec{\sigma}\cdot \hat{n}\right)=|-\hat{n}\rangle\langle-\hat{n}|
= \frac{1}{2}\begin{pmatrix}
 1-\cos\xi&
 -\sin\xi e^{-{\rm i} \tau}\\
 -\sin\xi e^{{\rm i} \tau} &
 1+\cos\xi
\end{pmatrix}=
{\frac{1}{2}\begin{pmatrix}
 1 &-1\\
 -1 & 1
\end{pmatrix},}\end{eqnarray}

\begin{eqnarray}
\tilde{\rho}_{0}^{\hat{n}}
=
\frac{1}{2}\begin{pmatrix}
\nu_1 \cos^2\theta + \nu_2 \sin^2\theta &
(\nu_1 - \nu_2)\sin\theta\cos\theta \\
(\nu_1 - \nu_2)\sin\theta\cos\theta &
\nu_1 \sin^2\theta + \nu_2 \cos^2\theta
\end{pmatrix},
\end{eqnarray}

\begin{eqnarray}
\tilde{\rho}_{1}^{\hat{n}}
=
\frac{1}{2}\begin{pmatrix}
\nu_1 \cos^2\theta + \nu_2 \sin^2\theta &
-(\nu_1 - \nu_2)\sin\theta\cos\theta \\
-(\nu_1 - \nu_2)\sin\theta\cos\theta &
\nu_1 \sin^2\theta + \nu_2 \cos^2\theta
\end{pmatrix},
\end{eqnarray}
\begin{eqnarray}
\rho_{B}=\tilde{\rho}_{0}^{\hat{n}}+\tilde{\rho}_{1}^{\hat{n}} =\begin{pmatrix}
\nu_1 \cos^2 \theta + \nu_2 \sin^2 \theta & 0 \\
0 & \nu_1 \sin^2 \theta + \nu_2 \cos^2 \theta
\end{pmatrix}.
\end{eqnarray}
Based on the above results, we can have the classical bound as
\begin{eqnarray}\label{eq:class-2}
    C_{\text{LHS}}
    &=&\frac{1}{2} \max_{\hat{n}_B} \left[ \mathrm{Tr} \left( |\hat{n}_B\rangle \langle \hat{n}_B| \rho_{B} \right) \right]\nonumber\\
    &=&\frac{1}{2} \max_{\hat{n}_B} \left[ \mathrm{Tr} \left[ \frac{1}{2}\begin{pmatrix}
 1+\cos\theta_B &
 \sin\theta_B e^{-{\rm i} \phi_B}\\
 \sin\theta_B e^{{\rm i} \phi_B} &
 1-\cos\theta_B
\end{pmatrix} \begin{pmatrix}
\nu_1 \cos^2 \theta  + \nu_2 \sin^2 \theta  &
0 \\
0 &
\nu_1 \sin^2 \theta + \nu_2 \cos^2 \theta
\end{pmatrix}\right] \right]\nonumber\\
&=& \frac{1}{4}\max_{\hat{n}_B} \left[(1+\cos\theta_B)(\nu_1 \cos^2 \theta  + \nu_2 \sin^2 \theta)+
(1-\cos\theta_B)(\nu_1 \sin^2 \theta + \nu_2 \cos^2 \theta)     \right]\nonumber\\
&=& \frac{1}{4}\max_{\hat{n}_B} \left[1+ (\nu_1  - \nu_2 )\cos2\theta \cos\theta_B     \right]\nonumber\\
&=& \frac{1+|(\nu_1 - \nu_2) \cos2\theta|}{4}.
\end{eqnarray}
The quantum expectation is given by (we take $\langle \mathcal{W} \rangle_1$ as $\langle \mathcal{W} \rangle$ and the parameter $\tau=0$)
\begin{eqnarray}
 \langle \mathcal{W} \rangle = \mathrm{Tr}[(|+\rangle\langle+|\otimes |\hat{n}_B \rangle\langle\hat{n}_B|) \rho],
\end{eqnarray}
i.e.,
\begin{eqnarray}
 \langle \mathcal{W} \rangle
  &=& \frac{1}{4}\mathrm{Tr}\biggr[\biggr(
 { \left(
  \begin{array}{cc}
   1 {+}\sin\xi \cos\delta  &
e^{-{\rm i}\tau} \left[-\cos\delta\cos\xi+\rm i\sin\delta  \right]\\
 e^{{\rm i}\tau}  \left[-\cos\delta\cos\xi-\rm i\sin\delta  \right] & 1 {-}\sin\xi \cos\delta  \\
  \end{array}
\right) }\otimes \begin{pmatrix}
 1+\cos\theta_B &
 \sin\theta_B e^{-{\rm i} \phi_B}\\
 \sin\theta_B e^{{\rm i} \phi_B} &
 1-\cos\theta_B
\end{pmatrix}\biggr) \rho\biggr],
\end{eqnarray}
i.e.,
\begin{eqnarray}
 \langle \mathcal{W} \rangle
&=& \frac{1}{4}\mathrm{Tr}\biggr[\biggr(
 { \frac{1}{2}
\left(
  \begin{array}{cc}
   1  &
 {\rm i}  \\
-{\rm i}    & 1  \\
  \end{array}
\right) }\otimes \begin{pmatrix}
 1+\cos\theta_B &
 \sin\theta_B e^{-{\rm i} \phi_B}\\
 \sin\theta_B e^{{\rm i} \phi_B} &
 1-\cos\theta_B
\end{pmatrix}\biggr) \rho\biggr]\nonumber\\
\end{eqnarray}
with
\begin{eqnarray}
\rho =
 \begin{bmatrix}
 \nu_1 \cos^2 \theta +\nu_2\sin ^2\theta    & 0  &  0  & (\nu_1-\nu_2) \sin \theta \cos \theta    \\
 0 & 0  & 0  & 0 \\
 0 & 0  & 0  & 0 \\
 (\nu_1-\nu_2) \sin \theta \cos \theta  & 0  &  0  &  \nu_1 {\sin^2 \theta} +\nu_2 {\cos ^2\theta}  \\
\end{bmatrix}.
\end{eqnarray}
After direct calculation, we obtain
\begin{eqnarray}
 \langle \mathcal{W} \rangle
&=&  \frac{1}{4} +
 \frac{1}{4} \biggr[\cos \theta_B [(\nu_1-\nu_2)\cos 2\theta]+\sin \theta_B [(\nu_1-\nu_2)\sin 2\theta] \sin \phi_B\biggr].
\end{eqnarray}
By taking $\sin\phi_B=1$, we have
\begin{eqnarray}
 \langle \mathcal{W} \rangle &=&  \frac{1}{4} +
 \frac{1}{4} \biggr[\cos \theta_B [(\nu_1-\nu_2)\cos 2\theta]+\sin \theta_B [(\nu_1-\nu_2)\sin 2\theta]\biggr]\nonumber\\
 &\leq & \frac{1}{4} +
 \frac{1}{4} \sqrt{[(\nu_1-\nu_2)\cos 2\theta]^2+[(\nu_1-\nu_2)\sin 2\theta]^2},
\end{eqnarray}
the equal sign occurs at
\begin{eqnarray}
 \tan \theta_B = \frac{\sin2\theta}{\cos2\theta}=\tan2\theta.
\end{eqnarray}
Thus
\begin{eqnarray}\label{eq:quatum-2}
 \langle \mathcal{W} \rangle^{\rm max}
 &=&  \frac{1}{4} + \frac{1}{4} \sqrt{[(\nu_1-\nu_2)\cos 2\theta]^2+[(\nu_1-\nu_2)\sin 2\theta]^2} \nonumber\\
 &=&  \frac{1}{4} + \frac{1}{4} \sqrt{[(\nu_1-\nu_2)\cos 2\theta]^2+\mathcal{C}^2}.
\end{eqnarray}
By comparing  the classical bound $C_{\text{LHS}}$ in  Eq. (\ref{eq:class-2}) and the maximal quantum expectation $\langle \mathcal{W} \rangle^{\rm max}$ in Eq. (\ref{eq:quatum-2}), one easily proves the relation (\ref{eq:sdsi-3}).

\begin{remark} For $\theta=\pi/4$, one has $\cos2\theta=0$, thus the classical bound as
\begin{eqnarray}
    C_{\text{LHS}} &=& \frac{1}{4},
\end{eqnarray}
and the maximal quantum expectation as
\begin{eqnarray}
 \langle \mathcal{W} \rangle^{\rm max}  &=&\frac{1+\left|\nu_1-\nu_2\right|}{4}
\end{eqnarray}
Obviously, the relation (\ref{eq:sdsi-3}) is valid for any $\nu_1-\nu_2\neq 0$ (i.e., the concurrence $\mathcal{C}=|\nu_1-\nu_2|> 0$). $\blacksquare$
\end{remark}

\begin{remark} By the way, for the quantum state $\rho$, the classical bound can be written as
\begin{eqnarray}
    C_{\text{LHS}}
&=& \frac{1+|(\nu_1 - \nu_2) \cos2\theta|}{4}=\frac{1+|\vec{F}|}{4}=\frac{1+\sqrt{\vec{F}^2}}{4},
\end{eqnarray}
where $\vec{F}$ is the Bloch vector of the state $\rho_{B}$. Similarly, the quantum expectation can be written as
\begin{eqnarray}
 \langle \mathcal{W} \rangle^{\rm max}
 &=&\frac{1+\sqrt{\vec{F}^2 +\mathcal{C}^2}}{4}.
\end{eqnarray}
$\blacksquare$
\end{remark}

\subsubsection{$0< \phi <\pi/2$}

\vspace{3mm}
\centerline{\emph{3.1. The Case of $\cos\beta=0$}}
\vspace{6mm}

In this situation, the state $\rho$ is given by
\begin{eqnarray}
\rho=\nu_1  \ket{\psi_1}\bra{\psi_1}+\nu_2 \ket{\psi_2}\bra{\psi_2},
\end{eqnarray}
where
\begin{eqnarray}
\ket{\psi_1} &=& \cos \theta \ket{00}+\sin \theta\ket{11}, \nonumber\\
 \ket{\psi_2} &=& \cos\phi(\cos \alpha \ket{01} +\sin\alpha\ket{10})+ e^{{\rm i}\beta}\sin\phi(\sin \theta \ket{00}-\cos \theta\ket{11}),
\end{eqnarray}
with $\beta=\pi/2$, $e^{{\rm i}\beta}={\rm i}$, $\theta\in[0, \pi/2]$, and $\alpha\in[0, \pi/2]$. For simplicity, in the following we have taken $\beta=\pi/2$ as example (for the case of $\beta=3\pi/2$, it will yield the similar result).

The concurrence of the state $\rho$ is given by
\begin{eqnarray}
\mathcal{C} &=& \sqrt{-s_2-2\sqrt{s_1}},
\end{eqnarray}
with
\begin{eqnarray}
 s_2 &=& \frac{1}{2} \nu_2^2 \sin 2 \alpha \sin 2 \theta  \sin ^2 2 \phi -\nu_2^2 \sin ^2 2 \alpha \cos ^4 \phi
 -\sin ^2 2 \theta \left(\nu_1^2 +\nu_2^2 \sin ^4\phi \right)- 2 \nu_1 \nu_2 \cos^2 2 \theta \sin ^2 \phi, \nonumber\\
s_1 &=& \frac{1}{2} \nu_1^2 \nu_2^2 \left(-\sin2 \alpha \sin 2 \theta \sin ^2 2 \phi +2 \sin ^2 2 \alpha
 \sin ^2 2 \theta \cos ^4\phi +2 \sin ^4\phi \right).
\end{eqnarray}
In the following, we would like to study the problem by two different cases: $\sin 2 \alpha= 0$ and $\sin 2 \alpha> 0$.


\textcolor{blue}{\emph{Example 3.---}} Let us consider a simple case of $\sin2\alpha=0$ with $\alpha=0$ (for $\alpha=\pi/2$, the result is similar). In this situation, the state $\rho$ is given by
\begin{eqnarray}
\rho=\nu_1  \ket{\psi_1}\bra{\psi_1}+\nu_2 \ket{\psi_2}\bra{\psi_2},
\end{eqnarray}
where
\begin{eqnarray}
\ket{\psi_1} &=& \cos \theta \ket{00}+\sin \theta\ket{11}, \nonumber\\
 \ket{\psi_2} &=& \cos\phi \ket{01} + e^{{\rm i}\beta}\sin\phi(\sin \theta \ket{00}-\cos \theta\ket{11}),
\end{eqnarray}
with $\beta=\pi/2$, $e^{{\rm i}\beta}={\rm i}$, $\theta\in[0, \pi/2]$. Actually, in this case, the concurrence of the state $\rho$ is equal to
\begin{eqnarray}\label{eq:concur-1}
\mathcal{C} &=& \sqrt{-s_2-2\sqrt{s_1}}=\sqrt{\sin^2 2\theta \times \left[\nu_1-\nu_2 \sin^2\phi\right]^2}=\left|\left(\nu_1-\nu_2 \sin^2\phi\right)\sin 2\theta\right|\nonumber\\
&=& |\nu_1-\nu_2 \sin^2\phi|\sin 2\theta.
\end{eqnarray}
{In Sec. \ref{sec:IIIB} we have known that
for the measurement parameters {$\tau=\beta=\pi/2$} and $\xi$ is arbitrary, one can have $\mathcal{M}=\mathcal{M}^\dagger$ for separable states.} In this situation, Alice's measurement direction is given by
\begin{eqnarray}
&& \hat{n}=(\sin\xi \cos\tau, \sin\xi \sin\tau, \cos\xi)=(0, \sin\xi, \cos\xi).
\end{eqnarray}
For convenience, we can adopt the following the measurement parameters
\begin{eqnarray}
&& {\tau=\beta=\frac{\pi}{2}}, \;\;\;\; \xi=\frac{\pi}{2}, \;\;\;\; \delta=\frac{\pi}{2},
\end{eqnarray}
then we obtain
\begin{eqnarray}
&& \hat{n}=(\sin\xi \cos\tau, \sin\xi \sin\tau, \cos\xi)=(0, \sin\xi, \cos\xi)=(0, 1, 0)=\hat{e}_y,
\end{eqnarray}
\begin{eqnarray}
&& \ket{+\hat{n}}=\begin{pmatrix}
\cos\frac{\xi}{2}\\
\sin\frac{\xi}{2}\, e^{{\rm i}\tau}\\
\end{pmatrix}=\cos\frac{\xi}{2} |0\rangle+\sin\frac{\xi}{2}\, e^{{\rm i}\tau} |1\rangle
=\frac{1}{\sqrt{2}} \left(|0\rangle+ {{\rm i}}\, |1\rangle\right),
\end{eqnarray}

\begin{eqnarray}
 \ket{-\hat{n}}=\begin{pmatrix}
\sin\frac{\xi}{2}\\
-\cos\frac{\xi}{2} \, e^{{\rm i}\tau} \\
\end{pmatrix}= \sin\frac{\xi}{2}|0\rangle -\cos\frac{\xi}{2} \, e^{{\rm i}\tau} |1\rangle=\frac{1}{\sqrt{2}} \left(|0\rangle- {{\rm i}}\,|1\rangle\right),
\end{eqnarray}

\begin{eqnarray}
P^{\hat{n}}_{0}&=&\frac{1}{2}\left(\openone+\vec{\sigma}\cdot \hat{n}\right)=|+\hat{n}\rangle\langle+\hat{n}|
= \frac{1}{2}\begin{pmatrix}
 1+\cos\xi&
 \sin\xi e^{-{\rm i} \tau}\\
 \sin\xi e^{{\rm i} \tau} &
 1-\cos\xi
\end{pmatrix}=
{\frac{1}{2}\begin{pmatrix}
 1 & -{\rm i}\\
 {\rm i} & 1
\end{pmatrix},}
\end{eqnarray}

\begin{eqnarray}
P^{\hat{n}}_{1}=\frac{1}{2}\left(\openone-\vec{\sigma}\cdot \hat{n}\right)=|-\hat{n}\rangle\langle-\hat{n}|
= \frac{1}{2}\begin{pmatrix}
 1-\cos\xi&
 -\sin\xi e^{-{\rm i} \tau}\\
 -\sin\xi e^{{\rm i} \tau} &
 1+\cos\xi
\end{pmatrix}=
{\frac{1}{2}\begin{pmatrix}
 1 & {\rm i}\\
 -{\rm i} & 1
\end{pmatrix},}
\end{eqnarray}

%
%
%
%
%

\begin{eqnarray}
\rho_{B}
& =&
\left(
\begin{array}{cc}
 \nu_2 \sin ^2\alpha \cos ^2\phi+\nu_1 \cos ^2\theta+\nu_2 \sin ^2\theta \sin ^2\phi
 & \nu_2 \sin \phi \cos\phi \left(e^{{\rm i} \beta } \cos \alpha \sin \theta-e^{-{\rm i} \beta }  \sin \alpha \cos\theta\right) \\
  \nu_2 \sin \phi \cos\phi \left(e^{-{\rm i} \beta } \cos \alpha \sin \theta-e^{{\rm i} \beta }  \sin \alpha \cos\theta\right) &
  \nu_2 \cos ^2\alpha \cos ^2\phi+\nu_1 \sin ^2\theta+\nu_2 \cos ^2\theta \sin ^2\phi \\
\end{array}
\right)\nonumber\\
& =&
\left(
\begin{array}{cc}
 \nu_2 \sin ^2\alpha \cos ^2\phi+\nu_1 \cos ^2\theta+\nu_2 \sin ^2\theta \sin ^2\phi
 & {{\rm i}} \, \nu_2 \sin \phi \cos\phi \left( \cos \alpha \sin \theta +  \sin \alpha \cos\theta\right) \\
  -{{\rm i}} \, \nu_2 \sin \phi \cos\phi \left( \cos \alpha \sin \theta+  \sin \alpha \cos\theta\right) &
  \nu_2 \cos ^2\alpha \cos ^2\phi+\nu_1 \sin ^2\theta+\nu_2 \cos ^2\theta \sin ^2\phi \\
\end{array}
\right).
\end{eqnarray}
We can have the classical bound as
\begin{eqnarray}
    C_{\text{LHS}}
    &=&\frac{1}{2} \max_{\hat{n}_B} \left[ \mathrm{Tr} \left( |\hat{n}_B\rangle \langle \hat{n}_B| \rho_{B} \right) \right]\nonumber\\
    &=& \frac{1}{2} \max_{\hat{n}_B} \biggr\{ \mathrm{Tr} \biggr[ \frac{1}{2}\begin{pmatrix}
 1+\cos\theta_B &
 \sin\theta_B e^{-{\rm i} \phi_B}\\
 \sin\theta_B e^{{\rm i} \phi_B} &
 1-\cos\theta_B
\end{pmatrix} \nonumber\\
&& \times
\left(
\begin{array}{cc}
 \nu_2 \sin ^2\alpha \cos ^2\phi+\nu_1 \cos ^2\theta+\nu_2 \sin ^2\theta \sin ^2\phi
 & {{\rm i}} \, \nu_2 \sin \phi \cos\phi \left( \cos \alpha \sin \theta +  \sin \alpha \cos\theta\right) \\
  -{{\rm i}} \, \nu_2 \sin \phi \cos\phi \left( \cos \alpha \sin \theta+  \sin \alpha \cos\theta\right) &
  \nu_2 \cos ^2\alpha \cos ^2\phi+\nu_1 \sin ^2\theta+\nu_2 \cos ^2\theta \sin ^2\phi \\
\end{array}
\right)\biggr] \biggr\}\nonumber\\
&=& \frac{1}{4}+ \frac{1}{4} \max_{\hat{n}_B} \biggr[-\nu_2 \sin \alpha \cos\theta \sin \theta_B \sin 2 \phi \sin \phi_B
-\nu_2 \cos\alpha \sin\theta \sin \theta_B \sin2 \phi \sin\phi_B\nonumber\\
&&-\nu_2 \cos 2 \alpha \cos \theta_B \cos ^2\phi+\cos ^2\theta \cos \theta_B \left(\nu_1-\nu_2 \sin ^2\phi\right)
-\nu_1 \sin ^2\theta\cos \theta_B+\nu_2 \sin ^2\theta \cos \theta_B \sin ^2\phi\biggr].
\end{eqnarray}
By taking $\sin\phi_B=-1$, we have
\begin{eqnarray}\label{eq:CLHS-1}
    C_{\text{LHS}}
&=& \frac{1}{4}+ \frac{1}{4} \max_{\hat{n}_B} \biggr\{
\sin \theta_B \left(\nu_2 \sin \alpha \cos\theta  \sin 2 \phi +\nu_2 \cos\alpha \sin\theta  \sin2 \phi\right)\nonumber\\
&&+ \cos \theta_B \left[-\nu_2 \cos 2 \alpha \cos ^2\phi+\cos ^2\theta  \left(\nu_1-\nu_2 \sin ^2\phi\right)
-\nu_1 \sin ^2\theta +\nu_2 \sin ^2\theta  \sin ^2\phi \right]\biggr\}\nonumber\\
&=& \frac{1}{4}+ \frac{1}{4} \max_{\hat{n}_B} \biggr\{
\sin \theta_B \left[\nu_2 \sin (\alpha+\theta) \sin 2 \phi\right]+\cos \theta_B \left[-\nu_2 \cos 2 \alpha \cos ^2\phi+ \nu_1\cos2\theta
- \nu_2 \cos2\theta  \sin ^2\phi\right]\biggr\}\nonumber\\
&=& \frac{1}{4}+ \frac{1}{4} \max_{\hat{n}_B} \biggr\{
\sin \theta_B \left[\nu_2 \sin (\alpha+\theta) \sin 2 \phi\right]+\cos \theta_B \left[-\nu_2 \cos 2 \alpha \cos ^2\phi
+ \cos2\theta (\nu_1- \nu_2  \sin ^2\phi)\right]\biggr\}\nonumber\\
&\leq & \frac{1}{4}+ \frac{1}{4}\sqrt{
[\nu_2 \sin (\alpha+\theta) \sin 2 \phi]^2+ \left[-\nu_2 \cos 2 \alpha \cos ^2\phi
+ \cos2\theta (\nu_1- \nu_2  \sin ^2\phi)\right]^2},
\end{eqnarray}
the equal sign occurs at
\begin{eqnarray}
 \tan \theta_B = \frac{\nu_2 \sin (\alpha+\theta) \sin 2 \phi}{-\nu_2 \cos 2 \alpha \cos ^2\phi
+ \cos2\theta (\nu_1- \nu_2  \sin ^2\phi)}.
\end{eqnarray}
Because $\sin 2\alpha =0$ with $\alpha=0$, we have the classical bound as
\begin{eqnarray}\label{eq:class-3}
    C_{\text{LHS}} &=& \frac{1}{4}+ \frac{1}{4}\sqrt{
(\nu_2 \sin \theta \sin 2 \phi)^2+ \left[-\nu_2 \cos ^2\phi
+ \cos2\theta (\nu_1- \nu_2  \sin ^2\phi)\right]^2}.
\end{eqnarray}
Accordingly, the quantum expectation is given by (we take $\langle \mathcal{W} \rangle_1$ as $\langle \mathcal{W} \rangle$ and the parameter $\tau=\beta=\pi/2$)
\begin{eqnarray}
 \langle \mathcal{W} \rangle = \mathrm{Tr}[(|+\rangle\langle+|\otimes |\hat{n}_B \rangle\langle\hat{n}_B|) \rho],
\end{eqnarray}
i.e.,
\begin{eqnarray}
 \langle \mathcal{W} \rangle
  &=& \frac{1}{4}\mathrm{Tr}\biggr[\biggr(
 { \left(
  \begin{array}{cc}
   1 {+}\sin\xi \cos\delta  &
e^{-{\rm i}\tau} \left[-\cos\delta\cos\xi+\rm i\sin\delta  \right]\\
 e^{{\rm i}\tau}  \left[-\cos\delta\cos\xi-\rm i\sin\delta  \right] & 1 {-}\sin\xi \cos\delta  \\
  \end{array}
\right) }\otimes \begin{pmatrix}
 1+\cos\theta_B &
 \sin\theta_B e^{-{\rm i} \phi_B}\\
 \sin\theta_B e^{{\rm i} \phi_B} &
 1-\cos\theta_B
\end{pmatrix}\biggr) \rho\biggr]\nonumber\\
  &=& \frac{1}{4}\mathrm{Tr}\biggr[\biggr(
 { \left(
  \begin{array}{cc}
   1  &
1 \\\
 1   & 1  \\
  \end{array}
\right) }\otimes \begin{pmatrix}
 1+\cos\theta_B &
 \sin\theta_B e^{-{\rm i} \phi_B}\\
 \sin\theta_B e^{{\rm i} \phi_B} &
 1-\cos\theta_B
\end{pmatrix}\biggr) \rho\biggr],
\end{eqnarray}
with
\begin{eqnarray}
\rho
&=&
{\tiny
 \begin{bmatrix}
  \nu_1 \cos^2 \theta +\nu_2\sin ^2\theta  \sin ^2\phi  & \nu_2e^{{\rm i} \beta } \cos \alpha \sin \theta  \sin \phi  \cos \phi  &  \nu_2e^{{\rm i} \beta } \sin \alpha  \sin \theta  \sin \phi \cos \phi  & \nu_1 \sin \theta \cos \theta-\nu_2\sin \theta  \cos \theta  \sin ^2\phi  \\
 \nu_2e^{-{\rm i} \beta } \cos \alpha \sin \theta \sin \phi \cos \phi & \nu_2\cos ^2\alpha \cos ^2\phi  & \nu_2  \sin \alpha \cos \alpha  \cos ^2\phi & -\nu_2e^{-{\rm i} \beta } \cos \alpha  \cos \theta  \sin \phi  \cos \phi \\
 \nu_2e^{-{\rm i} \beta }\sin \alpha  \sin \theta \sin \phi \cos \phi  & \nu_2  \sin \alpha \cos \alpha  \cos ^2\phi  & \nu_2\sin ^2\alpha  \cos ^2\phi  & -\nu_2e^{-{\rm i} \beta }\sin \alpha   \cos \theta \sin \phi  \cos \phi \\
 \nu_1 \sin \theta \cos \theta -\nu_2\sin \theta \cos \theta \sin ^2\phi  & -\nu_2e^{{\rm i} \beta } \cos \alpha  \cos \theta  \sin \phi  \cos \phi  & -\nu_2e^{{\rm i} \beta }\sin \alpha   \cos \theta  \sin \phi  \cos \phi  &  \nu_1 \sin^2 \theta+\nu_2\cos ^2\theta  \sin ^2\phi  \\
\end{bmatrix}}\nonumber\\
&=&
 \begin{bmatrix}
  \nu_1 \cos^2 \theta +\nu_2\sin ^2\theta  \sin ^2\phi  & \nu_2  {\rm i} \sin \theta  \sin \phi  \cos \phi
  &  0  & \nu_1 \sin \theta \cos \theta-\nu_2\sin \theta  \cos \theta  \sin ^2\phi  \\
 -{\rm i} \nu_2 \sin \theta \sin \phi \cos \phi & \nu_2 \cos ^2\phi
 &  & {\rm i}\nu_2 \cos \theta  \sin \phi  \cos \phi \\
 0  & 0  & 0  & 0 \\
 \nu_1 \sin \theta \cos \theta -\nu_2\sin \theta \cos \theta \sin ^2\phi  & -{\rm i} \nu_2 \cos \theta  \sin \phi  \cos \phi
 & 0  &  \nu_1 \sin^2 \theta+\nu_2\cos ^2\theta  \sin ^2\phi  \\
\end{bmatrix}
\end{eqnarray}
After direct calculation, we can have
\begin{eqnarray}
 \langle \mathcal{W} \rangle
 &=&\frac{1}{4}+ \frac{1}{4} \biggr[
  \cos\theta_B \left( \nu_1\cos2 \theta - \nu_2 \sin^2\theta \cos2 \phi - \nu_2 \cos^2\theta\right)
+ \sin \theta_B \cos\phi_B [\sin 2 \theta  (\nu_1-\nu_2 \sin^2\phi)] \nonumber\\
&& - \sin \theta_B \sin\phi_B(  \nu_2 \sin\theta  \sin2\phi) \biggr].
 \end{eqnarray}
Because
\begin{eqnarray}
 \nu_1\cos2 \theta - \nu_2 \sin^2\theta \cos2 \phi - \nu_2 \cos^2\theta &= &\nu_1 \cos2\theta -\nu_2 \cos ^2\phi
- \nu_2\cos2\theta   \sin ^2\phi \nonumber\\
&=& -\nu_2 \cos ^2\phi
+ \cos2\theta (\nu_1- \nu_2  \sin ^2\phi),
\end{eqnarray}
then we have
\begin{eqnarray}
 \langle \mathcal{W} \rangle
&= &\frac{1}{4}+ \frac{1}{4} \biggr[
  \cos\theta_B \left[ -\nu_2 \cos ^2\phi
+ \cos2\theta (\nu_1- \nu_2  \sin ^2\phi)\right]
+ \sin \theta_B \cos\phi_B [\sin 2 \theta  (\nu_1-\nu_2 \sin^2\phi)] \nonumber\\
&& - \sin \theta_B \sin\phi_B(  \nu_2 \sin\theta  \sin2\phi) \biggr]\nonumber\\
&\leq & \frac{1}{4}+ \frac{1}{4} \sqrt{
  \left[ -\nu_2 \cos ^2\phi
+ \cos2\theta (\nu_1- \nu_2  \sin ^2\phi)\right]^2
+ [\sin 2 \theta  (\nu_1-\nu_2 \sin^2\phi)]^2 + (  \nu_2 \sin\theta  \sin2\phi)^2 },
\end{eqnarray}
the equal sign occurs at
\begin{eqnarray}
&& \tan \phi_B = {-}\frac{\nu_2 \sin\theta  \sin2\phi}{{\sin2\theta} (\nu_1-\nu_2 \sin^2\phi)}, \nonumber\\
&& \tan \theta_B = \frac{\sqrt{
  [\sin 2 \theta  (\nu_1-\nu_2 \sin^2\phi)]^2 + (  \nu_2 \sin\theta  \sin2\phi)^2 }}{-\nu_2 \cos ^2\phi
+ \cos2\theta (\nu_1- \nu_2  \sin ^2\phi)},
\end{eqnarray}
and the maximal value is obtained due to the relation $(\cos\theta_B)^2+(\sin \theta_B \cos\phi_B)^2+(\sin \theta_B \sin\phi_B)^2=1$. Thus
\begin{eqnarray}\label{eq:quatum-3}
 \langle \mathcal{W} \rangle^{\rm max}
&=& \frac{1}{4}+ \frac{1}{4} \sqrt{
  \left[ -\nu_2 \cos ^2\phi
+ \cos2\theta (\nu_1- \nu_2  \sin ^2\phi)\right]^2
+  (  \nu_2 \sin\theta  \sin2\phi)^2 +\mathcal{C}^2}.
\end{eqnarray}
By comparing  the classical bound $C_{\text{LHS}}$ in  Eq. (\ref{eq:class-3}) and the maximal quantum expectation $\langle \mathcal{W} \rangle^{\rm max}$ in Eq. (\ref{eq:quatum-3}), one easily proves the relation (\ref{eq:sdsi-3}).


\textcolor{blue}{\emph{Example 4.---}} Let us consider the case of $\sin2\alpha >0$, which satisfies the condition $\sin^2\phi - \sin2\alpha \sin2\theta \cos^2\phi>0$.
In this case, the square of concurrence of the state $\rho$ is equal to
\begin{eqnarray}
\mathcal{C}^2 &=& -s_2-2\sqrt{s_1}= G_1-G_2,
\end{eqnarray}
with
\begin{eqnarray}
G_1 &=& -s_2= -\frac{1}{2} \nu_2^2 \sin 2 \alpha \sin 2 \theta  \sin ^2 2 \phi +\nu_2^2 \sin ^2 2 \alpha \cos ^4 \phi
 +\sin ^2 2 \theta \left(\nu_1^2 +\nu_2^2 \sin ^4\phi \right)+ 2 \nu_1 \nu_2 \cos^2 2 \theta \sin ^2 \phi\nonumber\\
&=& -2 \nu_2^2 \sin 2 \alpha \sin 2 \theta  \sin^2 \phi \cos^2 \phi +\nu_2^2 \sin ^2 2 \alpha \cos ^4 \phi
 +\sin ^2 2 \theta \left(\nu_1^2 +\nu_2^2 \sin ^4\phi \right)+ 2 \nu_1 \nu_2 \cos^2 2 \theta \sin ^2 \phi\nonumber\\
&=& \nu_2^2  \left( \cos^2\phi \sin2\alpha - \sin2\theta \sin^2\phi \right)^2  + 2 \nu_1 \nu_2 \cos^2 2\theta \sin^2\phi + \nu_1^2 \sin^2 2\theta,
\end{eqnarray}
\begin{eqnarray}
G_2&=& 2 \sqrt{s_1} =\sqrt{4\times \frac{1}{2} \nu_1^2 \nu_2^2 \left(-\sin2 \alpha \sin 2 \theta \sin ^2 2 \phi +2 \sin ^2 2 \alpha
 \sin ^2 2 \theta \cos ^4\phi +2 \sin ^4\phi \right)}\nonumber\\
&=& \sqrt{2 \nu_1^2 \nu_2^2 \left(-4\sin2 \alpha \sin 2 \theta \sin ^2 \phi \cos ^2 \phi+2 \sin ^2 2 \alpha
 \sin ^2 2 \theta \cos ^4\phi +2 \sin ^4\phi \right)}\nonumber\\
&=& \sqrt{4 \nu_1^2 \nu_2^2 \left(-2\sin2 \alpha \sin 2 \theta \sin ^2 \phi \cos ^2 \phi+ \sin ^2 2 \alpha
 \sin ^2 2 \theta \cos ^4\phi + \sin ^4\phi \right)}\nonumber\\
 &=&  \sqrt{4 \nu_1^2 \nu_2^2 \left( \sin^2\phi - \sin2\alpha \sin2\theta \cos^2\phi \right)^2}\nonumber\\
 &=& 2 \nu_1 \nu_2 \left( \sin^2\phi - \sin2\alpha \sin2\theta \cos^2\phi \right)>0.
\end{eqnarray}
Then we have
\begin{eqnarray}
\mathcal{C}^2 &=& \left[\nu_2^2  \left( \cos^2\phi \sin2\alpha - \sin2\theta \sin^2\phi \right)^2  + 2 \nu_1 \nu_2 \cos^2 2\theta \sin^2\phi + \nu_1^2 \sin^2 2\theta\right]- 2 \nu_1 \nu_2 \left( \sin^2\phi - \sin2\alpha \sin2\theta \cos^2\phi \right)\nonumber\\
&=& \nu_2^2 \left( \cos^2\phi \sin2\alpha - \sin2\theta \sin^2\phi \right)^2 + 2 \nu_1 \nu_2 \left( -\sin^2 2\theta  \sin^2\phi + \sin2\alpha \sin2\theta \cos^2\phi \right)+ \nu_1^2 \sin^2 2\theta \nonumber\\
&=& \nu_2^2 \left( \cos^2\phi \sin2\alpha - \sin2\theta \sin^2\phi \right)^2 + 2 (\nu_1 \sin 2\theta) \nu_2 \left( -\sin 2\theta  \sin^2\phi + \sin2\alpha \cos^2\phi \right)+ \nu_1^2 \sin^2 2\theta \nonumber\\
&=& \nu_2^2 \left( \sin2\alpha \cos^2\phi  - \sin2\theta \sin^2\phi \right)^2 + 2 (\nu_1 \sin 2\theta) \nu_2 \left( \sin2\alpha \cos^2\phi -\sin 2\theta  \sin^2\phi \right)+ \nu_1^2 \sin^2 2\theta \nonumber\\
&=& \left[ \nu_2 \left( \sin2\alpha \cos^2\phi  - \sin2\theta \sin^2\phi \right) + \nu_1 \sin 2\theta\right]^2.
\end{eqnarray}
{In Sec. \ref{sec:IIIB} we have known that
for the measurement parameters {$\tau=\beta=\pi/2$} and $\xi$ is arbitrary, one can have $\mathcal{M}=\mathcal{M}^\dagger$ for separable states.} In this situation, Alice's measurement direction is given by
\begin{eqnarray}
&& \hat{n}=(\sin\xi \cos\tau, \sin\xi \sin\tau, \cos\xi)=(0, \sin\xi, \cos\xi).
\end{eqnarray}
For convenience, we can adopt the following the measurement parameters
\begin{eqnarray}
&& {\tau=\beta=\frac{\pi}{2}}, \;\;\;\; \xi=\frac{\pi}{2}, \;\;\;\; \delta=\frac{\pi}{2},
\end{eqnarray}
then similarly we obtain the classical bound as
\begin{eqnarray}
    C_{\text{LHS}} &=& \frac{1}{4}+ \frac{1}{4} \max_{\hat{n}_B} \biggr[-\nu_2 \sin \alpha \cos\theta \sin \theta_B \sin 2 \phi \sin \phi_B
-\nu_2 \cos\alpha \sin\theta \sin \theta_B \sin2 \phi \sin\phi_B\nonumber\\
&&-\nu_2 \cos 2 \alpha \cos \theta_B \cos ^2\phi+\cos ^2\theta \cos \theta_B \left(\nu_1-\nu_2 \sin ^2\phi\right)
-\nu_1 \sin ^2\theta\cos \theta_B+\nu_2 \sin ^2\theta \cos \theta_B \sin ^2\phi\biggr]\nonumber\\
&=& \frac{1}{4}+ \frac{1}{4} \max_{\hat{n}_B} \biggr\{
\sin \theta_B \left(\nu_2 \sin \alpha \cos\theta  \sin 2 \phi +\nu_2 \cos\alpha \sin\theta  \sin2 \phi\right)\nonumber\\
&&+ \cos \theta_B \left[-\nu_2 \cos 2 \alpha \cos ^2\phi+\cos ^2\theta  \left(\nu_1-\nu_2 \sin ^2\phi\right)
-\nu_1 \sin ^2\theta +\nu_2 \sin ^2\theta  \sin ^2\phi \right]\biggr\}\nonumber\\
&=& \frac{1}{4}+ \frac{1}{4} \max_{\hat{n}_B} \biggr\{
\sin \theta_B \left[\nu_2 \sin (\alpha+\theta) \sin 2 \phi\right]+\cos \theta_B \left[-\nu_2 \cos 2 \alpha \cos ^2\phi+ \nu_1\cos2\theta
- \nu_2 \cos2\theta  \sin ^2\phi\right]\biggr\}\nonumber\\
&=& \frac{1}{4}+ \frac{1}{4} \max_{\hat{n}_B} \biggr\{
\sin \theta_B \left[\nu_2 \sin (\alpha+\theta) \sin 2 \phi\right]+\cos \theta_B \left[-\nu_2 \cos 2 \alpha \cos ^2\phi
+ \cos2\theta (\nu_1- \nu_2  \sin ^2\phi)\right]\biggr\}\nonumber\\
&\leq & \frac{1}{4}+ \frac{1}{4}\sqrt{
[\nu_2 \sin (\alpha+\theta) \sin 2 \phi]^2+ \left[-\nu_2 \cos 2 \alpha \cos ^2\phi
+ \cos2\theta (\nu_1- \nu_2  \sin ^2\phi)\right]^2}.
\end{eqnarray}
Thus the classical bound reads
\begin{eqnarray}\label{eq:class-4}
C_{\text{LHS}}
    &= & \frac{1}{4}+ \frac{1}{4}\sqrt{
[\nu_2 \sin (\alpha+\theta) \sin 2 \phi]^2+ \left[-\nu_2 \cos 2 \alpha \cos ^2\phi
+ \cos2\theta (\nu_1- \nu_2  \sin ^2\phi)\right]^2}.
\end{eqnarray}
Accordingly, the quantum expectation is given by
\begin{eqnarray}
 \langle \mathcal{W} \rangle = \mathrm{Tr}[(|+\rangle\langle+|\otimes |\hat{n}_B \rangle\langle\hat{n}_B|) \rho],
\end{eqnarray}
i.e.,
\begin{eqnarray}
 \langle \mathcal{W} \rangle
  &=& \frac{1}{4}\mathrm{Tr}\biggr[\biggr(
 { \left(
  \begin{array}{cc}
   1 {+}\sin\xi \cos\delta  &
e^{-{\rm i}\tau} \left[-\cos\delta\cos\xi+\rm i\sin\delta  \right]\\
 e^{{\rm i}\tau}  \left[-\cos\delta\cos\xi-\rm i\sin\delta  \right] & 1\hilight{-}\sin\xi \cos\delta  \\
  \end{array}
\right) }\otimes \begin{pmatrix}
 1+\cos\theta_B &
 \sin\theta_B e^{-{\rm i} \phi_B}\\
 \sin\theta_B e^{{\rm i} \phi_B} &
 1-\cos\theta_B
\end{pmatrix}\biggr) \rho\biggr]\nonumber\\
  &=& \frac{1}{4}\mathrm{Tr}\biggr[\biggr(
 { \left(
  \begin{array}{cc}
   1  &
1 \\\
 1   & 1  \\
  \end{array}
\right) }\otimes \begin{pmatrix}
 1+\cos\theta_B &
 \sin\theta_B e^{-{\rm i} \phi_B}\\
 \sin\theta_B e^{{\rm i} \phi_B} &
 1-\cos\theta_B
\end{pmatrix}\biggr) \rho\biggr].
\end{eqnarray}
After direct calculation, we can have
\begin{eqnarray}
 \langle \mathcal{W} \rangle
 &=&\frac{1}{4}+ \frac{1}{4} \biggr\{ \cos\theta_B \left[-\nu_2 \cos2\alpha \cos^2\phi
 +\cos2\theta \left(\nu_1-\nu_2 \sin^2\phi\right)\right] -  \sin\theta_B\sin\phi_B [ \nu_2\sin(\alpha +\theta)\sin2\phi]\nonumber\\
 &&+ \sin\theta_B \cos\phi_B  [\nu_2  \cos^2\phi( \sin2 \alpha+ \sin2 \theta)
 +\sin2\theta (\nu_1-\nu_2)]\biggr\},
 \end{eqnarray}
which can reach its maximum
\begin{eqnarray}\label{eq:quatum-4}
 \langle \mathcal{W} \rangle^{\rm max}
 &=&\frac{1}{4}+ \frac{1}{4} \sqrt{\Gamma^2},
 \end{eqnarray}
with
\begin{eqnarray}
\Gamma^2 &=&\left[-\nu_2 \cos2\alpha \cos^2\phi
 +\cos2\theta \left(\nu_1-\nu_2 \sin^2\phi\right)\right]^2 + [ \nu_2\sin(\alpha +\theta)\sin2\phi]^2\nonumber\\
&& +[\nu_2  \cos^2\phi( \sin2 \alpha+ \sin2 \theta)
 +\sin2\theta (\nu_1-\nu_2)]^2\nonumber\\
 &=&[ \nu_2\sin(\alpha +\theta)\sin2\phi]^2+\left[-\nu_2 \cos2\alpha \cos^2\phi
 +\cos2\theta \left(\nu_1-\nu_2 \sin^2\phi\right)\right]^2 \nonumber\\
 &&+ \left[ \nu_2 \left( \sin2\alpha \cos^2\phi  - \sin2\theta \sin^2\phi \right) + \nu_1 \sin 2\theta\right]^2\nonumber\\
&=&[ \nu_2\sin(\alpha +\theta)\sin2\phi]^2+\left[-\nu_2 \cos2\alpha \cos^2\phi
 +\cos2\theta \left(\nu_1-\nu_2 \sin^2\phi\right)\right]^2 + \mathcal{C}^2.
 \end{eqnarray}
By comparing  the classical bound $C_{\text{LHS}}$ in  Eq. (\ref{eq:class-4}) and the maximal quantum expectation $\langle \mathcal{W} \rangle^{\rm max}$ in Eq. (\ref{eq:quatum-4}), one easily proves the relation (\ref{eq:sdsi-3}).

\begin{remark} By the way, for the quantum state $\rho$, the classical bound can be written as
\begin{eqnarray}
    C_{\text{LHS}}
&=& \frac{1+|\vec{F}|}{4}=\frac{1+\sqrt{\vec{F}^2}}{4},
\end{eqnarray}
where $\vec{F}$ is the Bloch vector of the state $\rho_{B}$. Similarly, the quantum expectation can be written as
\begin{eqnarray}
 \langle \mathcal{W} \rangle^{\rm max}
 &=&\frac{1+\sqrt{\vec{F}^2 +\mathcal{C}^2}}{4}.
\end{eqnarray}
$\blacksquare$
\end{remark}


\vspace{6mm}
\centerline{\emph{3.2. The Case of $\cos\beta=1$}}
\vspace{6mm}

In this situation, the state $\rho$ is given by
\begin{eqnarray}
\rho=\nu_1  \ket{\psi_1}\bra{\psi_1}+\nu_2 \ket{\psi_2}\bra{\psi_2},
\end{eqnarray}
where
\begin{eqnarray}
\ket{\psi_1} &=& \cos \theta \ket{00}+\sin \theta\ket{11}, \nonumber\\
 \ket{\psi_2} &=& \cos\phi(\cos \alpha \ket{01} +\sin\alpha\ket{10})+ e^{{\rm i}\beta}\sin\phi(\sin \theta \ket{00}-\cos \theta\ket{11}),
\end{eqnarray}
with $\beta=0$, $\cos\beta=1$, $e^{{\rm i}\beta}=1$, $\theta\in[0, \pi/2]$, and $\alpha\in[0, \pi/2]$. For simplicity, in the following we have taken $\cos\beta=1$ as example (for the case of $\cos\beta=-1$ and $\beta=\pi$, it will yield the similar result). The concurrence of the state $\rho$ is given by
\begin{eqnarray}
\mathcal{C} &=& \sqrt{-s_2-2\sqrt{s_1}},
\end{eqnarray}
with
\begin{eqnarray}
 s_2 &=& -\frac{1}{2} \nu_2^2 \sin 2 \alpha \sin 2 \theta  \sin ^2 2 \phi \left( 2\cos^ 2 \beta-1\right)-\nu_2^2 \sin ^2 2 \alpha \cos ^4 \phi
 -\sin ^2 2 \theta \left(\nu_1^2 +\nu_2^2 \sin ^4\phi \right)- 2 \nu_1 \nu_2 \cos^2 2 \theta \sin ^2 \phi \nonumber\\
&=& -\frac{1}{2} \nu_2^2 \sin 2 \alpha \sin 2 \theta  \sin ^2 2 \phi -\nu_2^2 \sin ^2 2 \alpha \cos ^4 \phi
 -\sin ^2 2 \theta \left(\nu_1^2 +\nu_2^2 \sin ^4\phi \right)- 2 \nu_1 \nu_2 \cos^2 2 \theta \sin ^2 \phi, \nonumber\\
s_1 &=& \frac{1}{2} \nu_1^2 \nu_2^2 \left[\sin2 \alpha \sin 2 \theta \sin ^2 2 \phi \left( 2\cos^ 2 \beta-1\right) +2 \sin ^2 2 \alpha
 \sin ^2 2 \theta \cos ^4\phi +2 \sin ^4\phi \right]\nonumber\\
&=& \frac{1}{2} \nu_1^2 \nu_2^2 \left(\sin2 \alpha \sin 2 \theta \sin ^2 2 \phi +2 \sin ^2 2 \alpha
 \sin ^2 2 \theta \cos ^4\phi +2 \sin ^4\phi \right).
\end{eqnarray}
In the following, we would like to study the problem by two different cases: $\sin 2 \alpha= 0$ and $\sin 2 \alpha> 0$.


\textcolor{blue}{\emph{Example 5.---}} Let us consider a simple case of $\sin2\alpha=0$ with $\alpha=0$ (for $\alpha=\pi/2$, the result is similar). In this case, the concurrence of the state $\rho$ is equal to
\begin{eqnarray}
\mathcal{C} &=& \sqrt{-s_2-2\sqrt{s_1}}=\sqrt{\sin^2 2\theta \times \left[\nu_1-\nu_2 \sin^2\phi\right]^2}=\left|\left(\nu_1-\nu_2 \sin^2\phi\right)\sin 2\theta\right|\nonumber\\
&=& |\nu_1-\nu_2 \sin^2\phi|\sin 2\theta.
\end{eqnarray}
{In Sec. \ref{sec:IIIB} we have known that
for the measurement parameters {$\tau=\beta=0$} and $\xi$ is arbitrary, one can have $\mathcal{M}=\mathcal{M}^\dagger$ for separable states.} In this situation, Alice's measurement direction is given by
\begin{eqnarray}
&& \hat{n}=(\sin\xi \cos\tau, \sin\xi \sin\tau, \cos\xi)=(\sin\xi, 0, \cos\xi).
\end{eqnarray}
For convenience, we can adopt the following the measurement parameters
\begin{eqnarray}
&& {\tau=\beta=0}, \;\;\;\; \xi=\frac{\pi}{2}, \;\;\;\; \delta=\frac{\pi}{2},
\end{eqnarray}
then we obtain
\begin{eqnarray}
&& \hat{n}=(\sin\xi \cos\tau, \sin\xi \sin\tau, \cos\xi)=(1, 0, 0)=\hat{e}_x,
\end{eqnarray}
\begin{eqnarray}
&& \ket{+\hat{n}}=\begin{pmatrix}
\cos\frac{\xi}{2}\\
\sin\frac{\xi}{2}\, e^{{\rm i}\tau}\\
\end{pmatrix}=\cos\frac{\xi}{2} |0\rangle+\sin\frac{\xi}{2}\, e^{{\rm i}\tau} |1\rangle
=\frac{1}{\sqrt{2}} \left(|0\rangle+ |1\rangle\right),
\end{eqnarray}

\begin{eqnarray}
 \ket{-\hat{n}}=\begin{pmatrix}
\sin\frac{\xi}{2}\\
-\cos\frac{\xi}{2} \, e^{{\rm i}\tau} \\
\end{pmatrix}= \sin\frac{\xi}{2}|0\rangle -\cos\frac{\xi}{2} \, e^{{\rm i}\tau} |1\rangle=\frac{1}{\sqrt{2}} \left(|0\rangle- |1\rangle\right),
\end{eqnarray}

\begin{eqnarray}
P^{\hat{n}}_{0}&=&\frac{1}{2}\left(\openone+\vec{\sigma}\cdot \hat{n}\right)=|+\hat{n}\rangle\langle+\hat{n}|
= \frac{1}{2}\begin{pmatrix}
 1+\cos\xi&
 \sin\xi e^{-{\rm i} \tau}\\
 \sin\xi e^{{\rm i} \tau} &
 1-\cos\xi
\end{pmatrix}= {\frac{1}{2}\begin{pmatrix}
 1 &1\\
 1 & 1
\end{pmatrix},}
\end{eqnarray}

\begin{eqnarray}
P^{\hat{n}}_{1}=\frac{1}{2}\left(\openone-\vec{\sigma}\cdot \hat{n}\right)=|-\hat{n}\rangle\langle-\hat{n}|
= \frac{1}{2}\begin{pmatrix}
 1-\cos\xi&
 -\sin\xi e^{-{\rm i} \tau}\\
 -\sin\xi e^{{\rm i} \tau} &
 1+\cos\xi
\end{pmatrix}= {\frac{1}{2}\begin{pmatrix}
 1 & -1\\
 -1 & 1
\end{pmatrix},}
\end{eqnarray}

\begin{eqnarray}
\rho_{B}
& =&
\left(
\begin{array}{cc}
 \nu_2 \sin ^2\alpha \cos ^2\phi+\nu_1 \cos ^2\theta+\nu_2 \sin ^2\theta \sin ^2\phi
 & \nu_2 \sin \phi \cos\phi \left(e^{{\rm i} \beta } \cos \alpha \sin \theta-e^{-{\rm i} \beta }  \sin \alpha \cos\theta\right) \\
  \nu_2 \sin \phi \cos\phi \left(e^{-{\rm i} \beta } \cos \alpha \sin \theta-e^{{\rm i} \beta }  \sin \alpha \cos\theta\right) &
  \nu_2 \cos ^2\alpha \cos ^2\phi+\nu_1 \sin ^2\theta+\nu_2 \cos ^2\theta \sin ^2\phi \\
\end{array}
\right)\nonumber\\
& =&
\left(
\begin{array}{cc}
 \nu_2 \sin ^2\alpha \cos ^2\phi+\nu_1 \cos ^2\theta+\nu_2 \sin ^2\theta \sin ^2\phi
 &  \nu_2 \sin \phi \cos\phi \left( \cos \alpha \sin \theta {-}  \sin \alpha \cos\theta\right) \\
   \nu_2 \sin \phi \cos\phi \left( \cos \alpha \sin \theta  {-}  \sin \alpha \cos\theta\right) &
  \nu_2 \cos ^2\alpha \cos ^2\phi+\nu_1 \sin ^2\theta+\nu_2 \cos ^2\theta \sin ^2\phi \\
\end{array}
\right).
\end{eqnarray}
We can have the classical bound as
\begin{eqnarray}
    C_{\text{LHS}}
    &=&\frac{1}{2} \max_{\hat{n}_B} \left[ \mathrm{Tr} \left( |\hat{n}_B\rangle \langle \hat{n}_B| \rho_{B} \right) \right]\nonumber\\
    &=& \frac{1}{2} \max_{\hat{n}_B} \biggr[ \mathrm{Tr} \biggr[ \frac{1}{2}\begin{pmatrix}
 1+\cos\theta_B &
 \sin\theta_B e^{-{\rm i} \phi_B}\\
 \sin\theta_B e^{{\rm i} \phi_B} &
 1-\cos\theta_B
\end{pmatrix} \nonumber\\
&& \times\left(
\begin{array}{cc}
 \nu_2 \sin ^2\alpha \cos ^2\phi+\nu_1 \cos ^2\theta+\nu_2 \sin ^2\theta \sin ^2\phi
 &  \nu_2 \sin \phi \cos\phi \left( \cos \alpha \sin \theta \hilight{-}  \sin \alpha \cos\theta\right) \\
   \nu_2 \sin \phi \cos\phi \left( \cos \alpha \sin \theta  \hilight{-}  \sin \alpha \cos\theta\right) &
  \nu_2 \cos ^2\alpha \cos ^2\phi+\nu_1 \sin ^2\theta+\nu_2 \cos ^2\theta \sin ^2\phi \\
\end{array}
\right)
\biggr] \biggr]\nonumber\\
&=& \frac{1}{4}+ \frac{1}{4} \max_{\hat{n}_B} \biggr[-\nu_2 \sin \alpha \cos\theta \sin \theta_B \sin 2 \phi \cos \phi_B
+\nu_2 \cos\alpha \sin\theta \sin \theta_B \sin2 \phi \cos\phi_B\nonumber\\
&&-\nu_2 \cos 2 \alpha \cos \theta_B \cos ^2\phi+\cos ^2\theta \cos \theta_B \left(\nu_1-\nu_2 \sin ^2\phi\right)
-\nu_1 \sin ^2\theta\cos \theta_B+\nu_2 \sin ^2\theta \cos \theta_B \sin ^2\phi\biggr].
\end{eqnarray}
By taking $\cos\phi_B=1$, we have
\begin{eqnarray}
    C_{\text{LHS}}
&=& \frac{1}{4}+ \frac{1}{4} \max_{\hat{n}_B} \biggr\{
\sin \theta_B \left(-\nu_2 \sin \alpha \cos\theta  \sin 2 \phi +\nu_2 \cos\alpha \sin\theta  \sin2 \phi\right)\nonumber\\
&&+ \cos \theta_B \left[-\nu_2 \cos 2 \alpha \cos ^2\phi+\cos ^2\theta  \left(\nu_1-\nu_2 \sin ^2\phi\right)
-\nu_1 \sin ^2\theta +\nu_2 \sin ^2\theta  \sin ^2\phi \right]\biggr\}\nonumber\\
&=& \frac{1}{4}+ \frac{1}{4} \max_{\hat{n}_B} \biggr\{
\sin \theta_B \left[\nu_2 \sin (\theta-\alpha) \sin 2 \phi\right]+\cos \theta_B \left[-\nu_2 \cos 2 \alpha \cos ^2\phi+ \nu_1\cos2\theta
- \nu_2 \cos2\theta  \sin ^2\phi\right]\biggr\}\nonumber\\
&=& \frac{1}{4}+ \frac{1}{4} \max_{\hat{n}_B} \biggr\{
\sin \theta_B \left[\nu_2 \sin (\theta-\alpha) \sin 2 \phi\right]+\cos \theta_B \left[-\nu_2 \cos 2 \alpha \cos ^2\phi
+ \cos2\theta (\nu_1- \nu_2  \sin ^2\phi)\right]\biggr\}\nonumber\\
&\leq & \frac{1}{4}+ \frac{1}{4}\sqrt{
[\nu_2 \sin (\theta-\alpha) \sin 2 \phi]^2+ \left[-\nu_2 \cos 2 \alpha \cos ^2\phi
+ \cos2\theta (\nu_1- \nu_2  \sin ^2\phi)\right]^2},
\end{eqnarray}
the equal sign occurs at
\begin{eqnarray}
 \tan \theta_B = \frac{\nu_2 \sin (\theta-\alpha) \sin 2 \phi}{-\nu_2 \cos 2 \alpha \cos ^2\phi
+ \cos2\theta (\nu_1- \nu_2  \sin ^2\phi)}.
\end{eqnarray}
Because $\sin 2\alpha =0$ with $\alpha=0$, we have the classical bound as
\begin{eqnarray}\label{eq:class-5}
    C_{\text{LHS}} &=& \frac{1}{4}+ \frac{1}{4}\sqrt{
(\nu_2 \sin \theta \sin 2 \phi)^2+ \left[-\nu_2 \cos ^2\phi
+ \cos2\theta (\nu_1- \nu_2  \sin ^2\phi)\right]^2}.
\end{eqnarray}
Accordingly, the quantum expectation is given by
\begin{eqnarray}
 \langle \mathcal{W} \rangle = \mathrm{Tr}[(|+\rangle\langle+|\otimes |\hat{n}_B \rangle\langle\hat{n}_B|) \rho],
\end{eqnarray}
i.e.,
\begin{eqnarray}
 \langle \mathcal{W} \rangle
  &=& \frac{1}{4}\mathrm{Tr}\biggr[\biggr(
 { \left(
  \begin{array}{cc}
   1 {+}\sin\xi \cos\delta  &
e^{-{\rm i}\tau} \left[-\cos\delta\cos\xi+\rm i\sin\delta  \right]\\
 e^{{\rm i}\tau}  \left[-\cos\delta\cos\xi-\rm i\sin\delta  \right] & 1 {-}\sin\xi \cos\delta  \\
  \end{array}
\right) }\otimes \begin{pmatrix}
 1+\cos\theta_B &
 \sin\theta_B e^{-{\rm i} \phi_B}\\
 \sin\theta_B e^{{\rm i} \phi_B} &
 1-\cos\theta_B
\end{pmatrix}\biggr) \rho\biggr],
\end{eqnarray}
i.e.,
\begin{eqnarray}
 \langle \mathcal{W} \rangle
&=& \frac{1}{4}\mathrm{Tr}\biggr[\biggr(
 { \frac{1}{2}
\left(
  \begin{array}{cc}
   1  &
 {\rm i}  \\
-{\rm i}    & 1  \\
  \end{array}
\right) }\otimes \begin{pmatrix}
 1+\cos\theta_B &
 \sin\theta_B e^{-{\rm i} \phi_B}\\
 \sin\theta_B e^{{\rm i} \phi_B} &
 1-\cos\theta_B
\end{pmatrix}\biggr) \rho\biggr].
\end{eqnarray}
After direct calculation, we can have the general expression of $\langle \mathcal{W} \rangle$ as
\begin{eqnarray}
 \langle \mathcal{W} \rangle&=&\frac{1}{4}+ \frac{1}{4} \left\{ \nu_2 \left[ \cos(\alpha - \theta) \cos\tau \sin\beta - \cos\beta \cos(\alpha + \theta) \sin\tau \right] \sin2\phi \right\} \nonumber\\
&&+ \frac{1}{4} \cos\theta_B \biggr[ -\nu_2 \cos2\alpha \cos^2\phi + \cos2\theta \left( \nu_1 - \nu_2 \sin^2\phi \right) \nonumber\\
&& + \nu_2 \left[ -\sin\beta \cos(\alpha + \theta) \cos\tau + \cos\beta \cos(\alpha - \theta) \sin\tau \right] \sin2\phi \biggr] \nonumber\\
&&+ \frac{1}{4} \sin\theta_B \sin\phi_B \left\{ \cos\tau \left[ -\nu_2 \cos^2\phi \sin2\alpha + \sin2\theta \left( \nu_1 - \nu_2 \sin^2\phi \right) \right] - \nu_2 \sin\beta \sin(\alpha + \theta) \sin 2\phi \right\} \nonumber\\
&&+ \frac{1}{4} \sin\theta_B \cos\phi_B \left\{ \sin\tau \left[ \nu_2 \cos^2\phi \sin 2\alpha + \sin2\theta \left( \nu_1 - \nu_2 \sin^2\phi \right) \right] - \nu_2 \cos\beta \sin(\alpha - \theta) \sin 2\phi \right\}.
\end{eqnarray}
For $\tau=\beta=0$ and $\alpha=0$, we have
\begin{eqnarray}
 \langle \mathcal{W} \rangle
  &=&\frac{1}{4}+ \frac{1}{4} \biggr\{\cos\theta_B \left[ -\nu_2 \cos2\alpha \cos^2\phi + \cos2\theta \left( \nu_1 - \nu_2 \sin^2\phi \right) \right] \nonumber\\
&&+ \sin\theta_B \sin\phi_B \left[ -\nu_2 \cos^2\phi \sin2\alpha + \sin2\theta \left( \nu_1 - \nu_2 \sin^2\phi \right) \right]  +  \sin\theta_B \cos\phi_B \left[  \nu_2 \sin(\theta-\alpha) \sin 2\phi \right]\biggr\}\nonumber\\
&=&\frac{1}{4}+ \frac{1}{4} \biggr\{\cos\theta_B \left[ -\nu_2 \cos^2\phi + \cos2\theta \left( \nu_1 - \nu_2 \sin^2\phi \right) \right] \nonumber\\
&&+ \sin\theta_B \sin\phi_B \left[\sin2\theta \left( \nu_1 - \nu_2 \sin^2\phi \right) \right]  +  \sin\theta_B \cos\phi_B \left[  \nu_2 \sin\theta \sin 2\phi \right]\biggr\},
\end{eqnarray}
which can reach its maximum
\begin{eqnarray}\label{eq:quatum-5}
 \langle \mathcal{W} \rangle^{\rm max}
 &=&\frac{1}{4}+ \frac{1}{4} \sqrt{\Gamma^2},
 \end{eqnarray}
with
\begin{eqnarray}
\Gamma^2 &=&\left[-\nu_2 \cos^2\phi
 +\cos2\theta \left(\nu_1-\nu_2 \sin^2\phi\right)\right]^2 + ( \nu_2\sin\theta \sin2\phi)^2 +[\sin2\theta \left( \nu_1 - \nu_2 \sin^2\phi \right)]^2\nonumber\\
 &=& (\nu_2 \sin \theta \sin 2 \phi)^2+ \left[-\nu_2 \cos ^2\phi
+ \cos2\theta (\nu_1- \nu_2  \sin ^2\phi)\right]^2+\mathcal{C}^2,
\end{eqnarray}
By comparing  the classical bound $C_{\text{LHS}}$ in  Eq. (\ref{eq:class-5}) and the maximal quantum expectation $\langle \mathcal{W} \rangle^{\rm max}$ in Eq. (\ref{eq:quatum-5}), one easily proves the relation (\ref{eq:sdsi-3}).


\textcolor{blue}{\emph{Example 6.---}} Let us consider the case of $\sin2\alpha>0$. In this case, the square of concurrence of the state $\rho$ is equal to
\begin{eqnarray}
\mathcal{C}^2 &=& -s_2-2\sqrt{s_1}= G_1-G_2,
\end{eqnarray}
with
\begin{eqnarray}
 G_1 &=& -s_2 = \frac{1}{2} \nu_2^2 \sin 2 \alpha \sin 2 \theta  \sin ^2 2 \phi
  +\nu_2^2 \sin ^2 2 \alpha \cos ^4 \phi
 +\sin ^2 2 \theta \left(\nu_1^2 +\nu_2^2 \sin ^4\phi \right)
 + 2 \nu_1 \nu_2 \cos^2 2 \theta \sin ^2 \phi, \nonumber\\
G_2 &=& 2\sqrt{s_1} = 2\times
 \sqrt{\frac{1}{2} \nu_1^2 \nu_2^2 \left(\sin2 \alpha \sin 2 \theta \sin ^2 2 \phi +2 \sin ^2 2 \alpha
 \sin ^2 2 \theta \cos ^4\phi +2 \sin ^4\phi \right)}\nonumber\\
&=& 2\sqrt{s_1} = 2\times
 \sqrt{ \nu_1^2 \nu_2^2 \left( 2 \sin2 \alpha \sin 2 \theta \sin ^2 \phi \cos ^2 \phi + \sin ^2 2 \alpha
 \sin ^2 2 \theta \cos ^4\phi + \sin ^4\phi \right)}\nonumber\\
&=& 2\nu_1 \nu_2 \left(\sin^2\phi + \sin2\alpha \sin2\theta \cos^2\phi\right)>0.
\end{eqnarray}
One can have
\begin{eqnarray}
&& \mathcal{C}^2= [\nu_1 \sin 2\theta - \nu_2 \left( \sin 2\theta\sin^2\phi + \sin2\alpha \cos^2\phi \right)]^2.
\end{eqnarray}
{In Sec. \ref{sec:IIIB} we have known that
for the measurement parameters {$\tau=\beta=0$} and $\xi$ is arbitrary, one can have $\mathcal{M}=\mathcal{M}^\dagger$ for separable states.} In this situation, Alice's measurement direction is given by
\begin{eqnarray}
&& \hat{n}=(\sin\xi \cos\tau, \sin\xi \sin\tau, \cos\xi)=(\sin\xi, 0, \cos\xi).
\end{eqnarray}
For convenience, we can adopt the following the measurement parameters
\begin{eqnarray}
&& {\tau=\beta=0}, \;\;\;\; \xi=\frac{\pi}{2}, \;\;\;\; \delta=\frac{\pi}{2},
\end{eqnarray}
similarly we have the classical bound as
\begin{eqnarray}\label{eq:class-6}
    C_{\text{LHS}}
&= & \frac{1}{4}+ \frac{1}{4}\sqrt{
[\nu_2 \sin (\theta-\alpha) \sin 2 \phi]^2+ \left[-\nu_2 \cos 2 \alpha \cos ^2\phi
+ \cos2\theta (\nu_1- \nu_2  \sin ^2\phi)\right]^2}.
\end{eqnarray}
Accordingly, we have the quantum expectation as
\begin{eqnarray}
 \langle \mathcal{W} \rangle
  &=&\frac{1}{4}+ \frac{1}{4} \biggr\{\cos\theta_B \left[ -\nu_2 \cos2\alpha \cos^2\phi + \cos2\theta \left( \nu_1 - \nu_2 \sin^2\phi \right) \right] \nonumber\\
&&+ \sin\theta_B \sin\phi_B \left[ -\nu_2 \cos^2\phi \sin2\alpha + \sin2\theta \left( \nu_1 - \nu_2 \sin^2\phi \right) \right]  +  \sin\theta_B \cos\phi_B \left[  \nu_2 \sin(\theta-\alpha) \sin 2\phi \right]\biggr\},
\end{eqnarray}
which can reach its maximum
\begin{eqnarray}\label{eq:quatum-6}
 \langle \mathcal{W} \rangle^{\rm max}
 &=&\frac{1}{4}+ \frac{1}{4} \sqrt{\Gamma^2},
 \end{eqnarray}
with
\begin{eqnarray}
\Gamma^2 &=&  \left[ -\nu_2 \cos2\alpha \cos^2\phi + \cos2\theta \left( \nu_1 - \nu_2 \sin^2\phi \right) \right]^2
+ \left[ -\nu_2 \cos^2\phi \sin2\alpha + \sin2\theta \left( \nu_1 - \nu_2 \sin^2\phi \right) \right]^2 \nonumber\\
&& + \left[  \nu_2 \sin(\theta-\alpha) \sin 2\phi \right]^2 \nonumber\\
&=&  \left[  \nu_2 \sin(\theta-\alpha) \sin 2\phi \right]^2 +\left[ -\nu_2 \cos2\alpha \cos^2\phi + \cos2\theta \left( \nu_1 - \nu_2 \sin^2\phi \right) \right]^2 \nonumber\\
&& + [\nu_1 \sin 2\theta - \nu_2 \left( \sin 2\theta\sin^2\phi + \sin2\alpha \cos^2\phi \right)]^2 \nonumber\\
&=&  \left[  \nu_2 \sin(\theta-\alpha) \sin 2\phi \right]^2 +\left[ -\nu_2 \cos2\alpha \cos^2\phi + \cos2\theta \left( \nu_1 - \nu_2 \sin^2\phi \right) \right]^2 +\mathcal{C}^2.
\end{eqnarray}
By comparing  the classical bound $C_{\text{LHS}}$ in  Eq. (\ref{eq:class-6}) and the maximal quantum expectation $\langle \mathcal{W} \rangle^{\rm max}$ in Eq. (\ref{eq:quatum-6}), one easily proves the relation (\ref{eq:sdsi-3}).

\begin{remark} By the way, for the quantum state $\rho$, the classical bound can be written as
\begin{eqnarray}
    C_{\text{LHS}}
&=& \frac{1+|\vec{F}|}{4}=\frac{1+\sqrt{\vec{F}^2}}{4},
\end{eqnarray}
where $\vec{F}$ is the Bloch vector of the state $\rho_{B}$. Similarly, the quantum expectation can be written as
\begin{eqnarray}
 \langle \mathcal{W} \rangle^{\rm max}
 &=&\frac{1+\sqrt{\vec{F}^2 +\mathcal{C}^2}}{4}.
\end{eqnarray}
$\blacksquare$
\end{remark}

\vspace{8mm}
\centerline{\emph{3.3. The Case of General $\cos\beta$}}
\vspace{8mm}

In this situation, the state $\rho$ is given by
\begin{eqnarray}
\rho=\nu_1  \ket{\psi_1}\bra{\psi_1}+\nu_2 \ket{\psi_2}\bra{\psi_2},
\end{eqnarray}
where
\begin{eqnarray}
\ket{\psi_1} &=& \cos \theta \ket{00}+\sin \theta\ket{11}, \nonumber\\
 \ket{\psi_2} &=& \cos\phi(\cos \alpha \ket{01} +\sin\alpha\ket{10})+ e^{{\rm i}\beta}\sin\phi(\sin \theta \ket{00}-\cos \theta\ket{11}),
\end{eqnarray}
with $0<|\cos\beta|<1$, $\theta\in[0, \pi/2]$, and $\alpha\in[0, \pi/2]$. The square of concurrence of the state $\rho$ is given by
\begin{eqnarray}
\mathcal{C}^2 &=& -s_2-2\sqrt{s_1}=G_1-G_2,
\end{eqnarray}
with
\begin{eqnarray}
G_1 &=& \nu_1^2 \sin ^2 2 \theta + \nu_2^2 \left[ (\sin 2 \alpha \cos ^2 \phi -\sin 2 \theta \sin^2\phi)^2+ 4 \sin 2 \alpha \sin 2 \theta  \sin ^2 \phi \cos ^2 \phi \cos^ 2 \beta \right]+ 2 \nu_1 \nu_2 \cos^2 2 \theta \sin ^2 \phi,\nonumber\\
\end{eqnarray}
or
\begin{eqnarray}
G_1 &=& \nu_1^2 \sin ^2 2 \theta + \nu_2^2 \left[ (\sin 2 \alpha \cos ^2 \phi +\sin 2 \theta \sin^2\phi)^2-4 \sin 2 \alpha \sin 2 \theta  \sin ^2 \phi \cos ^2 \phi \sin^ 2 \beta \right]+ 2 \nu_1 \nu_2 \cos^2 2 \theta \sin ^2 \phi,\nonumber\\
\end{eqnarray}
and
\begin{eqnarray}
G_2 &=& 2 \nu_1 \nu_2  \sqrt{ \left( \sin^2\phi - \sin2\alpha \sin2\theta \cos^2\phi \right)^2+4 \sin2 \alpha \sin 2 \theta \sin^2 \phi \cos^2 \phi \cos^ 2 \beta },
\end{eqnarray}
or
\begin{eqnarray}
G_2
&=& 2 \nu_1 \nu_2 \sqrt{ \left( \sin^2\phi + \sin2\alpha \sin2\theta \cos^2\phi \right)^2-4 \sin2 \alpha \sin 2 \theta \sin^2 \phi \cos^2 \phi \sin^ 2 \beta }.
\end{eqnarray}
{In Sec. \ref{sec:IIIB} we have known that if one chooses the measurement parameters $\tau=\beta$ and $\xi$ is arbitrary, then he has $\mathcal{M}=\mathcal{M}^\dagger$ for separable states.}

\textcolor{blue}{\emph{Example 7.---}} Let us consider a simple case of $\sin2\alpha=0$ with $\alpha=0$ (for $\alpha=\pi/2$, the result is similar).
In this situation, Alice's measurement direction is given by
\begin{eqnarray}
&& \hat{n}=(\sin\xi \cos\tau, \sin\xi \sin\tau, \cos\xi)=(\sin\xi \cos\beta, \sin\xi \sin\beta, \cos\xi).
\end{eqnarray}
For convenience, we can adopt the following the measurement parameters
\begin{eqnarray}
&& {\tau=\beta}, \;\;\;\; \xi=\frac{\pi}{2}, \;\;\;\; \delta=\frac{\pi}{2},
\end{eqnarray}
then we obtain
\begin{eqnarray}
&& \hat{n}=(\sin\xi \cos\tau, \sin\xi \sin\tau, \cos\xi)=(\cos\beta, \sin\beta, 0),
\end{eqnarray}
\begin{eqnarray}
&& \ket{+\hat{n}}=\begin{pmatrix}
\cos\frac{\xi}{2}\\
\sin\frac{\xi}{2}\, e^{{\rm i}\tau}\\
\end{pmatrix}=\cos\frac{\xi}{2} |0\rangle+\sin\frac{\xi}{2}\, e^{{\rm i}\tau} |1\rangle
=\frac{1}{\sqrt{2}} \left(|0\rangle+ e^{{\rm i}\beta} |1\rangle\right),
\end{eqnarray}

\begin{eqnarray}
 \ket{-\hat{n}}=\begin{pmatrix}
\sin\frac{\xi}{2}\\
-\cos\frac{\xi}{2} \, e^{{\rm i}\tau} \\
\end{pmatrix}= \sin\frac{\xi}{2}|0\rangle -\cos\frac{\xi}{2} \, e^{{\rm i}\tau} |1\rangle=\frac{1}{\sqrt{2}} \left(|0\rangle- e^{{\rm i}\beta}|1\rangle\right),
\end{eqnarray}

\begin{eqnarray}
P^{\hat{n}}_{0}&=&\frac{1}{2}\left(\openone+\vec{\sigma}\cdot \hat{n}\right)=|+\hat{n}\rangle\langle+\hat{n}|
= \frac{1}{2}\begin{pmatrix}
 1+\cos\xi&
 \sin\xi e^{-{\rm i} \tau}\\
 \sin\xi e^{{\rm i} \tau} &
 1-\cos\xi
\end{pmatrix}= {\frac{1}{2}\begin{pmatrix}
 1 & e^{-{\rm i}\beta} \\
 e^{{\rm i}\beta}  & 1
\end{pmatrix},}
\end{eqnarray}

\begin{eqnarray}
P^{\hat{n}}_{1}=\frac{1}{2}\left(\openone-\vec{\sigma}\cdot \hat{n}\right)=|-\hat{n}\rangle\langle-\hat{n}|
= \frac{1}{2}\begin{pmatrix}
 1-\cos\xi&
 -\sin\xi e^{-{\rm i} \tau}\\
 -\sin\xi e^{{\rm i} \tau} &
 1+\cos\xi
\end{pmatrix}= {\frac{1}{2}\begin{pmatrix}
 1 & -e^{-{\rm i}\beta} \\
 -e^{{\rm i}\beta}  & 1
\end{pmatrix},}
\end{eqnarray}

\begin{eqnarray}
\rho_{B}
& =&
\left(
\begin{array}{cc}
 \nu_2 \sin ^2\alpha \cos ^2\phi+\nu_1 \cos ^2\theta+\nu_2 \sin ^2\theta \sin ^2\phi
 & \nu_2 \sin \phi \cos\phi \left(e^{{\rm i} \beta } \cos \alpha \sin \theta-e^{-{\rm i} \beta }  \sin \alpha \cos\theta\right) \\
  \nu_2 \sin \phi \cos\phi \left(e^{-{\rm i} \beta } \cos \alpha \sin \theta-e^{{\rm i} \beta }  \sin \alpha \cos\theta\right) &
  \nu_2 \cos ^2\alpha \cos ^2\phi+\nu_1 \sin ^2\theta+\nu_2 \cos ^2\theta \sin ^2\phi \\
\end{array}
\right).
\end{eqnarray}
For $\alpha=0$, we can have the classical bound as
\begin{eqnarray}
    C_{\text{LHS}}
    &=&\frac{1}{2} \max_{\hat{n}_B} \left[ \mathrm{Tr} \left( |\hat{n}_B\rangle \langle \hat{n}_B| \rho_{B} \right) \right]\nonumber\\
&=& \frac{1}{4}+ \frac{1}{4} \max_{\hat{n}_B}\biggr[-\nu_2 \sin \alpha \cos\theta \sin \theta_B \sin 2 \phi \cos (\beta -\phi_B)
+\nu_2 \cos\alpha \sin\theta \sin \theta_B \sin2 \phi \cos (\beta +\phi_B)\nonumber\\
&&-\nu_2 \cos 2 \alpha \cos \theta_B \cos ^2\phi+\cos ^2\theta \cos \theta_B \left(\nu_1-\nu_2 \sin ^2\phi\right)
-\nu_1 \sin ^2\theta\cos \theta_B+\nu_2 \sin ^2\theta \cos \theta_B \sin ^2\phi\biggr]\nonumber\\
&=& \frac{1}{4}+ \frac{1}{4} \max_{\hat{n}_B}
\biggr[\nu_2 \sin\theta \sin \theta_B \sin2 \phi \cos (\beta +\phi_B)\nonumber\\
&&-\nu_2 \cos \theta_B \cos ^2\phi+\cos ^2\theta \cos \theta_B \left(\nu_1-\nu_2 \sin ^2\phi\right)
-\nu_1 \sin ^2\theta \cos \theta_B+\nu_2 \sin ^2\theta \cos \theta_B \sin ^2\phi\biggr]\nonumber\\
&=& \frac{1}{4}+ \frac{1}{4} \max_{\hat{n}_B}
\biggr[\nu_2 \sin\theta \sin \theta_B \sin2 \phi (\cos \beta \cos\phi_B - \sin \beta \sin \phi_B)\nonumber\\
&&-\nu_2 \cos \theta_B \cos ^2\phi+\cos ^2\theta \cos \theta_B \left(\nu_1-\nu_2 \sin ^2\phi\right)
-\nu_1 \sin ^2\theta \cos \theta_B+\nu_2 \sin ^2\theta \cos \theta_B \sin ^2\phi\biggr]\nonumber\\
&=& \frac{1}{4}+ \frac{1}{4} \max_{\hat{n}_B}\biggr[  F_1 \sin \theta_B \cos\phi_B + F_2 \sin \theta_B \sin\phi_B+F_3 \cos \theta_B \biggr]\nonumber\\
&\leq &  \frac{1}{4}+ \frac{\sqrt{F_1^2+F_2^2+F_3^2}}{4}=\frac{1}{4}+ \frac{|\vec{F}|}{4},
\end{eqnarray}
with
\begin{eqnarray}
F_1 &=& \nu_2 \sin\theta  \sin2 \phi \cos \beta,  \nonumber\\
F_2 &=& -\nu_2 \sin\theta \sin2 \phi \sin \beta, \nonumber\\
F_3 &=& -\nu_2 \cos ^2\phi+\cos ^2\theta  \left(\nu_1-\nu_2 \sin ^2\phi\right)
-\nu_1 \sin ^2\theta + \nu_2 \sin ^2\theta \sin ^2\phi \nonumber\\
&=& \nu_1 (\cos ^2\theta - \sin ^2\theta) + \nu_2 (-\cos ^2\phi - \cos ^2\theta \sin ^2\phi +\sin ^2\theta \sin ^2\phi) \nonumber\\
&=& \nu_1 \cos 2\theta - \nu_2 (\cos ^2\phi + \cos 2\theta \sin ^2\phi ).
\end{eqnarray}
Then we have
\begin{eqnarray}\label{eq:class-7}
    C_{\text{LHS}}
&=&  \frac{1}{4}+ \frac{\sqrt{F_1^2+F_2^2+F_3^2}}{4} \nonumber\\
&=& \frac{1}{4}+\frac{1}{4} \sqrt{\left[\nu_1 \cos 2\theta - \nu_2 (\cos ^2\phi + \cos 2\theta \sin ^2\phi )\right]^2+( \nu_2 \sin\theta  \sin2 \phi \cos \beta)^2+( \nu_2 \sin\theta  \sin2 \phi \cos \beta)^2 } \nonumber\\
&=&\frac{1}{4}+\frac{1}{4}  \sqrt{\left[\nu_1 \cos 2\theta - \nu_2 (\cos ^2\phi + \cos 2\theta \sin ^2\phi )\right]^2+( \nu_2 \sin\theta  \sin2 \phi )^2}.
\end{eqnarray}
After direct calculation, we can have the general expression of $\langle \mathcal{W} \rangle$ as
\begin{eqnarray}
 \langle \mathcal{W} \rangle&=&\frac{1}{4}+ \frac{1}{4} \left\{ \nu_2 \left[ \cos(\alpha - \theta) \cos\tau \sin\beta - \cos\beta \cos(\alpha + \theta) \sin\tau \right] \sin2\phi \right\} \nonumber\\
&&+ \frac{1}{4} \cos\theta_B \biggr[ -\nu_2 \cos2\alpha \cos^2\phi + \cos2\theta \left( \nu_1 - \nu_2 \sin^2\phi \right) \nonumber\\
&& + \nu_2 \left[ -\sin\beta \cos(\alpha + \theta) \cos\tau + \cos\beta \cos(\alpha - \theta) \sin\tau \right] \sin2\phi \biggr] \nonumber\\
&&+ \frac{1}{4} \sin\theta_B \sin\phi_B \left\{ \cos\tau \left[ -\nu_2 \cos^2\phi \sin2\alpha + \sin2\theta \left( \nu_1 - \nu_2 \sin^2\phi \right) \right] - \nu_2 \sin\beta \sin(\alpha + \theta) \sin 2\phi \right\} \nonumber\\
&&+ \frac{1}{4} \sin\theta_B \cos\phi_B \left\{ \sin\tau \left[ \nu_2 \cos^2\phi \sin 2\alpha + \sin2\theta \left( \nu_1 - \nu_2 \sin^2\phi \right) \right] - \nu_2 \cos\beta \sin(\alpha - \theta) \sin 2\phi \right\}.
\end{eqnarray}
For $\tau=\beta$ and $\alpha=0$, we have
\begin{eqnarray}
 \langle \mathcal{W} \rangle &=&\frac{1}{4}+ \frac{1}{4} \left\{ \nu_2 \left[ \cos \theta \cos\beta \sin\beta - \cos\beta \cos \theta \sin\beta \right] \sin2\phi \right\} \nonumber\\
&&+ \frac{1}{4} \cos\theta_B \biggr[ -\nu_2 \cos^2\phi + \cos2\theta \left( \nu_1 - \nu_2 \sin^2\phi \right)  + \nu_2 \left[ -\sin\beta \cos\theta \cos\beta + \cos\beta \cos\theta \sin\beta \right] \sin2\phi \biggr] \nonumber\\
&&+ \frac{1}{4} \sin\theta_B \sin\phi_B \left\{ \cos\beta \left[ \sin2\theta \left( \nu_1 - \nu_2 \sin^2\phi \right) \right] - \nu_2 \sin\beta \sin\theta \sin 2\phi \right\} \nonumber\\
&&+ \frac{1}{4} \sin\theta_B \cos\phi_B \left\{ \sin\beta \left[\sin2\theta \left( \nu_1 - \nu_2 \sin^2\phi \right) \right]
+ \nu_2 \cos\beta \sin\theta \sin 2\phi \right\}\nonumber\\
&=&\frac{1}{4}+ \frac{1}{4} \cos\theta_B \left[ -\nu_2 \cos^2\phi + \cos2\theta \left( \nu_1 - \nu_2 \sin^2\phi \right) \right] \nonumber\\
&&+ \frac{1}{4} \sin\theta_B \sin\phi_B \left\{ \cos\beta \left[ \sin2\theta \left( \nu_1 - \nu_2 \sin^2\phi \right) \right] - \nu_2 \sin\beta \sin\theta \sin 2\phi \right\} \nonumber\\
&&+ \frac{1}{4} \sin\theta_B \cos\phi_B \left\{ \sin\beta \left[\sin2\theta \left( \nu_1 - \nu_2 \sin^2\phi \right) \right]
+ \nu_2 \cos\beta \sin\theta \sin 2\phi \right\},
\end{eqnarray}
which can reach its maximum
\begin{eqnarray}\label{eq:quatum-7}
 \langle \mathcal{W} \rangle^{\rm max}
 &=&\frac{1}{4}+ \frac{1}{4} \sqrt{\Gamma^2},
 \end{eqnarray}
with
\begin{eqnarray}
\Gamma^2 &=&  \left[ -\nu_2 \cos^2\phi + \cos2\theta \left( \nu_1 - \nu_2 \sin^2\phi \right) \right] ^2+ \left\{ \cos\beta \left[ \sin2\theta \left( \nu_1 - \nu_2 \sin^2\phi \right) \right] - \nu_2 \sin\beta \sin\theta \sin 2\phi \right\}^2\nonumber\\
&&+ \left\{ \sin\beta \left[\sin2\theta \left( \nu_1 - \nu_2 \sin^2\phi \right) \right]
+ \nu_2 \cos\beta \sin\theta \sin 2\phi \right\}^2\nonumber\\
&=& \left[ -\nu_2 \cos^2\phi + \cos2\theta \left( \nu_1 - \nu_2 \sin^2\phi \right) \right] ^2+  \cos^2\beta\left[ \sin2\theta \left( \nu_1 - \nu_2 \sin^2\phi \right) \right]^2 +\sin^2\beta (\nu_2  \sin\theta \sin 2\phi )^2\nonumber\\
&&+  \sin^2\beta\left[ \sin2\theta \left( \nu_1 - \nu_2 \sin^2\phi \right) \right]^2 +\cos^2\beta (\nu_2  \sin\theta \sin 2\phi )^2\nonumber\\
&=& \left[ -\nu_2 \cos^2\phi + \cos2\theta \left( \nu_1 - \nu_2 \sin^2\phi \right) \right] ^2+ \left[ \sin2\theta \left( \nu_1 - \nu_2 \sin^2\phi \right) \right]^2 + (\nu_2  \sin\theta \sin 2\phi )^2\nonumber\\
&=&\left[ -\nu_2 \cos^2\phi + \cos2\theta \left( \nu_1 - \nu_2 \sin^2\phi \right) \right] ^2+ (\nu_2  \sin\theta \sin 2\phi )^2+\mathcal{C}^2.
\end{eqnarray}
By comparing  the classical bound $C_{\text{LHS}}$ in  Eq. (\ref{eq:class-7}) and the maximal quantum expectation $\langle \mathcal{W} \rangle^{\rm max}$ in Eq. (\ref{eq:quatum-7}), one easily proves the relation (\ref{eq:sdsi-3}).

\begin{remark} By the way, for the quantum state $\rho$, the classical bound can be written as
\begin{eqnarray}
    C_{\text{LHS}}
&=& \frac{1+|\vec{F}|}{4}=\frac{1+\sqrt{\vec{F}^2}}{4},
\end{eqnarray}
where $\vec{F}$ is the Bloch vector of the state $\rho_{B}$. Similarly, the quantum expectation can be written as
\begin{eqnarray}
 \langle \mathcal{W} \rangle^{\rm max}
 &=&\frac{1+\sqrt{\vec{F}^2 +\mathcal{C}^2}}{4}.
\end{eqnarray}
$\blacksquare$
\end{remark}

\newpage

\section{Discussion}

In this section, we provide additional discussions to address several foundational questions regarding the physical scope, geometric boundaries, and methodological aspects of our steering theorem. These discussions are intended to clarify the conceptual implications of the theorem and its relation to existing results, while the key methodological derivations needed for the main result are reported in the METHODS section of the main manuscript.

The hierarchy of quantum nonlocality naturally raises the question of exact equivalence boundaries for mixed states. Bell's nonlocality is the strongest form of nonlocality, quantum entanglement is the weakest, and EPR steering lies between them. Gisin's theorem shows that all pure entangled states, namely rank-1 entangled states, possess Bell's nonlocality. Our theorem establishes the corresponding rank-2 boundary for EPR steering: all rank-2, including rank-1, entangled two-qubit states possess EPR steerability. Below, we elaborate on the theorem's relation to measurement-restricted scenarios, the geometric motivation for state-dependent inequalities, the breakdown of the equivalence in higher-rank systems, and the challenges of higher-dimensional generalizations.

\vspace{1em}

\begin{proof}[Comment (1).] Apparent contradiction with unsteerable rank-2 states in specific scenarios.

    \emph{Answer}:
    Certain existing studies have identified specific rank-2 entangled states that admit a local hidden state (LHS) model, thus appearing unsteerable. This does not contradict our theorem; rather, the two results address different physical limits.

    The unsteerability reported in such works is a \emph{measurement-restricted} phenomenon, evaluated under a constrained two-measurement-setting ($N=2$) scenario. When the number of measurement bases is restricted, the target party can often construct an LHS model to reproduce the conditional states. In contrast, our theorem evaluates the \emph{intrinsic steerability} in the asymptotic limit ($N \to \infty$). When finite-setting constraints are relaxed, these LHS models no longer hold. Thus, our theorem confirms that intrinsic steerability remains a universal property of all rank-2 entangled states once measurement constraints are removed.
\end{proof}

\begin{proof}[Comment (2).] Necessity of the state-dependent steering inequality.

    \emph{Answer}:
    Unlike Gisin's theorem which utilizes a fixed CHSH inequality, our approach relies on a \emph{state-dependent} steering bound. This is motivated strictly by the geometric complexity of mixed states.

    For pure states (rank-1), the state space geometry allows a fixed linear inequality to efficiently separate local models. However, classical mixing alters this geometry. Conventional fixed linear steering inequalities optimally probe only specific directions, inevitably leaving ``blind spots'' where weakly entangled mixed states go undetected. Meanwhile, numerical Semidefinite Programming (SDP) methods operate under finite measurement settings and may encounter precision issues near the separable threshold. By coupling the measurement operators directly with the internal parameters of the density matrix, our state-dependent boundary dynamically adapts to contour the threshold of steerability, providing an exact analytical condition free from finite-setting approximations.
\end{proof}

\begin{proof}[Comment (3).] The physical boundary of the equivalence in higher-rank systems.

    \emph{Answer}:
    It is a known fact that classical mixing can disrupt the equivalence between entanglement and steerability. Our theorem indicates that rank 2 is the highest boundary where this equivalence unconditionally holds for two qubits. To illustrate this point, we provide a concise comparison using Bell-diagonal states.

    For a generic two-qubit Bell-diagonal state with correlation matrix $T=\text{diag}(t_1,t_2,t_3)$, the condition
    \begin{eqnarray}
        \sum_{i=1}^3 t_i^2 \leq 1
    \end{eqnarray}
    gives a sufficient condition for the existence of a local hidden state model, and hence for unsteerability.

    \textbf{Case I: A rank-3 example.} \\
    Consider a rank-3 Bell-diagonal state with eigenvalues $\{\lambda_1,\lambda_2,\lambda_3,\lambda_4\}=\{0.51,0.245,0.245,0\}$. Since $\lambda_{\rm max}=0.51>0.5$, the state is entangled. Its correlation elements are
    \begin{eqnarray}
        t_1 &=& \lambda_1-\lambda_2+\lambda_3-\lambda_4=0.51,\nonumber\\
        t_2 &=& \lambda_1+\lambda_2-\lambda_3-\lambda_4=0.51,\nonumber\\
        t_3 &=& \lambda_1-\lambda_2-\lambda_3+\lambda_4=0.02.
    \end{eqnarray}
    Therefore,
    \begin{eqnarray}
        \sum_{i=1}^3 t_i^2=(0.51)^2+(0.51)^2+(0.02)^2=0.5206\leq1.
    \end{eqnarray}
    This state satisfies the above sufficient LHS condition and is therefore unsteerable, while it is still entangled. This provides a concrete rank-3 example showing that the universal equivalence between entanglement and EPR steerability no longer holds from rank 3 onward.

    \textbf{Case II: The rank-2 Bell-diagonal constraint.} \\
    For a rank-2 Bell-diagonal state, let $\lambda_3=\lambda_4=0$ and $\lambda_1+\lambda_2=1$. Up to local unitary transformations and relabelling of the Bell basis, the correlation elements can be written as
    \begin{eqnarray}
        t_1=\lambda_1-\lambda_2,\quad t_2=1,\quad t_3=\lambda_1-\lambda_2.
    \end{eqnarray}
    The Horodecki CHSH criterion states that a Bell-diagonal state violates the CHSH inequality whenever the sum of the two largest values among $\{t_1^2,t_2^2,t_3^2\}$ is larger than 1. For the above rank-2 Bell-diagonal state, this quantity is
    \begin{eqnarray}
        1+(\lambda_1-\lambda_2)^2.
    \end{eqnarray}
    Entanglement requires $\lambda_{\rm max}>0.5$, which is equivalent to $\lambda_1\neq\lambda_2$. Hence,
    \begin{eqnarray}
        1+(\lambda_1-\lambda_2)^2>1.
    \end{eqnarray}
    Therefore, every entangled rank-2 Bell-diagonal state violates the CHSH inequality and is Bell-nonlocal. Since Bell nonlocality implies EPR steerability, every entangled rank-2 Bell-diagonal state is necessarily steerable. This Bell-diagonal comparison illustrates why rank 2 is the sharp boundary for the Gisin-like equivalence in two-qubit systems, while the full theorem for arbitrary rank-2 two-qubit states is proven in the main manuscript.
\end{proof}

\begin{proof}[Comment (4).] Prospects and challenges for higher-dimensional and multipartite systems.

    \emph{Answer}:
    Generalizing this theorem to bipartite qudits ($d \times d$) or multipartite systems involves significant mathematical difficulties.

    For bipartite qudits, the primary challenge lies in the geometric complexity of the $SU(d)$ group. Our two-qubit proof benefits from the algebraic simplicity of Pauli matrices, which allows for a tractable parameterization of the density matrix. For generic $d$-dimensional systems, constructing an exact, state-dependent steering criterion requires managing a much larger set of coupled non-linear parameters, making the analytical derivation highly intricate.

    Furthermore, multipartite EPR steering involves complex correlation structures across different system partitions. A single bipartite analytical framework cannot straightforwardly capture these varied multipartite relationships. We leave the exploration of exact analytical criteria for these high-dimensional and multipartite scenarios as a direction for future research.
\end{proof}